\documentclass[12pt]{report}
\usepackage{graphicx}
\usepackage[centertags]{amsmath}
\usepackage{amsfonts}
\usepackage{amssymb}
\usepackage{amsthm}
\usepackage{hyperref}
\usepackage{mathrsfs}
\usepackage{bm}
\usepackage{amsfonts}
\usepackage{latexsym}
\usepackage[latin1,utf8]{inputenc}
\usepackage{amsmath}
\usepackage{palatino}
\usepackage{mathpazo}
\usepackage{textcomp}
\usepackage{float}
\usepackage{booktabs}
\usepackage{hyperref}
\hypersetup{colorlinks,citecolor=blue}
\hypersetup{colorlinks=true,linkcolor=red,filecolor=magenta,    urlcolor=blue}
\usepackage{amsmath}
\usepackage{graphicx}
\usepackage{makecell}
\usepackage{hyperref}
\usepackage[mathlines]{lineno}
\usepackage{amssymb}
\usepackage{enumitem}
\usepackage{mathtools}
\usepackage{mathrsfs}
\usepackage{setspace}
\usepackage{fancyhdr}
\usepackage{color}
\usepackage{graphicx}
\usepackage{tabularx}
\usepackage{longtable}
\usepackage[dvipsnames]{xcolor}
\usepackage{mathalfa}
\usepackage{calligra}
\usepackage[misc]{ifsym}
\DeclareMathAlphabet{\mathpzc}{OT1}{pzc}{m}{it}
\DeclareMathAlphabet{\mathcalligra}{T1}{calligra}{m}{n}
\hypersetup{colorlinks=true,citecolor=blue,linkcolor=red,urlcolor=magenta}
\usepackage{xcolor}
\usepackage[caption=false]{subfig}
\usepackage{commath}
\usepackage{lipsum}
\usepackage{amsmath}
\usepackage{amsthm}
\usepackage{hyperref}
\usepackage{amssymb}
\usepackage{enumitem}
\usepackage{mathtools}
\usepackage{mathrsfs}
\usepackage{setspace}
\usepackage[left=3.75cm, right = 2cm, top=2.45cm, bottom=1.5cm]{geometry}
\usepackage{parskip}
\usepackage{fancyhdr}
\usepackage{dashrule}
\usepackage{thmtools}
\usepackage{graphicx}
\usepackage[dvipsnames]{xcolor}
\usepackage{color}
\usepackage{multicol,caption}
\usepackage{subfig}
\usepackage{cleveref}
\usepackage{tabularx}
\usepackage{pdfpages}
\RequirePackage[bibstyle=ieee,citestyle=numeric-comp]{biblatex}
\definecolor{red}{rgb}{1,0,0}
\definecolor{maroon}{rgb}{0.5, 0.0, 0.0}
\definecolor{blue}{rgb}{0,0,1}
\definecolor{royalblue}{rgb}{0.0, 0.14, 0.4}
\definecolor{forestgreen}{rgb}{0.13, 0.55, 0.13}
\definecolor{halayaube}{rgb}{0.4, 0.22, 0.33}
\definecolor{lightbrown}{rgb}{0.71, 0.4, 0.11}
\definecolor{blue(ryb)}{rgb}{0.01, 0.28, 1.0}
\definecolor{blue-violet}{rgb}{0.54, 0.17, 0.89}
\definecolor{zaffre}{rgb}{0.0, 0.08, 0.66}
\definecolor{devpink}{rgb}{1,0,0.7}
\definecolor{devbrown}{rgb}{0.55, 0.1, 0.1}

\newlength{\defbaselineskip}
\newcommand{\setlinespacing}[1]%
           {\setlength{\baselineskip}{#1 \defbaselineskip}}

\theoremstyle{plain}

\newtheorem*{theorem*}{Theorem}

\newtheorem*{conjecture*}{Conjecture}
\theoremstyle{definition}
\newtheorem{definition}{Definition}[section]

\theoremstyle{remark}

\usepackage[T1]{fontenc}
\DeclareUnicodeCharacter{2212}{\textminus}
\DeclareUnicodeCharacter{0301}{\accent"00B4 }

\begin{document}
\title{\textsc{Study on Modifications of General Relativity: A Differential Geometric Approach}}
\author{Kavya N. S.}
	
{\thispagestyle{plain}\pagenumbering{gobble}

	\begin{center}
		\
		\\
		\color{black}
		{{\bf A Thesis Entitled}}
		\\[11 pt]
		\color{royalblue}
		{\Large{\bf THE STUDY ON MODIFIED THEORIES OF GENERAL RELATIVITY: A DIFFERENTIAL GEOMETRIC APPROACH
				\\[12 pt]
				 }
		} 
		\
		\\
		\color{black}
		{{\bf Submitted to the}}\\
		\
		\\
		\ \color{maroon}
		{\LARGE{\bf Faculty of Science and Technology}} \\
		
		\
		\\
		\color{forestgreen}
		{\Huge{\bf{\hskip .12 in } KUVEMPU\,\ \includegraphics[width=1in]{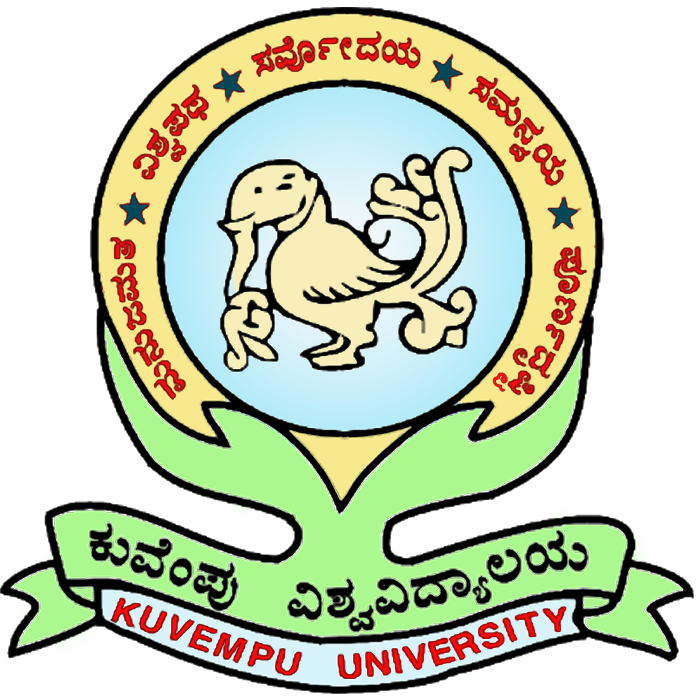}\,\ UNIVERSITY}}\\
		\
		\\
		\color{black}
		{{\bf For the Award of the Degree of}}\\
		\
		\\
		\color{lightbrown}
		{\Huge{\bf{ Doctor of Philosophy}}}\\
		\
		\\
		\color{black}
		{\bf in}\\
		\
		\\
		\color{blue(ryb)}
		{\Huge{\bf MATHEMATICS}}\\
		\
		\\
		\color{black}
		{\bf by}\\

		\color{maroon}
		{\Large \bf Ms. Kavya N. S.}\\890/31-08-2021\\
		\
		\\	\setstretch{1.2}
		\color{black}{\Large\bf Research Supervisor}\\[6 pt]
		\color{maroon}{\Large\bf Dr. Venkatesha, \normalsize M.Sc., Ph.D.} \\[3 pt]
		\color{black}{Professor }\\
		
		\color{black} \vspace{2pt} \hbox to \hsize{\hrulefill}
		\color{black}
	
		{{Department of P.G. Studies and Research in Mathematics,}}\\
		{{Kuvempu University,\, Jnanasahyadri,}}\\
		{{Shankaraghatta - 577 451,\, Shivamogga,}}\\
		{{Karnataka, INDIA.}}\\
		{\color{blue-violet} \large {\bf September-2024}}
	\end{center}
}
{\thispagestyle{plain}\def\baselinestretch{1}
\chapter*{\color{NavyBlue}\centering\textsc{\textbf{Declaration}}}
\def\baselinestretch{1.5}
\setstretch{1.5}

I, \textbf{\color{maroon}Kavya N. S.}, hereby declare that the thesis entitled \textbf{\color{maroon} The Study on Modified Theories of General Relativity: A Differential Geometric Approach}, submitted to the Faculty of Science and Technology, Kuvempu University for the award of the degree of Doctor of Philosophy in Mathematics is the result of research work carried out by me in the Department of P.G. Studies and Research in Mathematics, Kuvempu University under the guidance of \textbf{\color{maroon}Prof. Venkatesha}, Department of P.G. Studies and Research in Mathematics, Kuvempu University, Jnanasahyadri, Shankaraghatta.

I further declare that this thesis or part thereof has not been previously formed the basis of the award of any degree, associateship etc., of any other University or Institution.  \\
\
\\
\begin{multicols}{3}[]
\textbf{Place:} Shankaraghatta\\
\textbf{Date:} 24/09/2024\\
\
\\
\
\\
\
\\
{ Kavya N. S.}

\end{multicols}}
{\thispagestyle{plain}\def\baselinestretch{1}
\chapter*{\color{NavyBlue}\centering\textsc{\textbf{Certificate}}}
\def\baselinestretch{1.5}
\setstretch{1.5}
This is to certify that the thesis entitled \textbf{\color{maroon} The Study on Modified Theories of General Relativity: A Differential Geometric Approach}, submitted to the Faculty of Science and Technology, Kuvempu University for the award of the degree of Doctor of Philosophy in Mathematics by \textbf{\color{maroon}Ms. Kavya N. S.} is the result of bonafide research work carried out by her under my guidance in the Department of P.G. Studies and Research in Mathematics, Kuvempu University, Jnanasahyadri, Shankaraghatta.

This thesis or part thereof has not been previously formed the basis of the award of any degree, associateship etc., of any other University or Institution.  \\
\
\\
\begin{multicols}{3}[]
\textbf{Place:} Shankaraghatta\\
\textbf{Date:} 24/09/2024\\
\
\\
\
\\
\
\\
\begin{center}
     Prof. Venkatesha\\
     Research Supervisor
\end{center}

\end{multicols}}
{\thispagestyle{plain}\includepdf[pages={1}]{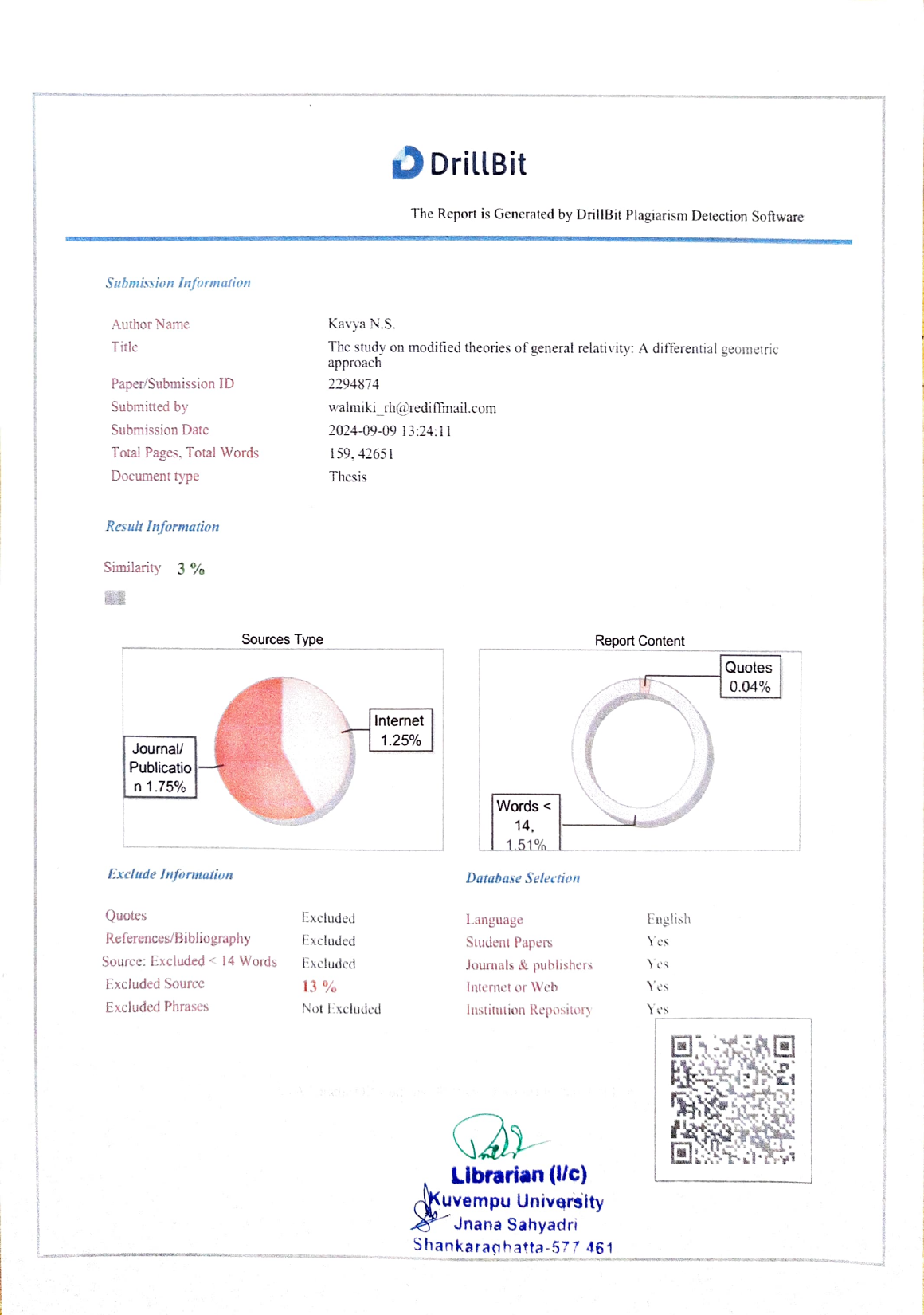}}
{\thispagestyle{plain}\def\baselinestretch{1}
\markboth{}{\small{{\it \hfill Acknowledgement}}}
\chapter*{\color{NavyBlue}\centering\textsc{\textbf{Acknowledgement}}}
\def\baselinestretch{1.5}

It gives me great pleasure to express my deep gratitude and sincere appreciation to everyone who assisted, guided, suggested, cooperated, and inspired me throughout this endeavor.

First and foremost, I am deeply indebted to my esteemed mentor and research supervisor, \textbf{Prof. Venkatesha}, Department of P.G. Studies and Research in Mathematics at Kuvempu University. His insightful guidance, invaluable advice, constant encouragement, constructive suggestions, and extensive discussions have been crucial during my research. His supportive attitude and personal attention have provided a solid foundation for my work. I am grateful for his dedication to reviewing my progress, making corrections, and finalizing my research papers and thesis. Beyond academic guidance, he has significantly contributed to my personal growth, teaching me resilience, ethical values, and life skills that have strengthened me as a person, for which I am deeply grateful. His care, keen involvement, and kind nature have always uplifted my spirits. I thank the Almighty for giving me such a mentor.

I extend my sincere thanks to \textbf{Prof. S. K. Narasimhamurthy}, Professor and Chairman of the Department of P.G. Studies and Research in Mathematics, Kuvempu University, for his cooperation, valuable suggestions, motivation, and support throughout my research.

I also wish to thank \textbf{Prof. B. J. Gireesha}, Professor in the Department of P.G. Studies and Research in Mathematics, Kuvempu University, for his moral and ethical advice, kind assistance, and encouragement throughout my research.

I would like to express my gratitude to \textbf{Dr. P. Venkatesh}, Associate Professor, Department of Mathematics, Sahyadri Science College, Kuvempu University, for his kind support and motivation.

I would like to thank \textbf{Dr. Pavithra G. M.}, Assistant Professor, Department of Mathematics, Sahyadri Science College, Kuvempu University, for supporting and encouraging me throughout my research.

I would like to express my sincere gratitude to \textbf{Dr. Devaraj Mallesha Naik}, my teacher and research senior, for his invaluable guidance in making important decisions throughout my research career, which has helped me achieve significant milestones.

I would like to express my deepest gratitude to \textbf{Prof. P. K. Sahoo}, Department of Mathematics, BITS-Pilani, Hyderabad Campus, Hyderabad, for his invaluable guidance and support throughout the research. His insightful feedback and continuous encouragement were crucial in upgrading my caliber. I am profoundly grateful for his kind support and time in my development as a researcher.

I am deeply appreciative and grateful to \textbf{Dr. G. Mustafa} from Zhejiang Normal University, China, \textbf{Prof. K. Bamba} and \textbf{Dr. S. Mandal} from Fukushima University, Japan, and \textbf{Mr. Sai Swagat Mishra} from BITS-Pilani, Hyderabad for their support, useful discussions, and collaboration during my Ph.D. work.

I would like to express my sincere thanks to my beloved seniors \textbf{Dr. C. Shruthi} and \textbf{Dr. M. L. Keerthi}, for their assistance, cooperation, valuable discussions, and the enjoyable moments we shared.

I am grateful to all my research colleagues for creating a friendly and supportive atmosphere in the Department of Mathematics at Kuvempu University.

I also wish to thank the non-teaching staff members during my Ph.D. days, \textbf{M. P. Latha}, \textbf{Ms. Ramya}, and \textbf{Mr. M. Shivakumar} in the Department of P.G. Studies and Research in Mathematics at Kuvempu University, for their assistance and support.

Special thanks to \textbf{Ms. L. Sudharani}, who is my friend, colleague, and collaborator. As a friend, she offered support during slow progress and low spirits; as a colleague, she encouraged and helped me; and as a collaborator, she spent countless hours discussing and suggesting solutions to research problems. Her continuous help and involvement in writing my thesis were invaluable.

I would also like to thank my friends and collaborators \textbf{Ms. Chaitra C. C} and \textbf{Ms. Varsha C. S.} for their constructive cooperation.   

I am thankful to Kuvempu University for providing financial assistance through the Junior Research Fellowship, which facilitated my research. I would like to acknowledge DST, New Delhi, India, for its financial support for research facilities under DST-FIST-2019.

I am deeply grateful to \textbf{K.H. Vasudeva Sir} and \textbf{Raghavendra Sir}, who ignited my passion for Mathematics during my early days, ultimately leading me to pursue a Ph.D.

Finally, I extend my heartfelt thanks and best regards to my wonderful family, who are my world: my dad, \textbf{Sri. Shashikantha N. C.}, and my mom, \textbf{Smt. Jyothi N. S.}; my younger brother, \textbf{Mr. Karthik Samak}; my ever-inspiring and lovely grandpa, \textbf{Sri. Chidambara Bhat}; and my best friend, \textbf{Mr. Srujan Ganesh Hegade}, for their encouragement, sacrifices, love, care, and support. I owe my accomplishments to my family, who stood by me throughout my research and during difficult times. They have been my biggest strength, not just during my Ph.D. journey, but throughout my entire life.

I am profoundly grateful to God for granting me strength, guiding me along the right path, and illuminating my decisions with an awareness of right and wrong. It is with heartfelt devotion that I dedicate everything to His lotus feet, acknowledging His boundless grace and wisdom that have shaped my journey.

Lastly, I thank everyone who helped me directly or indirectly in completing my research work, and I apologize if I could not mention each one personally.

\begin{multicols}{3}[]
\textbf{Place:} Shankaraghatta\\
\textbf{Date:}\\
\
\\
\
\\
\
\\
{ Kavya N. S.}

\end{multicols}}
{\thispagestyle{plain}
\vspace*{6cm}
\begin{center}
  \begin{figure}[h]
    \centering
        \includegraphics[width=0.2\linewidth]{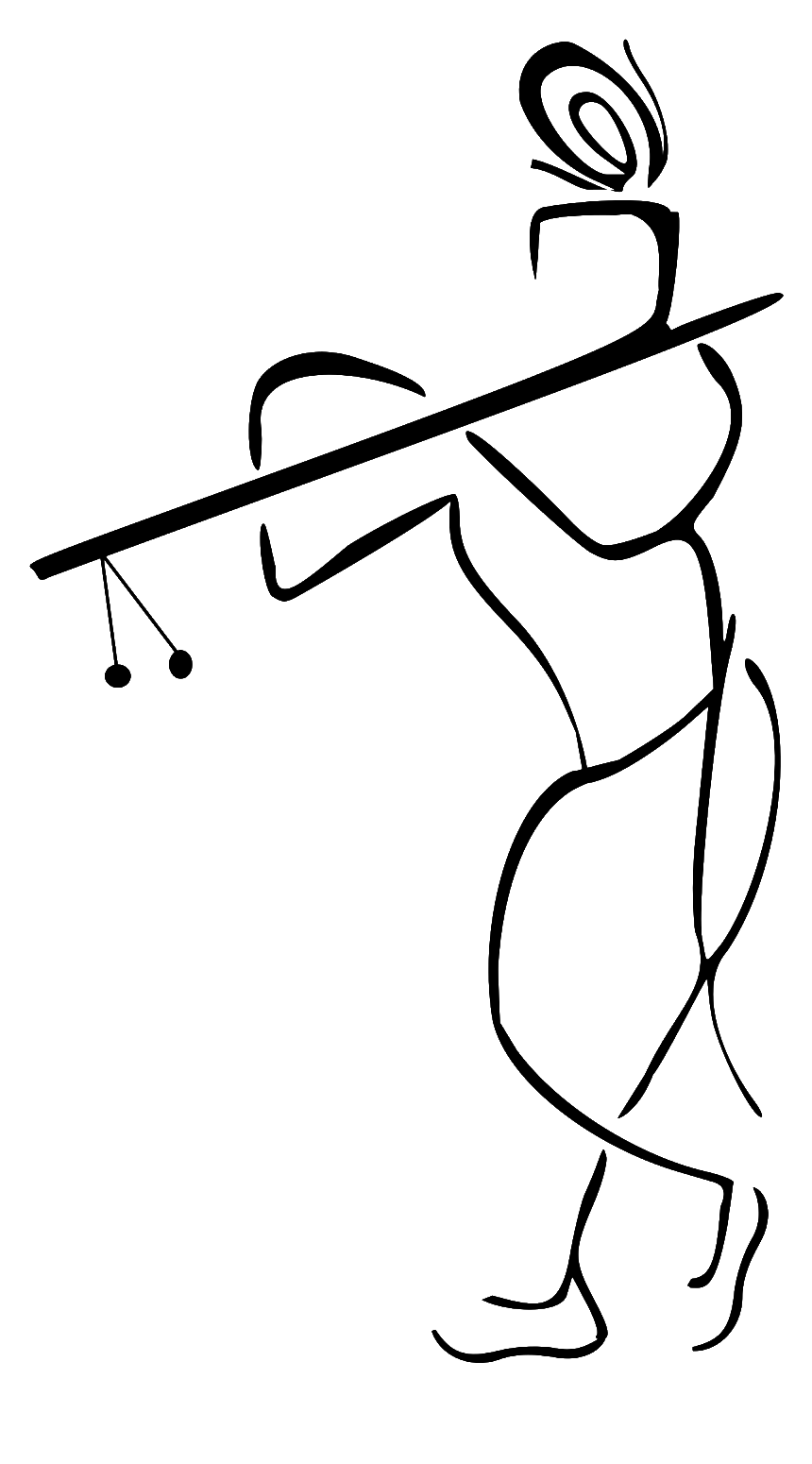}
    \end{figure}
    \textbf{\textit{Dedicated to Lord Krishna}}
\end{center}
\vspace*{\fill}}

\setlinespacing{1.5}
{\pagenumbering{roman}

\fancyhead[L]{}

\tableofcontents

}

\def\baselinestretch{1}
\markboth{}{\small{{\it \hfill Preface}}}
\chapter*{\textsc{Preface}}
\addcontentsline{toc}{chapter}{\textsc{Preface}}
\def\baselinestretch{1.5}
The introduction of General Relativity (GR) in 1915, has enhanced our understanding of the Universe and has undergone significant changes. While GR has consistently aligned with all observed data, concerns regarding its limitations have gradually surfaced. The notion of altering GR is not a recent development; in fact, modifications were suggested soon after the theory was introduced. The first significant modification came in 1919, when Weyl proposed adding higher-order invariants to the Einstein-Hilbert action, beyond the Ricci scalar. These early modifications were driven primarily by theoretical considerations rather than experimental findings. The first sign that GR might not be the ultimate theory of gravity emerged when it was discovered that GR is not renormalizable, leading to a stronger interest in incorporating higher-order terms into the action. 

Modifying GR to address its limitations, particularly in explaining dark matter, dark energy, and quantum gravity, has been a central focus of modern theoretical physics. Several approaches have been proposed, including extending the theory by introducing additional fields or adding higher-order curvature terms. Here, the main focus is to modify GR changing gravitational Lagrangian, thereby altering the geometric description. This leads to theories like teleparallel theory, which incorporates torsion, or symmetric teleparallel gravity, which uses a different connection than the Levi-Civita connection. Differential geometry plays a crucial role in these modifications, providing the mathematical framework to generalize GR's core concepts. Tools like differential forms, connections on tangent bundles, and tetrads allow for the exploration of higher geometric structures, enabling the proposal and analysis of new gravitational theories that extend or refine Einstein's original formulation. These contributions have been essential in exploring alternative models and testing the boundaries of GR in extreme conditions. 

This thesis contains seven chapters, followed by an appendix that provides relevant information regarding tools and techniques. As a foundational framework, \autoref{chap1} presents the fundamental definitions, concepts, and results essential for the subsequent chapters.

In \autoref{chap2}, we present a novel parametrization of the deceleration parameter to investigate the cosmological scenario which is both physically plausible and model-independent. Constrained by the observational dataset, we determine the model parameters using a Markov Chain Monte Carlo method. The results indicate that the Universe is currently in an accelerated phase. Furthermore, we apply the obtained parameter values to constrain $f(Q)$ gravity models and compare them with experimental results. This provides valuable insights into the accelerating Universe and underscores the importance and influence of parametrization on modified gravity models.

In exploring the dynamics of the Universe, modified gravity theories offer an alternative explanation for dark energy. Although the $\Lambda$CDM model has served as a foundational description for understanding the Universe, it also has its limitations. In \autoref{chap3}, the primary objective is to overcome these shortcomings and reconstruct the $\Lambda$CDM Universe. We aim to achieve this by embedding the $\Lambda$CDM scenario within a modified gravity framework. A very recently proposed gravity theory, known as $f(Q,\mathcal{L}_m)$ is set as background gravity theory. Here, the investigation is conducted within the framework of novel non-minimal coupling of $f(Q)$ gravity. A generic non-minimal form is considered, expressed in terms of two arbitrary functions $f_1(Q)$ and $f_2(Q)$. Analytic solutions are derived to mimic the $\Lambda$CDM paradigm. Through cosmographic analysis, the models are validated against observational results.

In \autoref{chap4}, we study locally rotationally symmetric homogeneous Bianchi-I spacetime with the isotropic matter distribution. This is done within the framework of $f(R,\mathcal{L}_m)$ gravity. Particularly, we consider a non-linear $f(R,\mathcal{L}_m)$ model, $f(R,\mathcal{L}_m)=\dfrac{1}{2}R+\mathcal{L}_m^{\,{\alpha}}$. Furthermore, $\omega$, the equation of state parameter, which is vital stuff in determining the present phase of the Universe is constrained. To constrain the model parameters and the equation of state parameter, we use 57 Hubble data points and 1048 Pantheon supernovae type Ia data samples. With the help of constrained values of parameters, we measure the anisotropy parameter. 

The \autoref{chap5} presents novel and physically plausible wormhole solutions within the framework of $f(Q,T)$ gravity theory, incorporating conformal symmetries. The investigation explores the feasibility of traversable wormholes under diverse scenarios, considering traceless, anisotropic, and barotropic equations of state. Additionally, the influence of model parameters on the existence and characteristics of these wormhole structures is thoroughly examined. 

Non-commutativity is a key feature of spacetime geometry. The \autoref{chap6} explores the traversable wormhole solutions in the framework of $f(R,\mathcal{L}_m)$ gravity within non-commutative geometry. By using the Gaussian and Lorentzian distributions, we construct tideless wormholes for the nonlinear $f(R,\mathcal{L}_m)$ model $f(R,\mathcal{L}_m)=\dfrac{R}{2}+\mathcal{L}_m^\alpha$. For both cases, we derive shape functions and discuss the required different properties with satisfying behavior. For the required wormhole properties, we develop some new constraints. The influence of the involved model parameter on energy conditions is analyzed graphically which provides a discussion about the nature of exotic matter. Further, we check the physical behavior regarding the stability of wormhole solutions through the TOV equation. An interesting feature regarding the stability of the obtained solutions via the speed of sound parameters within the scope of average pressure is discussed. 

Another well-known modified theory is $f(T)$ gravity, in which torsion is the basic geometric element of the gravity theory. in \autoref{chap7}, we study Big Bang Nucleosynthesis in the context of $f(T)$ gravity. The ability of Big Bang Nucleosynthesis theory to accurately predict the primordial abundances of helium and deuterium, as well as the baryon content of the Universe, is considered one of the most significant achievements in modern physics. We consider two highly motivated hybrid $f(T)$ models and constrain them using the observations from the Big Bang Nucleosynthesis era. In addition, using late-time observations of Cosmic Chronometers and Gamma-Ray-Bursts, the ranges of the model parameters are confined which are in good agreement with early time bounds. Subsequently, the common ranges obtained from the analysis for early and late times are summarized. Further, we verify the intermediating epochs by investigating the profiles of cosmographic parameters using the model parameter values from the common range. Here, we find the considered teleparallel models are viable candidates to explain the primordial-intermediating-present epochs.
\fancyhead[L]{}
\markboth{}{\small{{\it \hfill List of Figures}}}
\addcontentsline{toc}{chapter}{\textsc{List of Figures}}

\listoffigures
\addcontentsline{toc}{chapter}{\textsc{List of Tables}}
\listoftables

\newpage

\def\baselinestretch{1}
\markboth{}{\small{{\it \hfill Notations and Abbreviations}}}
\chapter*{\textsc{Notations and Abbreviations}}
\addcontentsline{toc}{chapter}{\textsc{Notations and Abbreviations}}
\def\baselinestretch{1.5}
\begin{multicols}{2}

\begin{itemize}
    \item $R$: Ricci scalar
    \item $Q$: Non-metricity
    \item $\mathcal{T}$: Torsion
    \item $\Lambda$: Cosmological constant
    \item $a$: Scale factor
    \item $z$: Redshift
    \item $q$: Deceleration parameter
    \item $G$: Newton's gravitational constant
    \item $R_{\mu \nu}$: Ricci tensor
    \item $T_{\mu \nu}$: Stress-energy tensor
    \item $R^{\sigma}_{\,\,\ \lambda \mu \nu}$: Riemann tensor
    \item $g_{\mu \nu}$: Metric tensor
    \item $g$: Determinant of Metric tensor
    \item $\omega$: EoS parameter
    \item $\Omega$: Density parameter
    \item $\chi^2$: Chi-Square
    \item $H_0$: Hubble Constant
\end{itemize}

\begin{itemize}
     \item \textbf{EoS}: Equation of State
    \item \textbf{DE}: Dark Energy
    \item \textbf{DM}: Dark Matter
    \item \textbf{LSS}: Large scale structure
    \item \textbf{GR}: General Relativity
    \item \textbf{CMB}: Cosmic Microwave Background
    \item \textbf{CDM}: Cold Dark Matter
    \item \textbf{ECs}: Energy conditions
    \item \textbf{BAO}: Baryonic Acoustic Oscillations
    \item \textbf{CC}: Cosmic Chronometer
    \item \textbf{GRBs}: Gamma Ray Bursts
    \item \textbf{SNeIa}: Type Ia supernovae
    \item \textbf{TEGR}: Teleparallel equivalent to GR
    \item \textbf{STEGR}: Symmetric teleparallel equivalent to GR
    \item \textbf{MCMC}:  Markov chain Monte Carlo
    
\end{itemize}

\end{multicols}

\setlinespacing{1.5}
\pagenumbering{arabic}
\newpage
\thispagestyle{empty}

\vspace*{\fill}
\begin{center}
    {\Huge \color{NavyBlue} \textbf{CHAPTER 1}}\
    \\
    \
    \\{\Huge\color{purple}\textsc{\textbf{Introduction}}}
\end{center}
\vspace*{\fill}

\pagebreak

\def\baselinestretch{1}
\markboth{}{\it\small{{Chapter 1 \hfill Introduction}}}
\chapter{\textsc{Introduction}}\label{chap1}
\def\baselinestretch{1.5}
\pagestyle{fancy}
\fancyhead[R]{\textit{Chapter 1}}
\markboth{Introduction}{}
\setstretch{1.5}

\section{Initiation}
During the 19th century, Newtonian mechanics was widely regarded as the most accurate theory for explaining gravity, as it effectively described both everyday phenomena and the motions of stars and planets. Nevertheless, this era also witnessed significant intellectual activity surrounding non-Euclidean geometries, spurred by foundational contributions from legendary scientists such as Bianchi, Riemann, Ricci-Curbastro, Lobachevsky, Gauss, Bolyai, and others. Euclidean geometry, which served as the foundation for classical physics, was eventually eclipsed by the rise of hyperbolic and elliptical geometries. These new geometrical concepts arose from the reexamination of the fundamentals of geometry, leading to two major approaches gaining traction: affine geometry, first introduced by Euler in 1748 and later further developed by Klein, Weyl, and  M\"obius.

In 1915, Albert Einstein drawing inspiration from these undertook an unconventional approach by employing the language of differential geometry to address the drawbacks of relativistic mechanics. The advent of General Relativity (GR) profoundly astonished the entire scientific community. This groundbreaking idea catalyzed numerous significant discoveries. However, for several years, Einstein's theory lacked experimental verification. This changed when Sir Arthur Eddington, a renowned astronomer, successfully demonstrated the correctness of Einstein's predictions during a total eclipse. This achievement marked not the end, but the beginning of an intensified pursuit for the theoretical completeness of GR. 

Even before GR was fully validated, some researchers were eager to propose extensions aimed at achieving broader objectives. In 1918, Weyl began exploring how to unify gravity and electromagnetism within a single, coherent geometric framework. To this end, he introduced an additional gauge field that establishes a unique length connection, with four extra degrees of freedom corresponding to electromagnetic potentials. In Weyl's geometry, alongside the GR connection, there exists a metric-incompatible, symmetric, and gauge-invariant length connection. This leads to the variation of both the length and direction of vectors during parallel transport. However, Weyl's theory encountered significant challenges, conflicting not only with certain experimental observations, such as the dependence of atomic clock spectral line frequencies on location and history, but also fundamentally with Quantum Mechanics, where particle masses depend on their past histories. 

In 1930, Einstein, continuing in the same vein of thought, suggested modifications to his theory. Drawn to the concepts of teleparallelism and tetrad formalism, he engaged in extensive and productive correspondence with Lanczos, Weitzenb\"ock, and Cartan. Recognizing that tetrad fields have sixteen independent components, Einstein associated ten of these components with the metric tensor. The remaining six were thought to correspond to a different connection, potentially representing electromagnetic potentials. A well-known extension of this theory is discussed in \autoref{chap7}, which can be found in the latter part of this thesis. 

In 1922, Cartan took a different approach by proposing a natural extension of GR that included not just the Levi-Civita connection, but also the torsion tensor, which is essentially the antisymmetric component of a metric-compatible affine connection. Building on this idea, he developed the corresponding geometric framework, suggesting that torsion could be physically linked to the intrinsic (quantum) angular momentum of matter. He also noted that torsion disappears in regions of vacuum. In the 1960s, Kibble and Sciama revisited the theory, reformulating it within the framework of gauge theory for the Poincar\'e group. This approach can be further extended to encompass the more general affine group, giving rise to metric-affine gauge gravity.

Several proposals and experiments have aimed to explore the fundamental nature of gravitation, particularly in establishing its geometric structure. During this process, there was a growing recognition that affinity and metricity could be viewed as distinct and independent concepts, with the affine connection potentially not adhering to the metric postulate. This idea is central to the Palatini approach, where GR is formulated using both a metric tensor and an affine connection, treated as separate geometric entities. By varying the Einstein-Hilbert action with respect to the metric, the Einstein field equations are derived; varying it with respect to the affine connection yields the metric compatibility condition, thereby restoring the Levi-Civita connection. This demonstrates that the structure of GR inherently involves metric compatibility and that the affine connection can be considered a true dynamical field. However, this alignment does not hold for extensions of GR, such as $f(R)$ theories.

These ideas contributed to the development of gravitational theories that extend beyond the Einstein framework, where field equations can be expressed not only in terms of scalar curvature but also other geometric invariants. In these theories, affine connections are no longer seen as secondary to the metric tensor; instead, they take on a dynamic and fundamental role. This shift in perspective has led to the diverse landscape of Modified Theories of Gravity we see today.

The main aim of the present chapter is to set some preliminary ideas for the study of modified theories. In this regard, we present some basic definitions in \autoref{sec2:chap1}. The concept of GR is discussed in \autoref{sec:chap1:GR}. The following section, \autoref{sec4:chap1}, introduces modified theories of gravity, and \autoref{sec5:chap1} outlines some equivalent formulations.

\section{Geometric Background}\label{sec2:chap1}

To further understand the concepts, it is necessary to provide some basic definitions of the terminologies used later in this thesis.

\begin{definition}[Curve]
    A \textbf{parametrized curve} in $\mathbb{R}^n$ is a map $\xi: (\alpha,\beta)\to\mathbb{R}^n$, for some some $\alpha$ and $\beta$ with $-\infty \le \alpha<\beta\le \infty$.
\end{definition}

\begin{definition}[Tangent Vector]
    If $\xi$ is a parameterized curve, the first derivative $\dot{\xi}(t)$ is called the \textbf{tangent vector} of $\xi$ at the point $\xi(t)$. 
\end{definition}



When geometrically analyzing the spatial structure and its behavior, viewing spacetime as a manifold is a fundamental concept in differential geometry. This perspective is crucial for understanding spacetime within the framework of gravity theory. A manifold is a mathematical structure that, while potentially intricate on a global scale, resembles Euclidean space in small, localized regions. This similarity to Euclidean space allows us to theoretically explain the kinematics of the Universe. One can define a manifold as

\begin{definition}[Manifold]
    Let $\mathcal{M}$ be a Hausdorff topological space. If each point $p\in \mathcal{M}$ has a neighborhood $\mathcal{U}$, which is homeomorphic to an open set $E$ in $\mathbb{R}^n$, where $\mathbb{R}^n$ is an $n$-dimensional Euclidean space, then $\mathcal{M}$ is said to be $n$-dimensional \textbf{topological manifold}. 
\end{definition}

In spacetime geometry, a four-dimensional differentiable manifold is used to represent spacetime, combining the three spatial dimensions with time. The curvature of this manifold, which is shaped by the presence of mass and energy, reflects the effects of gravity. 

\begin{definition}[Chart]
    Let $p\in\mathcal{M}$ and let $\mathcal{U}$ be a neighbourhood of $p$ in $\mathcal{M}$. If $E\subset \mathbb{R}^n$ is open, then the definition of topological manifold gives $\phi: \mathcal{U}\subset \mathcal{M}\to E\subset \mathbb{R}^n$ to be a homeomorphism. We call the pair $(\mathcal{U},\phi)$ a coordinate neighborhood of $p$ or \textbf{chart} on $\mathcal{M}$. 
\end{definition}

\begin{definition}[Tangent Space]
    If $p\in \mathcal{M}$, the set of all tangent vectors to $\mathcal{M}$ at $p$ is a vector space called the \textbf{tangent space} of $\mathcal{M}$ at $p$ and is denoted as $T_p(\mathcal{M})$. 
\end{definition}

\begin{definition}[Tangent Bundle]
    The \textbf{tangent bundle} $T\mathcal{M}$ of a manifold $\mathcal{M}$ is defined by $T\mathcal{M}=\cup _{p\in \mathcal{M}}T_p{\mathcal{M}}$.
\end{definition}

The point $p$ is uniquely determined by $n$-coordinates $(x^1(p),\dots,x^n(p))$ and is known as the local coordinates of the point $p$ in $\mathcal{U}$ with respect to the coordinate neighborhood $(\mathcal{U},\phi)$ and the $n$-tuple $(x^1,\dots,x^n)$ is called the \textbf{local coordinate system} on the neighbourhood $(\mathcal{U},\phi)$.

From the definition of the topological manifold, one can take an open set $E_\alpha$ in $\mathbb{R}^n$ such that $\phi_\alpha:\mathcal{U}_\alpha \to E_\alpha$ is a homeomorphism, then the collection $\left\{\mathcal{U}_\alpha, \phi_\alpha\right\}_{\alpha\in \mathcal{A}}$, where $\mathcal{A}$ is some index set, is called a \textbf{coordinate neighborhood system} or an \textbf{atlas} and is denoted by $S = \left\{\mathcal{U}_\alpha, \phi_\alpha\right\}_{\alpha\in \mathcal{A}}$.

\begin{definition}(Differentiable Manifold)
    A \textbf{differentiable manifold} is a topological space $\mathcal{M}$ that is locally equivalent to Euclidean space and equipped with a collection of \textit{charts}. Specifically,
    \begin{itemize}
        \item $\mathcal{M}$ is a Hausdorff and second-countable topological space.
        \item There exists an atlas $\{(U_\alpha, \varphi_\alpha)\}$, where $U_\alpha \subset\mathcal{M}$ is an open set and $\varphi_\alpha: U_\alpha \to \mathbb{R}^n$ is a homeomorphism onto its image, such that the collection of $U_\alpha$ covers $\mathcal{M}$.
        \item For any two overlapping charts $(U_\alpha, \varphi_\alpha)$ and $(U_\beta, \varphi_\beta)$, the transition maps
    \end{itemize} 
\[
\varphi_\beta \circ \varphi_\alpha^{-1} : \varphi_\alpha(U_\alpha \cap U_\beta) \to \varphi_\beta(U_\alpha \cap U_\beta)
\]
are smooth functions (infinitely differentiable). The pair $(M, \{(U_\alpha, \varphi_\alpha)\})$ is called a \textit{smooth manifold of dimension $n$}.

\end{definition}




The fundamental element in the description of the nature and dynamics of spacetime through mathematical language is tensors. It provides a mathematical approach to study and understand the properties of spaces that extend beyond basic Euclidean geometry. This framework is crucial in physics, particularly for analyzing complex systems within the fields of GR and beyond. Tensors serve to generalize vectors and scalars, facilitating the description of physical quantities in a manner that is independent of any specific coordinate system. Einstein's field equations, which illustrate how matter and energy shape the curvature of spacetime, are described using tensors. This makes tensor calculus indispensable for investigating astrophysical and cosmological phenomena. Some fundamentals regarding tensors and their properties are presented below.

\begin{definition}[Covarient and Contravariant Tensor] 
    Consider a set of \( n \) functions \( A_\mu \) defined over coordinates \( x^\mu \). These functions \( A_\mu \) are recognized as the components of a \textbf{covariant vector} if they adhere to the below-mentioned transformation 
    \begin{equation}
        \bar{A}_\mu = \frac{\partial x^\nu}{\partial \bar{x}^\mu} A_\nu.
    \end{equation}
    By multiplying both sides of this equation by \( \frac{\partial \bar{x}^\mu}{\partial x^\sigma} \) and using the identity \( \delta^\nu_\sigma A_\nu = A_\sigma \), we get
    \begin{equation*}
        A_\sigma = \frac{\partial \bar{x}^\mu}{\partial x^\sigma} \bar{A}_\mu.
    \end{equation*}
    Here, $A_\mu$ is the \textbf{covariant tensor} of the type $(0,1)$.

    The functions \( A^\mu \) are termed as the components of a \textbf{contravariant vector} if they transform according to 
    \begin{equation}
        \bar{A}^\mu = \frac{\partial \bar{x}^\mu}{\partial x^\nu} A^\nu.
    \end{equation}

By multiplying both sides of this equation by \( \frac{\partial x^\sigma}{\partial \bar{x}^\mu} \) and using the identity \( \delta^\sigma_\nu A^\nu = A^\sigma \), we get
\begin{equation*}
    A^\sigma = \frac{\partial x^\sigma}{\partial \bar{x}^\mu} \bar{A}^\mu.
\end{equation*}
In this context, \( A^\mu \) is referred to as a \textbf{contravariant tensor} of type \( (1,0) \).

It can also be written as a \textbf{mixed tensor} of the type $(m,n)$ as
\begin{equation}
    A_{\mu_1 \mu_2 \ldots \mu_n}^{\nu_1 \nu_2 \ldots \nu_m} = \frac{\partial \bar{x}^{\mu_1}}{\partial x^{\sigma_1}} \frac{\partial \bar{x}^{\mu_2}}{\partial x^{\sigma_2}} \cdots \frac{\partial \bar{x}^{\mu_m}}{\partial x^{\sigma_m}} \frac{\partial x^{\lambda_1}}{\partial \bar{x}^{\nu_1}} \frac{\partial x^{\lambda_2}}{\partial \bar{x}^{\nu_2}} \cdots \frac{\partial x^{\lambda_n}}{\partial \bar{x}^{\nu_n}} A_{\sigma_1 \sigma_2 \ldots \sigma_m}^{\lambda_1 \lambda_2 \ldots \lambda_n}.
\end{equation}
\end{definition}

\begin{definition}[Line-element]
    In an \( n \)-dimensional space, the distance between two points in the neighbourhood is expressed in the form of \textbf{line-elemen}t as
    \begin{equation}
        ds^2 =  g_{\mu\nu} dx^\mu dx^\nu.
    \end{equation}
    Here, $g_{\mu\nu}$ is called the \textbf{metric tensor}.
\end{definition}

With the signature $(+,-,-,-)$, a vector is said to be 
\begin{itemize}
    \item \textbf{Null}: when $g^{\mu\nu}A_\mu A_\nu = 0$.
    \item \textbf{Unit}: when $g^{\mu\nu}A_\mu A_\nu = 1$.
    \item \textbf{Time-like}: when $g^{\mu\nu}A_\mu A_\nu > 0$.
    \item \textbf{Space-like}: when $g^{\mu\nu}A_\mu A_\nu < 0$.
\end{itemize}

When we take the partial derivative of a tensor, it does not necessarily result in another tensor. However, when constructing expressions involving partial derivatives that form the components of a tensor, the metric is crucial. To facilitate this, E.B. Christoffel introduced notation derived from the partial derivatives of the metric tensors \(g_{\mu\nu}\). This notation is known as the \textbf{Christoffel symbol} and is defined as

\begin{equation}
    [\mu\nu; \kappa] = \Tilde{\Gamma}^\mu_{\nu\kappa} = \frac{1}{2} \left[\frac{\partial g_{\mu\kappa}}{\partial x^\nu} + \frac{\partial g_{\nu\kappa}}{\partial x^\mu} - \frac{\partial g_{\mu\nu}}{\partial x^\kappa}\right]
\end{equation}

\begin{definition}[Parallel Transport]
    If \(\xi: I \rightarrow \mathcal{M}\) is a smooth curve defined on an interval \([a, b]\), and \(\zeta \in T_x \mathcal{M}\), with \(x = \xi(a)\), then a vector field \(Y\) along \(\xi\) (specifically, the value of \(Y\) at \(y = \xi(b)\)) is referred to as the \textbf{parallel transport} of \(\zeta\) along \(\xi\) if the following conditions hold:
    \begin{gather}
        \nabla_{\xi'(t)} Y = 0, \quad \forall\; t \in [a, b],\\
        Y_{\xi(a)} = \zeta.
    \end{gather}
\end{definition}

\begin{definition}[Affine connection]
    Let $\mathcal{M}$ be a smooth manifold and let $V(\mathrm{T}\mathcal{M})$ be the space of vector fields on $\mathcal{M}$, that is, the space of smooth sections of the tangent bundle $\mathrm{T}\mathcal{M}$. Then an affine connection on $\mathcal{M}$ is a bilinear map
\begin{gather}
V(\mathrm{T}\mathcal{M}) \times V(\mathrm{T}\mathcal{M}) \rightarrow V(\mathrm{T}M), \\
(X, Y) \mapsto \nabla_X Y,
\end{gather}
such that for all $f$ in the set of smooth functions on $\mathcal{M}$, written $C^\infty(\mathcal{M}, \mathbb{R})$, and all vector fields $X, Y$ on $M$:

1. $\nabla_{fX} Y = f \nabla_X Y$, that is, $\nabla$ is $C^\infty(\mathcal{M}, \mathbb{R})$-linear in the first variable;

2. $\nabla_X(fY) = (\partial_X f) Y + f \nabla_X Y$, where $\partial_X$ denotes the directional derivative; that is, $\nabla$ satisfies the Leibniz rule in the second variable.

\end{definition}

\begin{definition}[Levi-civita connection]
    An affine connection \(\nabla\) is known as a \textbf{Levi-Civita connection} if it satisfies metric compatibility  The connection preserves the metric and torsion-free conditions, meaning \(\nabla g = 0\) and for any vector fields \(X\) and \(Y\), the relation \(\nabla_X Y - \nabla_Y X = [X, Y]\) holds, where \([X, Y]\) denotes the Lie bracket of the vector fields \(X\) and \(Y\).
\end{definition}

    The geodesic equation is expressed as
\begin{equation}
    \frac{d^2 x^\mu}{ds^2} + \Gamma^\mu_{\alpha \beta} \frac{dx^\alpha}{ds} \frac{dx^\beta}{ds} = 0
\end{equation}
where \( s \) is a scalar parameter, such as the proper time, and \( \Gamma^\mu_{\alpha \beta} \) represents the affine connection.

\section{General Relativity}\label{sec:chap1:GR}

To explain how matter and energy influence the shape of spacetime, Einstein introduced a groundbreaking theory of gravitation called GR. This theory is defined by a set of dynamic equations, known as the Einstein field equations, which describe how gravity arises from the curvature of spacetime caused by the presence of mass and energy. Einstein formulated these equations as tensor equations in 1915. The motion of objects in this curved spacetime is accurately described by geodesic equations.

Einstein undertook the development of GR on three key principles
\begin{enumerate}[]
    \item \textbf{Mach’s Principle}: Ernest Mach proposed ideas that influenced Einstein, but they lacked rigorous proof. Several interpretations of Mach's principle exist, some of which are incorporated into GR
    \begin{itemize}
        \item The geometry of space is influenced by the distribution of matter.
        \item In the absence of matter, geometry would not exist.
        \item The local inertial frame is determined entirely by the dynamics of the universe.
    \end{itemize}
    Although not all of Mach's ideas are fully applicable (since there are solutions to Einstein’s equations even in a vacuum) this principle was a crucial inspiration for Einstein.
   \item \textbf{Principle of Covariance}: In line with this principle, the governing laws of gravitation must be independent of the choice of coordinate system. This means that the gravitational field equations should remain invariant under any coordinate transformations, which requires that the equations be formulated in a tensorial form.
    \item \textbf{Causality Principle}: Every point in spacetime must have a universally applicable distinction between past, present, and future.
    
    \item \textbf{Equivalence Principle}: This principle asserts that the laws of physics are identical in all inertial reference frames, meaning no single inertial frame is preferred over another.
    \begin{itemize}
        \item The strong equivalence principle states that all laws of physics are consistent across all inertial frames.
        \item The weak equivalence principle, a more specific version, refers to the consistency of the laws governing the motion of freely falling particles.
    \end{itemize}
    The equivalence principle implies that gravity can be locally nullified by free-fall, leading to a situation where Special Relativity holds true.
\end{enumerate}

\begin{definition}[Einstein's Field Equations]
    Einstein's field equation is written in the form
    \begin{equation}\label{eq:EFE}
        G_{\mu\nu}=\kappa T_{\mu\nu}.
    \end{equation}
\end{definition} 

In the above equation, $G_{\mu\nu}$ represents the Einstein tensor, and $T_{\mu\nu}$ is the energy-momentum tensor. 

The Einstein tensor is defined as 
\begin{equation}\label{eq:einsteintensor}
    G_{\mu\nu}:= R_{\mu\nu} - \frac{1}{2} g_{\mu\nu} R.
\end{equation}

Here, $R_{\mu\nu}$ is called the \textbf{Ricci tensor} which is obtained by contacting the \textbf{Riemann curvature tenso}r:

\begin{equation}
    R^\lambda_{\mu\nu\kappa} =
\frac{\partial \Tilde{\Gamma}^\lambda_{\nu\kappa}}{\partial x^\mu} - \frac{\partial \Tilde{\Gamma}^\lambda_{\mu\kappa}}{\partial x^\nu} + \Tilde{\Gamma}^\rho_{\nu\kappa} \Tilde{\Gamma}^\lambda_{\mu\rho} - \Tilde{\Gamma}^\rho_{\mu\kappa} \Tilde{\Gamma}^\lambda_{\nu\rho}.
\end{equation}
The Riemann curvature tensor provides a mathematical description of how spacetime is warped in response to matter and energy. On contracting the Ricci tensor, one can get the \textbf{Ricci scalar} $R$. It contributes to the curvature term that describes the geometry of the Universe.

The energy-momentum tensor is defined as
\begin{equation}\label{eq:chap1:emt}
    T_{\mu\nu} = \frac{2}{\sqrt{-g}} \left[ \frac{\partial (\sqrt{-g} \mathcal{L}_F)}{\partial g^{\mu\nu}} - \frac{\partial}{\partial x^\alpha} \left(\frac{\partial (\mathcal{L}_F \sqrt{-g})}{\partial g^{\mu\nu, \alpha}}\right) \right].
\end{equation}

For isotropic matter distribution \eqref{eq:chap1:emt} can be written as
\begin{equation}\label{eq:chap1:isotropicT}
    T^{\mu\nu} = (\rho + p) u^\mu u^\nu + p g^{\mu\nu},
\end{equation}

where \( \rho \) represents the energy density, \( p \) denotes the isotropic pressure, \( u^\mu \) is the four-velocity of the fluid, and \( g^{\mu\nu} \) is the metric tensor of spacetime. This tensor describes a fluid that has uniform pressure in all directions, with no viscosity or heat conduction.

Further, the energy-momentum tensor for an anisotropic fluid, where pressures differ along different directions, is given by

\begin{equation}\label{eq:chap1:anisotropicT}
    T^{\mu\nu} = (\rho + p_t) u^\mu u^\nu + p_t g^{\mu\nu} + (p_r - p_t) v^\mu v^\nu.
\end{equation}
Here, \( p_r \) is the pressure along the direction of anisotropy, \( p_t \) is the pressure in the perpendicular directions, \( v^\mu \) is a unit spacelike vector indicating the direction of anisotropy, and the other symbols have their usual meanings. This tensor generalizes the concept of a perfect fluid by accounting for direction-dependent pressures, providing a more accurate description of fluids with anisotropic stress-energy distributions.

\subsection{Supportive Evidences for GR}

GR is founded on several key principles, each supported by observational tests that confirm its predictions.  The Weak Equivalence Principle, which indicates that all freely falling objects experience the same gravitational acceleration regardless of their composition, is validated by studies such as those by Schlamminger et al. \cite{Schlamminger:2007ht}. Local Lorentz Invariance, which means that the laws of physics are uniform regardless of the velocity of the freely falling reference frame, was demonstrated by Rossi and Hall \cite{Rossi:1941zz}. Local Position Invariance suggests that these laws remain unchanged by variations in the position or time of the freely falling frame, a principle explored by Fischer et al. \cite{Fischer:2004jt}. Evidence for GR includes the Gravitational Deflection of Light, where light bends around massive objects, affecting the apparent position of the light source, as shown by van der Wel et al. \cite{vanderWel:2013ffv}. The Perihelion Precession of Mercury, a shift in Mercury’s orbit accurately predicted by GR but not explained by Newtonian physics, has been confirmed through recent observations, including those from the Messenger spacecraft \cite{Park:2017zgd}. The Lense-Thirring Effect, where a rotating mass influences nearby spacetime, causing a precession in the spin of a gyroscope, has been observed in experiments such as those by Ciufolini et al. \cite{Ciufolini:2016ntr}. Gravitational Redshift, where light from a massive body appears redshifted depending on the observer’s distance, was confirmed in studies like those by Barstow et al. \cite{Barstow:2005mx}. The detection of Gravitational Waves, which are ripples in spacetime caused by time-varying mass distributions, was achieved by Abbott et al. \cite{LIGOScientific:2016aoc}. Lastly, the existence of Black Holes has been visually confirmed with the Event Horizon Telescope capturing an image of the black hole in galaxy M87 \cite{EventHorizonTelescope:2019ths}. These observations collectively reinforce the validity of GR.

\subsection{Need for Modifications}
For many decades, there was no need to doubt GR as it consistently provided accurate explanations for experimental and observational results. Nevertheless, both theoretical insights and observations indicate that GR may require significant modifications in regimes of strong and weak gravity. The Einstein equations, derived from the variation of the Einstein-Hilbert action—which is linearly dependent on the curvature scalar \( R \)—have served as the foundation of GR. However, in this section, we will explore reasons why the assumption of a linear relationship with the curvature scalar may not hold true.

\textbf{Lacks in analyzing gravity at all scales:} As we know, the principle of Equivalence is a foundational principle of GR. One notable implication of this principle is that all entities, including light, follow the same laws. Therefore, the Einstein Equivalence Principle (EEP) serves as the link between gravity and other areas of physics. Unlike the gauge invariance seen in the electromagnetic field, the EEP is not a fundamental symmetry but rather an experimentally observed fact. Einstein initially termed it the equivalence hypothesis, later promoting it to a principle as its essential role in extending special relativity to encompass gravity became apparent. It is remarkable that the EEP holds true, especially given the rigorous precision of contemporary tests.  Moreover, testing the viability of EEP is pretty much difficult when it comes to cosmic scales. 

\textbf{Problematic in strong gravity regime:} When considering strong gravity regimes, the extrapolated results from observational surveys suggest that one needs to go beyond GR. For instance, the insufficiency of GR in the description of primordial black holes and the inclusion of higher-order curvature terms in the early Universe. 

\textbf{Requiring quantum corrections:} Our current understanding identifies the four fundamental interactions of nature. The first three (strong, weak, electromagnetic) are successfully explained by the Standard Model of particle physics, which achieves a partial unification of the electromagnetic and weak interactions. Aside from the issue of non-zero neutrino masses, there is no experimental evidence that contradicts the Standard Model at this time. Gravity is described by GR. Consequently, there seems to be little empirical motivation to search for new physical laws. However, from a theoretical perspective, the situation is less satisfactory. GR is a classical theory, while the Standard Model is a quantum field theory that incompletely unifies the interactions.

We still lack a theory of quantum gravity that is both physically and mathematically consistent. As we examine smaller and smaller length scales, quantum effects become increasingly significant and cannot be ignored. Considering quantum matter within a classical gravitational framework introduces additional challenges, particularly the possibility that zero-point fluctuations of the matter fields could produce a non-zero vacuum energy density, \( \rho_{\text{vac}} \).

\textbf{Initial singularity problem:} The initial singularity issue is a fundamental challenge in cosmology, particularly concerning the standard cosmological model and Einstein's Field Equations. According to these models, the Universe began from an initial state of infinite density, temperature, and curvature, known as the singularity. This singularity represents a point where the laws of physics as we know them break down, leading to infinities that are problematic for our understanding of the Universe's origins. When standard governing equations are applied to the very early Universe, they predict a singularity where all physical quantities become infinite. This prediction suggests that GR cannot fully describe the conditions of the Universe at its inception.

\textbf{Cosmic coincidence:} Cosmological observations indicate that the current value of the dark energy (DE) density parameter, \( \Omega_{\Lambda 0} \), is approximately 0.7, which is comparable to the current value of the matter-energy density parameter, \( \Omega_{m0} \), around 0.3. This suggests that, despite the matter-energy density \( \rho_m \) varying over time while the DE density \( \rho_{\Lambda} \) remains constant, they happen to be of the same order in the present epoch. This apparent coincidence would require precise fine-tuning of parameters in the early Universe and is referred to as the cosmological coincidence problem. It is important to note that this issue arises in many other DE models as well.

\textbf{Fine-tuning problem:} Observations estimate that the current energy density of the cosmological constant is approximately \( \rho_{\Lambda} \sim 10^{-47} \) GeV. If the cosmological constant arises from vacuum energy density, this value is drastically different from the vacuum energy density predicted by quantum field theory, which is around \( \rho_{\text{vac}} \sim 10^{74} \) GeV. This discrepancy between \( \rho_{\Lambda} \) and the expected energy scales highlights a significant conflict, indicating that fine-tuning is required.

\textbf{Conservation of local energy-momentum tensor:}
In GR, the local energy-momentum tensor \( T_{\mu\nu} \) is not conserved in the same way as in flat spacetime. While in special relativity, the conservation law \( \partial^\mu T_{\mu\nu} = 0 \) holds, in curved spacetime, the appropriate expression is \( \nabla^\mu T_{\mu\nu} = 0 \), which accounts for spacetime curvature. This equation does not imply global conservation of energy and momentum as it does in flat spacetime, because the concept of global energy conservation is not generally applicable in curved spacetime. The energy-momentum of the gravitational field itself cannot be captured by a global conservation law, and while quasi-local quantities can be defined in specific spacetimes, they do not provide a general solution.

\textbf{Ambiguity in the nature of the Universe:}
Einstein's ``great blunder" refers to his introduction of the cosmological constant (\(\Lambda\)) into his field equations of GR to force a static Universe model, which was widely believed at the time. His original equations suggested that the Universe should be either expanding or contracting, but to counteract this and maintain a static Universe, Einstein added \(\Lambda\) as a repulsive force to balance gravity. However, in 1929, Edwin Hubble's discovery of the expanding Universe showed that this addition was unnecessary, leading Einstein to consider it a mistake. Ironically, the cosmological constant was later revived in modern cosmology to explain the accelerating expansion of the Universe, driven by DE, but Einstein's original motivation for \(\Lambda\) was indeed a misstep. This clearly puts forth the ambiguity present in the determination of the nature of the Universe.

\textbf{Exploring the dark sector of the Universe:} GR works well with visible matter but does not provide insights into what dark matter (DM) is composed of. While GR can describe how DM influences the curvature of spacetime and gravitational effects, it does not explain its nature or origin. DM does not interact with electromagnetic forces, making it invisible and undetectable by traditional methods that rely on light. This makes it difficult to confirm its presence directly using GR. 

GR can include a cosmological constant to account for the accelerated expansion of the universe, which is attributed to DE. However, the value of $\Lambda$ derived from observations is many orders of magnitude smaller than theoretical predictions from quantum field theory, leading to the so-called cosmological constant problem.

\textbf{Rising tensions:}
The distance ladder method yields local (low-redshift) measurements of the Hubble constant $(H_0)$ that are notably higher than those derived from the angular scale of cosmic microwave background (CMB) fluctuations within the $\Lambda$ Cold Dark Matter ($\Lambda$CDM) framework. When these local direct measurements of $H_0$ are combined, they exhibit a tension with CMB-based indirect measurements of $H_0$, reaching or exceeding the $5\sigma$ level, especially when various local measurement techniques are integrated.

Direct observations of the growth rate of cosmological perturbations—through methods such as Weak Lensing, Cluster Counts, and Redshift Space Distortions—suggest a slower growth rate compared to the predictions based on $\Lambda$CDM parameters, with a discrepancy of about $2-3\sigma$. Within the framework of GR, this slower growth rate would imply a lower matter density and a reduced amplitude of the primordial fluctuation spectrum than what is suggested by the $\Lambda$CDM model.

\textbf{Beyond Cosmological Principle:} 
The $\Lambda$CDM model adheres to the cosmological principle, which posits that, on sufficiently large scales, the Universe appears uniform in all directions and from every location. This principle was initially adopted during the development of its predecessor models due to a lack of data to differentiate between more complex anisotropic or inhomogeneous alternatives. As a result, these assumptions of homogeneity and isotropy were integrated into the standard cosmological model. Recent observations, however, have suggested possible violations of the cosmological principle, particularly concerning isotropy. Such deviations have led to scrutiny of the standard model of cosmology, and its predictions might fail to accurately describe the late Universe. These issues also affect the validity of the cosmological constant in the $\Lambda$CDM model, as the existence of DE is contingent upon the accuracy of the cosmological principle.

\textbf{Exotic matter issue:} In the context of GR, the existence of exotic matter cannot be ruled out when constructing a traversable wormhole solution. Keeping in mind the success of GR and to address its limitations, some equivalent formulations of GR have been proposed in the literature, which allow for further modifications of these theories. In the following section, we will discuss these modified theories in detail.

\section{Modified Theories} \label{sec4:chap1}

\begin{figure}[htbp]
    \centering
    \includegraphics[width=0.7\linewidth]{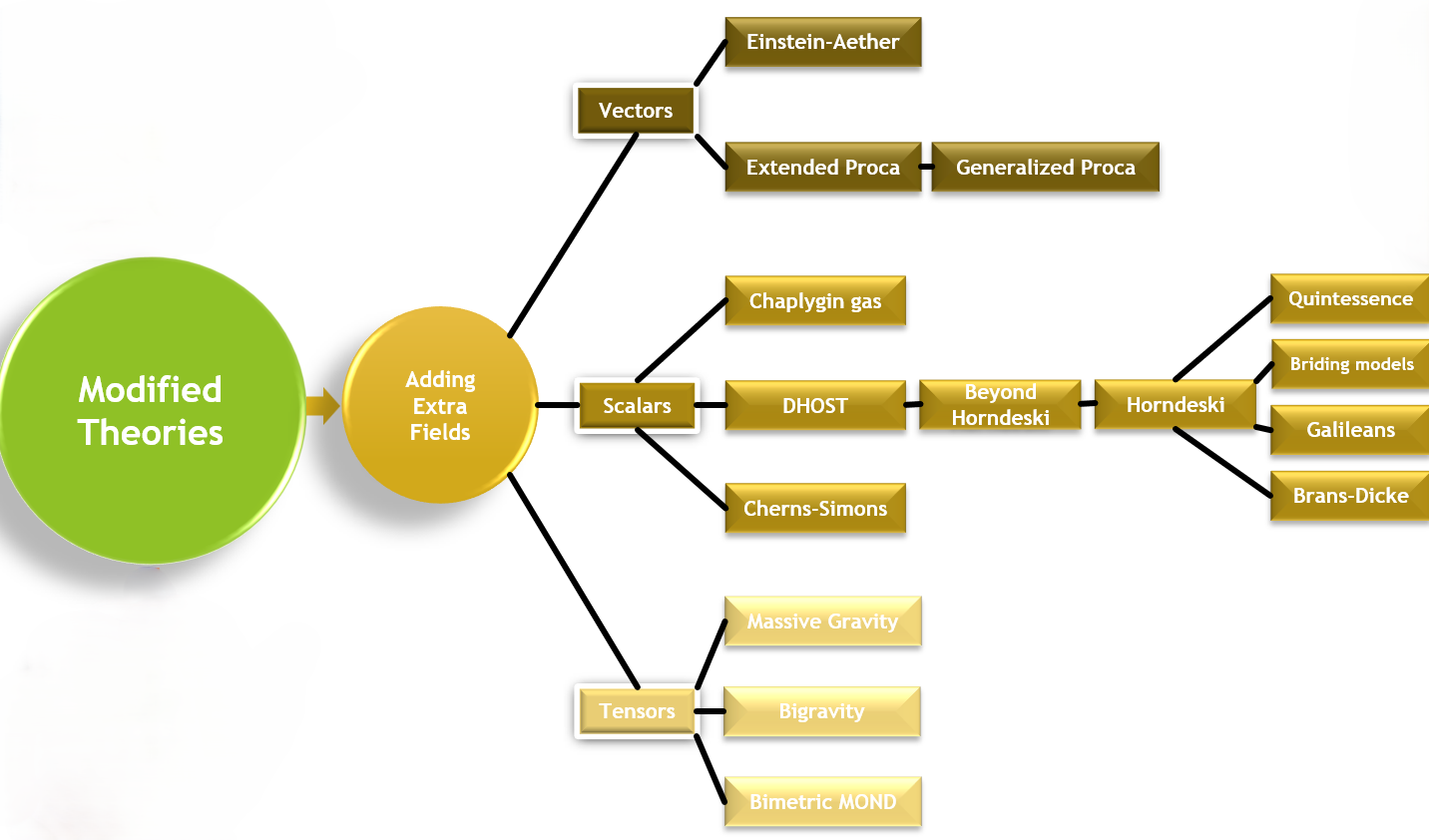}
    \caption{Schematic representation of the proposed modified gravity theories by adding extra fields.}
    \label{fig:chap1:fields}
\end{figure}

Given that Einstein's theory has been remarkably successful in explaining the majority of the physical phenomena that we witness, one must acknowledge its contribution. Yet we also have to admit that there are still a lot of unanswered questions, and any changes to the theory that try to solve these questions need to have a consistent limit that is consistent with GR. This idea is the basis of the scalar-tensor theories, which are among the most extensively studied methods for altering GR. Originally, these theories were proposed as a possible better way to integrate Mach's principle into gravitational theory. The idea was to create a residual effect that relies on a universal feature, like the total mass of the Universe so that the gravitational field could not be fully measured away. Generally, in these theories, the matter part of the Einstein field equation \eqref{eq:EFE} is modified by adding additional scalar, vector, or tensor fields.  \autoref{fig:chap1:fields} depicts the modified theories constructed using this approach.

\begin{figure}[htbp]
    \centering
    \includegraphics[width=0.7\linewidth]{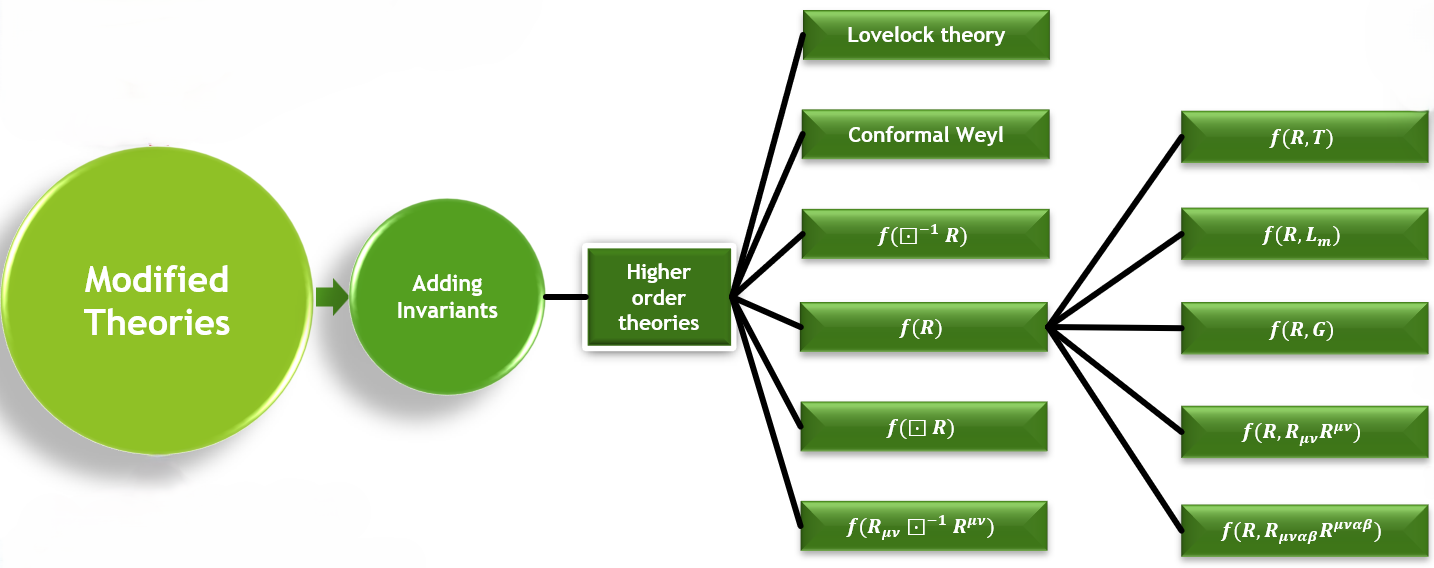}
    \caption{Schematic representation of the proposed modified gravity theories by adding invariant.}
    \label{fig:chap1:invarients}
\end{figure}

As we have seen in \autoref{sec:chap1:GR}, the metric-affine approach begins by constructing a curvature scalar, treating the connection and the metric tensor as independent entities. This framework can then be significantly expanded by formulating Lagrangians that incorporate additional scalar terms derived from the symmetric part of the Ricci tensor and its contractions with the metric tensor. Such an approach opens up several promising possibilities, including the potential to eliminate or smooth out cosmological singularities, which are typically unavoidable in standard GR scenarios. The modified theories developed through this approach are represented in \autoref{fig:chap1:invarients}.

In Quantum Mechanics, Zero Point Energy (ZPE) represents the lowest possible energy a quantum system can possess. Unlike classical mechanics, where systems remain at rest in their lowest energy state, quantum systems constantly fluctuate due to the Heisenberg uncertainty principle. This behavior is also observed in Quantum Field Theory. To extend quantum principles to GR, one would need a quantum theory of the gravitational field, commonly referred to as Quantum Gravity. Unfortunately, such a theory yet to be developed. However, the search for a viable ZPE candidate continues, despite the well-known challenge that ZPE calculations often encounter divergences.

\begin{figure}[htbp!]
    \centering
    \includegraphics[width=0.6\linewidth]{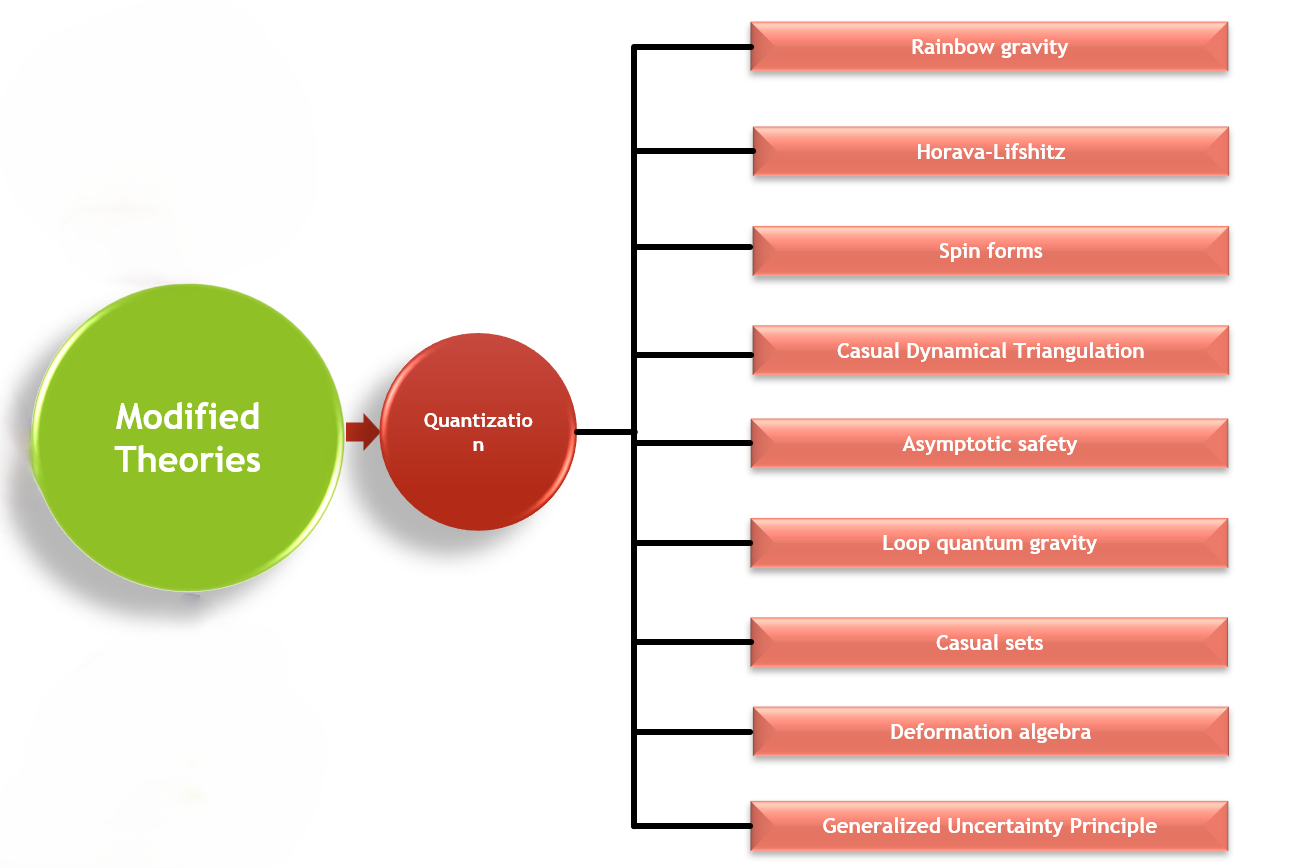}
    \caption{Schematic representation of the proposed modified gravity theories by adding the quantization effect.}
    \label{fig:chap1:quantum}
\end{figure}

These divergences are typically managed through regularization and renormalization techniques. In the context of ordinary gravity, the ZPE for quantum fluctuations of the gravitational field can be derived by reformulating the Wheeler-DeWitt equation in a manner resembling an expectation value computation. This derivation stems from the Arnowitt-Deser-Misner decomposition of spacetime, based on the modified line element. The theories proposed based on the incorporation of the quantization effect are schematically represented in \autoref{fig:chap1:quantum}

\begin{figure}[htbp]
    \centering
    \includegraphics[width=0.8\linewidth]{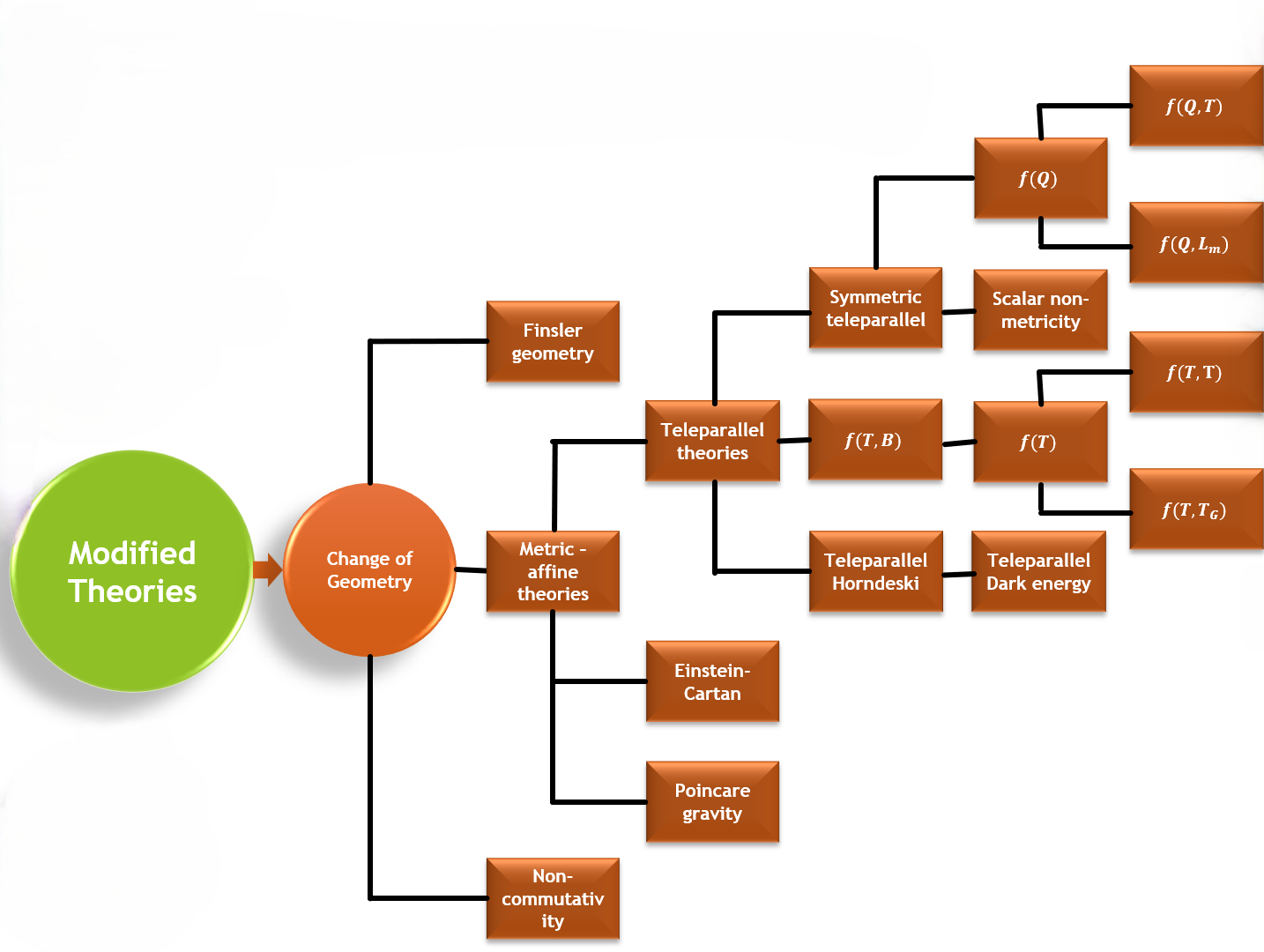}
    \caption{Schematic representation of the proposed modified gravity theories by changing geometric description}
    \label{fig:chap1:changeofgeometry}
\end{figure}

Alternatively, a distinct and somewhat unconventional approach to describing gravity as a manifestation of geometry can be explored through teleparallel gravity and its extensions. One of the appealing aspects of teleparallel gravity is the second-order nature of its equations and the flexibility of the framework, which allows for the reformulation of many theories that rely on the metric and the Levi-Civita connection. A particularly intriguing feature is the theory's ability to dispense with the equivalence principle, thereby reinterpreting gravitation as an interaction more akin to other fundamental forces. The framework offers a remarkable collection of new features, such as the potential to interpret gravity as a gauge theory of translations, which helps bridge the gap between gravity and other fundamental interactions. It also raises intellectually challenging questions regarding the definition and characterization of singularities within this alternative formulation of gravitational physics. \autoref{fig:chap1:changeofgeometry} provides a schematic categorization of the theories that emerge from modifying the geometry of GR.

Gravity theories are inherently considered within a geometric context. In the original formulation of GR, spacetime coordinates \( x^{\mu} \) and a metric tensor \( g_{\mu\nu} \) define a (semi-) Riemannian space. Cartan's work introduced a more appropriate, coordinate-independent framework for the geometric approach to gravity theories. In this framework, local frames are described by a set of vectors \( e_A \), whose components \( e_A^{\mu} \) are known as the vierbein, \( e_A = e_A^{\mu}\partial_{\mu} \). Provided the frame field is not degenerate, it has an inverse (the coframe one-form \( e^A \)) which is dual to the frame with respect to the inner product \( e^A \cdot e^B = \delta^A_B \). The concept of parallel transport is used to describe the relationship between frames at different points in spacetime. 

\begin{figure}[htbp!]
    \centering
    \includegraphics[width=\linewidth]{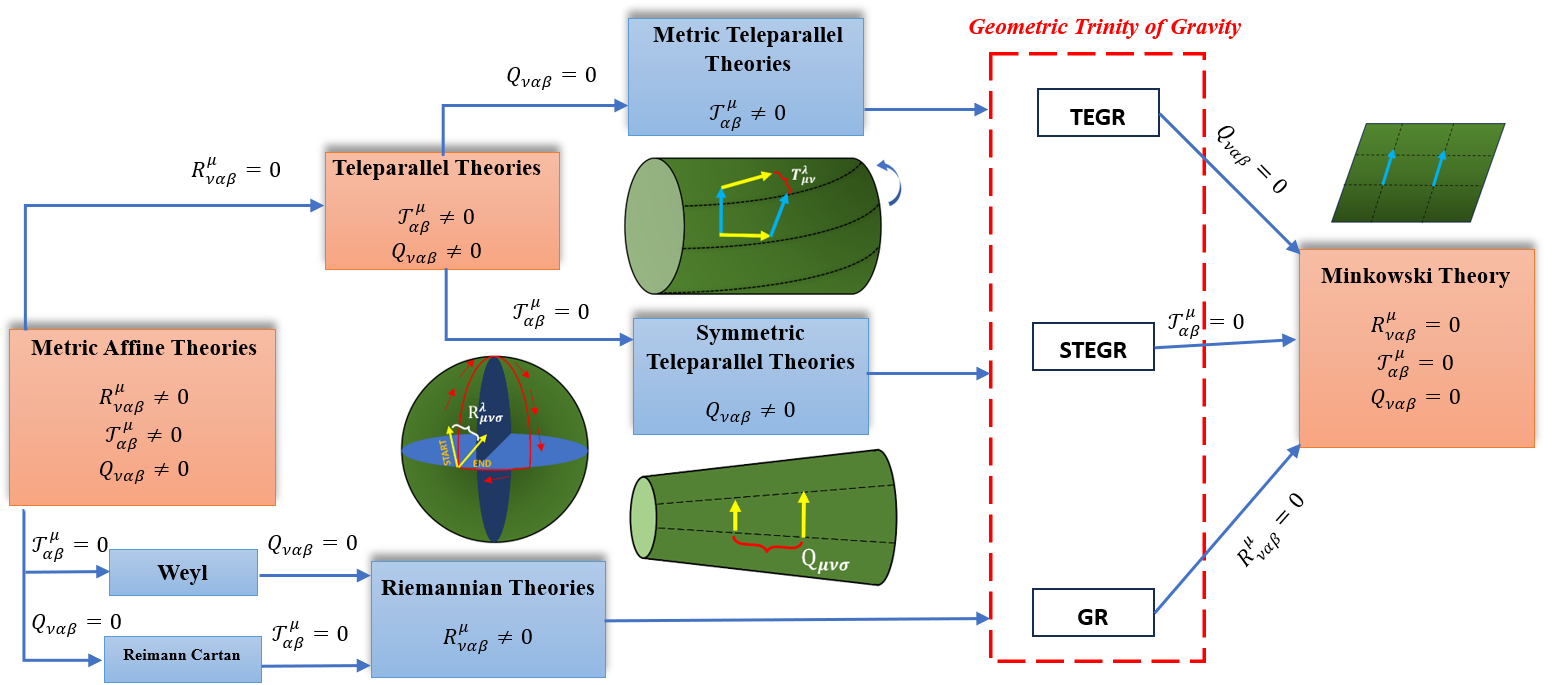}
    \caption{A potential categorization of gravity theories arising within the framework of metric-affine geometry}
    \label{fig:chap1:trinity}
\end{figure}

Since this study focuses on modified gravity through a differential geometric approach, particular emphasis is placed on exploring the dynamics of the Universe by adapting the technique of changing geometry. In the next section, we will examine how the equivalent formalism of GR can be presented by altering the underlying geometry.

\section{Equivalent Formulations}\label{sec5:chap1}

The progression of theories about gravity has evolved in tandem with advancements in differential geometry. According to these geometric models, spacetime is endowed with a metric framework within a broader context. This framework is described by the line element \( ds = F(x^1, \ldots, x^n; dx^1, \ldots, dx^n) \), where \( F(x; \xi) > 0 \) for \( \xi \neq 0 \), defined over the tangent bundle \( T\mathcal{M} \), and \( F \) is homogeneous of degree 1 in \( \xi \). In particular, when \( F^2 = g_{\mu\nu} dx^\mu dx^\nu \), it aligns with Riemannian geometry. An intriguing modification of GR involves adopting a more generalized geometric perspective, as proposed by Weyl and Cartan. This approach has led to the development of theories such as Riemann-Cartan and Weyl-Cartan.

An initial modification to GR involves generalizing the definition of affine connections, incorporating components beyond Levi-Civita \cite{Capozziello:2022zzh}. This leads to the formalism of a metric-affine theory, characterized by the triplet $\{\mathcal{M}, g_{\mu\nu}, \Gamma^{\rho}_{\mu\nu}\}$. Here, $\mathcal{M}$ represents a four-dimensional spacetime manifold, $g_{\mu\nu}$ denotes a rank-two symmetric tensor with 10 independent components, and $\Gamma^{\rho}_{\mu\nu}$ signifies the affine connection which can be uniquely decomposed as \cite{BeltranJimenez:2019esp}

\begin{equation}
    \Gamma^{\rho}_{\mu\nu} = \tilde{\Gamma}^{\rho}_{\mu\nu} + K^{\rho}_{\mu\nu} + L^{\rho}_{\mu\nu},
\end{equation}
where $\tilde{\Gamma}^{\rho}_{\mu\nu}:= \frac{1}{2} g^{\rho\lambda} (\partial_{\mu} g_{\lambda\nu} + \partial_{\nu} g_{\mu\lambda} - \partial_{\lambda} g_{\mu\nu}) $ is the Levi-Civita connection, $K^{\rho}_{\mu\nu}:=\frac{1}{2} (\mathcal{T}^{\rho}_{\mu \nu} + \mathcal{T}^{\rho}_{\nu \mu} - \mathcal{T}^{\rho}_{\mu\nu})$ is the contortion tensor, and $L^{\rho}_{\mu\nu}:=\frac{1}{2} (Q^{\rho}_{\mu\nu} - Q^{\rho}_{\mu \nu} - Q^{\rho}_{\nu \mu})$ is the disformation tensor. The corresponding geometric entities are represented by 
the curvature tensor $R^\mu_{\nu \alpha \beta}:= \partial_\rho \Gamma^\mu_{\nu \sigma} - \partial_\sigma \Gamma^\mu_{\nu \rho} + \Gamma^\mu_{\tau \rho} \Gamma^\tau_{\nu \sigma} - \Gamma^\mu_{\tau \sigma} \Gamma^\tau_{\nu \rho}$, the torsion tensor  $\mathcal{T}^\mu_{\nu \rho}:= 2 \Gamma^\mu_{[\rho \nu]} \equiv \Gamma^\mu_{\rho \nu} - \Gamma^\mu_{\nu \rho}$, and the non-metricity tensor  $Q_{\rho \mu \nu} := \nabla_\mu g_{\nu \rho} \equiv \partial_\mu g_{\nu \rho} - \Gamma^\lambda_{(\nu | \mu} g_{\rho) \lambda} \neq 0$.

The tensors exhibit symmetries as shown below
\begin{gather}
    R^{\mu}_{\nu \rho \sigma} = -R^{\mu}_{\nu \sigma \rho},\\
    \mathcal{T}^{\mu}_{\nu \rho} = -\mathcal{T}^{\mu}_{\rho \nu},\\
    Q_{\mu \nu \rho} = Q_{\mu \rho \nu}.
\end{gather}

The aforementioned geometric quantities influence the parallel transport of a vector on a manifold in different ways. Specifically

\begin{itemize}
    \item \textbf{Curvature} becomes evident when a vector is parallel transported along a closed loop on a non-flat surface, returning to its starting point at an angle different from its initial orientation.
    \item \textbf{Torsion} introduces a rotational geometry, causing the parallel transport of two vectors to be antisymmetric when the vectors and direction of transport are exchanged. This does not lead to the closed path of parallelograms.
    \item \textbf{Non-metricity} affects the length of vectors during transport.
\end{itemize}

In a general metric-affine theory, these effects can interact and may also correspond to physical quantities. For example, the torsion tensor is associated with spin in the Einstein-Cartan theory \cite{Hehl:1976kj}. Metric-affine geometries encompass a wide range of theories, with their dynamics connected to the tensors  $R^\mu_{\nu \rho \sigma}, \mathcal{T}^\mu_{\nu \rho}$,  and  $Q_{\mu \nu \rho}$. These theories can be broadly categorized as shown in \autoref{fig:chap1:trinity}.

GR is formulated using the metric  $g_{\mu \nu}$ whereas TEGR employs the tetrads $e^A_{\mu}$ for the dynamic description of gravity and the spin connection $\omega^A_{B \mu}$ to represent inertial effects. The STEGR, on the other hand, incorporates the Palatini approach, treating the metric  $g_{\mu \nu}$ and the affine connection $ \Gamma^\mu_{\alpha \beta}$ as two distinct dynamical entities. Similar to other fundamental forces, gravity can be reinterpreted as a gauge theory through the frameworks of TEGR and STEGR. What sets gravitation apart is its universal nature; it influences all objects, regardless of their internal composition, a concept encapsulated by the Equivalence Principle in GR. 

\begin{figure}[htbp]
    \centering
    \includegraphics[width=\linewidth]{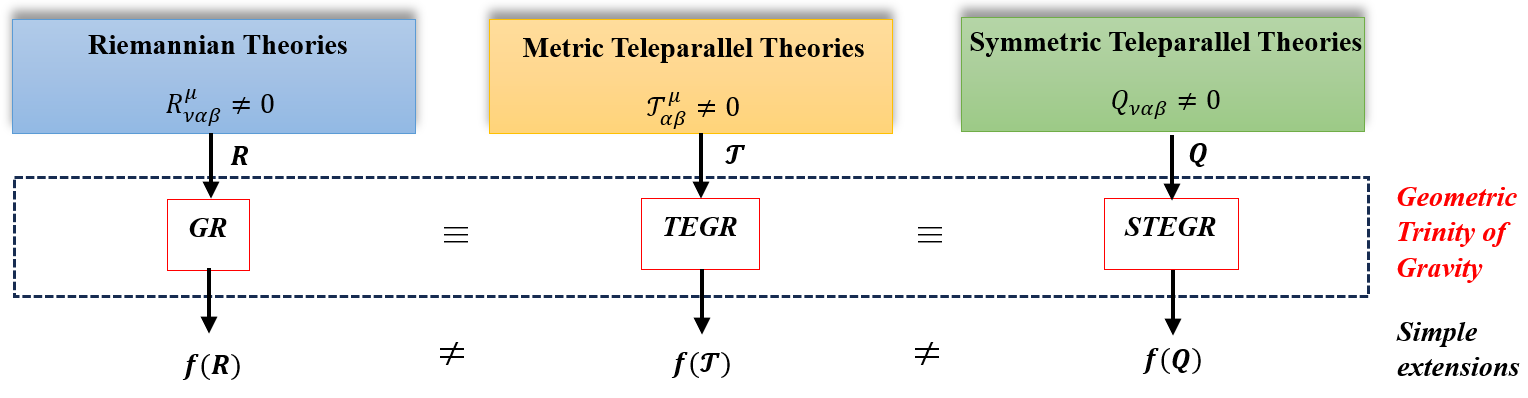}
    \caption{Schematic representation of equivalent theories and their extensions}
    \label{fig:cha1:trinity1}
\end{figure}

\begin{figure}[htbp]
    \centering
    \includegraphics[width=0.7\linewidth]{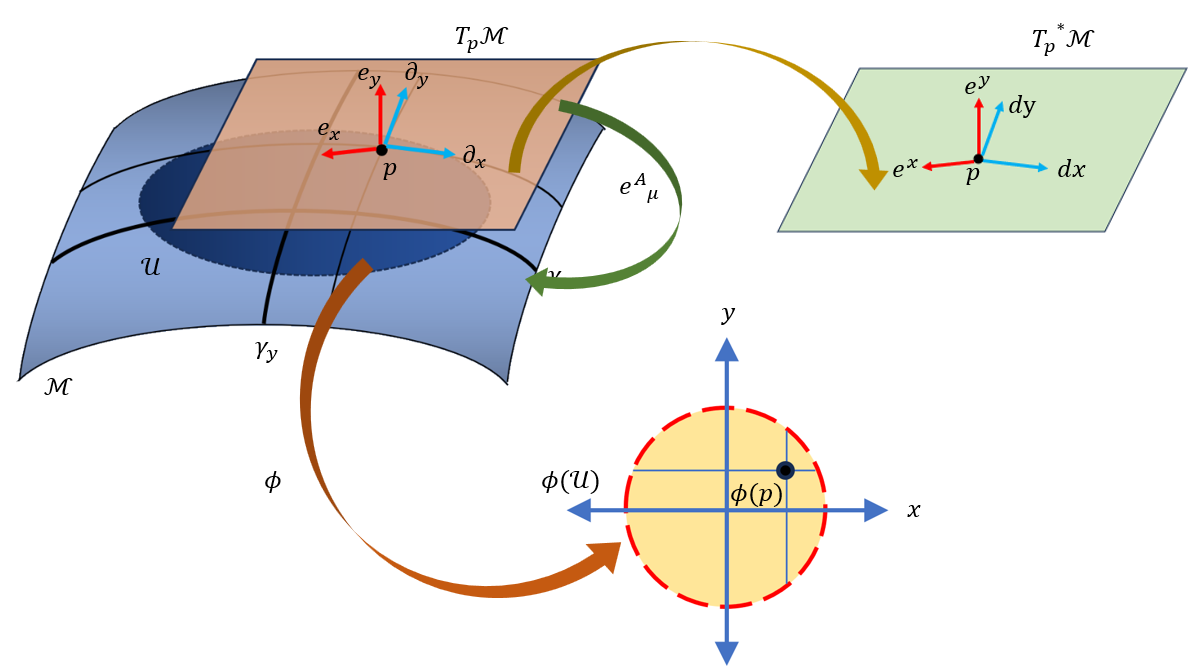}
    \caption{Pictorial representation of the tetrad formalism}
    \label{fig:chap1:manifold}
\end{figure}

\begin{figure}[htbp!]
    \centering
    \includegraphics[width=0.5\linewidth]{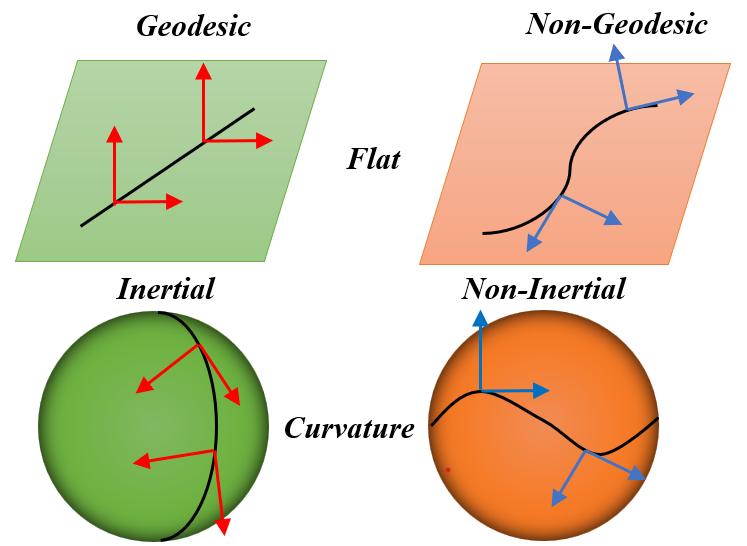}
    \caption{Pictorial representation of the behavior of tetrads under the absence and the influence of gravity.}
    \label{fig:chap1:geodesic}
\end{figure}

In some discussions of torsion-based formulations, the validity of the Equivalence Principle is questioned. However, we will emphasize how it can still be upheld within these theories, even though it is not their foundational principle. This point is particularly significant because if the Equivalence Principle were proven to be fundamentally violated, it might suggest that the ultimate theory of gravitation could be non-metric.

Within these equivalent frameworks, we can establish alternative representations of the gravitational field, each reflecting the same degrees of freedom (DoFs) and corresponding to specific geometric invariants: the Ricci curvature scalar $R$, the torsion scalar $\mathcal{T}$, and the non-metricity scalar $Q$. This leads to the concept of the Geometric Trinity of Gravity, encompassing GR, TEGR, and STEGR. Just as GR can be extended to \(f(R)\) gravity, TEGR and STEGR can be extended to \(f(\mathcal{T})\) and \(f(Q)\) gravity, respectively, where \(f\) is a smooth arbitrary function. It is important to note that the equivalence among the three original theories generally does not hold in their extended versions, as they result in different dynamics and DoFs (see \autoref{fig:cha1:trinity1}). Specifically, in \(f(R)\) gravity, the governing equations are of order four in the metric representation, whereas in \(f(\mathcal{T})\) and \(f(Q)\) theories, they remain in order two. Additionally, unlike TEGR and STEGR, where a gauge choice can simplify the calculations, \(f(\mathcal{T})\) and \(f(Q)\) generally do not allow for such a simplification. Instead, one must address the field equations for the spin connection in \(f(\mathcal{T})\) and the affine field equations in \(f(Q)\).

\subsection{Formalism of tetrads}

The geometric framework for any gravitational theory is based on the tangent bundle, a fundamental structure inherent in any smooth spacetime. At every point in spacetime, one can define a tangent vector space associated with that location. Within this framework, vectors and covectors can be represented using a general linear orthonormal basis, known as tetrads or vielbeine, in their respective domains.

Consider a general metric spacetime \((\mathcal{M}, g_{\mu\nu})\), where \(\mathcal{M}\) is a four-dimensional differentiable manifold of class \(C^\infty\), and at each point \(p \in \mathcal{M}\), the tangent spaces \(T_p \mathcal{M}\) are Minkowski spacetimes with metric \(\eta_{AB}\). The symmetric metric tensor \(g_{\mu\nu}\) defines the geometry of \(\mathcal{M}\). Under these conditions, there is a compatible atlas of charts \(A\), which provides an open cover of \(\mathcal{M}\). For each point \(p \in \mathcal{M}\), $\exists$ a chart \((\mathcal{U}, \phi)\) where \(\mathcal{U}\) is an open neighborhood of \(p\) and \(\phi: \mathcal{U} \to \phi(\mathcal{U}) \subset \mathbb{R}^4\) is a homeomorphism. Additionally, for any two charts \((\mathcal{U}, \phi)\) and \((\mathcal{V}, \psi)\) in \(A\), the map \(\psi \circ \phi^{-1}: \phi(\mathcal{U} \cap \mathcal{V}) \to \psi(\mathcal{U} \cap \mathcal{V})\) is a \(C^\infty\)-diffeomorphism known as the coordinate transformation. Thus, each point \(p \in \mathcal{M}\) can be assigned coordinates \((x^0, x^1, x^2, x^3) := \phi(p) \in \mathbb{R}^4\).

By defining the coordinate axes \(x^\mu\) in \(\mathbb{R}^4\), we can establish the corresponding coordinate curves \(\gamma_{x^\mu}\) on \(\mathcal{M}\) using the charts. Consequently, all curves parallel to the coordinate axes in \(\mathbb{R}^4\) create a grid on \(\mathcal{M}\) that allows for the precise identification of points in spacetime. A natural or holonomic basis for each tangent space \(T_p \mathcal{M}\) at point \(p\) consists of vectors tangent to the coordinate lines at \(p\), denoted as \(\partial_\mu := \left.\frac{\partial}{\partial x^\mu}\right|_p\). For covector fields defined on the cotangent space \(T_p^* \mathcal{M}\) (the space of all linear maps \(\alpha: T_p \mathcal{M} \to \mathbb{R}\)), the basis \(\{dx^\mu\}\) at point \(p\) holds the orthonormality condition
\begin{equation}
    dx^\mu \partial_\nu = \delta^\mu_\nu.
\end{equation}

The tangent space \(T_p \mathcal{M}\) and the cotangent space \(T_p^* \mathcal{M}\) at \(p\) are interconnected through the metrics \(g_{\mu \nu}\) and \(\eta_{AB}\).

Vectors and covectors at \(p\) can be expressed in terms of this natural basis. We define a set of orthonormal vectors and covectors related to the natural basis through:
\[ e^A := e^A_\mu \partial_\mu, \quad e_A := e_{A \mu} dx^\mu, \]
where the coefficients \(\{e^A_\mu\}\) are known as the tetrad transformation matrix and belong to the general linear group of real \(4 \times 4\) invertible matrices \(\text{GL}(4, \mathbb{R})\). The tetrads serve as a bridge between the general manifold (represented by $\mu,\nu$) and Minkowski spacetime (with represented by $A,B$), as shown by
\begin{equation}\label{eq:chap1:gmunu}
    g_{\mu \nu} = \eta_{AB} e^A_\mu e^B_\nu, \quad \eta_{AB} = g_{\mu \nu} e^\mu_A e^\nu_B.
\end{equation}
Thus, a tetrad field provides a local frame that connects the coordinate charts on \(\mathcal{M}\) to the preferred orthonormal basis \(e^A\) in the tangent space, simplifying calculations significantly.

As \(g_{\mu \nu}\) varies across the manifold \(\mathcal{M}\), the vierbeine \(e^A_\mu\) vary accordingly. The determinant of the relation given in \eqref{eq:chap1:gmunu} is \(-g = e^2\), where \(e\) is the determinant of \(e^A_\mu\), and the negative sign reflects the signature of \(\eta_{AB}\). Thus, the vierbeine represents the square root of the metric. The tetrads and their properties are illustrated in \autoref{fig:chap1:manifold}.

A particular category of frames is the inertial frames, denoted as \(\{e^0_A\}\), where the coefficients of anholonomy \(f^0_{CAB}\) satisfy \(f^0_{CAB} = 0\). Consequently, these frames are holonomic. This characteristic is not just local but is true for all frames within this inertial class everywhere. In the absence of gravitational effects, the anholonomy arises solely from inertial forces present in these frames. The metric \(g_{\mu\nu}\) then simplifies to the Minkowski metric \(\eta_{\mu\nu}\). The metric \(\eta_{\mu\nu}\) is a function of the spacetime point and is independent of whether \(\{e^A\}\) is holonomic (inertial) or not. In this case, tetrads consistently relate the tangent Minkowski space to a Minkowski spacetime as given by \(\eta_{AB} = \eta_{\mu\nu} e^\mu_A e^\nu_B\). These are the frames found in Special Relativity, often referred to as trivial frames or trivial tetrads, and they are particularly useful in contexts involving torsion. In the absence of inertial forces, the inertial frame class is characterized by vanishing structure coefficients (See \autoref{fig:chap1:geodesic}).

\subsection{Governing Equations of Equivalent Theories}

\subsubsection{Formulation of TEGR}\label{sub:chap1:TEGR}

In the previous section, we explained how tetrads are formed. TEGR theory uses the Weitzenb\"{o}ck connection instead of the usual torsion-less Levi-Civita connection of GR. Using the connection, the non-null torsion tensor can be obtained as
\begin{equation}
    \mathcal{T}_{\mu \nu}^{\lambda}=\overset{w}{\Gamma}_{\nu \mu}^\lambda-\overset{w}{\Gamma}_{\mu \nu}^\lambda=e_{\gamma}^\lambda(\partial_{\mu}e_{\nu}^\gamma-\partial_{\nu}e_{\mu}^\gamma).
\end{equation}
Here $\overset{w}{\Gamma}_{\nu \mu}^\lambda$ is the Weitzenb\"{o}ck connection and the tetrad fields $e_{\mu}^\gamma$ or $e_{\mu}(x^\gamma)$ create a tangent space at each point $x^\gamma$ of the manifold. The metric tensor of the manifold can be written in terms of the tetrad fields as $g_{\mu \nu}(x)=\eta_{\alpha \beta} \, e_\mu^{\alpha}(x) \, e_\nu^{\beta}(x)$, with the pseudo-Riemannian metric $\eta_{\alpha \beta}=diag(1,-1,-1,-1)$.

 The torsion scalar can be defined as 
\begin{equation}
    T\equiv {S_{\lambda}}^{\mu \nu} T_{\mu \nu}^\lambda
\end{equation}
where the superpotential tensor is defined as ${S_{\lambda}}^{\mu \nu} \equiv \frac{1}{2}({K^{\mu \nu}}_{\lambda}+\delta_\lambda^\mu {T^{\alpha \nu}}_{\alpha}-\delta_\lambda^\nu {T^{\alpha \mu}}_{\alpha})$ and the contorsion tensor is defined as ${K^{\mu \nu}}_{\lambda} \equiv -\frac{1}{2}({T^{\mu \nu}}_{\lambda}-{T^{\nu \mu}}_{\lambda}-{T_{\lambda}}^{\mu \nu})$. Using the torsion scalar one can modify the GR action to teleparallel action as
\begin{equation}\label{eq:chap1:actionT}
    S=\frac{1}{2k^2} \int d^4xe[T+f(T)]
\end{equation}
where $k=\sqrt{8\pi G}$ and $e=det(e_\mu^A)=\sqrt{-g}$. 

Variation of the above action with respect to the tetrads gives rise to the motion equation
\begin{equation}
   \left[e^{-1}\partial_\mu\,(e\,{e^\lambda_{\,A}} \,S_\lambda^{\,\, \nu \mu})-e^\alpha_{\,A} \,T^{\lambda}_{\,\, \mu \alpha}\, S_{\lambda}^{\,\, \mu \nu} \right]+\frac{1}{4} \,e^\nu_{\,A} \,T=4 \pi G \,e^\lambda_{\,A} \stackrel{em}{T}_\lambda^{\,\,\, \nu},
\end{equation}
 where $\stackrel{em}{T}_\lambda^{\,\,\, \nu}$ denotes the usual energy momentum tensor.

\subsubsection{Formulation of STEGR}\label{sub:chap1:STEGR}

In general, for theories based on nonmetricity, the first two geometric quantities, namely the curvature tensor and the torsion tensor, are zero. The disformation tensor is defined as

\begin{equation}
    {L^\mu}_{\rho\beta} \equiv -\frac{1}{2} g^{\mu\nu} \left( \nabla_{\rho} g_{\nu\beta} + \nabla_{\beta} g_{\nu\rho} - \nabla_{\nu} g_{\rho\beta} \right).
\end{equation}

Using this disformation tensor, the nonmetricity scalar \( Q \) can be expressed as

\begin{equation}
    Q \equiv -g^{\rho\beta} \left( {L^{\mu}}_{\nu\rho} {L^{\nu}}_{\beta\mu} - {L^{\mu}}_{\nu\mu} {L^{\nu}}_{\rho\beta} \right).
\end{equation}

Additionally, the relation \(\nabla_{\lambda} \overset{\circ}{=} \partial_{\lambda}\) leads to \( Q \overset{\circ}{=} -\mathcal{L}_E \), where the symbol '◦' represents gauge equivalence and 

\begin{equation}
    \mathcal{L}_E = g^{\alpha\beta}\left(\left\{{^{\lambda}}_{\mu\alpha}\right\}\left\{{^{\mu}}_{\beta\lambda}\right\} - \left\{{^{\lambda}}_{\mu\lambda}\right\}\left\{{^{\mu}}_{\alpha\beta}\right\}\right)
\end{equation}

is the Lagrangian for the equations of motion originally proposed by Einstein with $\left\{{^{\mu}}_{\alpha\beta}\right\}$ being the Christeffol symbol of the second kind. The nonmetricity tensor \( Q_{\lambda\rho\beta} \) has two important traces, defined as

\begin{equation}
    Q_\lambda = Q^{\;\;\;\rho}_{\lambda\;\;\rho}, \quad \Tilde{Q}_\lambda = {Q^{\rho}}_{\lambda\rho}.
\end{equation}

Nonmetricity leads to a failure in preserving the length of vectors. To maintain length conservation, the conditions \( Q_{(\lambda \mu \nu)} = 0 \) and \( Q_{\lambda (\mu \nu)} = 0 \) are imposed. The introduction of nonmetricity brings about specific geometric effects, resulting in unique differences compared to GR. The way indices change under the covariant derivative is also described differently. The deviation of the anomalous acceleration implies that the four-velocity is no longer orthogonal to the four-acceleration. However, one can recover a scenario equivalent to GR, STGER. The action for STEGR is given by

\begin{equation}\label{eq:chap1:actionQ}
    S_{\text{STEGR}}=\int \left(\frac{1}{2}\, \mathcal{L}_{\text{STEGR}} +\mathcal{L}_m\right)	\sqrt{-g}\, d^4x,
\end{equation}

where \( \mathcal{L}_{\text{STEGR}} \) represents the nonmetricity scalar \( Q \), \( g \) is the determinant of the metric tensor, and \( \mathcal{L}_m \) is the matter Lagrangian. Using the relations

\begin{equation}
    Q = R + \nabla_{\mu} (Q^{\mu} - \Tilde{Q}^{\mu}),
\end{equation}

and 

\begin{equation}
    \nabla_{\mu} (Q^{\mu} - \Tilde{Q}^{\mu}) \equiv \frac{1}{\sqrt{-g}} \partial_{\mu} \left( \sqrt{-g} \left( Q^{\mu} - \Tilde{Q}^{\mu} \right) \right),
\end{equation}
where \( R \) is the Ricci scalar, the STEGR action is dynamically equivalent to GR, with the exception of a boundary term. This boundary term vanishes because the boundary is fixed and the metric variation at the boundary is zero. Importantly, this framework involves second-order field equations, unlike the gravitational field equations in theories that only use the Levi-Civita connection, which is of fourth order.

Further, one can express the nonmetricity conjugate \({P^\gamma} _{\alpha\beta}\) as

\begin{equation}
    {P^\gamma} _{\alpha\beta} = \frac{1}{4}\left[-Q^\gamma_{\;\;\alpha\beta} + 2Q^{\;\;\;\gamma}_{\left(\alpha\;\;\beta\right)} + Q^\gamma g_{\alpha\beta} - \Tilde{Q}^\gamma g_{\alpha\beta} - \delta^\gamma_{(\alpha} Q_{\beta)}\right].
\end{equation}

From the above equation, the nonmetricity scalar takes the form

\begin{equation}
    Q = -Q_{\gamma \alpha \beta} P^{\gamma \alpha \beta}.
\end{equation}

 Varying the action \eqref{eq:chap1:actionQ} with respect to $g_{\alpha\beta}$ yields the following expression for the governing field equation
	   \begin{equation}
		    \frac{-2}{\sqrt{-g}} \nabla_\gamma \left(\sqrt{-g} \, P^\gamma_{\;\;\alpha \beta}\right) - \frac{1}{2} g_{\alpha\beta} Q - Q \left(P_{\alpha \gamma \delta} Q^{\;\;\gamma \delta}_{\beta} - 2Q^{\gamma \delta}_{\;\;\alpha} P_{\gamma \delta \beta}\right) = T_{\alpha\beta}.
		\end{equation}

In the upcoming chapters, we will explore how the kinematics of the universe can be understood under the influence of these modifications.

\newpage
\thispagestyle{empty}

\vspace*{\fill}
\begin{center}
    {\Huge \color{NavyBlue} \textbf{CHAPTER 2}}\\
    \
    \\
    {\Large\color{purple}\textsc{\textbf{Reconstruction of \texorpdfstring{$f(Q)$}{f(Q)} gravity via Newly Parameterized Deceleration Parameter}}}\\
    \ 
    \\
    \textbf{Publication based on this chapter}\\
\end{center}
\textsc{Impact of a Newly Parametrized Deceleration parameter on the Accelerating Universe and the Reconstruction of $f(Q)$ Non-metric Gravity Models}, DM Naik, \textbf{NS Kavya}, L Sudharani, V Venkatesha, \textit{The European Physical Journal C} \textbf{83}, 9, 1-15, 2023 (Springer, Q1, IF - 4.4) DOI: \href{ https://doi.org/10.1140/epjc/s10052-023-12029-1}{10.1140/epjc/s10052-023-12029-1}
\vspace*{\fill}

\pagebreak

\def\baselinestretch{1}
\chapter{\textsc{Reconstruction of \texorpdfstring{$f(Q)$}{f(Q)} gravity via Newly Parameterized Deceleration Parameter}}\label{chap2}
\def\baselinestretch{1.5}
\pagestyle{fancy}
\fancyhead[R]{\textit{Chapter 2}}

\textbf{Highlights:}
{\textit{
\begin{itemize}
    \item The current chapter presents a novel parametrization of the deceleration parameter to investigate the cosmological scenario.
    \item The newly proposed parametric form of the deceleration parameter is both physically plausible and model-independent.
    \item Constrained by a combined dataset of 31 cosmic chronometers (CC) data points, 26 non-correlated baryonic acoustic oscillations (BAO) points, and 1701 Pantheon$+$ data points from supernovae type Ia (SNeIa), we determine the model parameters using a Markov Chain Monte Carlo (MCMC) method.
    \item The analysis explores the kinematic behavior of the model, including the transition from deceleration to acceleration. 
    \item The results indicate that the Universe is currently in an accelerated phase.
    \item We apply the obtained parameter values to constrain $f(Q)$ gravity models and compare them with observations.
\end{itemize}}}

\section{Introduction}\label{sec:chap2:intro}


Recent studies have explored alternative frameworks, modifying GR or introducing new fields, to build on $\Lambda$CDM's success while addressing its limitations. In light of these considerations, a new approach known as "reconstruction" has emerged as a response, whereby observational data is directly incorporated into the process of constructing cosmological models. This approach holds significant promise in enriching our understanding of the Universe and enhancing the accuracy and efficiency of future cosmological surveys.

The reconstruction approach offers the key advantage of being independent of the specific gravity model underlying cosmological studies. It encompasses two methods: non-parametric reconstruction, which involves deriving models directly from observational data through statistical procedures, and parametric reconstruction, which establishes a kinematic model with free parameters and subsequently constrains these parameters through statistical analysis of observational data. The parametric reconstruction approach has been successfully employed in studying the accelerating Universe through various cosmological entities, including the jerk parameter, the deceleration parameter, and the Hubble parameter, shedding light on the behavior of DE. Also, numerous works have focused on parametrizing physical parameters, such as pressure, energy density, and equation of state (EoS) parameters. These parametrizations aid in obtaining precise solutions to the Einstein field equations (see \cite{Roy:2022fif,Akarsu:2013lya,Gong:2006gs,Capozziello:2022jbw,Koussour:2022jyk,Mukherjee:2016trt,Mukherjee:2016eqj,Pantazis:2016nky,Jaime:2018ftn,Nair:2011tg,Sudharani:2023vhv,Sudharani:2023ywc,Koussour:2024xeh}).

Furthermore, the remarkable observations from cosmological surveys, including those by Riess et al. \cite{SupernovaSearchTeam:1998fmf} and Perlmutter et al. \cite{SupernovaCosmologyProject:1998vns}, have confirmed the current acceleration of the Universe's expansion following a period of deceleration characterized by structure formation. The deceleration parameter is a vital quantity in explaining this phenomenon, and parametrizing it offers an appropriate approach compared to other kinematic quantities. Several works have proposed parametrizations of the deceleration parameter \cite{Nair:2011tg,Akarsu:2013lya,Gong:2006gs,delCampo:2012ya,Mamon:2016dlv,Naik:2023gma} (one can see the references therein). In addition, more prominently, the redshift-based deceleration parameter is advantageous as it directly relates to the Universe's expansion rate and facilitates comparisons among different observational datasets.


Additionally, investigating $f(Q)$ theories can provide insights into the cosmic acceleration resulting from different geometries compared to Riemannian geometry. The connection between the disformation tensor and the Levi-Civita connection in $f(Q)$ gravity highlights the intricate interplay between non-metricity and the geometry of spacetime. The current chapter aims to present the consequences and potential applications of $f(Q)$ gravity in deepening our understanding of the fundamental nature of gravity and its effects on the large-scale structure of the Universe. In the literature, one can see several prominent studies on $f(Q)$ gravity \cite{Lazkoz:2019sjl,Capozziello:2022wgl,Capozziello:2022tvv,Esposito:2021ect,Harko:2018gxr}. 

Here, we propose a newly reconstructed form of the deceleration parameter that aligns with both physical considerations and theoretical arguments. This parameterization is independent of the specific gravity model. The free parameters are constrained via statistical analysis of observational data using the Bayesian approach and Markov Chain Monte Carlo (MCMC) analysis. Specifically, we utilize data sets including Cosmic Chronometers (CC), Baryonic Acoustic Oscillations (BAO), and the latest Pantheon$^+$ (SNeIa) dataset. Based on these constraints, we study the kinematics of the model. More interestingly, with these obtained parameters, we constrain $f(Q)$ gravity models and compare the results with observational data. 

This chapter is structured as follows:
The basic field equations in $f(Q)$ gravity are discussed in \autoref{chap2:sectionII}. The newly proposed parameterization and its characteristics are explained in \autoref{chap2:sectionIII}. The results obtained from statistical analysis of the data sets are given in \autoref{chap2:sectionIV}. Further, using the results obtained from the statistical method, we interpret the behavior of the Universe for our newly defined deceleration parameter model in \autoref{chap2:sectionVI}. With the obtained parameter values we constrain $f(Q)$ gravity models in \autoref{chap2:sectionVII}. Finally, we conclude our results in \autoref {chap2:sectionVIII}.

	
\section{Geometric Formulation of \texorpdfstring{$f(Q)$}{f(Q)} Gravity}\label{chap2:sectionII}

Recently, a research paper by J. B. Jimenez et al. \cite{BeltranJimenez:2019tme} introduced a novel approach to gravity known as $f(Q)$ gravity. This theory is distinctive as it is solely defined by the non-metricity $\nabla_\alpha g_{\mu\nu}\neq 0$ of the spacetime, where both curvature and torsion vanish and $\Gamma^\alpha_{\;\;\mu\nu}=0$. 


In $f(Q)$ gravity, the gravitational interactions are described by the following action
\begin{equation}\label{c2:action}
S=\int \left[ -\frac{1}{2}f(Q)+\mathcal{L}_m\right]\sqrt{-g}d^4x.
\end{equation}

The selection of the non-metricity scalar and the action described above in $f(Q)$ gravity is motivated by the desire to reproduce GR in a certain limit  \cite{BeltranJimenez:2019tme}. Specifically, when the function $f$ is chosen as $f=Q$, the action \eqref{c2:action} yields, up to a boundary term STEGR as discussed in \autoref{sub:chap1:STEGR}. This classical correspondence with GR provides a significant motivation for this particular choice. This allows for a seamless transition between the two theories and ensures that $f(Q)$ gravity remains consistent with the well-established framework of GR. The action in equation \eqref{c2:action} incorporates both the gravitational and matter sectors, where the former is described by the function $f(Q)$ and the latter by the matter Lagrangian $\mathcal{L}_m$.




The field equation for $f(Q)$ gravity, which describes the gravitational dynamics in this modified theory, can be obtained by varying the action integral \eqref{c2:action} with respect to the metric tensor $g_{\mu\nu}$. The resulting equation is given by
\begin{equation}\label{fieldequation1}
    \begin{split}
        f_Q(P_{\mu \beta \gamma}Q_{\nu}^{\;\;\beta\gamma}
        -2Q_{\beta\gamma\mu}P^{\beta\gamma}_{\;\;\;\;\;\nu})+\frac{1}{2}g_{\mu\nu}f
        +\frac{2}{\sqrt{-g}}\nabla_\alpha (\sqrt{-g}f_Q P^\alpha_{\;\;\mu\nu})=T_{\mu\nu},
    \end{split}
\end{equation}
where $f_Q$ represents the partial derivative of the function $f$ with respect to $Q$. These field equations play a fundamental role in $f(Q)$ gravity, governing the behavior of the metric tensor and the connection. They establish the relationship between the non-metricity of spacetime, the matter-energy distribution encoded in the energy-momentum tensor $T_{\mu\nu}$, and the functional form of $f(Q)$. In addition to varying the action with respect to the metric tensor, we can also vary it with respect to the connection. This variation yields the equation
\begin{equation}
    \nabla_\mu\nabla_\nu (\sqrt{-g}f_Q P^{\mu\nu}_{\;\;\;\;\;\gamma})=0.
\end{equation}

\section{Flat FLRW Cosmology}

To investigate the evolution of the Universe, it is often useful to make the assumption that the background spacetime is isotropic and homogeneous. This assumption allows us to employ the Friedmann-Lemaitre-Robertson-Walker (FLRW) metric, which describes a homogeneous and isotropic Universe. In the case of an isotropic and homogeneous Universe, we specifically consider the flat FLRW metric, given by
\begin{equation}\label{eq:chap2:flrw}
    ds^2=-dt^2+a^2(t)(dx^2+dy^2+dz^2),
\end{equation}
where $a(t)$ is the scale factor that quantifies the size of the expanding Universe and is related to the Hubble parameter as $H=\frac{\dot{a}}{a}$. Here, the overhead dot indicates derivative with cosmic time $t$, and the same convention is used throughout the article. This framework allows us to study the behavior of cosmological models, analyze the dynamics of matter and energy within the Universe, and investigate phenomena such as cosmic expansion, the age of the Universe, and the behavior of different components like DM and DE. In this case, the non-metricity scalar is given by 
\begin{equation}\label{eq:chap2:nonmetricity}
    Q = 6H^2.
\end{equation}

By employing the FLRW metric, the equations governing the dynamics of the Universe can be derived. These equations are known as the cosmological equations of motion. In this context, for isotropic matter distribution \eqref{eq:chap1:isotropicT}, the cosmological equations of motion are given by
\begin{gather}
    6f_{Q} H^2-\frac{1}{2}f=\rho, \text{and}\label{eq:friedman1}\\
    \dot{H}(12 f_{QQ}H^2+f_{Q})=-\frac{1}{2}(\rho + p).\label{eq:friedman2}
\end{gather}
The variables $\rho$ and $p$ correspond to the energy density and the pressure of the matter fluid, respectively. In the absence of any interaction, the conservation equation $\dot{\rho}+3H(\rho + p)=0$ holds true, ensuring consistency with the aforementioned cosmological equations.  To provide a more familiar form, these cosmological equations can be written in the standard Friedman equations format
\begin{gather}
    3H^2=\rho_{eff},\label{eq:eff1}\\
    2\dot{H}+3H^2=-p_{eff},\label{eq:eff2}
\end{gather}
by defining the effective pressure $p_{eff}$ and effective density $\rho_{eff}$ of the total fluid as
\begin{gather}
    \rho_{eff}=\frac{1}{2f_Q}\left(\frac{f}{2}+\rho \right),\label{eq:rhoeff}\\
    p_{eff}=\frac{1}{f_Q}\left(p-\frac{f}{2}+3H^2(f_Q+8\dot{H} f_{QQ}) \right).\label{eq:peff}
\end{gather}
An important quantity of interest is the effective EoS, denoted as $\omega_{eff}$. It is defined as the ratio of the effective pressure $p_{eff}$ to the effective density $\rho_{eff}$,
\begin{equation}
    \omega_{eff}=\frac{p_{eff}}{\rho_{eff}}=\frac{4p-2f+12H^2(f_Q+8\dot{H}f_{QQ})}{f+2\rho}.
\end{equation}
The value of $\omega_{eff}$ in $f(Q)$ gravity can play a significant role in addressing issues related to DE. It allows for the investigation of the nature of DE and its impact on the expansion of the Universe. It is worth noting that for an accelerating Universe, one requires $\omega_{eff} < -\frac{1}{3}$. These equations provide a comprehensive framework for studying the behavior of the Universe within the context of $f(Q)$ gravity. They offer insights into the effective energy density, pressure, and EoS, which play essential roles in understanding the dynamics and properties of DE.

\section{Newly Proposed Deceleration
Parameter}\label{chap2:sectionIII}
The deceleration parameter plays a fundamental role in characterizing the evolution of the homogeneous and isotropic cosmos, making it a crucial parameter in cosmology. It is directly associated with the second time derivative of the scale factor in the FLRW metric and can be expressed as
\begin{equation}
    q=-\frac{\ddot{a}}{H^2a}=-\left(\frac{\dot{H}}{H^2}+1\right).
\end{equation}
The deceleration parameter provides insight into the rate at which the Universe's expansion is accelerating or decelerating. A positive value of $q$ indicates a decelerating expansion, while a negative value signifies an accelerating expansion. The relationship between the Hubble parameter and the deceleration parameter is given by
\begin{equation}\label{Eq:Hztoq}
    H(z)=H_0 \exp \left(\int_0^z (q(\xi)+1)d\ln (\xi+1)\right),
\end{equation}
where $H_0$ denotes the value of the Hubble parameter at $z=0$, and $z$ represents the redshift, which is related to the scale factor $a$ through $z=-1+\frac{1}{a}$.

The choice of a specific parametric form for the deceleration parameter $q(z)$ allows us to investigate various cosmological models and explore the properties of DE and modified gravity theories. The deceleration parameter plays a crucial role in understanding the dynamics of the Universe, and therefore, several authors have proposed parametrized forms of $q(z)$ based on practical and theoretical considerations, as discussed in \autoref{sec:chap2:intro}. It is important to note that different parametric forms of $q(z)$ have different ranges of applicability. Some parametrizations work well when the redshift $z$ is much smaller than 1 ($z \ll 1$), while others may not accurately predict the future evolution of the Universe.  Ideally, the chosen parametric form should be valid and predictive for all ranges of redshifts, encompassing both the early and late stages of the Universe's evolution. This ensures that the parametrization captures the full dynamics and behavior of DE and provides reliable insights into the expansion history of the Universe.

The parametrization of the deceleration parameter offers several advantages over other kinematic models, making it a valuable tool for studying the dynamics of the Universe. The deceleration parameter provides a direct physical interpretation, allowing us to understand the formation and evolution of cosmic structures. By choosing a suitable form for $q(z)$, we can capture the behavior of cosmic expansion across different epochs, shedding light on the intricate processes involved in the growth of large-scale structures. Crucially, the deceleration parameter must approach $1/2$ at high redshifts to meet the requirements of cosmic structure formation, ensuring that the Universe undergoes the necessary deceleration for structures to develop over cosmic timescales. The adherence of the Universe to the second law of thermodynamics is another important aspect that the parametrization of $q(z)$ can address. The second law imposes constraints on the dynamics of the Universe, and incorporating thermodynamic considerations into the parametrization allows us to satisfy these constraints. One such constraint is that the FLRW Universe approaches thermodynamic equilibrium in the distant future, ensuring the consistency of the cosmological model. By incorporating these thermodynamic aspects, we can refine the parametrization of $q(z)$ and ensure its adherence to the underlying principles of thermodynamics. Moreover, the parametrization of $q(z)$ offers predictive power, enabling us to accurately describe the past and present cosmic expansion and make reliable predictions about the future evolution of the Universe. By carefully selecting an appropriate parametric form, we can extrapolate the behavior of $q(z)$ beyond the range of observed redshifts, providing insights into the long-term fate of the Universe. This predictive capability is particularly valuable in understanding whether the cosmic expansion will continue to accelerate, decelerate, or undergo a transition to a different regime in the future.

Considering the factors discussed above, we propose a new parametric expression for the deceleration parameter $q(z)$ that satisfy several important criteria. Firstly, we ensure that $q(z)$ remains finite for all redshifts within the range $z\in [-1,\infty]$. This condition guarantees that the parametrized form remains well-defined and applicable across a wide range of cosmic epochs. Additionally, we impose the requirement that $q(z)\geq -1$ for all $z\in [-1,\infty]$, $q(z)\to -1$ and $dq/dz>0$ as $z\to -1$
indicating that the FLRW Universe approaches thermodynamic equilibrium in the distant future, as discussed in a previous study by Campo et al. \cite{delCampo:2012ya} and Capozziello et al. \cite{Capozziello:2022jbw}. To ensure an effective and reliable analysis, we focus on just two parameters, motivated by the challenges of higher-dimensional spaces and the potential for degeneracies that hinder parameter determination and increase uncertainties. Exploring high-dimensional spaces becomes computationally demanding and time consuming, while a smaller parameter space enables more efficient sampling and convergence checks. Taking into account these considerations, we introduce a parametrized model for the deceleration parameter $q(z)$ as given by the expression
\begin{equation}\label{Eq:qz-I}
    q(z)=-1+a\left( \frac{(1+z)^3}{z^3+5z^2+b}\right),
\end{equation}
where $a$ and $b$ are model parameters. These parameters control the shape and behavior of the deceleration parameter and allow for customization based on specific cosmological scenarios. By adopting this parametric form for $q(z)$, we meet the requirements of finiteness, positivity of $q(z)+1$, and convergence of the FLRW Universe to thermodynamic equilibrium in the distant future. The next step involves comparing these parametric expressions with observational data to determine the best-fit values for the model parameters and assess their agreement with empirical evidence. This analysis will provide insights into the dynamics and future evolution of the Universe within the framework of this parametrized model.

It is important to note that the cubic polynomial $z^3+5z^2+b$ exhibits distinct root characteristics depending on the value of $b$. When $b>0$, the polynomial will possess one negative root (which is less than -5) and two complex roots. On the other hand, when $b<0$, it can have at most one positive real root. Consequently, when $b>0$, we observe that the equation $z^3+5z^2+b\neq 0$ holds true for any $z\in [-1,\infty)$. In order to ensure the finite nature of the deceleration parameter across all redshifts, it becomes necessary to impose the condition $b>0$ in the parametric form. By adhering to this condition, we guarantee that the deceleration parameter remains free from divergences, thereby upholding the validity of the model throughout the entire range of redshifts.
This provides a solid foundation for using this parametrized model to analyze and interpret observational data on the expansion history of the Universe.

\section{Observational Constraints}\label{chap2:sectionIV}
In order to build a cosmological model that accurately describes the behavior of our Universe, it is crucial to rely on robust observational data and employ appropriate methodologies for parameter estimation. The data utilized in this analysis consists of CC, BAO, and the Pantheon$+$ sample derived from observations of SNeIa. By leveraging these diverse and complementary data sets, we can effectively constrain the model parameters $a$ and $b$, enabling a comprehensive analysis of the Universe's evolution and providing valuable insights into its underlying dynamics and properties (see Appendix for more details regarding datasets and methodology used). 

The results of our analysis are presented in the form of contour plots, which illustrate the joint constraints on the model parameters. These contour plots depict regions in the parameter space that are consistent with the observational data at different confidence levels. In particular, we present the contours up to $3\sigma$ ($99.7\%$) confidence level, indicating the regions where the model is in good agreement with the observed data.

\textbf{Results:}
The dynamics of the Universe can be better understood with the help of cosmological surveys and observational data. In order to derive physically meaningful parameter space for the free parameters, statistical analysis of the observed datasets is performed in the context of parametric reconstruction. In this section, we will use these datasets to constrain the model parameters $a$, $b$, and $H_0$ for the parametrization of deceleration parameter presented earlier. Furthermore, we will attempt to reconstruct the deceleration parameter $q$ by finding the best-fit parameter values. To achieve this, we will use the expression of deceleration parameter \eqref{Eq:qz-I} and numerically compute the Hubble parameter using equation \eqref{Eq:Hztoq}, where $H_0$ is also a free parameter.

\begin{figure}[htbp]
\centering
    \includegraphics[width=0.7\linewidth]{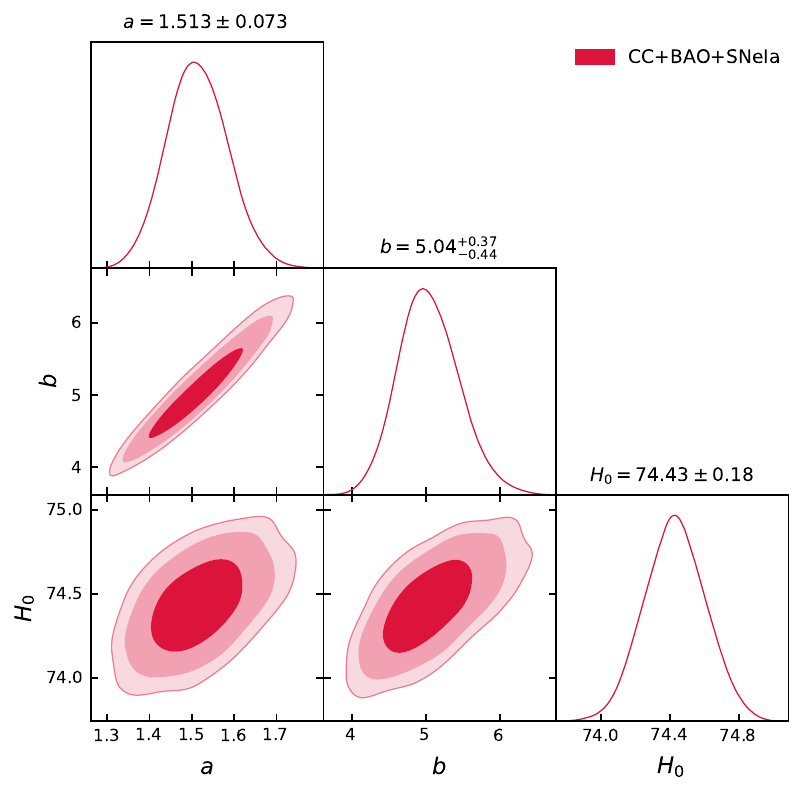}
    \caption{2D-contour plot of the model parameters  $a$, $b$, and $H_0$, indicating the most likely values and confidence regions up to $3\sigma$ obtained from the combined analysis of CC, BAO and SNeIa datasets.}
    \label{fig:param-I}
\end{figure}

The MCMC method is employed for the analysis, and the results are presented in the form of contour plots in \autoref{fig:param-I}. These contour plots illustrate regions in the parameter space that are consistent with the observational data at various confidence levels. Specifically, we present the contours up to $3\sigma$ ($99.7\%$) confidence level, indicating the regions where the model agrees well with the observed data.
Based on our analysis, the mean values of the model parameters $a$, $b$, and $H_0$ are determined to be $1.513^{+0.073}_{-0.073}$, $5.04^{+0.44}_{-0.37}$, and $74.43^{+0.18}_{-0.18}$ (with 1$\sigma$ error), respectively. Notably, the obtained value of $a=1.513^{+0.073}_{-0.073}$ leads to an interesting consequence. At high redshifts, the deceleration parameter approaches nearly $1/2$, which is significant for the cosmic structure formation. This finding suggests that the Universe undergoes the necessary deceleration required for the development of cosmic structures over vast timescales. The agreement between the obtained value of $a$ and the requirements for cosmic structure formation further strengthens the consistency of our model with observational data.

\begin{figure*}[htbp!]
     \centering
     \subfloat[Comparing observed CC and BAO data with theoretical curves for Hubble function.\label{fig:hcf1}]{\includegraphics[width=0.8\linewidth]{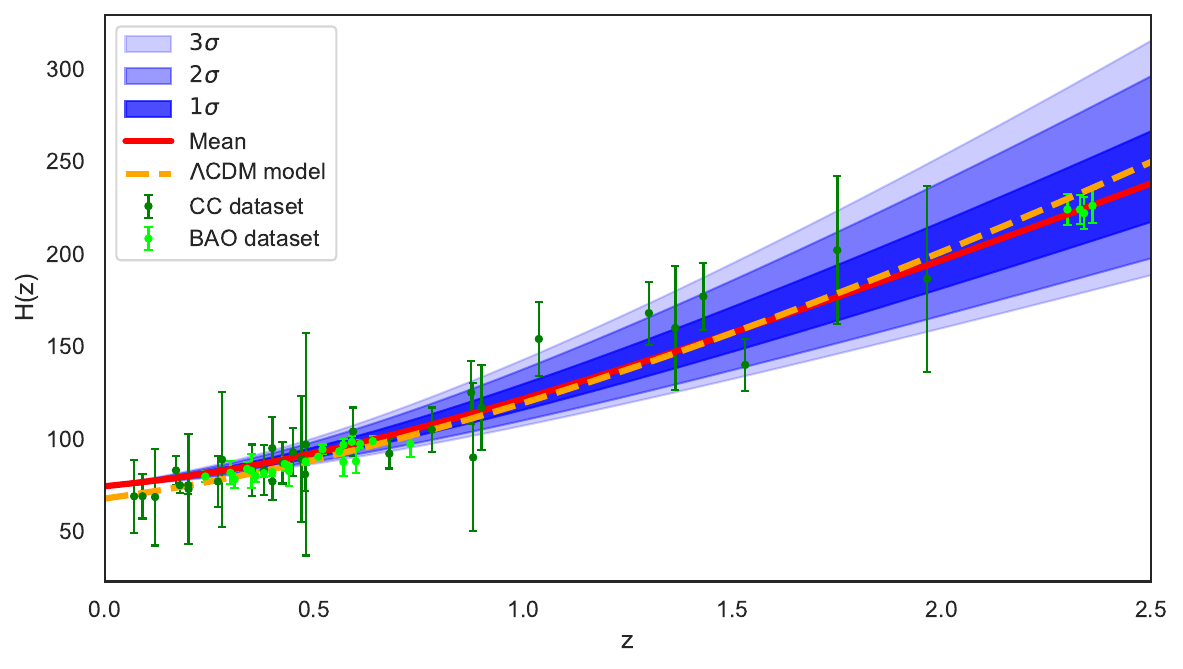}}\\
    \subfloat[Plot of observed distance modulus along with the corresponding error bars and the best-fit theoretical curves for distance modulus function.\label{fig:mucf1}]{\includegraphics[width=0.8\linewidth]{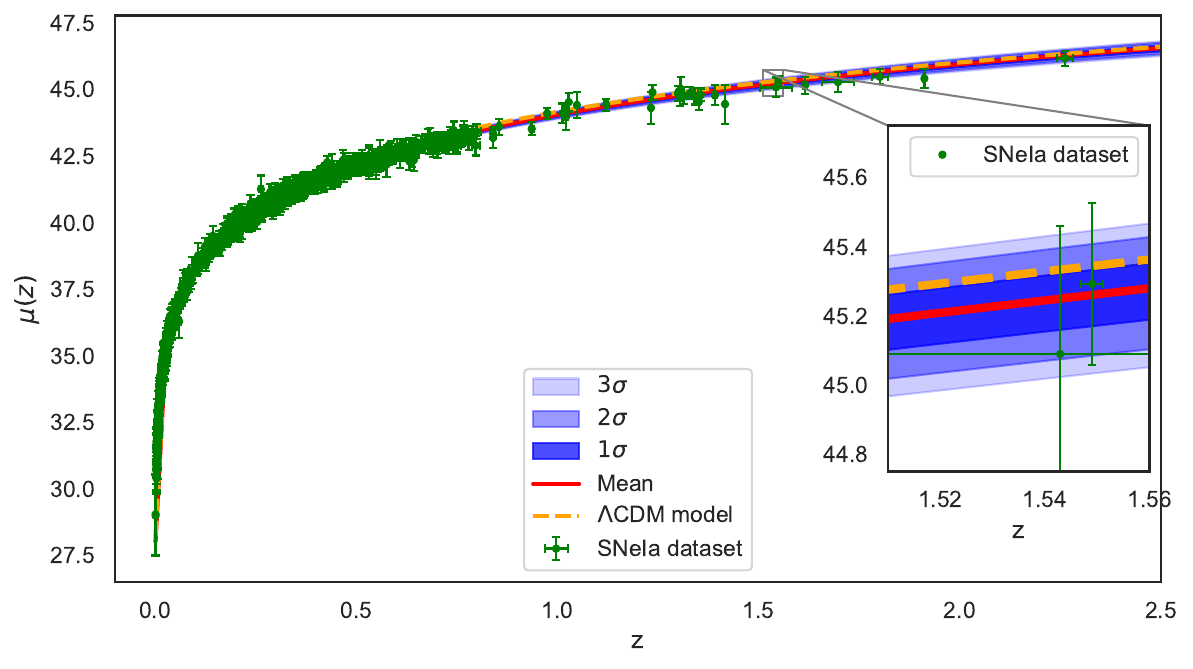}}
     \label{fig:errorplot}
     \caption{Comparison of the obtained best-fit theoretical curves (red line) of the Hubble function $H(z)$ and distance modulus function $\mu(z)$  with their corresponding $1\sigma$, $2\sigma$, and $3\sigma$ error bands (blue shaded regions) against the $\Lambda$CDM model (orange dotted line) with $\Omega_\Lambda=0.7$, $\Omega_m=0.3$, and $H_0=67.8 \,km\;s^{-1}Mpc^{-1}$.}
 \end{figure*}

In Figure~\autoref{fig:hcf1}, we present the observed data on $H(z)$, accompanied by error bars, as well as the best-fit theoretical curves represented by a red line. The shaded regions in blue indicate $1\sigma$, $2\sigma$, and $3\sigma$ error bands of the Hubble function $H(z)$. The agreement between the model predictions and the observed data is evident from the consistency of the error bars with the shaded regions. This visual representation confirms the accuracy of our model in capturing the observed behavior of the Hubble function.
Furthermore, in Figure~\autoref{fig:mucf1}, we display the error plot of the distance modulus. The observed distance modulus of the 1701 SNeIa dataset is depicted, along with the best-fit theoretical curves of the distance modulus function $\mu(z)$ shown as a red line. The blue-shaded regions correspond to the error bands at a confidence level of up to 99.7\%. The consistency observed in these plots further strengthens the confidence in the reliability of our results.

In addition to analyzing the model parameters, we have successfully constrained the present value of the Hubble parameter $H_0$. Utilizing the combined CC+BAO+SNeIa dataset, we determine the value of $H_0$ to be $74.43^{+0.18}_{-0.18}$ $km\,s^{-1}Mpc^{-1}$, accompanied by a 1-$\sigma$ error. The results of this constrained $H_0$ value are presented in \autoref{fig:comph}, where we compare our findings with the outcomes of previous studies.

\begin{table*}[t]
    \caption{Results of MCMC for parameters $a$, $b$ and $H_0$ (km/s/Mpc) with $1\sigma$-$3\sigma$ errors}
    \label{tab:tab_1}
     \setstretch{1.4}
	\centering
	\begin{tabularx}{\linewidth}{ >{\centering\arraybackslash}X >{\centering\arraybackslash}X >{\centering\arraybackslash}X >{\centering\arraybackslash}X}
        \hline
        \hline
		\multicolumn{1}{c}{Confidence Level} &  $a$ & $b$ & $H_0$\\ 
		\hline
              $1\sigma$ &$1.513^{+0.073}_{-0.073}$ & $5.04^{+0.37}_{-0.44}$ & $74.43^{+0.18}_{-0.18}$ \\
             $2\sigma$ & $1.51^{+0.15}_{-0.14}$ & $5.04^{+0.87}_{-0.77}$ & $74.43^{+0.35}_{-0.34}$ \\
             $3\sigma$ &$1.51^{+0.19}_{-0.17}$ & $5.04^{+1.2}_{-0.96}$ & $74.43^{+0.45}_{-0.46}$ \\
            \hline
            \hline
	\end{tabularx}
    \end{table*}

\section{ Dynamics of the Model}\label{chap2:sectionVI}
The dynamics of a cosmological model provide valuable insights into the behavior and evolution of the Universe. In this section, we explore the dynamics of our model through three key aspects: the transition from a deceleration to an acceleration phase, the analysis of the jerk parameter, and the examination of the Om diagnostics. Each of these subsections sheds light on different aspects of the model's behavior and helps us better understand the underlying dynamics of the Universe. By studying these aspects, we gain a deeper understanding of the fundamental processes that govern the expansion and evolution of our Universe.

\subsection{Transition from Deceleration to Acceleration Phase}
The transition from a deceleration to an acceleration phase holds significant importance in understanding the dynamics of the Universe. Prior to this transition, the Universe was characterized by a decelerating expansion, driven by the gravitational interaction between matter and radiation. However, as the cosmos continued to expand and matter became more dispersed, the gravitational force gradually weakened, leading to a shift in the cosmic dynamics. This transition marks a turning point in the evolution of the Universe, as it entered a phase of cosmic acceleration. The current accelerating action of the Universe can be quantified by estimating negative values of the deceleration parameter. Exploring this transition is essential for comprehending the underlying mechanisms driving the expansion and evolution of the cosmos. By studying the transition from deceleration to acceleration, we gain valuable insights into the dynamic nature of our Universe.

\begin{figure}[htbp!]
    \centering
    \label{fig:D-I}{\includegraphics[width=0.8\linewidth]{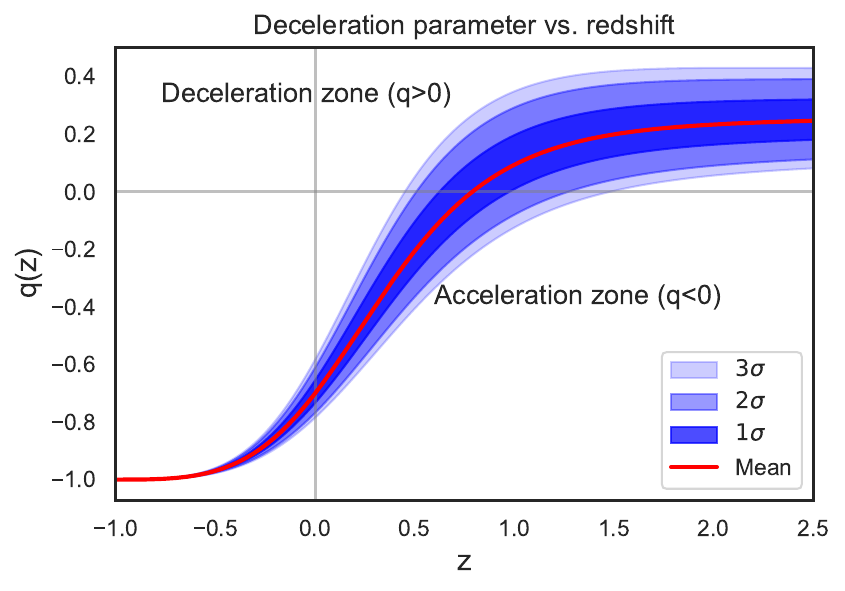}}
    \caption{Deceleration parameters as a function of redshift for proposed parametric  model, obtained from  the combined CC, BAO, and SNeIa datasets, with shaded zones representing $68\%$, $95\%$, and $99.7\%$ confidence levels.}
    \label{fig:dcp}
\end{figure}

In \autoref{fig:dcp}, we present the results, showcasing three deceleration parameter curves accompanied by their corresponding $1\sigma$, $2\sigma$, and $3\sigma$ error bands. The displayed curves clearly depict a transition of the Universe from a decelerating phase to an accelerating phase. Notably, we observe that the occurrence of this phase transition is influenced by the variations in the model parameters $a$ and $b$. The specific redshift value at which the transition takes place is determined as $z_t = 0.789^{+0.186}_{-0.165}$, based on the constrained parameters derived from the combined CC+BAO+SNeIa dataset, with a 1-$\sigma$ error. Remarkably, these values align with those reported by several other researchers in diverse scenarios \cite{delCampo:2012ya,Mamon:2016dlv,AlMamon:2015ali}. Furthermore, the current estimate of the deceleration parameter is $q_0=-0.7^{+0.045}_{-0.033}$, with a 1-$\sigma$ error. These findings consistently agree with values previously reported in the literature \cite{Lu:2011ue,Banerjee:2005ef}. To facilitate reference, these values are compiled in \autoref{tab:tab_2}. Further, we present the obtained $q_0$ in \autoref{fig:compq} and compare them with the results of previous studies.

\subsection{Jerk Parameter}
The jerk parameter (j) serves as a significant quantity in our understanding of the accelerating Universe. It is defined as the dimensionless third-order derivative of the cosmic scale factor $a(t)$. Furthermore, we can express the jerk parameter as a function of redshift $z(t)$ using the deceleration parameter $q(z)$, which can be given by the equation
\begin{equation}\label{eq:jerk}
j(z)=q(z)(2q(z)+1)+\frac{dq}{dz}(1+z).
\end{equation}
The jerk parameter plays a crucial role in discerning various DE models \cite{BOSS:2016wmc}, as deviations from the value of $j=1$ would favor non-$\Lambda$CDM models. For our specific model, the expression for the jerk parameter can be derived from the expression for the deceleration parameter. It takes the form
\begin{equation}\label{j(z)}
   j(z)= \frac{2 a^2 (z+1)^6-a z (3 z+10) (z+1)^4}{\left(b+(z+5) z^2\right)^2}+1.
\end{equation}
The equation provides insights into the behavior of the jerk parameter within the framework of our model.

\begin{figure}[htbp]
    \centering
    \label{fig:Jerk1}{\includegraphics[width=0.8\linewidth]{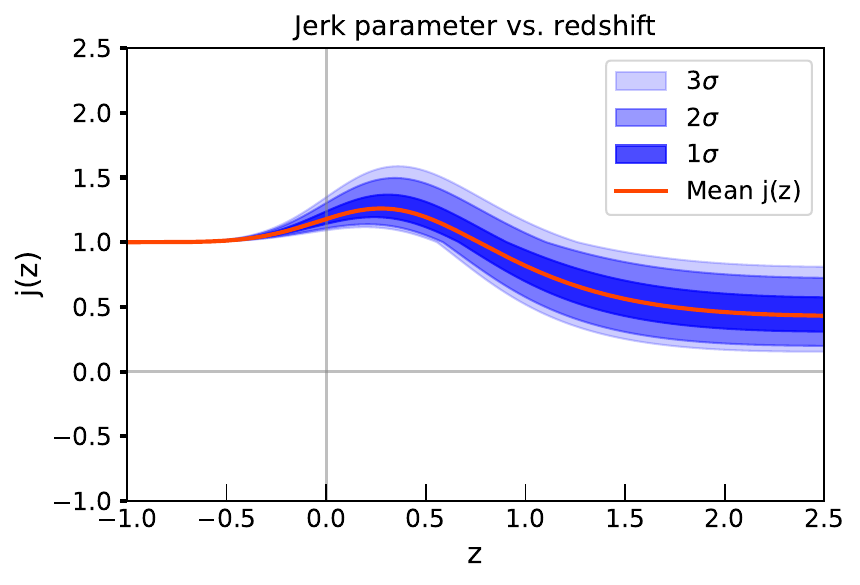}}
    \caption{Jerk parameters as a function of redshift for the proposed model, obtained from  the combined CC, BAO, and SNeIa datasets, with shaded zones representing $68\%$, $95\%$, and $99.7\%$ confidence levels.}
    \label{fig:Jerk}
\end{figure}

\autoref{fig:Jerk} presents the evolution of the jerk parameter, $j(z)$, within the $3\sigma$ error regions for the combined CC+BAO+SNeIa dataset. The results depicted in \autoref{fig:Jerk} indicates that our model only exhibits slight deviations from the concordance $\Lambda$CDM model at the present epoch. These deviations, observed in the value of $j_0$, prompt further investigation as the underlying cause of cosmic acceleration remains unknown. Furthermore, our analysis demonstrates that the future behavior of our model is marginally consistent with the $\Lambda$CDM model. The current value of the jerk parameter, along with its corresponding $1\sigma$ error, is estimated as $j_0=1.18^{+0.058}_{-0.038}$, which is consistent with the results obtained in \cite{Mamon:2016dlv,Mehrabi:2021cob}. These findings, presented in \autoref{tab:tab_2}, provide validation for our model and lend support to the notion of an accelerating Universe.

\subsection{\texorpdfstring{$Om(z)$}{Om(z)} Diagnostic}
The $Om(z)$ diagnostic is an effective tool in differentiating between different DE or cosmological models from the standard $\Lambda$CDM model. Sahni et al. introduced this diagnostic in 2008 \cite{Sahni:2008xx}, and since it has been studied extensively by numerous researchers. The function $Om(z)$ relates the observed Hubble parameter, which is a measure of the rate of expansion of the Universe, to the density of matter in the Universe. A constant value of $Om(z)$ at any redshift indicates that the DE behaves like a cosmological constant. However, if $Om(z)$ varies with redshift, it suggests that the DE is dynamic and changes its form over time. Furthermore, the slope of $Om(z)$ can distinguish between two distinct types of dynamic DE models: quintessence and phantom. A positive slope in $Om(z)$ implies a phantom phase, while a negative slope implies a quintessence phase. In general, the $Om(z)$ diagnostic is an influential tool for studying various DE and cosmological models and provides important insights into the expansion of the Universe.

\begin{figure}[htbp!]
    \centering
    \includegraphics[scale=0.6]{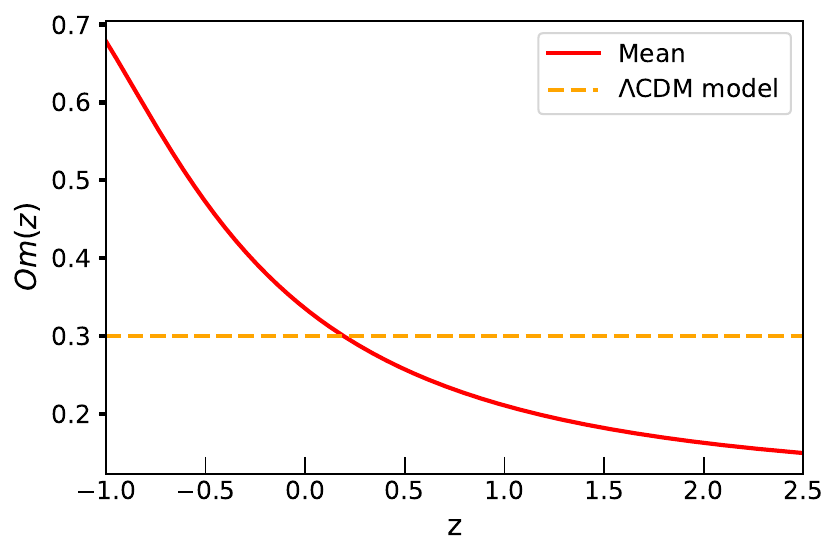}
    \caption{The behavior of the $Om(z)$ diagnostic vs. redshift $z.$}
    \label{fig:Om}
\end{figure}

In a Universe with flat spatial geometry, the Om(z) diagnostic can be expressed as 
\begin{equation}\label{Eq:Om}
    Om(z)=\frac{E^2(z)-1}{(1+z)^3-1},
\end{equation}
where $E(z)=H(z)/H_0$ and $H(z)$ is obtained numerically using the equation \eqref{Eq:Hztoq}. By utilizing the mean values of the constrained parameters derived from the combined CC+BAO+SNeIa dataset, we plot the evolution of $Om(z)$ with respect to $z$ in \autoref{fig:Om}. The resulting graph illustrates that $Om(z)$ exhibits a negative slope for all redshift ranges, indicating a quintessence phase. This suggests that our model displays distinct behavior compared to the standard $\Lambda$CDM model.

    \begin{table*}[!]
    \caption{Summary of the results for deceleration parameter, effective EoS, $z_t$, and jerk parameter obtained from constrained values of model parameter with $1\sigma$-$3\sigma$ errors}
    \label{tab:tab_2}
     \setstretch{1.4}
	\centering
	\begin{tabularx}{\linewidth}{>{\centering\arraybackslash}X >{\centering\arraybackslash}X >{\centering\arraybackslash}X >{\centering\arraybackslash}X >{\centering\arraybackslash}X}
        \hline
        \hline
		\multicolumn{1}{c} {Confidence Level} &  $q_0$ & $\omega^{eff}_0$ & $z_t$ & $j_0$\\ 
		\hline
              $1\sigma$ &$-0.7^{+0.045}_{-0.033}$ & $-0.8^{+0.03}_{-0.085}$ & $0.789^{+0.186}_{-0.165}$ & $1.18^{+0.058}_{-0.038}$ \\
             $2\sigma$ & $-0.7^{+0.089}_{-0.068}$ & $-0.8^{+0.059}_{-0.045}$ & $0.789^{+0.481}_{-0.281}$ & $1.18^{+0.123}_{-0.072}$ \\
             $3\sigma$ &$-0.7^{+0.117}_{-0.085}$ & $-0.8^{+0.078}_{-0.057}$ & $0.789^{+0.690}_{-0.338}$& $1.18^{+0.168}_{-0.087}$ \\
            \hline
            \hline
	\end{tabularx}
    \end{table*}

 \section{Constraints on \texorpdfstring{$f(Q)$}{f(Q)} Models}\label{chap2:sectionVII}
The exploration of alternative theories of gravity, such as $f(Q)$ gravity models, provides valuable insights into the fundamental nature of the Universe. In this section, we aim to constrain the parameters of two specific $f(Q)$ gravity models using the observed values of the deceleration parameter $q_0$ and the Hubble constant $H_0$. The first model we consider is the power-law form of $f(Q)$ gravity and the second model is the logarithmic form of $f(Q)$ gravity. To derive constraints on these models, we utilize the generalized Friedmann equations in the present-day values, taking into account the contribution of pressureless matter while neglecting the influence of radiation. By analyzing the available observational data, we aim to determine the parameter values that best describe the dynamics of the Universe within the framework of $f(Q)$ gravity models. Thus, \eqref{eq:friedman1} and \eqref{eq:friedman2} take the forms
\begin{gather}
    6f_{Q} H_0^2-\frac{1}{2}f_0=\rho_0,\label{Eq:present-01}\\
    (12 f_{QQ_0}H_0^2+f_{Q_0})\dot{H_0}=-\frac{1}{2}\rho_0,\label{Eq:present-02}
\end{gather}
where a subscript $(_0)$ indicates the present-day value of the corresponding parameter. We make use of \eqref{Eq:present-01} and \eqref{Eq:present-02} to set constraints on the parameter of $f(Q)$ model. 

\subsection{Power-law Form of \texorpdfstring{$f(Q)$}{f(Q)} Gravity Model}
In this subsection, we focus on the power-law form of the $f(Q)$ gravity model, given by the equation
\begin{equation} \label{Eq:Model-01}
    f(Q)=Q+\alpha Q_0\left(\frac{Q}{Q_0}\right)^\beta,
\end{equation}
where $\alpha$ and $\beta$ are scalars. Though there are models like $f(Q)=Q+\alpha Q^\beta$ in the literature \cite{Mandal:2020lyq,Capozziello:2022tvv}, we chose the above specific form because it allows us to express the parameters $\alpha$ and $\beta$ as dimensionless quantities. By substituting the equation \eqref{Eq:Model-01} into the equations  \eqref{Eq:present-01} and \eqref{Eq:present-02}, we obtain
\begin{gather}
    \alpha  (2 \beta -1)-\Omega_{m_0}+1=0, \text{and}\label{Eq1:Model_01}\\
    2 (q_0+1) (\alpha  \beta  (2 \beta -1)+1)-3 \Omega_{m_0}=0, \text{respectively,}\label{Eq2:Model_01}
\end{gather}
where $\Omega_{m_0}$ is the dimensionless matter density parameter.  
By substituting the values $q_0=-0.7$, $H_0=74.43$ obtained in this study along with $\Omega_{m_0}=0.315$ \cite{Planck:2018vyg}, into Equations~\eqref{Eq1:Model_01} and \eqref{Eq2:Model_01}, we find the corresponding parameter values for the power-law $f(Q)$ gravity model as $\alpha =0.255708$ and $\beta =-0.839416$. Alternatively, considering the observational constraints from Planck2018 \cite{Planck:2018vyg} results, with values $q_0=-0.5275$, $H_0=67.4$, and $\Omega_{m_0}=0.315$, we obtain parameter values of $\alpha = 0.685$ and $\beta=3.09284\times 10^{-16}$. Similarly, by using the observational constraints from the SH0ES team \cite{Riess:2021jrx} with values $q_0=-0.51$, $H_0=73.3$, and $\Omega_{m_0}=0.326$, we find parameter values of $\alpha =0.678107$ and $\beta =0.00302792$. These parameter values are summarized in \autoref{tab:tab_fQ}.

\begin{figure}[htbp!]
    \centering
    {\includegraphics[width=0.8\linewidth]{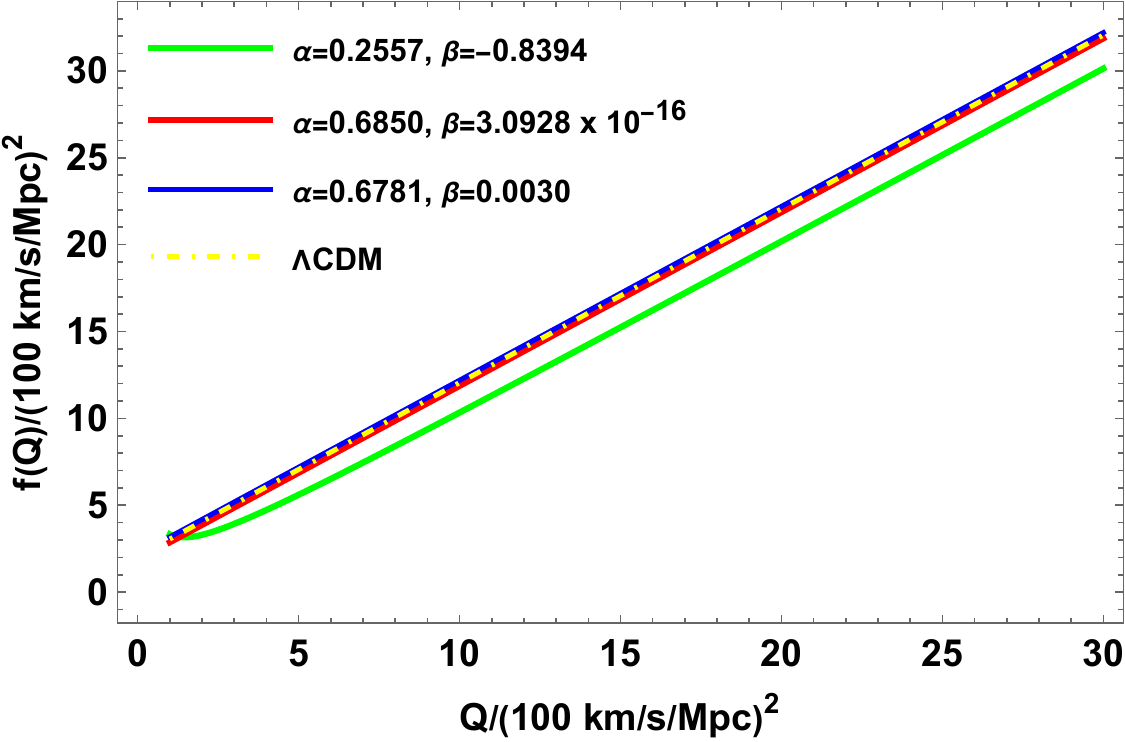}}
    \caption{Comparison between the reconstructed $f(Q)=Q+\alpha Q_0\left(\frac{Q}{Q_0}\right)^\beta$ and $\Lambda$CDM model.}
    \label{fig:fQ1}
\end{figure}

\autoref{fig:fQ1} displays the reconstructed $f(Q)$ functions for the power-law $f(Q)$ gravity model in comparison to the $\Lambda$CDM model with $\Omega_{m_0}=0.3$ and $H_0=70$. The $f(Q)$ curves corresponding to the parameter sets $\alpha=0.685$, $\beta=3.09284\times 10^{-16}$ and $\alpha=0.678107$, $\beta=0.00302792$ closely align with the $\Lambda$CDM model. These curves exhibit a strong similarity to the standard cosmological scenario. However, the $f(Q)$ curve corresponding to $\alpha =0.255708$ and $\beta =-0.839416$ shows a small deviation from the $\Lambda$CDM model. While still resembling the standard cosmological scenario, this particular parameter set exhibits a slight departure from the expected behavior.

\begin{table*}
    \caption{Summary of the Findings for Constraints on $f(Q)$ Gravity Models}
    \label{tab:tab_fQ}
     \setstretch{1.4}
	\centering
     \begin{tabular}{cl cl cl cl}
     &\multicolumn{2}{c}{\textbf{Power-law $f(Q)$ gravity}} & \multicolumn{2}{c}{\textbf{Logarithmic $f(Q)$ gravity}}\\
     \cmidrule{2-5}\addlinespace[0pt]\cmidrule{2-5}
		 & $\alpha$ & $\beta$ &  $\alpha$ & $\beta$\\ 
		\cmidrule{2-5}
              $Present\; work$ &$0.255708$ & $-0.839416$ & $1.155$ & $-0.42$\\
             $Planck2018$ & $0.685$ & $3.09284\times 10^{-16}$ &$0.771667$ &$-0.228333$ \\
             $ SH0ES\; team$ &$0.678107$ & $0.00302792$ &$0.773973$ & $-0.223986$\\
             \cmidrule{2-5} \addlinespace[0pt]\cmidrule{2-5}
	\end{tabular}
\end{table*}

\subsection{Logarithmic Form of \texorpdfstring{$f(Q)$}{f(Q)} Gravity Model}
As our second specific example, we examine the $f(Q)$ gravity model expressed as
\begin{equation}
    f(Q)=\alpha Q+\beta Q_0\log(Q/Q_0).
\end{equation}
A similar kind of model has also been explored in recent studies, such as the work by  \cite{Capozziello:2022tvv}. By substituting this model into the equations \eqref{Eq:present-01} and \eqref{Eq:present-02}, we obtain the following 
\begin{gather}
    \alpha +2 \beta -\Omega_{m_0}=0, \label{Eq1:Model_03}\\
    2 (q_0+1) (\alpha -\beta )-3 \Omega_{m_0}=0.\label{Eq2:Model_03}
\end{gather}

By employing the values $q_0=-0.7$ and $H_0=74.43$ obtained in this study, along with $\Omega_{m_0}=0.315$, we derive the parameter values for the logarithmic $f(Q)$ gravity model as $\alpha =1.155$ and $\beta =-0.42$. On the other hand, when considering the values $q_0=-0.5275$, $H_0=67.4$, and $\Omega_{m_0}=0.315$ based on the observational constraints from Planck Collaboration \cite{Planck:2018vyg}, the resulting parameter values are $\alpha = 0.771667$ and $\beta=-0.228333$. Additionally, adopting the values $q_0=-0.51$, $H_0=73.3$, and $\Omega_{m_0}=0.326$ following the observational constraints by the SH0ES team \cite{Riess:2021jrx}, yields the parameter values of $\alpha =0.773973$ and $\beta =-0.223986$. These parameter values are summarized in \autoref{tab:tab_fQ}.

\begin{figure}[htbp!]
    \centering
    {\includegraphics[width=0.8\linewidth]{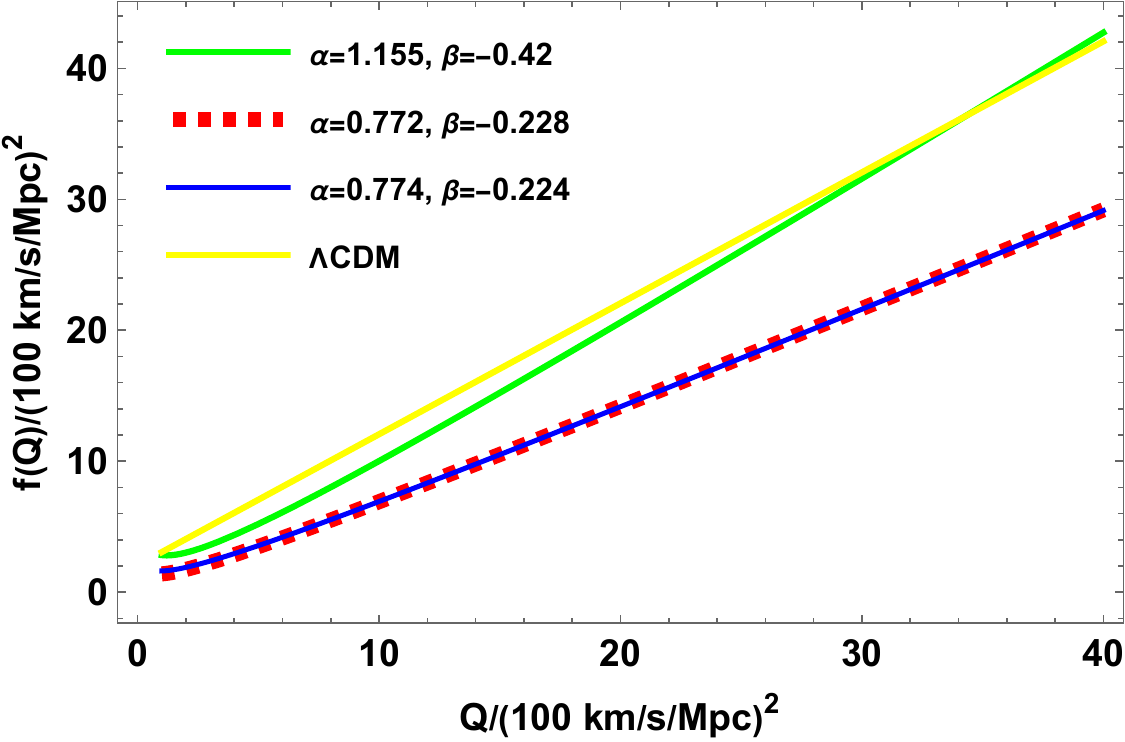}}
    \caption{Graph of $f(Q)=\alpha Q+\beta Q_0\log(Q/Q_0)$  against $\Lambda$CDM model.}
    \label{fig:fQ2}
\end{figure}

\autoref{fig:fQ2} depicts the reconstructed $f(Q)$ function for the logarithmic form of the $f(Q)$ gravity model, compared to the $\Lambda$CDM model with $\Omega_{m_0}=0.3$ and $H_0=70$. The plot showcases the minor deviations of the logarithmic form of the $f(Q)$ model from the standard cosmological scenario. Notably, the $f(Q)$ curve corresponding to $\alpha =1.155$ and $\beta =-0.42$ closely resembles the $\Lambda$CDM model. However, the $f(Q)$ curves associated with the other two parameter sets, $\alpha = 0.771667$, $\beta=-0.228333$, and $\alpha =0.773973$, $\beta =-0.223986$, exhibit increasing deviations from the $\Lambda$CDM model as the redshift increases. These findings underscore the capability of the logarithmic form of the $f(Q)$ gravity model to reproduce the standard cosmological scenario while also manifesting deviations from it within specific parameter regimes.

Overall, our analysis of both the power-law and logarithmic forms of $f(Q)$ gravity models reveals their ability to reproduce the standard cosmological scenario while also exhibiting departures from it in specific parameter regimes. The derived parameter constraints and the corresponding reconstructed $f(Q)$ functions provide insights into the behavior of these models and their compatibility with observational data. These findings contribute to our understanding of modified gravity theories and their implications for cosmology.

\section{Concluding Remarks}\label{chap2:sectionVIII}
The investigation of cosmic evolution through reconstruction methods has shed light on the dynamics of the Universe, providing valuable insights into its expansion history and the nature of DE and modified gravity. Building upon the advancements in parametric reconstruction and considering the importance of a suitable deceleration parameter model, this study has made significant contributions to our understanding of the cosmological scenario. The choice of a parametric form for $q(z)$ is a critical step in cosmological analysis. It must be done thoughtfully, considering both observational constraints and theoretical considerations, to ensure that the parametrization is valid and predictive across all ranges of redshifts. This ensures robust investigations of the Universe's dynamics and facilitates the exploration of DE and modified gravity theories. In this chapter, we introduced a new parametrization for the deceleration parameter, offering a more flexible and model-independent approach to studying the dynamics of the Universe.  The proposed deceleration parameter model in this article is capable of explaining several physical phenomena, such as satisfying the second law of thermodynamics and describing the entropy of the Universe.

Here, the utilization of observational probes including 31 data sets of CC, 26 non-correlated points from BAO, and 1701 newly updated SNeIa data points has enabled the constraint of cosmological parameters and validation of the viability of the proposed model. The Bayesian statistical inference techniques and MCMC methods used to constrain the model parameters have allowed us to accurately analyze the data and derive meaningful conclusions. The best fits (see \autoref{tab:tab_1}) obtained from this procedure are used to analyze the kinematic behavior of the Universe. Our findings demonstrate that the best-fit parameters of our models align well with the observed data. This success in constraining the model's parameters showcases the importance of utilizing observational data and statistical methods to improve our understanding of the Universe's behavior.

\begin{figure}[htbp]
\centering
    \includegraphics[width=0.5\linewidth]{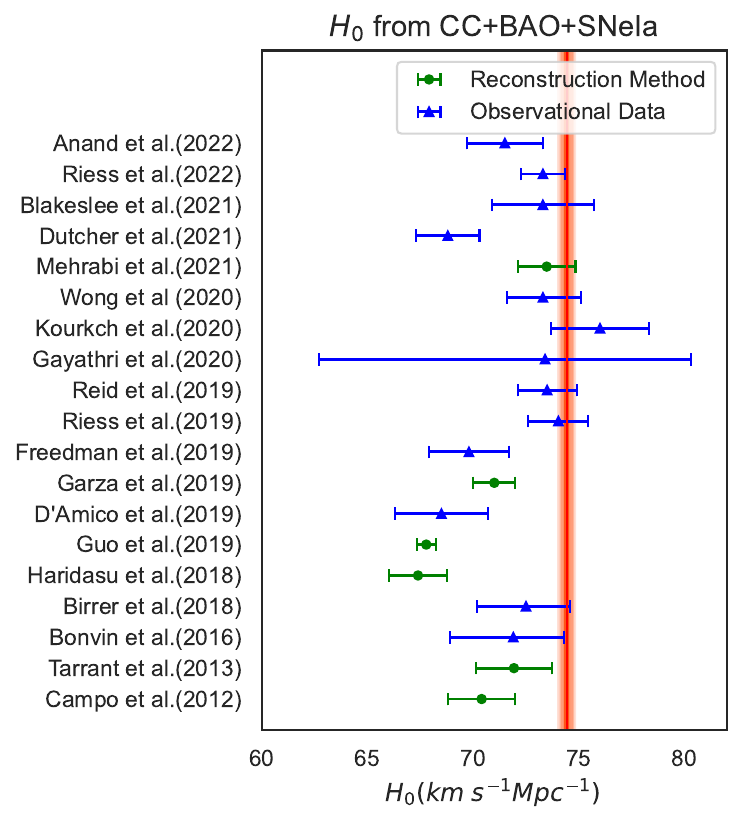}
    \caption{A plot illustrating the current values of $H_0$ along with their corresponding error bars obtained from various studies is presented. The orange-red shaded strips represent the $H_0$ value with $1\sigma$, $2\sigma$, and $3\sigma$ errors obtained in this work.}
    \label{fig:comph}
\end{figure}

\begin{figure}[htbp]
\centering
    \includegraphics[width=0.5\linewidth]{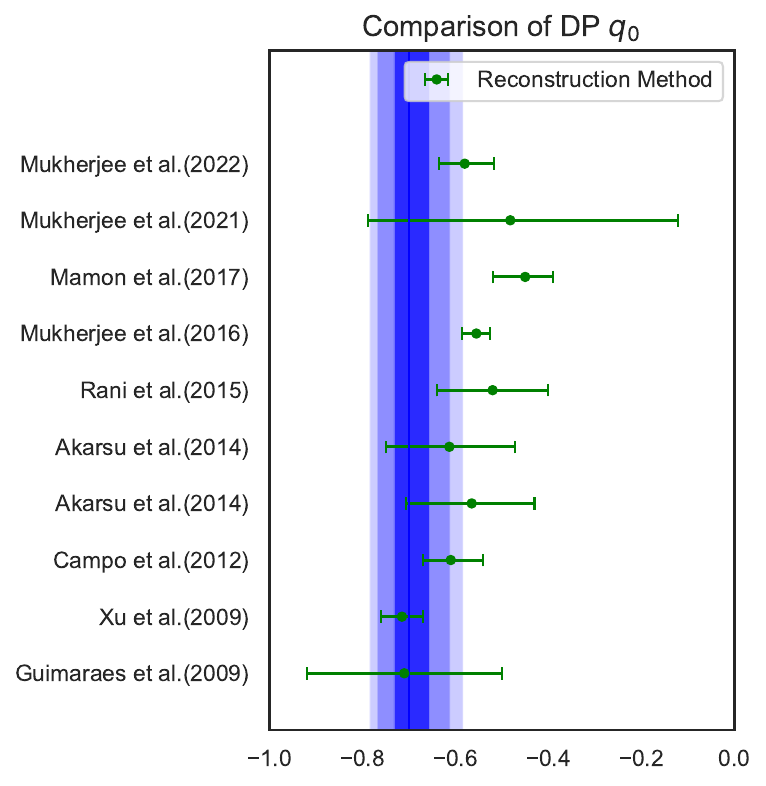}
    \caption{A graph displaying the present values of deceleration parameter, accompanied by their corresponding error bars, as obtained from several studies is presented \cite{Mukherjee:2016trt,delCampo:2012ya,Akarsu:2013lya,Mamon:2016dlv,Mukherjee:2020vkx,Mukherjee:2020ytg,Guimaraes:2009mp,Rani:2015lia,Xu:2009ct}. The blue shaded regions on the graph indicate the $q_0$ value, along with their $1\sigma$, $2\sigma$, and $3\sigma$ errors obtained in this study.}
    \label{fig:compq}
\end{figure}

Through the investigation of the proposed model's dynamics, which includes examining the transition from deceleration to acceleration, analyzing the Jerk Parameter $(j)$, and utilizing the $Om(z)$ diagnostic, we gain a better understanding of the Universe's behavior and its evolutionary processes.  The results of the cosmological quantities for the statistically estimated values of model parameters are in \autoref{tab:tab_2}.  The current deceleration parameter value, up to $3\sigma$ error, is presented and compared with earlier studies in \autoref{fig:compq}. Moreover, our study effectively constrained both the model parameters and the Hubble constant $H_0$. Notably, our obtained Hubble constant values align with those derived from other studies that employed reconstruction methods (both parametric and non-parametric) \cite{Haridasu:2018gqm,Roman-Garza:2018cxf,Mehrabi:2021cob,delCampo:2012ya,Tarrant:2013xka,Guo:2018ans} and observational data (including the distance ladder method, Tip of the Red Giant Branch (TRGB) technique, and H0LiCOW) \cite{Riess:2019cxk,Freedman:2019jwv,Bonvin:2016crt,Birrer:2018vtm,Gayathri:2020mra,DAmico:2019fhj,SPT-3G:2021eoc,Blakeslee:2021rqi,Kourkchi:2020iyz,Reid:2019tiq,Wong:2019kwg,Riess:2021jrx,Anand:2021sum}. We present our findings of the constrained $H_0$ in \autoref{fig:comph} and compare them with the results of previous studies.

By employing a careful analysis of the observational constraints, we obtained substantial findings into the behavior of the power-law and logarithmic $f(Q)$ gravity models.
We presented the parameter values of $f(Q)$ gravity models for three different sets of observational constraints: one obtained in the present work, one following the Planck2018 observational constraints, and one based on the SH0ES team's observational constraints. We summarized these parameter values in \autoref{tab:tab_fQ} and demonstrated their impact on the reconstructed $f(Q)$ function. Notably, we found that the power-law $f(Q)$ model closely resembles the $\Lambda$CDM model for certain parameter sets, while others exhibit increasing deviations as redshift increases.

Overall, this chapter offers valuable cosmological perspectives utilizing observational data and a novel deceleration parameter, providing meaningful astrophysical insights. These findings will aid in the continued exploration of the nature of DE and the future trajectory of the cosmos. Further, we reconstructed $f(Q)$ gravity models and their implications for the nature of the Universe.

\newpage
\thispagestyle{empty}

\vspace*{\fill}
\begin{center}
    {\Huge \color{NavyBlue} \textbf{CHAPTER 3}}\\
    \
    \\
    {\Large\color{purple}\textsc{\textbf{Embedding \texorpdfstring{$\Lambda$}{Lambda}CDM Universe in \texorpdfstring{$f(Q)$}{f(Q)} Gravity with Matter Coupling}}}\\
    \ 
    \\
    \textbf{Publication based on this chapter}\\
\end{center}
\textsc{Embedding the $\Lambda$CDM Framework in Non-minimal $f(Q)$ Gravity with Matter-Coupling}, \textbf{NS Kavya}, V Venkatesha, \textit{Physics Letters B} \textbf{856}, 138927, 2024 (Elsevier, Q1, IF – 4.3)  DOI: \href{https://doi.org/10.1016/j.physletb.2024.138927}{10.1016/j.physletb.2024.138927}
\vspace*{\fill}

\pagebreak

\def\baselinestretch{1}
\chapter{\textsc{Embedding \texorpdfstring{$\Lambda$}{Lambda}CDM Universe in \texorpdfstring{$f(Q)$}{f(Q)} Gravity with Matter Coupling}}\label{chap3}
\def\baselinestretch{1.5}
\pagestyle{fancy}
\fancyhead[R]{\textit{Chapter 3}}

\textbf{Highlights:}
{\textit{
\begin{itemize}
    \item The $\Lambda$CDM model has served as a foundational description for understanding the Universe, but it also has its limitations. 
    \item The primary objective of this chapter is to overcome these shortcomings and reconstruct the $\Lambda$CDM Universe. 
    \item We aim to achieve this by embedding the $\Lambda$CDM scenario within a modified gravity framework. 
    \item Here, the investigation is conducted within the framework of novel non-minimal coupling of $f(Q)$ gravity. A generic non-minimal form is considered, expressed in terms of two arbitrary functions $f_1(Q)$ and $f_2(Q)$. 
    \item Analytic solutions are derived to mimic the $\Lambda$CDM paradigm. 
    \item Finally, through cosmographic analysis, the models are validated against observational results. 
\end{itemize}}}

\section{Introduction}

A prominent challenge in contemporary cosmology is the need for a profound explanation of the origin and existence of the dark components of the Universe, which are accountable for the observed acceleration of the Universe supported by recent observations \cite{SupernovaCosmologyProject:1997czu,SupernovaSearchTeam:1998fmf}. To address this, introducing a cosmological constant $\Lambda$ into Einstein's equations emerges as the most straightforward method to induce accelerated cosmological expansion. This leads to the $\Lambda$CDM model which predicts a contemporary acceleration in cosmic expansion. For several decades, the $\Lambda$CDM model has served as a foundational framework for comprehending the dynamics of the Universe. Its ability to explain various cosmological phenomena using only a few fundamental parameters has established it as an essential foundation of cosmology. Despite the significance of the $\Lambda$CDM model in cosmological analyses, it is evident that further refinement is essential to deepen our comprehension of the Universe.

In this regard, in the previous chapter, we successfully reconstructed $f(Q)$ gravity models. This chapter explores a novel extension of the $f(Q)$ theory coupled with a matter Lagrangian. This new notion was first proposed by Harko et al in \cite{Harko:2018gxr}, and currently offers a broader opportunity for more detailed exploration. Interconnecting geometry with matter leads to a non-zero covariant derivative of the energy-momentum tensor, causing deviations from geodesic motion and violating the equivalence principle. Various expressions of $\mathcal{L}_m$, which represent matter sources, introduce supplementary forces orthogonal to the four-velocity \cite{Bertolami:2008ab,Bertolami:2007gv}. Though the STEGR produces GR-like scenarios \cite{BeltranJimenez:2017tkd,Nester:1998mp}, extensions of corresponding theories lead to different explanations for the Universe's dynamics (see \autoref{fig:cha1:trinity1}). On this matter, this newly proposed notable matter-extension of $f(Q)$ theory can illuminate probing cosmic evolution and provide viable insights into the geometry-matter interactions. 

In this chapter, we reconstruct the $\Lambda$CDM equivalent scenario within matter-Lagrangian coupled gravity using metric-affine formalism. The coupling is expressed in a generic non-minimal form. Here, we consider a barotropic fluid composition for the $\Lambda$CDM formulation within the modified theory at hand. The non-minimal coupling is addressed with two arbitrary functions of the nonmetricity scalar, with one term representing pure dependence on geometry and the other reflecting the influence of matter coupling. First, we derive analytical solutions for the presumed functional form of either of the arbitrary functions and assess the viability of the obtained scenario through cosmographic techniques.

The present chapter is organized as follows:  In \autoref{chap3:secII}, presents an overview of non-minimal $f(Q)$ gravity and \autoref{chap3:secIII} explores its implications for the FLRW Universe. \autoref{chap3:secIV} presents the $\Lambda$CDM model, considering barotropic fluid compositions. In \autoref{chap3:secV}, we investigate different scenarios that replicate the $\Lambda$CDM model using various non-minimal functional forms. Next, \autoref{chap3:secVI} outlines the cosmographic aspects of non-minimal $f(Q)$ gravity and we test the viability of models with the cosmographic approach. Finally, our conclusions are summarized in \autoref{chap3:secVII}.
 
\section{Geometric Formulation of Non-minimal \texorpdfstring{$f(Q)$}{f(Q)} Gravity} \label{chap3:secII}
    In \autoref{chap2:sectionII}, we explained the geometric formulation of $f(Q)$ theory. Here, we shall see the influence of matter coupling on the modified field equations. 
    By coupling matter and geometry, it is possible to achieve a more comprehensive formulation of $f(Q)$ theory. To achieve this objective, the nonminimal coupling in $f(Q)$ gravity is characterized by two arbitrary functions of $Q$, $f_1$ and $f_2$. These functions play a crucial role in the generalization of $f(Q)$ theory by governing the interaction between matter and geometry. The action integral for the theory is denoted as 
    \begin{equation}\label{c3:action}
			S=\int \left[\frac{1}{2} f_1(Q)+f_2(Q)\mathcal{L}_m \right]	\sqrt{-g}\, d^4x.
	\end{equation} 
  For the functional form $f_1(Q)=-Q$ and $f_2(Q)=1$, the study reduces to STEGR.

	    
	    
     

    By varying the action integral \eqref{c3:action} with respect to the metric tensor, we can get the governing equation for the non-minimal $f(Q)$ gravity, given by

          \begin{equation}\label{feq}
		  \begin{split}
		     \frac{2}{\sqrt{-g}} \nabla_\mu \left[\sqrt{-g} \left( \mathcal{F}_1+2\mathcal{F}_2 \mathcal{L}_m \right) P^\mu _{\;\;\alpha \beta}\right]  + \left( \mathcal{F}_1+2\mathcal{F}_2 \mathcal{L}_m \right) \left(P_{\alpha \mu \nu} Q^{\;\;\mu \nu} _{\beta} -2Q^{\mu \nu} _{\;\;\alpha} P_{\mu \nu\beta}\right)\\ + \frac{1}{2} g_{\alpha\beta} f_1 = -f_2 T_{\alpha\beta},
		  \end{split}
		\end{equation}

        where $\mathcal{F}_i\equiv \frac{\partial f_i }{\partial Q}$.


    Further, the variation of the field equation with respect to the connection gives
        \begin{equation}\label{conservation}
            \nabla_{\alpha}\nabla_{\beta}\left[\sqrt{-g} \left( \mathcal{F}_1+2 \mathcal{F}_2 \mathcal{L}_m \right) {P^{\alpha\beta}}_{\mu} - f_2 {\mathcal{H}_{\mu}}^{\alpha\beta}\right]=0.
        \end{equation}

    \section{Flat FLRW Cosmology in Non-minimal \texorpdfstring{$f(Q)$}{f(Q)} Gravity}\label{chap3:secIII}
    Our aim is to investigate the $\Lambda$CDM model within the framework of nonminimal $f(Q)$ gravity. For this purpose, we use flat FLRW metric \eqref{eq:chap2:flrw}. From the straightforward calculation, we can get the expression for nonmetricity scalar as \eqref{eq:chap2:nonmetricity}. 
    In order to establish the continuity equation, we make the natural choice for the matter Lagrangian $\mathcal{L}_m=p$. However, it is also possible to choose $\mathcal{L}_m=-\rho$. This would result in the existence of the extra force generated by the coupling mechanism. For $\mathcal{L}_m=p$, the energy-momentum source term would still persist, and the gravity would be non-conservative. 
    Now, for the current choice of $\mathcal{L}_m$ the energy-momentum tensor takes the form \eqref{eq:chap1:isotropicT}. Thus, we can get the field equations
    \begin{gather}
     \label{friedmann1} 3H^2=\gamma_1\left(-\rho+\gamma_2\right),\\
     \label{friedmann2}  \dot{H}+3H^2+\gamma_3 H=\gamma_1\left(p+\gamma_2\right),
    \end{gather}
		
    where $\gamma_1=\frac{f_2 }{2(f'_1+2 f'_2 p)}$, $  \gamma_2 = \frac{f_1}{2f_2}$ and $    \gamma_3 = \frac{2f_2' \dot{p}+\dot{Q}(f_1''+2 f_2''p)}{f_1'+2f_2' p}$. Here, the overhead dot represents the derivative with respect to cosmic time $t$, and the prime represents the derivative with respect to $Q$.
    Using the above equations, we have
    \begin{equation}
        \dot{H}+\gamma_3 H=\gamma_1\left(p+\rho\right).
    \end{equation}
    
    Further, the equation of continuity can be expressed as
    \begin{equation}\label{eq:continuity}
        \dot{\rho}+3H(\rho+p)=-\frac{6 f'_2 H \dot{H}}{f_2}(p+\rho).
    \end{equation}

    \section{\texorpdfstring{$\Lambda$}{Lambda}CDM model}\label{chap3:secIV}
    The present section deals with the reconstruction of the non-minimal $f(Q)$ gravity model, which can closely resemble the behavior of the well-known $\Lambda$CDM model. To accomplish this, we obtain real-valued functions of the non-metricity scalar that enable a precise description of the cosmic evolution in accordance with the $\Lambda$CDM. The Hubble function expressed in terms of redshift for the $\Lambda$CDM model is given by

    \begin{equation}\label{hofz}
        H(z)=\sqrt{\frac{\rho_0(1+z)^3+\Lambda}{3}},
    \end{equation}
    where $\rho_0$ is the present energy density $(>0)$ and $\Lambda$ is the cosmological constant. The scale factor $a(t)$ with $a(t_0)=1$ is related to the redshift $z$ by the expression $a = 1/(1+z)$.  Thus, the equation presented above can be rewritten in the following manner

    \begin{equation}\label{adot}
        \dot{a}=\sqrt{\frac{\rho_0+\Lambda a^3}{3a}}.
    \end{equation}

    Now, with the help of equation \eqref{eq:chap2:nonmetricity}, one can express equation \eqref{adot} in terms of the nonmetricity scalar as the function of the scale factor 
    \begin{equation}
        Q(a)=\dfrac{2(\rho_0+\Lambda a^3)}{a^3}.
    \end{equation}
   
    From this, we have
    
    \begin{equation}\label{aofQ}
        a(Q)=\left(\dfrac{2\rho_0}{Q-2\Lambda}\right)^{1/3}.
    \end{equation}

    In the view of equations \eqref{adot} and \eqref{aofQ}, we can express the Hubble parameter equation \eqref{hofz} in terms of the nonmetricity scalar as
    
    \begin{equation}
        H(Q)=\frac{1}{3a(Q)^{\frac{3}{2}}}\sqrt{\rho_0+\Lambda a^2(Q)}.
    \end{equation}

    To derive $f_i$'s that replicate the $\Lambda$CDM expansion, it is necessary to incorporate all of the relevant parameters expressed as functions of the non-metricity scalar into the Friedmann equation \eqref{friedmann1}. Now, substituting all the parameters in terms of  $Q$ into \eqref{friedmann1}, we obtain a first-order inhomogeneous differential equation, expressed as follows

    \begin{equation}\label{ode}
        \frac{Q}{2}=\frac{f_2(Q) \left(\frac{f_1(Q)}{2 f_2(Q)}-\rho (Q)\right)}{2 \left(f_1'(Q)+2 p(Q) f_2'(Q)\right)}.
    \end{equation}
    In the case of a vacuum, the aforementioned equation can be simplified to a first-order homogeneous linear differential equation, with $f_1$ being the only dependent variable. The solution for this equation is given by $f_1=k\sqrt{Q}$, where $k$ is the integrating constant. 
    
    In the present work, we assume a barotropic EoS where pressure is represented as $ p =  w \rho$. It is to be noted that the geometry of the Universe still adheres to the $\Lambda$CDM model. Using the aforementioned relation, we can express the energy density as
    \begin{equation}
        \rho(a)=\frac{\rho_0}{a^{3(1+w)}}.
    \end{equation}

     Here, by plugging equation \eqref{aofQ}, we can represent the energy density in terms of the nonmetricity scalar $Q$ as

    \begin{equation}\label{eos}
        \rho(Q)=\left[\frac{1}{2} (Q-2 \Lambda ) \rho_0^{-\frac{w }{w +1}}\right]^{w +1}.
    \end{equation}

    \section{Different non-minimal \texorpdfstring{$f(Q)$}{f(Q)} models}\label{chap3:secV}
        In the non-minimal generalization of $f(Q)$ gravity models, the functions $f_1$ and $f_2$ play a vital role. So, we try to find the unknown function by fixing either of them. The determined function along with the former one describes the modified gravity equivalent $\Lambda$CDM Universe.
        
    \subsection{Power-law form of \texorpdfstring{$f_2$}{f 2}}

    In this section, we shall consider a power-law form of $f_2(Q)$ model, given by 

    \begin{equation}\label{powerlaw}
        f_2(Q)=\xi  \left(\frac{Q}{Q_0}\right)^{\nu },
    \end{equation}
    where $\xi$ and $\nu$ are scalars and the constant $Q_0$ represents the present value of the non-metricity scalar.

    For barotropic fluid having the energy density \eqref{eos}, the inhomogeneous differential equation \eqref{ode} with a fixed power-law, $f_1$ polynomial can be represented as

    \begin{equation}
    \begin{split}
        -\frac{1}{2}\left(\xi  (Q-2 \Lambda ) \left(\frac{Q}{Q_0}\right)^{\nu } \left((Q-2 \Lambda ) \rho_0^{\frac{1}{w +1}-1}\right)^{w }-2^{w } f_1(Q) \rho_0^{\frac{w }{w +1}}\right)\\=Q 2^{w } \rho_0^{\frac{w }{w +1}} f_1'(Q)+\nu  \xi  w  (Q-2 \Lambda ) \left(\frac{Q}{Q_0}\right)^{\nu } \left((Q-2 \Lambda ) \rho_0^{\frac{1}{w +1}-1}\right)^{w }.
    \end{split}
    \end{equation}

    Finding the analytical solution for the obtained differential equation, we can get an expression for the unknown function $f_1$ as
    
    \begin{equation}
    \begin{split}
        f_1(Q)= &k_1 \sqrt{Q}+\Lambda  \xi  (2 \nu  w +1) \Gamma \left(\nu -\frac{1}{2}\right) \rho_0^{\frac{1}{w +1}-1} \left(2-\frac{Q}{\Lambda }\right)^{-w } \left(\frac{Q}{Q_0}\right)^{\nu } \\&\times\left((Q-2 \Lambda ) \rho_0^{\frac{1}{w +1}-1}\right)^{w } \, _2\tilde{F}_1\left(\nu -\frac{1}{2},-w -1;\nu +\frac{1}{2};\frac{Q}{2 \Lambda }\right),
    \end{split}  
    \end{equation}
    where $k_{1}$ is the constant of integration, $\Gamma(z)=\int_0^\infty t^{z-1}e^{-t} dt$ is the Gamma function and $_2\tilde{F}_1(a,b;c;z)=\frac{_2F_1(a,b;c;z)}{\Gamma(c)}$ is the regularized hypergeometric function.

    If the Universe is filled only with cold DM and ordinary matter then the solution reduces to
    \begin{equation}\label{eq:solf1_1}
        f_1(Q)=-\xi  \left(\frac{2 \Lambda }{1-2 \nu }+\frac{Q}{2 \nu +1}\right) \left(\frac{Q}{Q_0}\right)^{\nu }+k_1 \sqrt{Q}.
    \end{equation}
    For $\xi=1$ and $\nu=0$, GR can be retained. 
    One can examine another interesting scenario wherein matter fields reside at the boundary of a specific set that adheres to the strong energy condition. The boundary position suggests that these matter fields may exhibit some exceptional characteristics. In the framework of GR, these fields give rise to the concept of the Milne Universe, which undergoes a uniform and linear expansion over time. In this case, the $f_1$ polynomial becomes,

    \begin{equation}
    \begin{split}
         f_1(Q)=&\frac{1}{5 \Lambda }\left(2^{\nu -\frac{13}{6}} (2 \nu -3) \xi  (Q-2 \Lambda ) \left(\frac{Q}{\Lambda }\right)^{\frac{1}{2}-\nu } \right.\\&\left.\left(\sqrt{\rho_0} (Q-2 \Lambda )\right)^{2/3} \left(\frac{Q}{Q_0}\right)^{\nu } \, _2F_1\left(\frac{5}{3},\frac{3}{2}-\nu ;\frac{8}{3};1-\frac{Q}{2 \Lambda }\right)\right)+k_1 \sqrt{Q}.
    \end{split}
    \end{equation}

    Consequently, we deduced that the gravity theory described above, having the power-law functional form of matter coupling, would accurately replicate the expansion history of the $\Lambda$CDM model if the Universe is filled with barotropic fluid.

    \subsection{Logarithmic form of \texorpdfstring{$f_2$}{f 2}}

    Let us now presume another form of $f_2$ model, having by logarithmic function of non-metricity scalar,
    
    \begin{equation}\label{logmodel}
        f_2(Q)=\xi  \log \left(\frac{Q}{Q_0}\right).
    \end{equation}

    When considering a barotropic perfect fluid characterized by the energy density given in equation \eqref{eos} together with the inhomogeneous differential equation \eqref{ode}, subject to the above logarithmic form of $f_2$ polynomial, we get

    \begin{equation}
        \frac{Q}{2}=\frac{Q \rho_0 2^{w } f_1(Q)-\xi  Q \rho_0 \log \left(\frac{Q}{Q_0}\right) \left((Q-2 \Lambda ) \rho_0^{\frac{1}{w +1}-1}\right)^{w +1}}{4 \left(Q \rho_0 2^{w } f_1'(Q)+\xi  \rho_0 w  \left((Q-2 \Lambda ) \rho_0^{\frac{1}{w +1}-1}\right)^{w +1}\right)}.
    \end{equation}
    
    Thus, the unknown function $f_1$ obtained by fixing $f_2$ for barotropic fluid takes the following polynomial form
    
    \begin{equation}
    \begin{split}
         f_1(Q)=&-\frac{1}{Q-2 \Lambda }\left[\xi  \left(2-\frac{Q}{\Lambda }\right)^{-w } \left((Q-2 \Lambda ) \rho_0^{\frac{1}{w +1}-1}\right)^{w +1}\right.\\&\left. \left(-2 Q \, _3F_2\left(\frac{1}{2},\frac{1}{2},-w ;\frac{3}{2},\frac{3}{2};\frac{Q}{2 \Lambda }\right)\right.\right.\left.\left.+4 \Lambda  \, _3F_2\left(-\frac{1}{2},-\frac{1}{2},-w ;\frac{1}{2},\frac{1}{2};\frac{Q}{2 \Lambda }\right)\right.\right.\\&\left.\left.+\left(Q \, _2F_1\left(\frac{1}{2},-w ;\frac{3}{2};\frac{Q}{2 \Lambda }\right)+2 \Lambda  \, _2F_1\left(-\frac{1}{2},-w ;\frac{1}{2};\frac{Q}{2 \Lambda }\right)\right)\right.\right.\\&\left.\left.\times \left(\log \left(\frac{Q}{Q_0}\right)+2 w \right)\right)\right]+k_2 \sqrt{Q},
    \end{split}
    \end{equation}

    where $_p F_q(a_1,...,a_p,b_1,...,b_q;z)$ is hypergeometric function and $k_{2}$ is a constant.

    For $w=0$, the dust case can be recovered and the corresponding $f_1$ polynomial takes the form

    \begin{equation}\label{eq:solf1_2}
       f_1(Q)= -4 \Lambda  \xi +2 \xi  Q-\xi  (2 \Lambda +Q) \log \left(\frac{Q}{Q_0}\right)+k_2 \sqrt{Q}.
    \end{equation}

    Similarly, for $w=-1/3$, it can be expressed as,

    \begin{equation}
        \begin{split}
            f_1(Q)= &\frac{1}{9 \sqrt[3]{\sqrt{\rho_0} (Q-2 \Lambda )}}\left[\xi  \sqrt{\rho_0} \left(-144 \Lambda  \sqrt[3]{2-\frac{Q}{\Lambda }} \, _3F_2\left(-\frac{1}{2},-\frac{1}{2},\frac{1}{3};\frac{1}{2},\frac{1}{2};\frac{Q}{2 \Lambda }\right)\right.\right.\\&\left.\left.+16 \Lambda  \sqrt[3]{2-\frac{Q}{\Lambda }} \, _2F_1\left(-\frac{2}{3},-\frac{1}{2};\frac{1}{2};\frac{Q}{2 \Lambda }\right)\right.\right.\\&\left.\left.-\left(8 \Lambda  \sqrt[3]{2-\frac{Q}{\Lambda }} \, _2F_1\left(-\frac{1}{2},\frac{1}{3};\frac{1}{2};\frac{Q}{2 \Lambda }\right)+3 \sqrt[3]{2} (Q-2 \Lambda )\right)\right.\right.\\&\left.\left. \times\left(9 \log \left(\frac{Q}{Q_0}\right)-52\right)\right)\right]+k_2 \sqrt{Q}.
        \end{split}
    \end{equation}

    Therefore, our derived results lead to the conclusion that the non-minimal $f(Q)$ gravity, with its logarithmic functional form of matter coupling, can faithfully reproduce the expansion history of the $\Lambda$CDM model when the Universe is dominated by barotropic fluid.
    
    \subsection{Specific form of \texorpdfstring{$f_1$}{f 1}}

    So far, our examination has been focused on the reconstructed polynomials corresponding to a particular formulation of $f_2$. The function $f_1$ delineates a purely geometric component within the context of the nonminimally coupled Universe, while the introduction of coupling arises from the behavior of the function $f_2$. In this section, we explore the $\Lambda$CDM equivalent by choosing a specific form of the polynomial $f_1$, thus obtaining the suitable form of $f_2$. This is characterized by a power law expression

    \begin{equation}\label{f1}
        f_1(Q)=2 \Lambda -Q \left(\frac{Q}{Q_0}\right)^{\nu }.
    \end{equation}
    
   In this scenario, let us assume that the Universe is filled with fluid for which energy density purely depends on pressure. Then with the aid of the relation \eqref{eos} and the specific form of $f_1$ \eqref{f1}, the differential equation \eqref{ode} becomes

\begin{equation}
    \frac{Q}{2}=\frac{f_2(Q) (Q-2 \Lambda ) A+2^{w } \rho_0^{\frac{w }{w +1}} \left(Q \left(\frac{Q}{Q_0}\right)^{\nu }-2 \Lambda \right)}{(\nu +1) 2^{w +2} \rho_0^{\frac{w }{w +1}} \left(\frac{Q}{Q_0}\right)^{\nu }-4 w  (Q-2 \Lambda ) f_2'(Q) A},
\end{equation}
where $A=\left((Q-2 \Lambda ) \rho_0^{\frac{1}{w +1}-1}\right)^{w }$.
Upon solving the differential equation, we get the following functional form for $f_2$

    \begin{equation}\label{eq:f2}
    \begin{split}
        f_2(Q)=& 2^{w -1} (Q-2 \Lambda ) \left((Q-2 \Lambda ) \rho_0^{\frac{1}{w +1}-1}\right)^{-w -1}\\& \times\left(-2 \, _2F_1\left(1,\frac{1}{2 w }-w ;\frac{1}{2} \left(2+\frac{1}{w }\right);\frac{Q}{2 \Lambda }\right)\right.\\&\left.-\frac{1}{w ^2}\left((2 \nu +1) 2^{\nu +\frac{1}{2 w }} \left(\frac{Q}{Q_0}\right)^{\nu } \left(\frac{Q}{\Lambda }\right)^{-\nu -\frac{1}{2 w }}\right.\right.\\&\left.\left. \, _2F_1\left(-\nu -\frac{1}{2 w },-w ;1-w ;1-\frac{Q}{2 \Lambda }\right)\right)\right)+k_3 Q^{-\frac{1}{2 w }},
    \end{split}
    \end{equation}
where $k_{3}$ is constant of integration. 

For dust-like stuff scenarios, the functional form becomes

    \begin{equation}
        \begin{split}
            f_2(Q)= & \frac{2 \Lambda +(2 \nu +1) Q \left(\frac{Q}{Q_0}\right)^{\nu }}{Q-2 \Lambda }.
        \end{split}
    \end{equation}

In a similar manner, when the EoS takes the value of $-1/3$, the function \eqref{eq:f2} can be expressed as

    \begin{equation}
        \begin{split}
            f_2(Q)= & \frac{1}{4 \left(\sqrt{\rho_0} (Q-2 \Lambda )\right)^{2/3}}\left[\left(2-\frac{Q}{\Lambda }\right)^{2/3} \left(4 \Lambda  \, _2F_1\left(-\frac{3}{2},\frac{2}{3};-\frac{1}{2};\frac{Q}{2 \Lambda }\right)\right.\right.\\&\left.\left.-3 (2 \nu +1) Q \Gamma \left(\nu -\frac{1}{2}\right) \left(\frac{Q}{Q_0}\right)^{\nu } \, _2\tilde{F}_1\left(\frac{2}{3},\nu -\frac{1}{2};\nu +\frac{1}{2};\frac{Q}{2 \Lambda }\right)\right)\right]+k_3 Q^{3/2}.
        \end{split}
    \end{equation}

Hence, our derived outcomes indicate that non-minimal $f(Q)$ gravity, characterized by its specific functional form of geometry component, is capable of accurately reconstructing the expansion history observed in the $\Lambda$CDM model during a period of dominance by barotropic fluid within the Universe.

\section{Cosmography in non-minimal \texorpdfstring{$f(Q)$}{f(Q)} gravity}\label{chap3:secVI}

Cosmography offers a versatile approach for understanding cosmological parameters solely through kinematic properties of the Universe. Notably, this treatment remains model-independent, distinguishing it as an advantageous feature. Current advancements in the field of cosmography primarily concentrate on the late time dynamics of the Universe. Typically, cosmographic analyses rely on Taylor expansions, particularly applicable within the observable range where $z\ll 1$, providing insights into the present Universe. In this section, we employ the cosmographic framework outlined in \cite{Capozziello:2008qc,Capozziello:2011hj,Mandal:2020buf} and describe this in non-minimally coupled $f(Q)$ gravity.

Here, we consider two arbitrary functions, $f_1$ and $f_2$, which are dependent on the non-metricity scalar. In the previous section, we analytically derived the $\Lambda$CDM scenario by selecting a specific model and determining the complementary functional form. In this section, we repeat a similar procedure but with fixed forms for $f_2$, aiming to find the corresponding $f_1$ through a cosmographic approach. The main advantage of this approach is to connect the analytically obtained solution with the reconstructed model using empirical data. For this purpose, we explore the Taylor series expansion of $f_1(Q)$ as 

\begin{equation}\label{eq:series}
\begin{split}
    f_1(Q) =& F +\frac{1}{2!} F_1 (Q-Q_0) +\frac{1}{3!} F_2 (Q-Q_0)^2 + \frac{1}{4!} F_3 (Q-Q_0)^3 + \frac{1}{5!} F_4 (Q-Q_0)^4 + \mathcal{O},
\end{split} 
\end{equation}
where $F=f_1(Q_0)$, and $F_n=f_1^{(n)}(Q_0)$, $n$ representing $n^{th}$ order derivative of the function.

For the dust case, the Friedmann equations can be written as
\begin{gather}
    12 H^2=\frac{f_1-2 \rho  f_2}{f_1'},\\
    \dot{H}=\frac{f_2 \rho -2 H \left(2 \dot{p} f_2'+\dot{Q} \left(f_1''\right)\right)}{2 f_1'},
\end{gather}

Here, the overhead dot represents a derivative with respect to time $t$, and the prime represents a derivative with respect to non-metricity scalar $Q$. The successive derivatives of the second field equation lead to
\begin{align}
&\begin{split}
    \ddot{H}=\frac{1}{2 f_1'{}^2}& \left[-2 H f_1' \left(f_1'' \ddot{Q}+f_1''' \dot{Q}^2\right)-2 \dot{H} \dot{Q} f_1' f_1''\right.\\&\left.+2 H \dot{Q}^2 \left(f_1''\right){}^2+f_2 \dot{\rho } f_1'+\rho  \dot{Q} f_1' f_2'-f_2 \rho  \dot{Q} f_1''\right],
\end{split}
    \\
    &\begin{split}
        \dddot{H}=\frac{1}{2 f_1'{}^3}&\left[-2 H f_1'{}^2 \left(\dddot{Q} f_1''+3 f_1''' \dot{Q} \ddot{Q}+f_1'''' \dot{Q}^3\right)\right.\\&\left.-2 \dot{Q} \left(f_1'\right){}^2 f_1'' \ddot{H}+6 H \dot{Q} f_1' f_1'' \left(f_1'' \ddot{Q}+f_1''' \dot{Q}^2\right)\right.\\&\left.-4 \dot{H} \left(f_1'\right){}^2 \left(f_1'' \ddot{Q}+f_1''' \dot{Q}^2\right)+4 \dot{H} \dot{Q}^2 f_1' \left(f_1''\right){}^2\right.\\&\left.-4 H \dot{Q}^3 \left(f_1''\right){}^3+f_2 \left(f_1'\right){}^2 \ddot{\rho }-f_2 \rho  f_1' \left(f_1'' \ddot{Q}+f_1''' \dot{Q}^2\right)\right.\\&\left.+\rho  \left(f_1'\right){}^2 f_2' \ddot{Q}-2 f_2 \dot{\rho } \dot{Q} f_1' f_1''-2 \rho  \dot{Q}^2 f_1' f_2' f_1''\right.\\&\left.+\rho  \dot{Q}^2 \left(f_1'\right){}^2 f_2''+2 \dot{\rho } \dot{Q} \left(f_1'\right){}^2 f_2'+2 f_2 \rho  \dot{Q}^2 \left(f_1''\right){}^2\right].
    \end{split}
\end{align}

The derivatives of $H$ can be expressed in terms of cosmographic parameters. These expressions for the present time can be represented as

\begin{equation}\label{Hderivatives}
\begin{aligned}
    \dot{H}_0&=-H_0^2 \left(q_0+1\right),\\
    \ddot{H}_0&=H_0^3 \left(j_0+3 q_0+2\right),\\
    \dddot{H}_0&=H_0^4 \left(-4 j_0-3 q_0 \left(q_0+4\right)+s_0-6\right).
\end{aligned}
\end{equation}

These can be substituted into the following expressions to obtain derivatives of the non-metricity scalar at $t=t_0$:

\begin{gather}\label{qderivatives}
\begin{aligned}
    Q_0&=6 H_0^2, \\
    \dot{Q}_0&=12 H_0 \dot{H}_0, \\
    \ddot{Q}_0&=12 H_0 \ddot{H}_0+12 \dot{H}_0^2,\\
    \dddot{Q}_0&=12 H \dddot{H}_0+36 \dot{H}_0 \ddot{H}_0. 
\end{aligned}
\end{gather}

Let us consider the scenario where $f_2=\xi \left(\frac{Q}{Q_0}\right)^{\nu }$. Computing the aforementioned Friedmann equations and their derivatives, along with equations \eqref{Hderivatives} and \eqref{qderivatives}, at the present time yields

\begin{gather}
    \label{eq:m1_1}H_0^2=\frac{F-6 H_0^2 \xi  \Omega_{m0}}{12 F_1},
\end{gather}
\begin{gather}
    \label{eq:m1_2}H_0 \left(2 \left(q_0+1\right) \left(12 F_2 H_0^2+F_1\right)+3 \xi  \Omega_{m0}\right)=0,
    \end{gather}
\begin{gather}
    \begin{split}\label{eq:m1_3}
         &24 F_2 H_0^2 \left(j_0+q_0 \left(2 q_0+7\right)+4\right)+\frac{3}{F_1}\left(\xi  \Omega_{m0} \left(F_1 \left(2 \nu  \left(q_0+1\right)+3\right)-12 F_2 H_0^2 \left(q_0+1\right)\right)\right.\\&\left.-96 F_2^2 H_0^4 \left(q_0+1\right){}^2\right)+288 F_3 H_0^4 \left(q_0+1\right){}^2+2 F_1 \left(j_0+3 q_0+2\right)=0,
    \end{split}
    \end{gather}
\begin{gather}
    \begin{split}\label{eq:m1_4}
        &\frac{H_0 }{F_1}\left(3 \xi  \Omega_{m0} \left(-12 F_1 H_0^2 \left(12 F_3 H_0^2 \left(q_0+1\right){}^2\right.\right.\right.\\&\left.\left.\left.+F_2 \left(j_0+4 \nu  \left(q_0+1\right){}^2+q_0^2+11 q_0+9\right)\right)+288 F_2^2 H_0^4 \left(q_0+1\right){}^2\right.\right.\\&\left.\left.+F_1^2 \left(2 \nu  \left(j_0-\left(q_0-7\right) q_0+7\right)+4 \nu ^2 \left(q_0+1\right){}^2+3 \left(q_0+4\right)\right)\right)\right.\\&\left.+24 F_1^2 H_0^2 \left(12 H_0^2 \left(q_0+1\right) \left(12 F_4 H_0^2 \left(q_0+1\right){}^2\right.\right.\right.\\&\left.\left.\left.+F_3 \left(3 j_0+q_0 \left(5 q_0+19\right)+11\right)\right)+F_2 \left(2 j_0 \left(3 q_0+5\right)\right.\right.\right.\\&\left.\left.\left.+q_0 \left(q_0 \left(2 q_0+27\right)+48\right)-s_0+20\right)\right)-288 F_1 F_2 H_0^4\right.\\&\left.\times \left(q_0+1\right) \left(36 F_3 H_0^2 \left(q_0+1\right){}^2+F_2 \left(3 j_0+q_0 \left(5 q_0+19\right)+11\right)\right)\right.\\&\left.+6912 F_2^3 H_0^6 \left(q_0+1\right){}^3+2 F_1^3 \left(4 j_0+3 q_0 \left(q_0+4\right)-s_0+6\right)\right)=0.
    \end{split}
\end{gather}

Using equations \eqref{eq:m1_2}, \eqref{eq:m1_3}, and \eqref{eq:m1_4}, we can compute $q_0$, $j_0$, and $s_0$ as follows

\begin{gather}
    q_0= -\frac{3 \xi  \Omega_{m0}}{2 \left(12 F_2 H_0^2+F_1\right)}-1,
    \end{gather}
    \begin{gather}
    \begin{split}
        j_0= &\frac{1}{2 F_1 \left(12 F_2 H_0^2+F_1\right)}\left[-3 F_1 \left(96 F_3 H_0^4 \left(q_0+1\right){}^2\right.\right.\\&\left.\left.+8 F_2 H_0^2 \left(q_0 \left(2 q_0+7\right)+4\right)+\xi  \Omega_{m0} \left(2 \nu +2 \nu  q_0+3\right)\right)\right.\\&\left.+36 F_2 H_0^2 \left(q_0+1\right) \left(8 F_2 H_0^2 \left(q_0+1\right)+\xi  \Omega_{m0}\right)\right.\left.-2 F_1^2 \left(3 q_0+2\right)\right],
    \end{split}
    \end{gather}
    \begin{gather}
    \begin{split}
        s_0&= \frac{1}{2 F_1^2 \left(12 F_2 H_0^2+F_1\right)}\left[36 F_1 H_0^2 \left(-\xi  \Omega_{m0} \left(12 F_3 H_0^2 \left(q_0+1\right){}^2\right.\right.\right.\\&\left.\left.\left.+F_2 \left(j_0+4 \nu +q_0 \left(8 \nu +4 \nu  q_0+q_0+11\right)+9\right)\right)\right.\right.\\&\left.\left.-8 F_2 H_0^2 \left(q_0+1\right) \left(36 F_3 H_0^2 \left(q_0+1\right){}^2+F_2 \left(3 j_0\right.\right.\right.\right.\\&\left.\left.\left.\left.+q_0 \left(5 q_0+19\right)+11\right)\right)\right)+3 F_1^2 \left(96 H_0^4 \left(q_0+1\right) \right.\right.\\&\left.\left.\times\left(12 F_4 H_0^2 \left(q_0+1\right){}^2+F_3 \left(3 j_0+q_0 \left(5 q_0+19\right)+11\right)\right)\right.\right.\\&\left.\left.+8 F_2 H_0^2 \left(2 j_0 \left(3 q_0+5\right)+q_0 \left(q_0 \left(2 q_0+27\right)+48\right)+20\right)\right.\right.\\&\left.\left.+\xi  \Omega_{m0} \left(2 \nu  \left(j_0+q_0 \left(4 \nu +(2 \nu -1) q_0+7\right)\right)\right.\right.\right.\\&\left.\left.\left.+2 (\nu +2) (2 \nu +3)+3 q_0\right)\right)+864 F_2^2 H_0^4 \left(q_0+1\right){}^2 \right.\\&\left.\times\left(8 F_2 H_0^2 \left(q_0+1\right)+\xi  \Omega_{m0}\right)+2 F_1^3 \left(4 j_0+3 q_0 \left(q_0+4\right)+6\right)\right].
    \end{split}
\end{gather}

\begin{figure}[hbtp]
     \centering
     \includegraphics[width=0.9\linewidth]{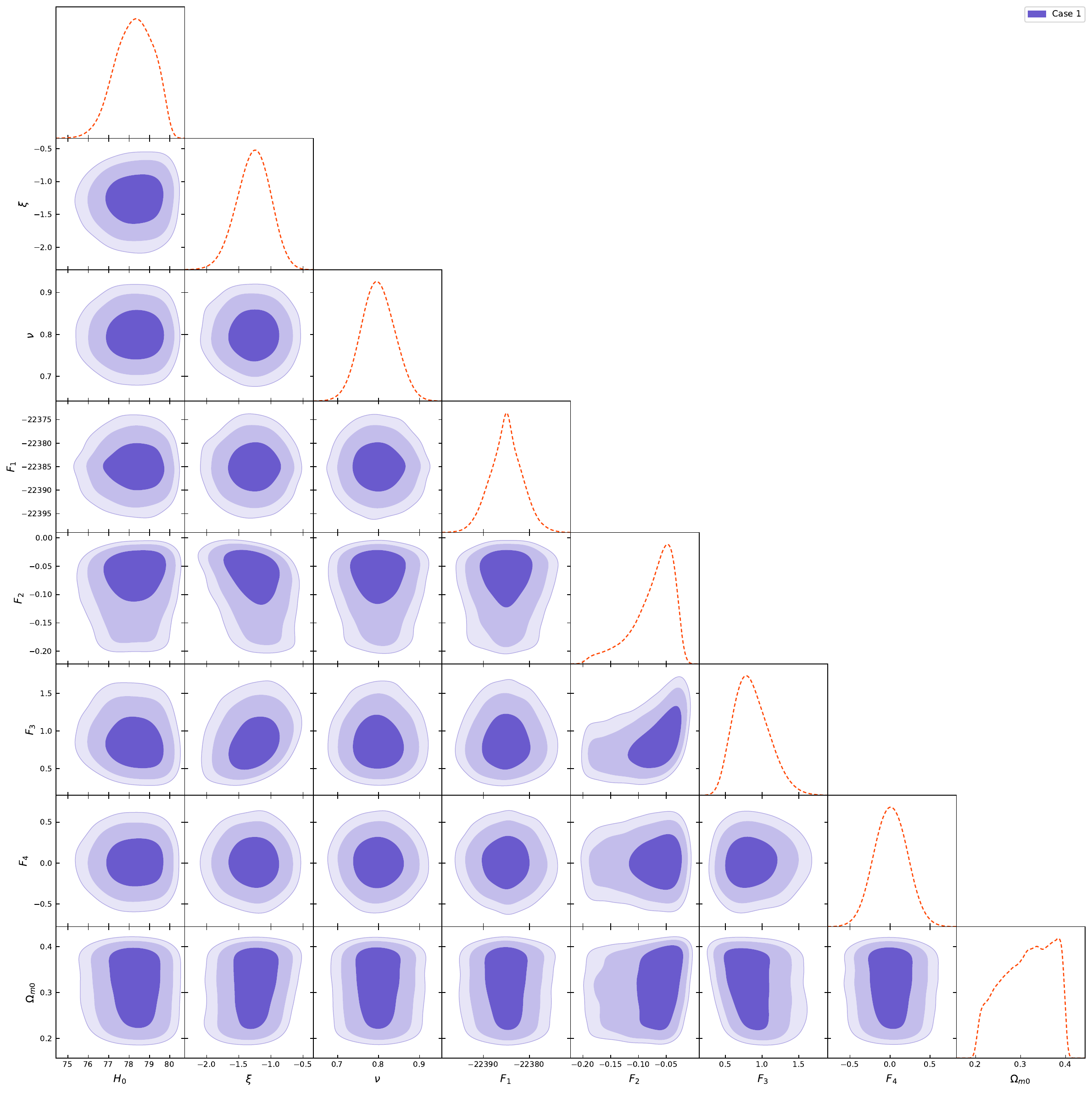}
     \caption{2D likelihood contours obtained from MCMC with power-law form}
     \label{fig:contour1}
 \end{figure}

Here, $F_1$, $F_2$, $F_3$, and $F_4$ represent free parameters, constrained using MCMC along with $\xi$, $\nu$, $\Omega_{m0}$, and $H_0$. For this purpose, we utilize 1701 Type Ia supernovae samples (see Appendix). The plot obtained through MCMC is provided in \autoref{fig:contour1}. The best fits for the constrained parameters from statistical analysis are as follows: $H_0 = 78.3$, $\xi = -1.28$, $\nu = 0.801$, $F_1 = 2.23849 \times 10^4$, $F_2 = 0.073$, $F_3 = 0.85$, $F_4 = 0.01$, and $\Omega_{m0} = 0.316$.

In the previous section, for the dust case with a power-law form of the $f_2$ function, we derived the analytical solution for the functional form of $f_1$ (see Eq. \eqref{eq:solf1_1}). This expression for the present scenario can be expressed as

\begin{equation}
    F= H_0 \left(\sqrt{6} k_1-\frac{6 H_0 \xi  \left((2 \nu +1) \Omega _{m0}-2\right)}{4 \nu ^2-1}\right).
\end{equation}

By comparing the above equation with equation \eqref{eq:m1_1}, we can determine the value of the integrating constant $k_1$. Therefore, for the mean values obtained from MCMC, the value of $k$ becomes $8.58673\times 10^6$. Consequently, $F =1.64686\times 10^9$ . One can then substitute these values into \eqref{eq:series} to obtain the functional form of $f_1(Q)$.

Now, repeating a similar procedure for the previously defined logarithmic form of $f_2(Q)$, we obtain the Friedman equations and their derivatives at the present time as

\begin{gather}
    \label{eq:m2_1}\frac{F}{F_1}=12 H_0^2,\\
    \label{eq:m2_2}
    \frac{H_0 \left(q_0+1\right) \left(12 F_2 H_0^2+F_1\right)}{F_1}=0,\\
    \label{eq:m2_3}\begin{split}
        &12 F_2 H_0^3 \left(j_0+q_0 \left(2 q_0+7\right)+4\right)+F_1 H_0 \left(j_0+3 q_0+2\right)\\&+144 F_3 H_0^5 \left(q_0+1\right){}^2+3 H_0 \xi  \left(q_0+1\right) \Omega_{m0}=\frac{144 F_2^2 H_0^5 \left(q_0+1\right){}^2}{F_1},
    \end{split}
    \\
    \label{eq:m2_4}\begin{split}
        &3 H_0 \xi  \Omega_{m0} \left(F_1 \left(j_0-\left(q_0-7\right) q_0+7\right)-24 F_2 H_0^2 \left(q_0+1\right){}^2\right)\\&+12 F_1 H_0^3 \left(12 H_0^2 \left(q_0+1\right) \left(12 F_4 H_0^2 \left(q_0+1\right){}^2\right.\right.\\&\left.\left.+F_3 \left(3 j_0+q_0 \left(5 q_0+19\right)+11\right)\right)+F_2 \left(2 j_0 \left(3 q_0+5\right)\right.\right.\\&\left.\left.+q_0 \left(q_0 \left(2 q_0+27\right)+48\right)-s_0+20\right)\right)+F_1^2 H_0 \left(4 j_0\right.\\&\left.+3 q_0 \left(q_0+4\right)-s_0+6\right)+\frac{3456 F_2^3 H_0^7 \left(q_0+1\right){}^3}{F_1}\\&=144 F_2 H_0^5 \left(q_0+1\right) \left(36 F_3 H_0^2 \left(q_0+1\right){}^2+F_2 \left(3 j_0\right.\right.\\&\left.\left.+q_0 \left(5 q_0+19\right)+11\right)\right).
    \end{split}
\end{gather}

From equations \eqref{eq:m2_2}-\eqref{eq:m2_4}, one can derive expressions for cosmographic parameters as 

\begin{gather}
    q_0= -1,\\
    \begin{split}
        j_0=& \frac{1}{F_1 \left(12 F_2 H_0^2+F_1\right)}\left[-3 F_1 \left(\left(q_0+1\right) \left(48 F_3 H_0^4 \left(q_0+1\right)+\xi  \Omega_{m0}\right)\right.\right.\\&\left.\left.+4 F_2 H_0^2 \left(q_0 \left(2 q_0+7\right)+4\right)\right)+144 F_2^2 H_0^4 \left(q_0+1\right){}^2\right.\left.-F_1^2 \left(3 q_0+2\right)\right],
    \end{split}
    \\
    \begin{split}
        s_0=& -\frac{1}{F_1 \left(12 F_2 H_0^2+F_1\right)}\left[72 F_2 H_0^2 \left(q_0+1\right) \left(2 F_2 H_0^2 \left(3 j_0\right.\right.\right.\\&\left.\left.\left.+q_0 \left(5 q_0+19\right)+11\right)+\left(q_0+1\right) \left(72 F_3 H_0^4 \left(q_0+1\right)+\xi  \Omega_{m0}\right)\right)\right.\\&\left.+3 F_1 \left(-48 H_0^4 \left(q_0+1\right) \left(12 F_4 H_0^2 \left(q_0+1\right){}^2+F_3 \left(3 j_0\right.\right.\right.\right.\\&\left.\left.\left.\left.+q_0 \left(5 q_0+19\right)+11\right)\right)-4 F_2 H_0^2 \left(2 j_0 \left(3 q_0+5\right)\right.\right.\right.\\&\left.\left.\left.+q_0 \left(q_0 \left(2 q_0+27\right)+48\right)+20\right)-\xi  \left(j_0-\left(q_0-7\right) q_0+7\right) \Omega_{m0}\right)\right.\\&\left.-\frac{3456 F_2^3 H_0^6 \left(q_0+1\right){}^3}{F_1}-F_1^2 \left(4 j_0+3 q_0 \left(q_0+4\right)+6\right)\right].
    \end{split}
\end{gather}

\begin{figure}[!]
     \centering
     \includegraphics[width=0.9\linewidth]{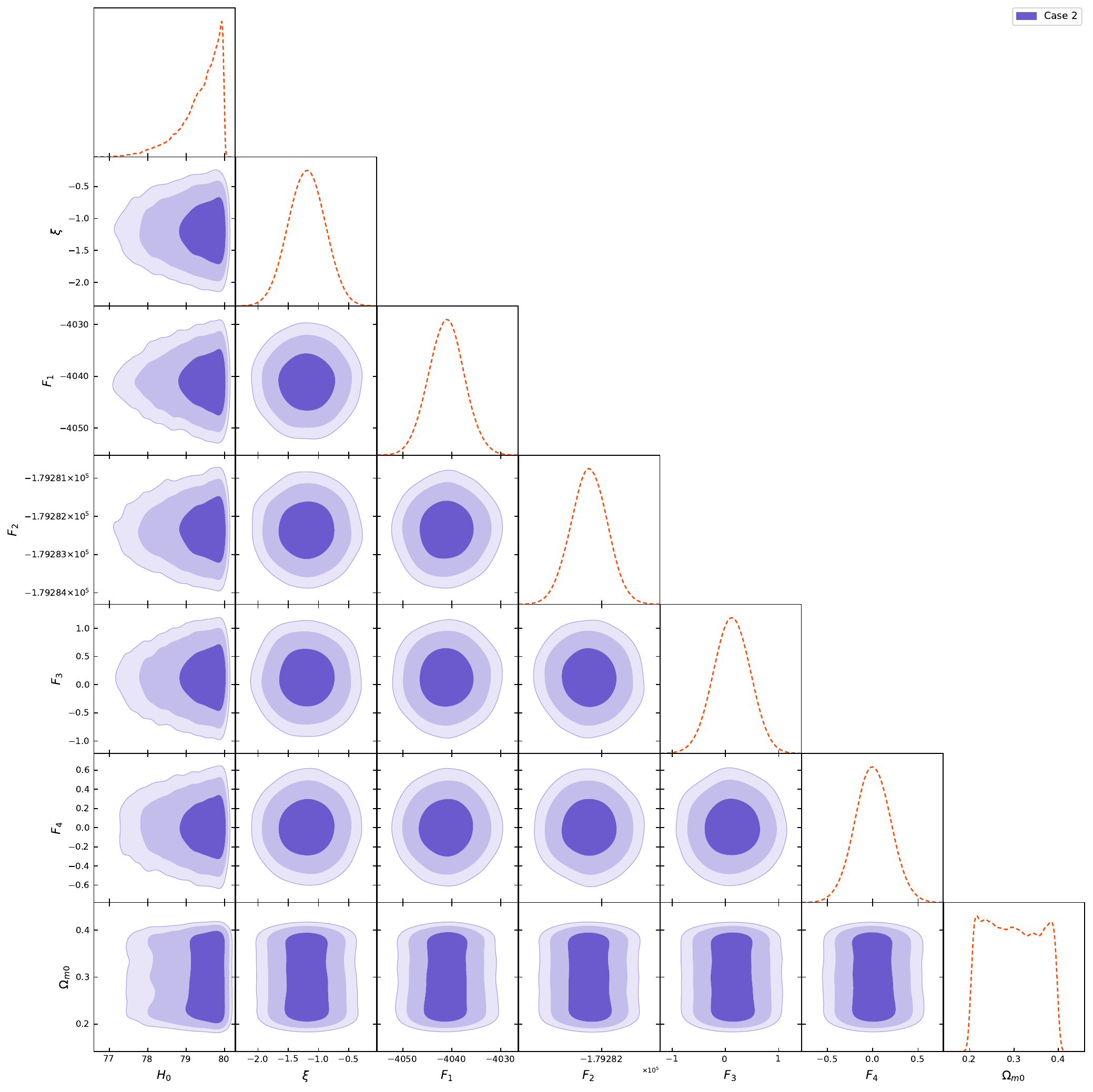}
     \caption{2D likelihood contours obtained from MCMC with logarithmic form}
     \label{fig:contour2}
 \end{figure}

After performing MCMC analysis (see \autoref{fig:contour2}), the free parameters $F_1$, $F_2$, $F_3$, $F_4$, $\xi$, $H_0$, and $\Omega_{m0}$ take the best fit values of $-4.0411\times10^3$, $-0.073$, $0.85$, $0.012$, $-1.19$, $79.35$, and $0.299$, respectively.

In the view of the present scenario, the equation \eqref{eq:solf1_2} yields
\begin{equation}
    F=\sqrt{6} H_0 k_2+12 H_0^2 \xi  \Omega_{m0}.
\end{equation}
Therefore, for the obtained values of parameters, the integrating constant $k_2$ takes the value $-1.57077\times 10^6$ and $F=-3.84794\times10^6$, when compared with equation \eqref{eq:m2_1}.

\paragraph{Discussion:}
The coupling of matter to geometry is achieved by introducing a coupling term $f_2$ in the action integral.
Generally, a non-zero functional value of \( f_2 \) results in a deviation from standard geodesic motion, causing the particle's motion to deviate from the regular path. This deviation becomes evident when the distance indicator, entering the luminosity distance, is altered. Therefore, to effectively constrain non-metric models, the data set used for model optimization plays a crucial role. Astronomical surveys employ different methodologies for this purpose. For instance, standard candles (such as supernova surveys) rely on known luminosities to gauge distances, while standard rulers (like BAO) use known physical sizes. BAO, as standard rulers, are less affected by the complexities and uncertainties inherent in stellar explosions that characterize standard candles. Hence, utilizing these data to constrain the gravity model is beneficial, especially when dealing with early Universe probes, as they offer the still unproven potential to constrain growth by observing changes in the amplitude of the power spectrum.

In this chapter, we utilized a supernova dataset featuring standard candles due to their standardized luminosities. They were instrumental in discovering the Universe's accelerating expansion \cite{SupernovaCosmologyProject:1997czu,SupernovaSearchTeam:1998fmf} and remain widely employed as cosmological tools. Over the last two decades, numerous surveys have concentrated on detecting these supernovae across a broad spectrum of redshifts (which include HF, CSP, LOSS, SWIFT, ASASSN, Pan-STRARRS, SDSS, GOODS, SCP, HST, etc). The methodology adopted in our study for the analysis with Pantheon+SH0ES data is in the appendix.

Another interesting fact to discuss is the influence of non-metricity-based formalism on distance indicators. Here, we have not considered the effect of non-metricity on distance indicators like BAO. This is because the BAO data comes from baryonic acoustic oscillations or the CMB anisotropy caused by baryon perturbations. The theoretical calculations \cite{Weinberg:2008zzc} show that non-relativistic effects (Jeans instability) and relativistic effects (gauge-invariant Bardeen potential) cause the BAO imprint in CMB data. It is important to note that one only uses the FLRW metric in the background, so the background manifold (diffeomorphic to \(\mathbb{R}^4\)) and the Riemannian metric, which is the standard flat FLRW metric, are the same throughout. In our treatment of non-metricity, as done in \eqref{eq:nonmetricity}, we merely present a new way of doing Riemannian geometry without using the standard Levi-Civita connection. Therefore, at the background level, all the relevant tensorial quantities, like curvature, would have different formulations \cite{Heisenberg:2023lru}. However, note that in no case does the background metric (the FLRW metric) change, nor do the physical properties. For example, the timelike geodesic of a particle or the null geodesic of photons remains the same (similar to the geodesic equation for the Levi-Civita connection), as discussed in \cite{Xu:2019sbp}. So, non-metricity only appears as an alternative geometry calculation (without the Levi-Civita connection), but it does not affect the background metric or the perturbation analysis done using the FLRW metric. The non-metricity based models offer promising approach in the investigation of dynamics of the Universe \cite{nojiri2024fqgravitygaussbonnetcorrections,Capozziello:2024vix,Mishra:2024rut,Naik:2023ykt,Hu:2023gui}. There are several interesting works in the literature that focus on analyzing physically viable $f(Q)$ models and constraining them with observational data. For instance, one can see in \cite{Nojiri:2024zab}, authors have focused on formulating convenient and mimetic \( f(Q) \) gravities, which are described in terms of the metric and four scalar fields. This formulation results in well-defined field equations. By using these equations, they have demonstrated how to reconstruct \( f(Q) \) models that can realize any given FLRW spacetime. These developed models, including convenient \( f(Q) \) gravity and mimetic \( f(Q) \) gravity, successfully describe both the inflationary and DE epochs, and they even have the potential to unify these cosmic phases. Additionally, it has been shown that these theories can easily include a radiation-dominated epoch and early DE. They also discuss potential reconstruction methods for spherically symmetric spacetimes, highlighting the flexibility of \( f(Q) \) gravity theories. The primary methodology adopted in their work is the analytical approach. Other than this technique, one can also consider several ways to reconstruct $f(Q)$ models like the one which is employed in \cite{Capozziello:2022wgl}. A key idea in this work is the method for reconstructing the \( f(Q) \) action without making prior assumptions about the cosmological model. Authors have utilized rational Pad\'e approximations, which ensure stable behavior of the cosmographic series at high redshifts, thereby mitigating the convergence issues of traditional methods. The reconstruction process involves a numerical inversion based on current observational data on cosmographic parameters.  Additionally, the study explores potential deviations from the standard $\Lambda$CDM model, highlighting the flexibility and relevance of \( f(Q) \) gravity in cosmological probes. In our present chapter, we have made use of both these techniques of finding solutions to reconstruct $\Lambda$CDM Universe analytically and validating the result using a cosmographic approach. 

\section{Concluding Remarks}\label{chap3:secVII}

Our primary goal is to reconstruct the $\Lambda$CDM Universe by addressing these limitations. A promising approach for achieving this is embedding the $\Lambda$CDM scenario within a modified gravity framework \cite{Salako:2013gka,Fay:2007uy,Astorga-Moreno:2024okq,Gadbail:2022jco,Ortiz-Banos:2021jgg,Paliathanasis:2023dfz,Goncalves:2023umv,Junior:2015bva}. Also, there are purely model-independent approaches to reconstruct the $\Lambda$CDM Universe in modified theories using cosmographic techniques (for instance, \cite{Capozziello:2008qc,Capozziello:2011hj,Mandal:2020buf}). In our work, we employ a blended procedure involving the derivation of solutions for functional forms of non-metricity scalar analytically and then validating the observationally agreeable model obtained through cosmography.

We investigate this within the framework of novel non-minimally coupled $f(Q)$ gravity. A generic coupling form is considered in the action, given by $\frac{1}{2}f_1(Q)+f_2(Q) \mathcal{L}_m$. Due to the existence of coupling, the underlying theoretical explanation for the corresponding dynamics deviates from that of $f(Q)$ gravity. The expression \eqref{conservation} indicates that the theory is non-conservative in general. Thus, one can have $\nabla_a{T^a}_b = \mathcal{E}$, where $\mathcal{E}$ represents the residual energy source. A gravity theory is considered conserved if $\mathcal{E}=0$. Since the energy balancing term pertains, this indicates the energy interactions within the system. Entities contributing to the energy balance term stem from both the energy-momentum tensor and the hyper-momentum tensor. The effect of coupling is also reflected in the trajectory of massive test particles, leading to an additional force $\mathcal{F}^a$. This force causes particles to deviate from their usual geodesic path and is orthogonal to the four-velocity. Similar to the energy balancing term, the extra force is contributed by both the energy-momentum and hyper-momentum tensor.

At this point, it is noteworthy that the choice of matter Lagrangian is crucial in analyzing the residual energy term and the motion trajectory of test particles. For example, $\mathcal{L}_m=-\rho$ nullifies the energy source arising from the energy-momentum tensor while retaining a non-zero extra force. Conversely, $\mathcal{L}_m=p$ eliminates the energy-momentum component of the extra force, suggesting the possibility of an energy transfer process within the system. In our current study, we adopt $\mathcal{L}_m=p$, ensuring that the extra force due to energy-momentum, given by $(-p+\mathcal{L}_m) h^a_b \nabla^b \log f_2$, vanishes. Here, $h^a_b$ denotes the projection operator. The corresponding continuity equation for this case is given in the equation \eqref{eq:continuity}.

For the purpose of embedding $\Lambda$CDM in non-minimally coupled $f(Q)$ gravity, we consider the flat FLRW metric \eqref{eq:chap2:flrw}. In dealing with non-metricity-based gravity, we establish the coincident gauge by employing the diffeomorphisms gauge equality. Given that we cannot select any specific lapse function, this flexibility is allowed by special cases in non-metricity-based theories, where the non-metricity scalar maintains a residual invariance for reparametrizing time. Therefore, the choice of lapse function $N(t) = 1$ is made to use such symmetry.

Among the two arbitrary functions of $Q$, namely $f_1$ and $f_2$, we analytically derived the functional form of one by setting the other, ensuring that both functions together replicate the $\Lambda$CDM Universe. When analytically solving for the functional forms, we initially set $f_2$ in power-law and logarithmic forms and constructed the $f_1$ function accordingly. Then, a specific choice of $f_1$ is made, and $f_2$ is reconstructed. The former choice involves fixing the coupling term and reconstructing the purely geometric part, while the latter method fixes the geometric component and finds the non-minimal potential.

In cosmographical analysis, the $f_1$ function is expressed in Taylor series form, with the coefficients computed at the present value. These coefficients depend on the cosmographic parameters $q_0$, $j_0$, and $s_0$. The free parameters are constrained using the statistical procedures of MCMC, which utilized 1701 data points of the Pantheon+SH0ES sample. The $H_0$ value calibrated from MCMC is slightly higher than that of the concordance model, consistent with results obtained in \cite{Mandal:2020buf}. The present value of $\Omega_{m0}$ aligns perfectly with recent observations. Through cosmography, we constructed the functional form of $f_1$ and integrated the results of MCMC with the analytic solutions of $f_1$ for both power-law and logarithmic cases by determining the suitable values of arbitrary constants.

In all, we have successfully reconstructed $\Lambda$CDM Universe in the framework of non-minimally coupled $f(Q)$ gravity. Since this gravity theory has been recently proposed, there are ample opportunities for future exploration to further investigate various scenarios within this framework and explore new possibilities at the theoretical level.

\newpage
\thispagestyle{empty}

\vspace*{\fill}
\begin{center}
    {\Huge \color{NavyBlue} \textbf{CHAPTER 4}}\\
    \
    \\
    {\Large\color{purple}\textsc{\textbf{Anisotropic Cosmological Model in \texorpdfstring{$f(R,\mathcal{L}_m)$}{f(R,L m)} Gravity}}}\\
    \ 
    \\
    \textbf{Publication based on this chapter}\\
\end{center}
\textsc{Constraining Anisotropic Cosmological Model in $f(R, L_m)$ Gravity}, \textbf{NS Kavya}, V Venkatesha, S Mandal, PK Sahoo, \textit{Physics of the Dark Universe} \textbf{38}, 101126, 2022 (Elsevier, Q1, IF - 5.5) DOI: \href{https://doi.org/10.1016/j.dark.2022.101126}{10.1016/j.dark.2022.101126}
\vspace*{\fill}

\pagebreak

\def\baselinestretch{1}
\chapter{\textsc{Anisotropic Cosmological Model in \texorpdfstring{$f(R,\mathcal{L}_m)$}{f(R,L m)} Gravity}}\label{chap4}
\def\baselinestretch{1.5}
\setstretch{1.5}
\pagestyle{fancy}
\fancyhead[R]{\textit{Chapter 4}}
\markboth{Anisotropic Cosmological Model in $f(R,\mathcal{L}_m)$ Gravity }{}

\textbf{Highlights:}
{\textit{
\begin{itemize}
    \item The observational evidence regarding the present cosmological aspects tells us about the presence of very little anisotropy in the Universe on a large scale.
    \item Here,  we attempt to study locally rotationally symmetric (LRS) homogeneous Bianchi-I spacetime with the isotropic matter distribution.
    \item This is done within the framework of $f(R,\mathcal{L}_m)$ gravity. Particularly, we consider a non-linear $f(R,\mathcal{L}_m)$ model, $f(R,\mathcal{L}_m)=\dfrac{1}{2}R+\mathcal{L}_m^{\,{\alpha}}$. 
    \item Furthermore, $w$, the EoS parameter, which is vital stuff in determining the present phase of the Universe is constrained.
    \item To constrain the model parameters and the EoS parameter, we use 57 Hubble data points and 1048 Pantheon supernovae type Ia data samples. 
    \item  Moreover, with the help of obtained values of parameters, we measure the anisotropy parameter for our model.
\end{itemize}}}

\section{Introduction}\label{chap4sec1}
		\par Over the past few decades, many scientific explorations have been taking place to decipher the mystic behavior of the Universe. Right from the early time inflation to the late time acceleration, from the black holes to the wormholes, from the DE to the gravitational waves, their entire course has been probing the very nature of the Universe. Just to look into the cosmological principle, the Universe on a large scale, was presumed to be both isotropic and homogeneous. But in 1992, Cosmic Background Explorer (COBE) successfully made a significant assertion about the existence of a small anisotropy in the large-scale cosmic microwave background \cite{Bennett:1996ce}. Moreover, in the later years, this was further supported by the measurements made by Balloon Observations of Millimetric Extragalactic Radiation and Geophysics (BOOMERanG) \cite{Boomerang:2000egs}, Cosmic Background Imager (CBI) \cite{Mason:2002tm}, Wilkinson Microwave Anisotropy Probe (WMAP) \cite{WMAP:2012nax}, and the Plank collaborations \cite{Planck:2018vyg}. Furthermore, intriguing advancements in the field of cosmology took place through the observational results of the two teams led by Perlmutter and Riess \cite{SupernovaSearchTeam:1998fmf,SupernovaCosmologyProject:1997czu}. These studies strive to endorse that the Universe is currently in the phase of accelerated expansion. To this point, there arose a question regarding the isotropic nature of the expansion of the Universe. Interestingly, recent developments suggest that the Universe tends to expand at a different rate in different directions \cite{Migkas:2020fza}. Though FLRW cosmology is the most successful, it is built based on cosmological principles. However, the observational evidence attempts to elucidate the presence of a slight difference in the strengths of microwaves coming from different axes. For this reason, the spacetime that can appropriately describe anisotropic and homogeneous geometry is Bianchi cosmology. Several works on such Bianchi cosmology with different modified gravity frameworks can be found in the literature. (See ref \cite{Amirhashchi:2017kbb,Bali:2008zzc,Koussour:2022wbi,Saha:2011azb,Akarsu:2009bhh,Shamir:2015rva,Fadragas:2013ina,HamaniDaouda:2011ydu,Rodrigues:2012qua,Rodrigues:2014xam, Capozziello:2002rd,Pradhan:2010dm})
	
		\par In the present scenario, to deal with the study of such aspects, the modified theoretic approach sounds more potent. Among these, the $f(R)$ theory of gravity has produced a reliable framework for evaluating the current cosmic evolution \cite{Carroll:2003wy}. Indeed, $f(R)$ theories can adequately explain the interpretations of late-time acceleration \cite{Capozziello:2006uv,Capozziello:2002rd}, the exclusion of the DM entity in the analysis of the dynamics of massive test particles \cite{Starobinsky:1980te}, and the unification of inflation with DE \cite{Nojiri:2007as}. Furthermore, numerous justifications indicate that higher-order theories, like $f(R)$ gravity, are capable of explaining the flatness of galaxies' rotational curves \cite{Capozziello:2004us}. With these motivations, several coupling theories came into existence \cite{Harko:2011kv, Houndjo:2016rlg, Odintsov:2013iba}. One such theory is the  $f(R,\mathcal{L}_m)$ theory of gravity \cite{Harko:2010mv}. Notably, this favors the occurrence of an extra force that is orthogonal to four-velocity. In addition, the so-called `extra' force accounts for the non-geodesic motions of the test particle. Consequently, a violation of the equivalence principle can be observed. Numerous contributions to this theory can be seen in the literature \cite{Bertolami:2006js,Bertolami:2007gv,Harko:2008qz,Harko:2010zi,Wu:2014yya,Goncalves:2021vim,Carvalho:2019ert,Jaybhaye:2021jwn,Lobo:2022aop}. Recently, Jaybhaye et al have studied cosmology in $f(R,\mathcal{L}_m)$ gravity \cite{Jaybhaye:2022gxq}.
		
		\par In the present chapter, we center on the study of theoretical exploration and observational validation of the LRS Bianchi type I spacetime and effectuate this in terms of $f(R,\mathcal{L}_m)$  formalism. Moreover, in assessing the expanding Universe the EoS parameter plays a prominent role. This predicts the fluid type in spacetime. In our work, we emphasize constraining this cosmological parameter $w$ and obtaining the best-fit values as per the observational measurements. This is accomplished with a statistical approach for incorporated sets of data samples. We use two types of data samples such as Hubble measurements and Pantheon SNe Ia sample. Further, with the anisotropy parameter, we measure anisotropy in spacetime. 
		
	This chapter is structured as follows: in  \autoref{chap4sec2}, the basic formulation of $f(R,\mathcal{L}_m)$ gravity is presented. The analysis of LRS Bianchi I within the framework of the $f(R,\mathcal{L}_m)$ gravity is made in  \autoref{chap4sec3}. The \autoref{chap4sec4} is brought with the examination of observational constraints and discussion of results. Finally, the last  \autoref{chap4sec5}, gives some concluding remarks. 

\section[Geometric Formulation of \texorpdfstring{$f(R,\mathcal{L}_m)$}{f(R,L m)} Gravity]{Geometric Formulation of $f(R,\mathcal{L}_m)$ Gravity}\label{chap4sec2}			
			
		\par With the matter lagrangian density $\mathcal{L}_m$ and the Ricci scalar $R$, the action integral for $ f(R,\mathcal{L}_m)$ theory reads,
		\begin{equation}\label{chap4:eq:action}
			S=\int f(R,\mathcal{L}_m)	\sqrt{-g}\, d^4x,
		\end{equation}
		where $f$ represents an arbitrary function of $R$ and $\mathcal{L}_m$. 
		
		\par The field equation for the $f(R,\mathcal{L}_m)$ gravity \cite{Harko:2010mv}, obtained by varying the action integral \eqref{chap4:eq:action} with respect to the metric tensor $g_{\mu\nu}$ is given by, 
	   \begin{equation}\label{chap4:eq:fieldequation1}
			\begin{split}
				(g_{\mu\nu}\nabla_\mu\nabla^{\mu}-\nabla_\mu\nabla_\nu)f_R(R,\mathcal{L}_m)-\dfrac{1}{2}\left[f(R,\mathcal{L}_m)- f_{\mathcal{L}_m}(R,\mathcal{L}_m)\mathcal{L}_m\right]g_{\mu\nu}\\+f_R(R,\mathcal{L}_m)R_{\mu\nu}=\dfrac{1}{2}f_{\mathcal{L}_m}(R,\mathcal{L}_m)T_{\mu\nu}.
			\end{split}
		\end{equation}
		Here, $f_R(R,\mathcal{L}_m)\equiv\frac{\partial f(R,\mathcal{L}_m)}{\partial R}$, $f_{\mathcal{L}_m}(R,\mathcal{L}_m)\equiv\frac{\partial f(R,\mathcal{L}_m)}{\partial \mathcal{L}_m}$. 
	From the explicit form of the field equation \eqref{chap4:eq:fieldequation1}, the covariant divergence of energy-momentum tensor $T_{\mu\nu}$ can be obtained as, 
		\begin{equation}
			\nabla^\mu T_{\mu\nu}=2\left\lbrace \nabla^\mu \text{ln}\left[f_{\mathcal{L}_m}(R,\mathcal{L}_m) \right]\right\rbrace \dfrac{\partial \mathcal{L}_m }{\partial g^{\mu\nu}}. 
		\end{equation}
		\par Furthermore, on contracting the field equation \eqref{chap4:eq:fieldequation1} we get,
		\begin{equation}\label{chap4:eq:traceoffieldequation}
			\begin{split}
				3\nabla_\mu\nabla^{\mu}f_R(R,\mathcal{L}_m)+f_R(R,\mathcal{L}_m)R-2\left[f(R,\mathcal{L}_m)\right.\left. -f_{\mathcal{L}_m}(R,\mathcal{L}_m)\mathcal{L}_m\right]=\dfrac{1}{2}f_{\mathcal{L}_m}(R,\mathcal{L}_m)T.
			\end{split}
		\end{equation}
		
		\section[LRS Bianchi-I Cosmology]{LRS Bianchi-I Cosmology in $f(R,\mathcal{L}_m)$ Gravity}\label{chap4sec3}
  
		\par For anisotropic and spatially homogeneous LRS Bianchi-I spacetime, the metric is described by, 
		\begin{equation}
			ds^2=-dt^2+\mathcal{A}^2(t)\;dx^2+\mathcal{B}^2(t)\;dy^2+\mathcal{B}^2(t)\;dz^2,
		\end{equation}
		where $\mathcal{A}$ and $\mathcal{B}$ are metric potentials that are the functions of (cosmic) time $t$ alone. If $\mathcal{A}(t)=\mathcal{B}(t)=a(t)$, then one can analyze the scenarios in flat FLRW spacetime.
		Now, the Ricci scalar for LRS Bianchi-I spacetime can be expressed as,
		\begin{equation}
			R=2\left[\dfrac{\ddot{\mathcal{A}}}{\mathcal{A}}+\dfrac{2\ddot{\mathcal{B}}}{\mathcal{B}}+\dfrac{2\dot{\mathcal{A}}\dot{\mathcal{B}}}{\mathcal{A}\mathcal{B}}+\dfrac{\dot{\mathcal{B}}^2}{\mathcal{B}^2}\right].
		\end{equation}
		
		With the directional Hubble parameters $H_x$, $H_y$, and $H_z$, the Ricci scalar for the corresponding metric is given by,
		
		\begin{equation}
			R=2(\dot{H}_x+2\dot{H}_y)+2(H_x^2+3H_y^2)+4H_xH_y.
		\end{equation}
		
		Here, $H_x=\frac{\dot{\mathcal{A}}}{\mathcal{A}}$ and $H_y= \frac{\dot{\mathcal{B}}}{\mathcal{B}}=H_z$ indicate the directional Hubble parameters along the corresponding coordinate axes. For $H_x=H_y=H$, i.e., for FLRW cosmology, the equation $R=2(\dot{H}_x+2\dot{H}_y)+2(H_x^2+3H_y^2)+4H_xH_y$ reduces to $R=6(2 H^2+\dot{H})$. In the present work, we are supposing the matter distribution to be described by the energy-momentum tensor of a perfect fluid \eqref{eq:chap1:isotropicT},
	Thus, the field equation \eqref{chap4:eq:fieldequation1} takes the form,
		
		\begin{align}
		\label{chap4:eq:fe1}\begin{split}
			-\dot{f}_R(H_{x}+2 H_{y})+(\dot{H}_{x}(t)+2 \dot{H}_{y}+H_{x}^{2}+2 H_{y}^{2})\, f_R-\dfrac{1}{2}\left(f-f_{\mathcal{L}_m} \mathcal{L}_m\right)=-\dfrac{\rho f_{\mathcal{L}_m}}{2},&
		\end{split}\\
		\label{chap4:eq:fe2}\begin{split}
			-\ddot{f}_R-2 H_{y}\, \dot{f}_R+(\dot{H}_{x}+H_{x}^{2}+2 H_{x} H_{y}) f_R-\dfrac{1}{2}\left(f-f_{\mathcal{L}_m} \mathcal{L}_m\right)=\dfrac{p f_{\mathcal{L}_m}}{2},&
		\end{split}\\
		\label{chap4:eq:fe3}\begin{split}
			-\ddot{f}_R-2 H_{x}\, \dot{f}_R+(\dot{H}_{y}+2 H_{x} H_{y}) f_R-\dfrac{1}{2}\left(f-f_{\mathcal{L}_m} \mathcal{L}_m\right)=\dfrac{p f_{\mathcal{L}_m}}{2}.&
		\end{split}
		\end{align}
		The dot $(\cdot)$ here represents the derivative with respect to the time $t$ and $f\equiv f(R,\mathcal{L}_m)$.
		
		\par Further, one can express the spatial volume $\mathcal{V}$ of the spacetime as,  
		\begin{equation}
			\mathcal{V}=a^3=AB^2.
		\end{equation}
		Thus the mean value of the Hubble parameter is given by,
		\begin{align}
			H=\dfrac{\dot{a}}{a}=\dfrac{1}{3}(H_x+2H_y).
		\end{align} 
	In further study, we are going to investigate the physical cosmological model and its application in the context of $f(R,\mathcal{L}_m)$ gravity using the above set of equations.	
    
    \section{\texorpdfstring{$f(R,\mathcal{L}_m)$}{f(R,L m)} Gravity Model}
		\par In the present section, we shall focus on the cosmological aspects of $f(R,\mathcal{L}_m)$ theory, with the relation between $R$ and $\mathcal{L}_m$ being
 		\begin{equation}
			f(R,\mathcal{L}_m)=\dfrac{1}{2}R+\mathcal{L}_m^{\,{\alpha}}	,
		\end{equation}
		where $\alpha\ne 0$ is a model parameter and one can retain GR for $\alpha=1$.
		
		\par Now, to find an exact solution to the field equations \eqref{chap4:eq:fe1}-\eqref{chap4:eq:fe3}, we have to consider the constraining relation. To this point, we shall presume the anisotropic relation that can be written in terms of shear $(\sigma)$ and expansion scalar $(\theta)$ as, 
		$$\theta^2 \propto \sigma^2, $$
		so that, for constant $\frac{\sigma}{\theta}$, the Hubble expansion can achieve isotropy \cite{De:2022shr,Koussour:2022nsc}. This condition gives rise to
		\begin{equation}
		    \mathcal{A}(t)=\mathcal{B}(t)^n,
		\end{equation}
		for some real non-zero $n$, and for $n=1$, we can retrieve flat FLRW cosmology. With this, one can get the relation between directional Hubble parameters as, 
		\begin{equation}\label{chap4:eq:hxnhy}
		    H_x=n H_y.
		\end{equation}
		Therefore, the averaged Hubble parameter can take the form,
		\begin{equation}\label{chap4:eq:hhy}
		    H=\dfrac{n+2}{3} H_y.
		\end{equation}
		Additionally, we shall relate the the pressure $p$ and energy density $\rho$ by,
		\begin{equation}
			p=w\rho.
		\end{equation}
		Now, we have two choices for the Lagrangian to proceed further such as $\mathcal{L}_m=-\rho$ or $\mathcal{L}_m=p$ \cite{Bertolami:2008ab}. But, in our study, we consider $\mathcal{L}_m=-\rho$ because it is the most adequate choice presented in \cite{Bertolami:2008ab}. In literature, there are many studies that have explored the choices for $\mathcal{L}_m$ and their applications [to see more details please check the references therein \cite{Bertolami:2008ab,Faraoni:2009rk}.
		
		Applying the above conditions, the field equations \eqref{chap4:eq:fe1}-\eqref{chap4:eq:fe3} become,
  
		\begin{align}
			\dfrac{9(2n+1) }{(n+2)^{2}}H^{2}&=-(-\rho)^\alpha,\\
			\dfrac{6\dot{H}}{(n+2)}+\dfrac{27 H^2}{(n+2)^2}&=-(-\rho)^\alpha\left[\alpha(1+w)+1\right],\\
			3\dot{H}\left(\dfrac{n+1}{n+2}\right)+9H^2\dfrac{n^2+3}{(n+2)^2}&= - (-\rho)^\alpha \left[\alpha(1+w)+1\right].
		\end{align}
		
		\par With the help of the aforementioned equations we can obtain an expression for the Hubble parameter $H$ in terms of redshift $z$ as,
		\begin{equation}
		    H(z)=\gamma_1\left(z+1\right)^{\gamma_1/\gamma_2},
		\end{equation}
		where, $\gamma_1=2(n+2)$ and $\gamma_2= -3[\alpha(w+1)(2n+1)+2(n-1)]$. Here, we used the scale factor $a(t)$ and redshift relation, which is given by
		\begin{equation}
		a(t)=\frac{1}{1+z}.
		\end{equation}
		
		We now aim to constrain the model parameters ($\alpha,\,\, n$) and cosmological parameters ($w$) using various observational measurements. Doing this helps us to present a physically realistic cosmological model, which can obey the astrophysical observations.
		\section[Observational Constraints]{Observational Constraints}\label{chap4sec4}
		\par So far, we looked into the formulation of LRS Bianchi I cosmology in $f(R,\mathcal{L}_m)$ gravity. It is to be noted that the substantial validation of the values for the parameters is significant in analyzing the cosmological aspects. In this regard, the present section puts forth the observational interpretations of the current scenario. The statistical technique we adopted assists us to constrain the parameters such as $w$, $\alpha$, and $n$. In particular, we have opted for the MCMC with a standard Bayesian technique. To achieve this, we center on two datasets, namely, $H(z)$ and Pantheon data. To proceed further, we consider the priors on parameters, which are $(-2.0<w<1.0)$ to keep in mind all the possible cases of the EoS parameter, $(-10<\alpha<+ 10)$ as it is a free model parameter, and $(0<n<100)$ to measure the anisotropy and also to keep in mind that our model should fit the observational dataset as well.

    \subsection{Anisotropy Parameter}

        \begin{table}
            \setstretch{1.5}
		    \centering
            \caption{Marginalized constrained data of the parameters $w$, $\alpha$ and $n$ and corresponding anisotropy measure $\Delta$ for different data samples with 68\% confidence level.}
		    \begin{tabular*}{\linewidth}{@{\extracolsep{\fill}}c c c c c}
                \hline\hline
				Dataset & $w$ & $\alpha$ & $n$ & $\Delta$\\ 
				\hline
				$H(z)$ & $-1.245^{+0.034}_{-0.031}$ & $5.57^{+0.61}_{-0.85}$ & $26.73\pm 0.36$ & $1.604\pm 0.005$\\
				Pantheon & $-1.175^{+0.030}_{-0.020}$  & $6.4\pm 0.93$ & $31.09\pm0.54$ & $1.626^{+0.003}_{-0.004}$\\
				Pantheon+$H(z)$ & $-1.281^{+0.052}_{-0.037}$ & $4.76^{+0.68}_{-0.84}$ & $28.65\pm 0.29$ & $1.654^{+0.005}_{-0.006}$\\
                \hline\hline\\
			\end{tabular*}
            \label{chap4:tab:mcmcresult}
		\end{table}

        \par It is known that the Universe is expanding at a different pace in different directions. This certainly leads to the anisotropy in the geometric structure of spacetime. A physical quantity that measures the amount of anisotropy that arises due to the expanding Universe is the anisotropic parameter. Mathematically, this can be expressed as, 
        \begin{equation}
    		\Delta=\dfrac{1}{3}\sum\limits_{i=1}^3\left(\dfrac{H_i-H}{H} \right)^2.
		\end{equation}
        For the present problem, this takes the form 
        \begin{equation}
		\Delta=\dfrac{2}{9H^2}\left(H_x-H_y\right)^2.
		\end{equation}
        From \eqref{chap4:eq:hxnhy} and \eqref{chap4:eq:hhy} we have, 
        \begin{equation}
		\Delta=2\left(\dfrac{n-1}{n+2}\right)^2.
		\end{equation}
        In the previous section, from the distinct observational datasets, we have obtained the value for n. Now, the corresponding value of the anisotropy measure $\Delta$ is given in the  \autoref{chap4:tab:mcmcresult}.

		\subsection{Results}
		\par
\begin{figure}[H]
			\centering
			\includegraphics[width=0.7\linewidth]{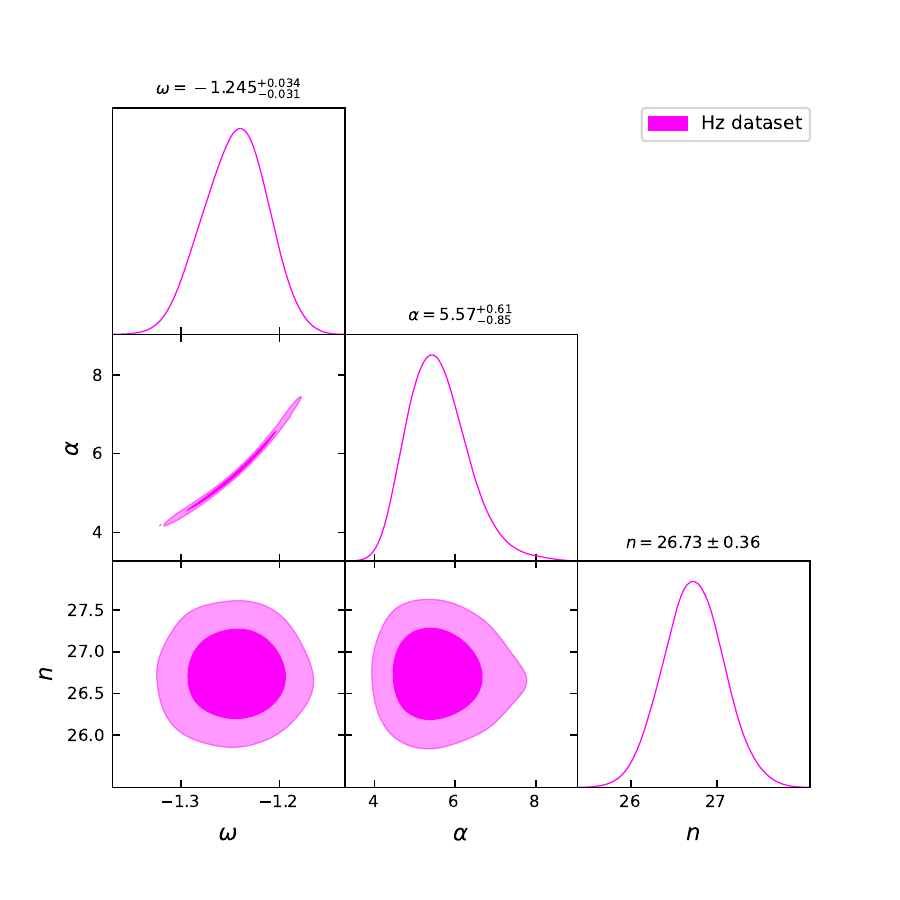}
			\caption{Contour plot with $1-\sigma$ and $2-\sigma$ errors for the parameters $w$, $\alpha$ and $n$ along with the constraint values for Hubble dataset.}
			\label{chap4:fig:kh}
		\end{figure}

\begin{figure}[H]
			\centering
			\includegraphics[width=\linewidth]{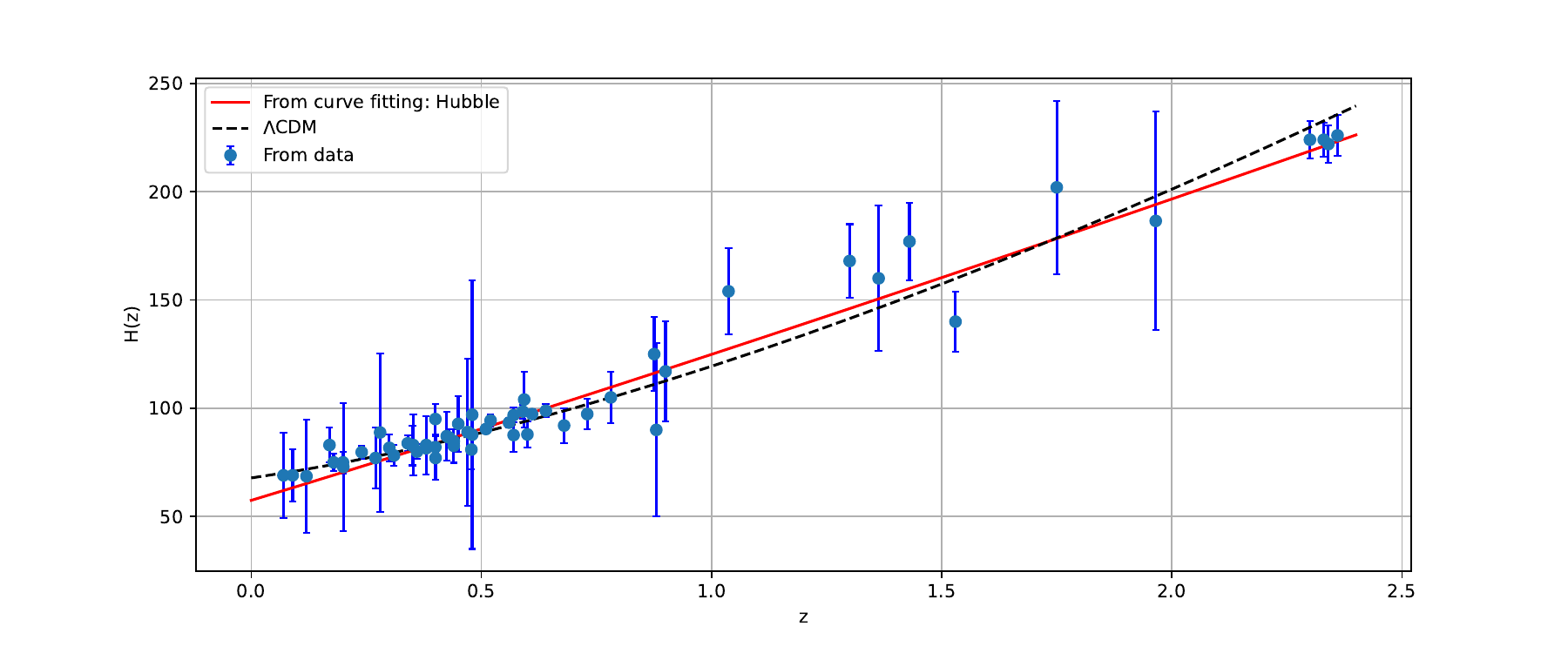}
			\caption{The profile of Hubble parameter versus redshift $z$. The line in red shows the curve for the model and the dotted line in black represents the $\Lambda$CDM model with $\Omega_{m0}$, $\Omega_{\Lambda 0}$ having the values 0.3 and 0.7 respectively. The dots with error bars in blue depicts 57 $H(z)$ sample points.
			}
			
			\label{chap4:fig:khf}
		\end{figure}

\begin{figure}[H]
			\centering
			\includegraphics[width=0.7\linewidth]{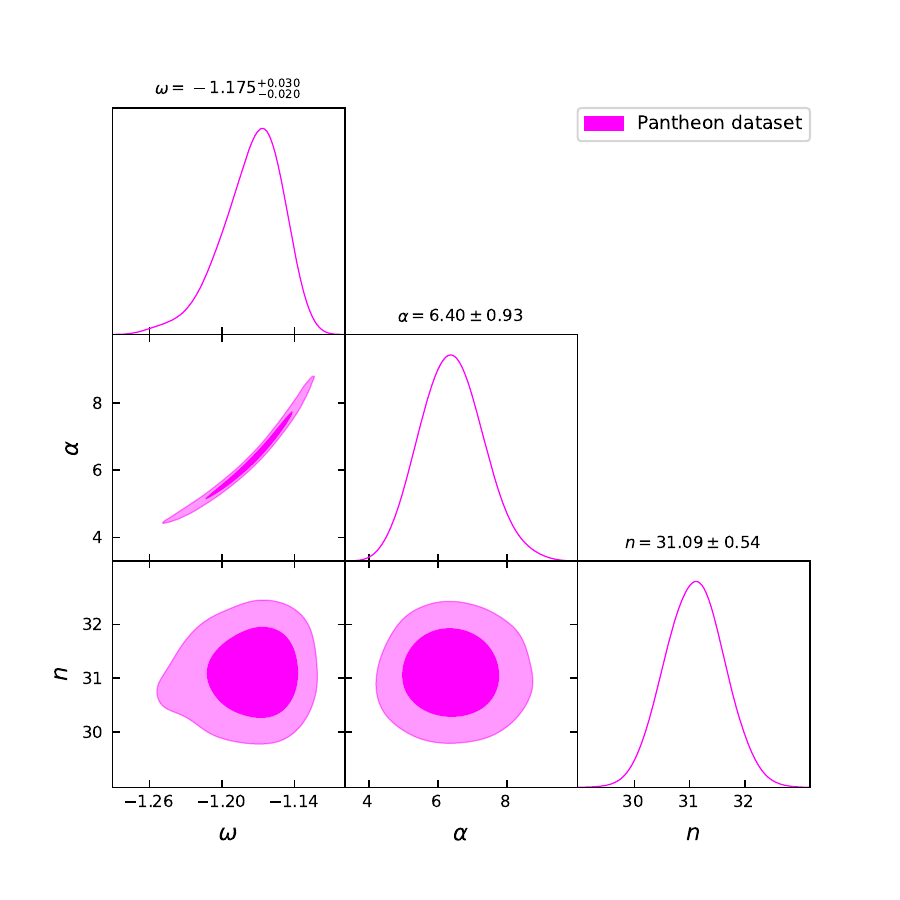}
			\caption{Contour plot with $1-\sigma$ and $2-\sigma$ errors for the parameters $w$, $\alpha$ and $n$ along with the constraint values for pantheon dataset.}
			\label{chap4:fig:kp}
		\end{figure}

\begin{figure}[H]
			\centering
	   	\includegraphics[width=1.0\linewidth]{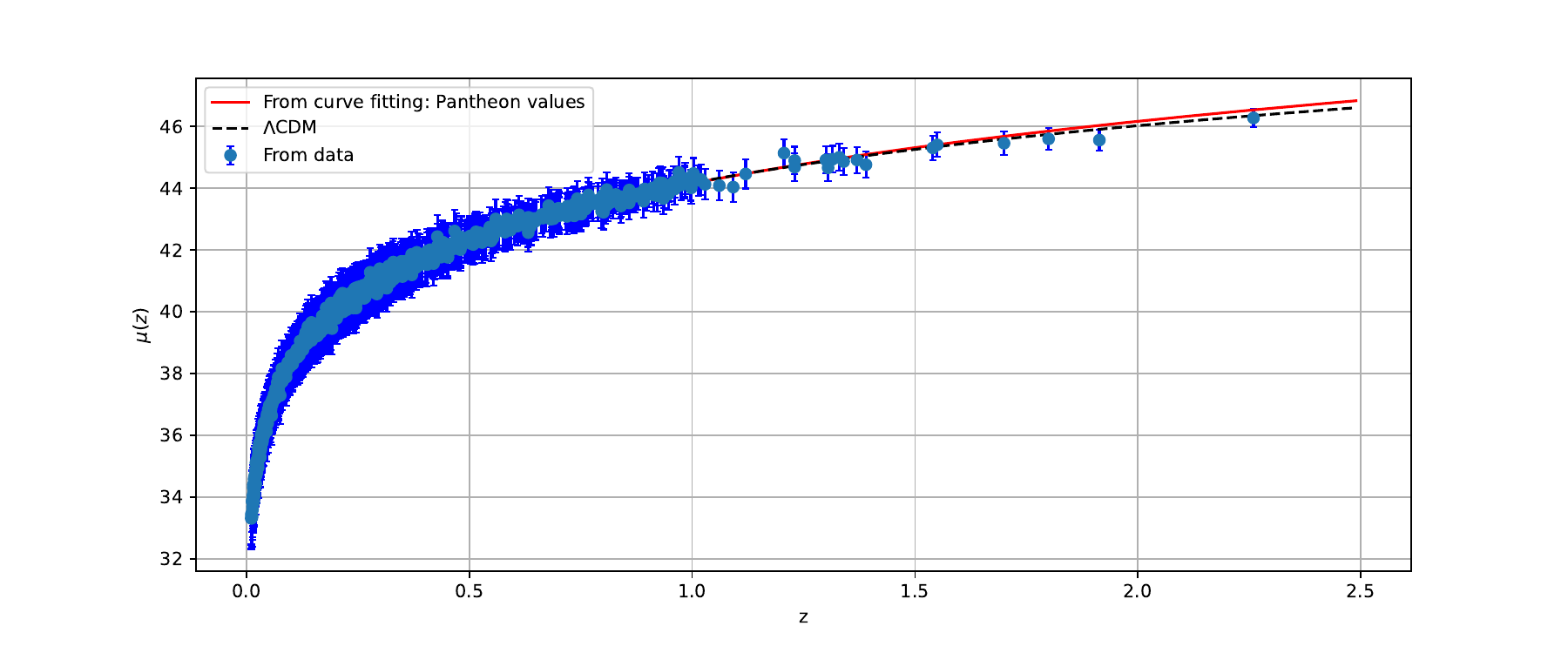}
			\caption{The profile of distance modulus versus redshift $z$. The line in red shows the curve for the model and the dotted line in black represents the $\Lambda$CDM model with $\Omega_{m0}$, $\Omega_{\Lambda 0}$ with the values 0.3 and 0.7, respectively. The dots with error bars in blue depict 1048 Pantheon sample points.}
			\label{chap4:fig:kpf}
		\end{figure}

\begin{figure}[H]
			\centering
			\includegraphics[width=0.7\linewidth]{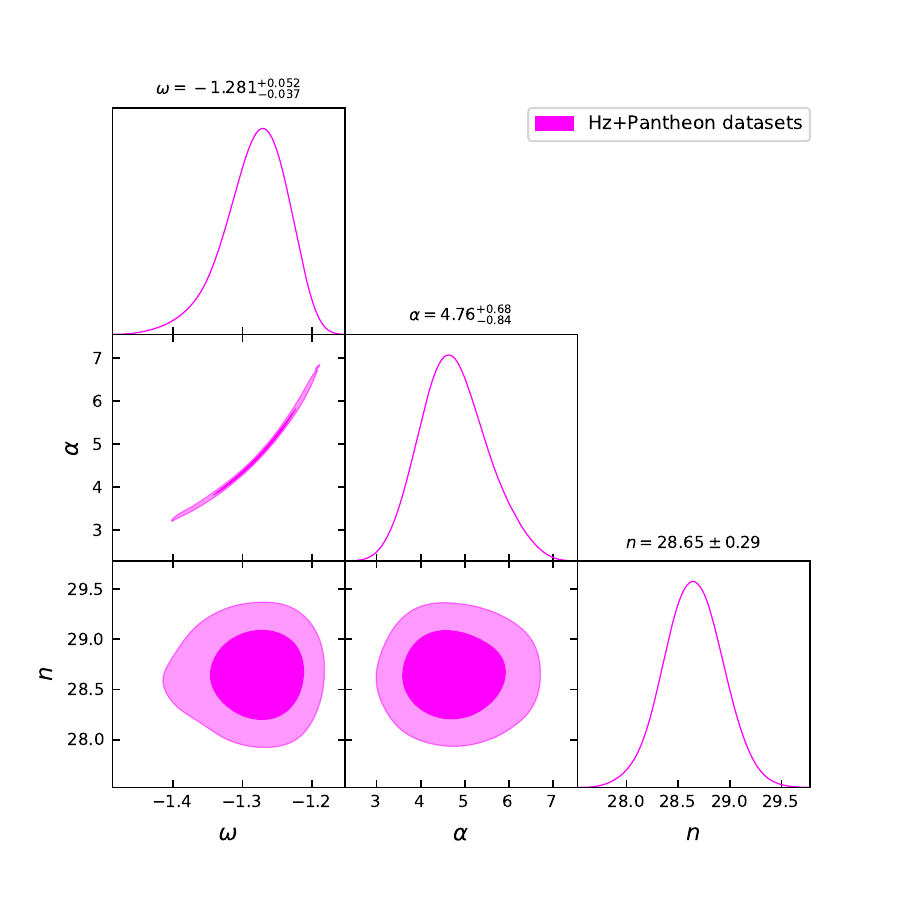}
			
			\caption{Contour plot with $1-\sigma$ and $2-\sigma$ errors for the parameters $w$, $\alpha$ and $n$ along with the constraint values for $H(z)+$Pantheon dataset.}
			\label{chap4:fig:Kcombine}
		\end{figure}

 Heretofore, we have checked over different data samples and have obtained the constraint values for the unknown parameters $w$, $\alpha$, and $n$. Further, we obtained the two-dimensional likelihood contours with $1-\sigma$ and $2-\sigma$ errors that are equipped with 68\% and 95\% confidence levels for Hubble, Pantheon, and Hubble+Pantheon data samples. These are depicted in \autoref{chap4:fig:kh}, \ref{chap4:fig:kp}, and \ref{chap4:fig:Kcombine}, respectively. First, we considered the $H(z)$ dataset with 57 data points. Here, for the model parameter $\alpha$, we have obtained the value $5.57^{+0.61}_{-0.85}$ and for the parameter $n$, which gives the relation between the directional Hubble parameters, the constrained value turns out to be $26.73\pm 0.36$. Next, for the SNe Ia Pantheon data sets with 1048 sample points, it yields, $\alpha=6.4\pm 0.93$ and $n=31.09\pm0.54$. Finally, for combined data sets in the last section, they attain the values, $\alpha=4.76^{+0.68}_{-0.84}$ and $n=28.65\pm 0.29$. Along with these, to compare our model with the $\Lambda$CDM model, we checked the Hubble parameter, $H(z)$ profile and distance modulus, $\mu(z)$ profile with the constraint values of unknown parameters $w$, $\alpha$, and $n$ for $H(z)$ and pantheon samples and illustrated subsequently in \autoref{chap4:fig:khf} and \ref{chap4:fig:kpf}. It is observed that, for both cases, our $f(R,\mathcal{L}_m)$ model fits nicely with the observational results. Moreover, it is also seen that our model is quite close to the $\Lambda$CDM model's profile. Moreover, it is well-known that the EoS parameter $w$ also plays a crucial role in describing the different energy-dominated evolution processes of the Universe. The present scenario of the Universe can be predicted by either quintessence phase $\left(-1<w<-\frac{1}{3}\right)$ or phantom phase $(w<-1)$. Now, for the present model, we found $w=-1.245^{+0.034}_{-0.031}$, $w=-1.175^{+0.030}_{-0.020}$ and $w=-1.281^{+0.052}_{-0.037}$, for $H(z)$, Pantheon, and $H(z)$+ Pantheon samples, respectively. Our results on $w$ align with the outputs of some observational studies (please see \cite{Pan-STARRS1:2017jku,Scolnic:2013efb,WMAP:2008ydk}). It is worthy to mention here that, our $f(R,\mathcal{L}_m)$ model admits phantom behavior for each data analysis. The constrained values so obtained are summarized in  \autoref{chap4:tab:mcmcresult}.
		
		\section{Concluding Remarks}\label{chap4sec5}
		\par The never-ending curiosity of the scientific community about the current cosmological aspects fosters us to look into the Universe beyond the standard gravity models. In this direction, $f(R,\mathcal{L}_m)$ formalism works pretty well. In the present article, we investigated the accelerated expansion of the Universe in the realm of $f(R,\mathcal{L}_m)$ gravity. In particular, we adopted a non-linear $f(R,\mathcal{L}_m)$ model $f(R,\mathcal{L}_m)=\dfrac{1}{2}R+\mathcal{L}_m^{\,{\alpha}}$. Further, in this, we focused on Bianchi I cosmology which is locally rotationally symmetric. Also, we considered expansion and shear scalar to vary proportionally which can lead to the isotropization of hubble expansion. 
        \par Then, to find the constraint values for the parameters we used the statistical MCMC approach with the Bayesian technique. Further, we analyzed the result for two different observational samples such as Hubble data and Pantheon data.  Furthermore, the EoS parameter, which is significant in explaining the behavior of the Universe, has been constrained. The constraint value so-obtained for $w$ ($-1.245^{+0.034}_{-0.031}$,  $-1.175^{+0.030}_{-0.020}$, $ -1.281^{+0.052}_{-0.037}$), suggests the phantom behavior of the Universe. In addition, with these values for the parameters, we compared our model with the $\Lambda$CDM model.
        
        \par Together with this, we can correlate the obtained outcomes with the existing results to assess the present aspects of the Universe. Besides, in dealing with a modified theoretic approach, to discuss these scenarios, we use cosmographic treatments and observational constraints. The usage of the former technique has led to numerous interesting investigations within the framework of several modified theories. For instance, appraising the cosmographic parameters such as the deceleration parameter, and EoS parameter with their present data helps us to examine the cosmic evolution \cite{Capozziello:2014zda,Capozziello:2019cav,Capozziello:2011hj}. Also, the latter approach of observational studies has been extensively done over the past few years \cite{Lazkoz:2019sjl,Mandal:2021bpd,Anagnostopoulos:2021ydo,Ayuso:2020dcu,Atayde:2021pgb}. Moreover, to focus on the EoS parameter $w$, we can see numerous works with a fixed value of $w$, say 1/3, 0, -1/3 so on, depending on the fluid dominated in the spacetime. Interestingly, in our work, such supposition for the $w$ value has not been admitted. Instead, more advantageously, its value has been constrained against observational results. As per the obtained outcome for $w$, we can infer the cosmic acceleration.
        
        \par Finally, for our model, we examined the nature of anisotropy with the aid of the anisotropy parameter. The anisotropy measure for $H(z)$, pantheon, and $H(z)$+pantheon is found as $1.604\pm 0.005$, $1.626^{+0.003}_{-0.004}$ and $1.654^{+0.005}_{-0.006}$ respectively. This model, under all the assumptions made, predicts an anisotropy that is in agreement with the dataset used.
        \par In all, these results could motivate us to further explore the studies in $f(R,\mathcal{L}_m)$ theory as this obeys the observational data.

\newpage
\thispagestyle{empty}

\vspace*{\fill}
\begin{center}
    {\Huge \color{NavyBlue} \textbf{CHAPTER 5}}\\
    \
    \\
    {\Large\color{purple}\textsc{\textbf{Wormhole Solutions in \texorpdfstring{$f(Q,T)$}{f(Q,T)} gravity with Conformal Symmetry}}}\\
    \ 
    \\
    \textbf{Publication based on this chapter}\\
\end{center}
\textsc{Probing the Existence of Wormhole Solutions in $f(Q,T)$ Gravity with Conformal Symmetry}, \textbf{NS Kavya}, G Mustafa, V Venkatesha, \textit{Annals of Physics} \textbf{468}, 169723, 2024 (Elsevier, Q1, IF - 3) DOI: \href{https://doi.org/10.1016/j.aop.2024.169723}{10.1016/j.aop.2024.169723}
\vspace*{\fill}

\pagebreak

\def\baselinestretch{1}
\chapter{\textsc{Wormhole Solutions in \texorpdfstring{$f(Q,T)$}{f(Q,T)} gravity with Conformal Symmetry}}\label{chap5}
\def\baselinestretch{1.5}
\pagestyle{fancy}
\fancyhead[R]{\textit{Chapter 5}}
\textbf{Highlights:}
{\textit{
\begin{itemize}
    \item This chapter presents novel and physically plausible wormhole solutions within the framework of $f(Q,T)$ gravity theory, incorporating conformal symmetries. 
    \item The investigation explores the feasibility of traversable wormholes under diverse scenarios, considering traceless, anisotropic, and barotropic equations of state (EoS). 
    \item Additionally, the influence of model parameters on the existence and characteristics of these wormhole structures is thoroughly examined. 
    \item Notably, the derived shape function satisfies all the necessary criteria. 
    \item Furthermore, in one of the cases, the presence of non-exotic fluid is confirmed, while in others, exotic matter is only required near the wormhole throat.
\end{itemize}}}

\section{Introduction}
Wormholes are unique geometric structures in the realm of theoretical physics, representing classical or quantum solutions to gravitational field equations. They offer the possibility of connecting two distant regions through a bridge-like pathway. In classical terms, they can be likened to instantons, representing tunneling events between asymptotic manifolds. A seminal work by Morris and Thorne \cite{Morris:1988cz} unveiled a critical insight into wormholes, indicating that their geometry can only be a viable solution to the Einstein equation if the energy-momentum tensor violates the null energy condition (NEC). This violation necessitates the presence of exotic matter, characterized by properties that challenge established laws of physics. However, the NEC, stated as $T^{ab}k_a k_b > 0$, holds true for all known forms of matter, $k^a$ is any null vector. In light of this condition, researchers \cite{Visser:1995cc,Gao:2016bin,Caceres:2019giy,Armendariz-Picon:2002gjc,Nicolis:2009qm,Kuhfittig:1999nd} have delved into different approaches to create wormhole-like spacetime, either by accommodating ordinary matter or by reducing the need for exotic matter, while still supporting the violation of effective NEC. One of the approaches for generating negative energy density (and therefore violating NEC) is through the Casimir effect. In \cite{Casimir:1948dh}, the author demonstrated that the Casimir effect could be employed to stabilize traversable wormholes. To achieve this, the placement of two sufficiently charged superconducting spheres at the mouths of the wormhole is sufficient. This innovative method offers a potential pathway to overcome the exotic matter requirement and provides possibilities for traversable wormholes.

Another compelling avenue for addressing the issue of the exotic matter lies in the realm of Modified theories. In the literature, extensive research has been conducted on various aspects of traversable wormhole geometries in the context of modified gravitational theories \cite{Boehmer:2012uyw,Lobo:2009ip,Lobo:2020kxn,Harko:2013yb,Kanti:2011jz,Usmani:2010cd,Rahaman:2005qi,Rahaman:2012pg,Zubair:2016cde,Ovgun:2018fnk,Mustafa:2021vqz,Capozziello:2018mqy,Capozziello:2020ncr,Capozziello:2020zbx,Capozziello:2022zoz,Capozziello:2012hr,Kuhfittig:2013hva,Kuhfittig:2005gy,Nada:2020jay,Venkatesha:2023zkm,Hassan:2022hcb,Kavya:2023mwv,Kavya:2023tjf,Chalavadi:2023sif,Chalavadi:2023zcw}. In recent years, in the context of $f(Q)$ gravity, numerous solutions have been extensively explored in both cosmological and astrophysical scenarios \cite{Capozziello:2022wgl,Lazkoz:2019sjl,Anagnostopoulos:2021ydo,Mustafa:2021ykn,Wang:2021zaz}. A viable extension of $f(Q)$ gravity known as $f(Q,T)$ was proposed in \cite{Xu:2019sbp}, that couples the nonmetricity $Q$ and the trace of the energy-momentum tensor $T$. This introduces a novel perspective on gravitational interactions and their implications. In this chapter, we investigate wormhole solutions within this framework of $f(Q,T)$ gravity. 

Furthermore, in our investigation, to establish a natural correlation between geometry and matter with the governing equations, we use inheritance symmetry that serves as a valuable technique \cite{Boehmer:2007md,Caceres:2019giy,Mustafa:2023hlu,Boehmer:2007rm,Mars:1994td,Maartens:1996gd,Harko:2004ui,Mak:2004hv,Kuhfittig:1999nd,Kavya:2023lms}. The conformal symmetry arises from a set of Conformal Killing Vectors (CKVs). Rahaman et al. \cite{Rahaman:2014dpa} highlight the significance of this mathematical technique, citing its ability to provide a deeper understanding of spacetime geometry and facilitate the derivation of exact solutions to the Einstein field equations in more comprehensive forms. Additionally, the study of this specific symmetry in spacetime holds great physical importance as it plays a crucial role in discovering conservation laws and devising spacetime classification schemes. Moreover, owing to the highly non-linearity of the Einstein field equations, CKVs enable the conversion of partial differential equations to ordinary differential equations, simplifying the mathematical analysis of the system. In the context of GR, these find numerous applications in geometric configurations, kinematics, and dynamics based on structural theory. They are used to explore unexpected expansion and shear phenomena.

This chapter aims to address the fundamental issue concerning the physical viability of a traversable wormhole within the framework of GR, the requirement of exotic fluids. We explore solutions in the context of $f(Q,T)$ gravity theory, seeking to find physically plausible wormholes that do not necessitate the presence of exotic matter. By investigating various scenarios and considering conformal symmetries, we strive to present wormhole solutions that offer a more realistic perspective and potentially pave the way for traversable wormholes without the need for exotic fluids.

Following is the chapter layout: \autoref{sec:chap5:II} presents the geometric perspectives of the theory, including the governing field equations, wormhole metric, and their corresponding properties and conformal killing vectors. In \autoref{sec:chap5:III}, the energy conditions are examined, and in \autoref{sec:chap5:IV}, three different wormhole solutions are presented along with their embedding diagram. In \autoref{sec:chap5:V} the equilibrium condition and anisotropy analysis are discussed. Finally, in \autoref{sec:chap5:VI} we concluded our results.
	
\section{Geometric Formulation of \texorpdfstring{$f(Q,T)$}{f(Q,T)} Gravity}\label{sec:chap5:II}

The action integral for $f(Q,T)$ theory, where $f$ is an arbitrary function of nonmetricity $Q$ and the trace of the energy-momentum tensor, can be expressed as
		\begin{equation}\label{eq:action}
			S=\int \frac{1}{2}\, f(Q,T)	\sqrt{-g}\, d^4x +S_M.
		\end{equation}
		
        Here,  $S_M=\int \mathcal{L}_m \sqrt{-g}\, d^4x$ $f(Q,T)$.
                               
     
	    

  
In addition, we have 
  
		\begin{equation} 
		    \Theta _{a b} = g^{c d} \frac{\delta T_{c d}}{\delta g^{a b}}.
		\end{equation}

        Varying the action \eqref{eq:action} with respect to $g_{ab}$ yields the following expression for the governing field equation
	   \begin{equation}\label{fd}
		     \frac{-2}{\sqrt-g} \nabla_c \left(\sqrt -g f_Q P^c _{\;\;a b}\right) - \frac{1}{2} g_{ab} f + f_T\left( T_{a b} + \Theta_{a b}\right)  - f_Q \left(P_{a c d} Q^{\;\;c d} _{b} -2Q^{c d} _{\;\;a} P_{c d b}\right) = T_{ab},
		\end{equation}
		where $f_Q = \frac{df}{dQ}$ and $f_T = \frac{df}{dT}$.

\section{Wormhole Geometry in \texorpdfstring{$f(Q,T)$}{f(Q,T)} Gravity }

The Morris-Thorne traversable wormhole metric is given by,

 \begin{equation}\label{eq:whmetric}
		ds^2=e^{2\mathcal{R}_f(r)}dt^2-\dfrac{dr^2}{1-\dfrac{S_f(r)}{r} }  - r^2\left(d\theta^2+\text{sin}^2\theta \,d\phi^2\right). 
	\end{equation}
 
    Here, both $\mathcal{R}_f$, the redshift function, and $S_f$, the shape function, are functions of the radial coordinate $r$. To ensure the absence of a spacetime horizon, the function $\mathcal{R}_f$ should remain finite throughout the entire domain. The radial coordinate $r$ extends from $r_0$, the throat radius, to $\infty$. Additionally, we introduce another function, $L_f$, the proper radial distance function, to describe the spacetime geometry. It can be expressed as
    
    \begin{equation}\label{prd}
        L_f(r)=\pm \int_{r_0}^r \sqrt{\dfrac{r}{r-S_f(r)}}dr.
    \end{equation}
    
    To achieve traversability, the shape function $S_f$ should obey the following conditions:
    \begin{itemize}
        \item The shape function $S_f$ has a fixed point at the minimum value $r_0$, meaning that $S_f(r_0) = r_0$. 
        \item  It is a monotonically increasing function and ensures that the spacetime becomes asymptotically flat, as $\frac{S_f(r)}{r}$ approaches zero for infinitely large radial coordinates.
        \item  The shape function satisfies the flaring-out condition, given by $\frac{S_f(r) - rS_f'(r)}{S_f(r)^2} > 0$. At the throat, this condition becomes $S_f'(r_0) < 1$.
        \item To maintain a finite value for the proper radial distance function $f_L$, the shape function must satisfy the inequality $S_f < r$ for all values of $r$.
    \end{itemize}

\subsection{Conformal Killing Vector (CKVs)}

Collineations encompass a variety of symmetries arising from both geometrical considerations and physically relevant quantities. Among these symmetries, CKVs stand out as particularly advantageous, offering a deeper insight into spacetime geometry. From a mathematical perspective, conformal motions or CKVs are motions along which the metric tensor of spacetime remains invariant up to a scale factor i.e., the metric can undergo a conformal mapping onto itself along the vector $\eta$ through the action of the Lie derivative operator, denoted as
\begin{equation}\label{eq:ckv}
        \mathcal{L}_{\eta}g_{ab}=g_{na}\eta^{n}_{;b}+g_{an}\eta^{n}_{;b}=\Psi(r) g_{ab}.
    \end{equation}
Here, $\Psi$, $\eta^n$, and $g_{ab}$ represent the conformal factor, CKVs, and metric tensor, respectively. These CKVs provide valuable tools for exploring the intricate relationships between geometry and physical properties within the spacetime framework. By plugging equation \eqref{eq:ckv} into \eqref{eq:whmetric} yields the following expressions

\begin{gather}
    \eta^1 \mathcal{R}_f'(r)=\frac{\psi(r)}{2},\\
    \eta^1 = \frac{r\psi(r)}{2},\\
    \eta^1 \left(-\frac{S_f(r)-rS_f'(r)}{r^2-rS_f(r)}\right)+2\eta^1_{,1}=\psi(r).
\end{gather}

As a result, from the aforementioned system of equations, we obtain

\begin{equation}\label{eq:ckvsol}
     e^{2\mathcal{R}_f(r)}=C_1 r^2,~\text{and}~
    \left(1-\frac{S_f(r)}{r}\right)^{-1}=\frac{C_2}{\psi^2(r)}.
\end{equation}

For simplicity, let us assume $\xi(r) = \psi^2(r)$. Consequently, the expression for the shape function $S_f(r)$ is then given by

\begin{equation}\label{eq:main}
    S_f(r)=r\left(1-\frac{\xi(r)}{C_2}\right).
\end{equation}

\section{Energy Conditions}\label{sec:chap5:III}

In the analysis of wormhole solutions, a notable aspect is the violation of energy conditions. Within the context of modified theories of gravity, due to the change in geometries, there is a deviation from the traditional relativistic governing equations. For this purpose, we shall consider the effective gravitational equation for the wormhole metric \eqref{eq:whmetric}, represented by

\begin{gather}
    \rho^{eff}=\dfrac{S_f'}{r^2},\\
    p_r^{eff}=2\left(1-\frac{S_f}{r}\right)\frac{\mathcal{R}_f'}{r}-\frac{S_f}{r^3},\\
    p_{t}^{eff}=\left(1-\frac{S_f}{r}\right)\Bigg[\mathcal{R}_f''+\mathcal{R}_f'^2-\dfrac{(rS_f'-S_f)\mathcal{R}_f'}{2r(r-S_f)}-\dfrac{(rS_f'-S_f)}{2r^2(r-S_f)}+\dfrac{\mathcal{R}_f'}{r}\Bigg],
\end{gather}

where primes ($'$) represent the derivative of the function with respect to $r$, $\rho$ is energy density and $p_r$ and $p_t$ are radial and tangential pressures. Basically, energy conditions are obtained as a consequence of the Raychaudhuri equation. These provide the interpretations related to physical aspects associated with the motion of energy and matter. For anisotropic matter distribution \eqref{eq:chap1:anisotropicT}

		\begin{enumerate}[label=$\circ$,leftmargin=*]
			\setlength{\itemsep}{4pt}
			\setlength{\parskip}{4pt}
			\setlength{\parsep}{4pt}
			\item \textit{Null Energy Conditions (NECs)}: $\rho+p_t\ge0$ and $\rho+p_r\ge0$.
			\item \textit{Weak Energy Conditions (WECs)}: $\rho\ge0\implies$  $\rho+p_t\ge0$ and $\rho+p_r\ge0$.
			\item \textit{Strong Energy Conditions (SECs)}: $\rho+p_j\ge0\implies$  $\rho+\sum_j p_j\ge0   \ \forall\ j$.
			\item \textit{Dominant Energy Conditions (DECs)}: $\rho\ge0\implies$ $\rho-|p_r|\ge0$ and $\rho-|p_t|\ge0$.
		\end{enumerate}

\section{Conformally Symmetric Wormhole Models}\label{sec:chap5:IV}

In this section, we shall consider a physically plausible $f(Q,T)$ gravity model, given by $ f(Q,T)=\alpha Q+\beta T$ \cite{Xu:2019sbp}. Here, $\alpha$ and $\beta$ are model parameters. For $\alpha=-1$ and $\beta=0$, the scenario reduces to STEGR. For the metric \eqref{eq:whmetric}, the nonmetricity scalar takes the form
\begin{equation}\label{eq:nonmetricity}
    Q=-\frac{S_f}{r^2} \left[ \frac{rS_f'-S_f}{r(r-S_f)} + 2 \mathcal{R}_f'\right].
\end{equation}

For the model in hand, the gravitational field equations can be expressed as

    \begin{align}
     \rho =& \frac{\alpha  \left[S_f' \left(4 \beta -5 \beta  r \mathcal{R}_f'-6\right)+5 \beta  \left(2 r (r-S_f) \mathcal{R}_f''+\mathcal{R}_f' \left(2 r (r-S_f) \mathcal{R}_f'-3 S_f+4 r\right)\right)\right]}{6 \left(2 \beta ^2+\beta -1\right) r^2},
     \end{align}
     \begin{align}
     \begin{split}
        p_r=&\frac{\alpha}{6 \left(2 \beta ^2+\beta -1\right) r^3}\left[r \left(\beta  S_f' \left(5 r \mathcal{R}_f'+8\right)+2 r \left(\mathcal{R}_f' \left(2 \beta -5 \beta  r \mathcal{R}_f'-6\right)-5 \beta  r \mathcal{R}_f''\right)\right)\right.\\&\left.+S_f \left(-12 \beta +r \left(10 \beta  r \mathcal{R}_f''+\mathcal{R}_f' \left(-9 \beta +10 \beta  r \mathcal{R}_f'+12\right)\right)+6\right)\right],   
     \end{split}
    \end{align}
     \begin{align}
    \begin{split}
        p_t=& \frac{\alpha}{6 \left(2 \beta ^2+\beta -1\right) r^3}  \left[r \left(S_f' \left(2 \beta -(\beta -3) r \mathcal{R}_f'+3\right)+2 r \left((\beta -3) r \mathcal{R}_f''+\mathcal{R}_f' \left(-4 \beta\right.\right.\right.\right.\\&\left.\left.\left.\left. +(\beta -3) r \mathcal{R}_f'-3\right)\right)\right)\right.\left.+S_f \left(6 \beta +r \left(\mathcal{R}_f' \left(9 \beta -2 (\beta -3) r \mathcal{R}_f'+3\right)\right.\right.\right.\\&\left.\left.\left.-2 (\beta -3) r \mathcal{R}_f''\right)-3\right)\right].
    \end{split}
\end{align}

Using CKVs the above set of equations can be rewritten as

\begin{gather}
     \label{eq:rho}\rho =\frac{\alpha  \left((\beta +6) r \xi'(r)+2 (8 \beta +3) \xi(r)+2 (2 \beta -3)C_2\right)}{6 (\beta +1) (2 \beta -1)C_2 r^2},\\
    \label{eq:pr}p_r=-\frac{\alpha  \left(13 \beta  r \xi'(r)+(18-8 \beta ) \xi(r)+4 \beta C_2-6C_2\right)}{6 (\beta +1) (2 \beta -1)C_2 r^2},\\
    \label{eq:pt}p_t=\frac{\alpha  \left(-(\beta +6) r \xi'(r)-2 (8 \beta +3) \xi(r)+8 \beta C_2\right)}{6 (\beta +1) (2 \beta -1)C_2 r^2}.
\end{gather}

\subsection{Case 1: Wormhole admitting traceless fluid}

In this section, we explore the dynamics of a traceless fluid \cite{Boehmer:2012uyw,Lobo:2009ip,Mustafa:2019oiy} governed by a specific EoS, namely $\rho - p_r - 2p_t= 0$. It is worth noting that in systems involving traceless fluids, the Casimir effect can also play a significant role. The Casimir effect arises from quantum field theory and gives rise to a measurable force between closely spaced parallel plates or boundaries due to the influence of vacuum fluctuations. To investigate the properties of this fluid, we substitute the expressions for the energy density \eqref{eq:rho} as well as the pressures \eqref{eq:pr} and \eqref{eq:pt} into the equation $\rho - p_r - 2p_t= 0$. This results in obtaining the differential equation that describes the behavior of the traceless fluid. It is given by,

\begin{equation}
    \frac{\alpha  \left[-(8 \beta +9) r \xi'(r)-2 (10 \beta +9) \xi(r)+2 (2 \beta +3)C_2\right]}{(\beta +1) (2 \beta -1)C_2 r}=0.
\end{equation}

To satisfy the throat condition for the shape function $S_f$, we impose the initial condition $\xi(r)=0$ using the relation \eqref{eq:main}. The solution of the above equation is obtained as,

\begin{equation}
    \xi(r)=-\frac{(2 \beta +3)C_2 (8 \beta  r+9 r)^{-\frac{20 \beta }{8 \beta +9}-\frac{18}{8 \beta +9}} \left(((8 \beta +9) r_0)^{\frac{20 \beta }{8 \beta +9}+\frac{18}{8 \beta +9}}-((8 \beta +9) r)^{\frac{20 \beta }{8 \beta +9}+\frac{18}{8 \beta +9}}\right)}{10 \beta +9}.
\end{equation}

Thus, the shape function $S_f$ takes the form,

\begin{equation}\label{eq:sf}
    S_f(r)=r-\frac{(2 \beta +3) r ((8 \beta +9) r)^{-\frac{2 (10 \beta +9)}{8 \beta +9}} \left(((8 \beta +9) r)^{\frac{20 \beta +18}{8 \beta +9}}-((8 \beta +9) r_0)^{\frac{20 \beta +18}{8 \beta +9}}\right)}{10 \beta +9}.
\end{equation}

\begin{figure}
            \centering
            \subfloat[$S_f$\label{fig:asf1}]{\includegraphics[width=0.49\linewidth]{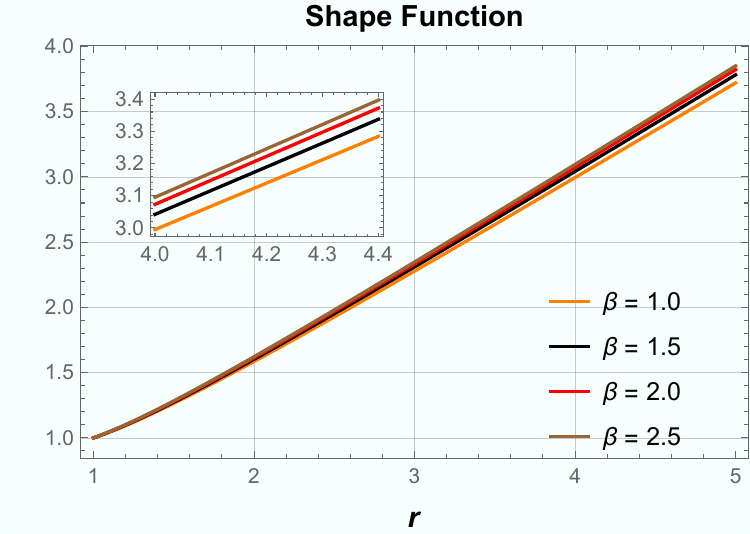}}
    	    \subfloat[$\frac{S_f}{r}$\label{fig:asf2}]{\includegraphics[width=0.49\linewidth]{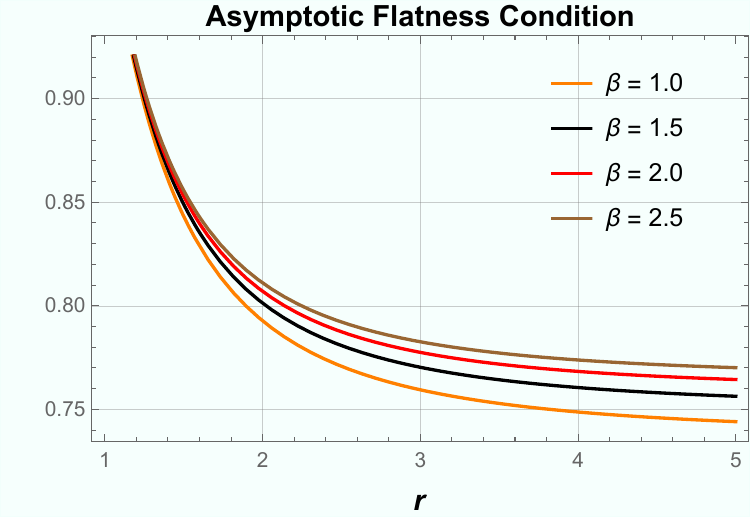}}\\
    	    \subfloat[$S_f'$\label{fig:asf3}]{\includegraphics[width=0.49\linewidth]{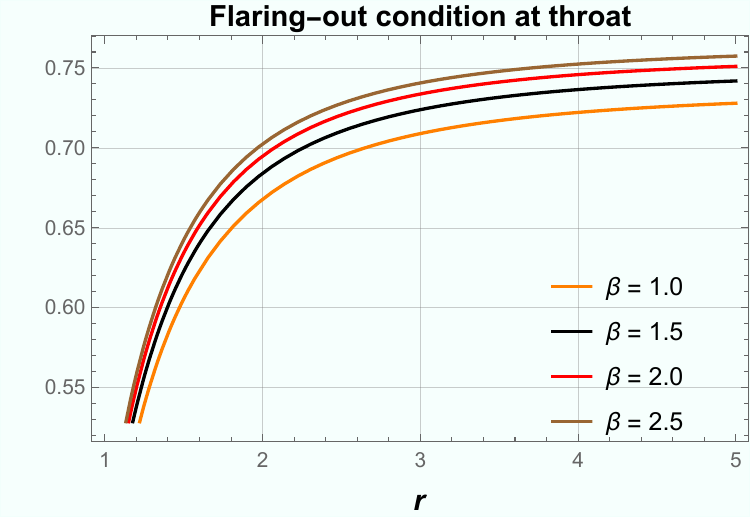}}
            \subfloat[$\frac{S_f-r S_f'}{S_f^2}$\label{fig:asf4}]{\includegraphics[width=0.49\linewidth]{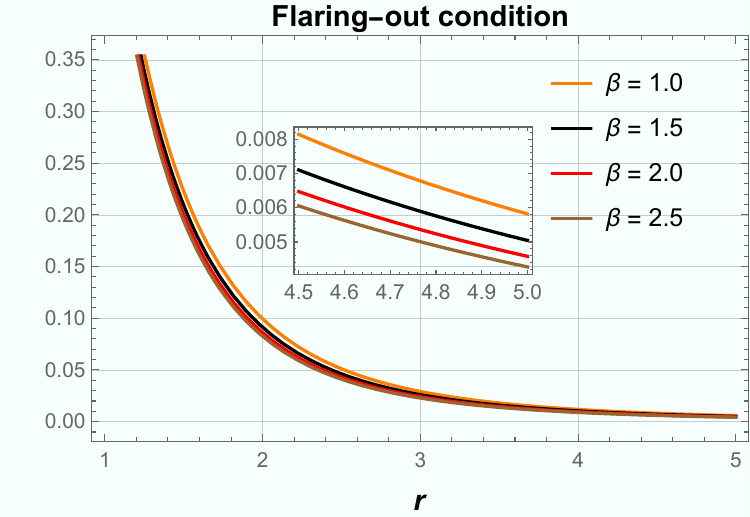}}
            \caption{Case 1: Plot showing the characteristics of $S_f$ for $\alpha=1.5$, $C_2=4$, $r_0=1$. }
            \label{fig:Gsf}
        \end{figure}

\begin{figure*}
	    \centering
	    \subfloat[ $\rho$\label{fig:arho}]{\includegraphics[width=0.49\linewidth]{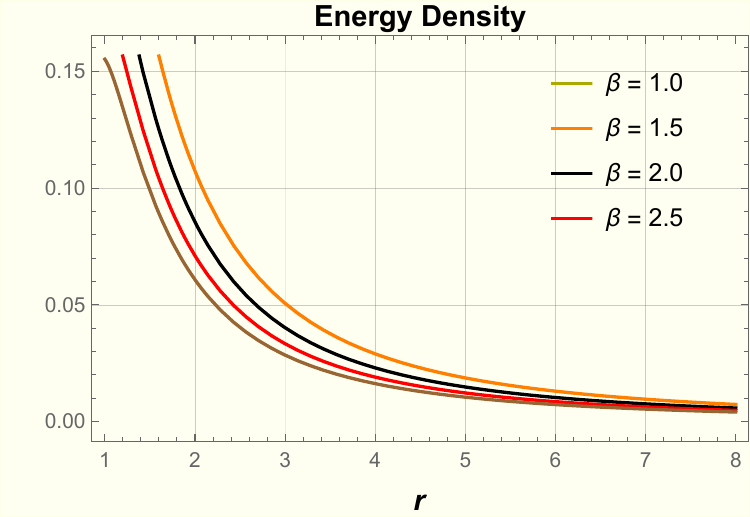}}
	    \subfloat[$\rho+p_{r}$\label{fig:ae1}]{\includegraphics[width=0.5\linewidth]{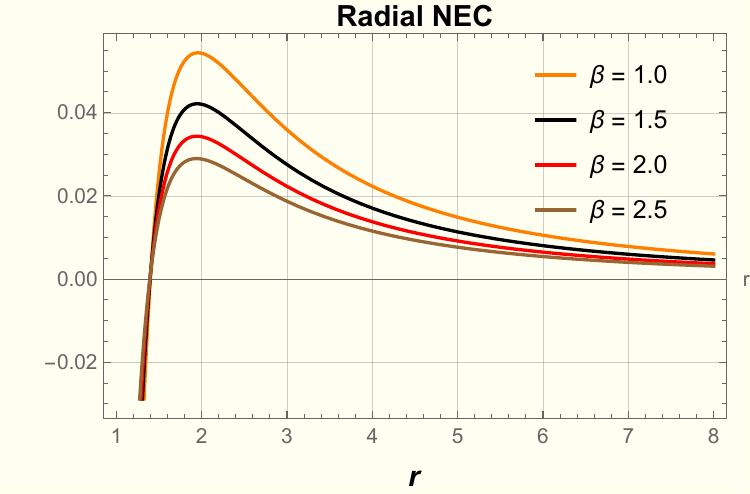}}\\
	    \subfloat[$\rho+p_t$\label{fig:ae2}]{\includegraphics[width=0.49\linewidth]{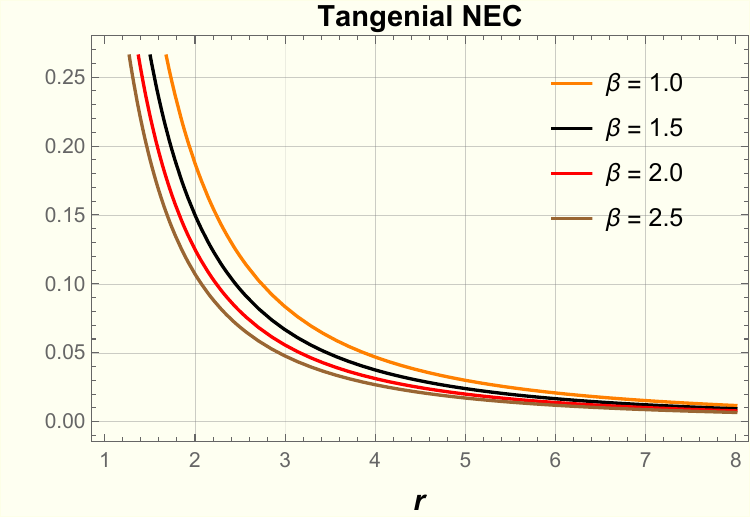}}
	    \subfloat[$\rho-|p_{r}|$\label{fig:ae3}]{\includegraphics[width=0.5\linewidth]{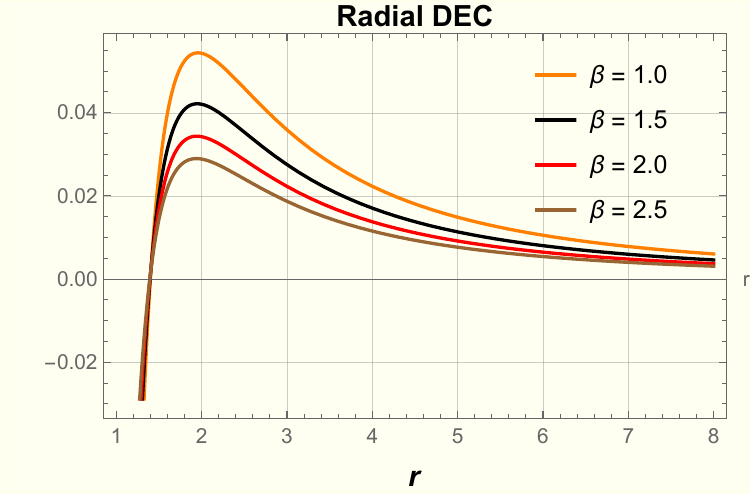}}\\
	    \subfloat[$\rho-|p_{t}|$\label{fig:ae4}]{\includegraphics[width=0.5\linewidth]{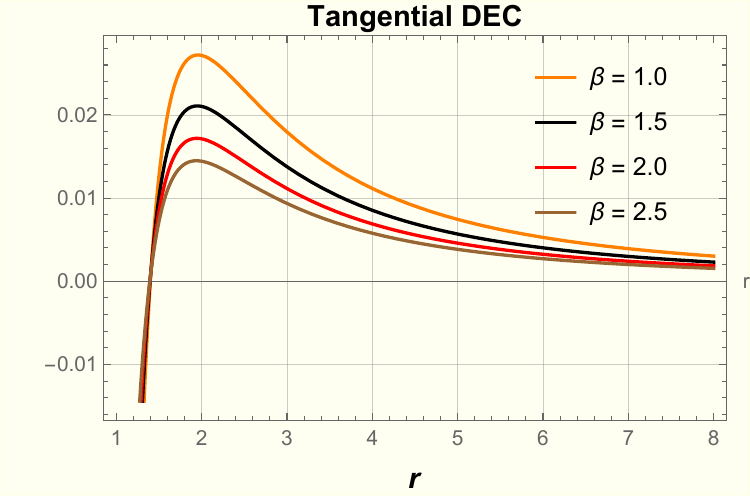}}
	    \subfloat[SEC $\rho+p_r+2p_t$\label{fig:ae5}]{\includegraphics[width=0.49\linewidth]{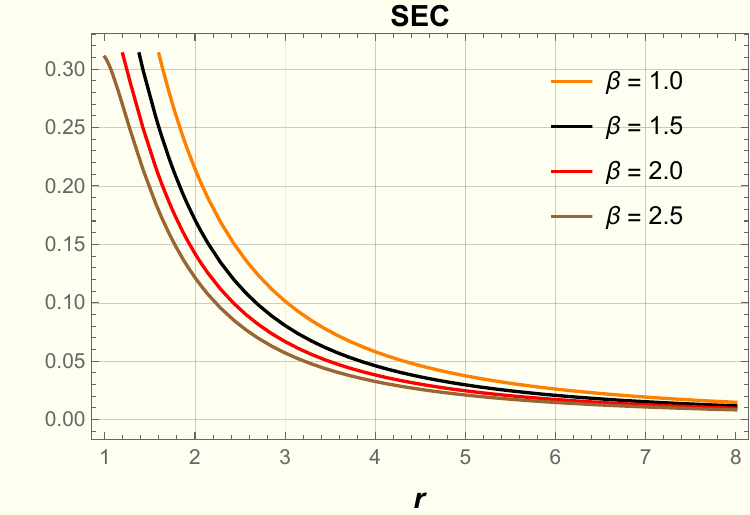}}
	    \caption{Case 1: Plot showing the profile of energy density and various ECs for $\alpha=1.5$, $C_2=4$, $r_0=1$.}
	    \label{fig:1caseec}
	\end{figure*}

This function is a monotonically increasing and $S_f(r)<r$ for $\beta\in\mathbb{R}-[-1.5,-0.75]$ [\figureautorefname~\ref{fig:asf1}]. In order to flaring-out condition $S_f'$ at the wormhole throat, $\beta$ should take the value greater than $-1.5$. \figureautorefname~\ref{fig:asf2} and \ref{fig:asf3} represent the satisfying behavior of flaring-out condition. Consequently, the effective NEC is violated. Due to CKVs, the asymptotic flatness condition cannot be achieved \cite{Rahaman:2014dpa}. The function $S_f/r$ approaches to $\frac{8 \beta +6}{10 \beta +9}$ for infinitely large $r$ [\figureautorefname~\ref{fig:asf2}]. From \eqref{eq:sf}, equations \eqref{eq:rho}, \eqref{eq:pr} and \eqref{eq:pt} can be rewritten as

\begin{gather}
    \rho=\frac{\delta\left(6 \alpha  (8 \beta +9) ((8 \beta +9) r)^{\frac{20 \beta +18}{8 \beta +9}}-9 \alpha  (2 \beta +3) ((8 \beta +9) r_0)^{\frac{20 \beta +18}{8 \beta +9}}\right)}{(10 \beta +9) r},\\
    p_r=\frac{\delta \left(-2 \alpha  \beta  (8 \beta +9) ((8 \beta +9) r)^{\frac{20 \beta +18}{8 \beta +9}}-27 \alpha  (\beta +1) (2 \beta +3) ((8 \beta +9) r_0)^{\frac{20 \beta +18}{8 \beta +9}}\right)}{(\beta +1) (10 \beta +9) r},\\
    p_t=\frac{\delta \left(\alpha  (4 \beta +3) (8 \beta +9) ((8 \beta +9) r)^{\frac{20 \beta +18}{8 \beta +9}}+9 \alpha  (\beta +1) (2 \beta +3) ((8 \beta +9) r_0)^{\frac{20 \beta +18}{8 \beta +9}}\right)}{(\beta +1) (10 \beta +9) r}.
\end{gather}
Here, $\delta = ((8 \beta +9) r)^{\frac{9}{2 (8 \beta +9)}-\frac{7}{2}}.$
The energy density remains positive throughout the domain [\figureautorefname~\ref{fig:arho}]. The tangential NEC and SEC are satisfied. The radial NEC and DECs are violated for $r\approx r_0$, whereas for $r\gg r_0$, these ECs are satisfied.  In this scenario, the violation of radial NEC indicates the requirement of hypothetical fluid at the wormhole throat.

 \subsection{Case 2: Wormhole admitting anisotropy relation}

In this section, we shall examine a scenario in which the radial and tangential pressures are connected by the equation $p_t=A p_r$. By using equations \eqref{eq:pr} and \eqref{eq:pt}, we obtain the relation given below

\begin{equation}
    \frac{\alpha  \left(r (13 A \beta -\beta -6) \xi'(r)-2 (A (4 \beta -9)+8 \beta +3) \xi(r)+4 (A+2) \beta  C_2-6 A C_2\right)}{(\beta +1) (2 \beta -1) C_2 r}=0.
\end{equation}

Solving the above differential equation for $\xi(r)$ with the initial condition $\xi(r_0)=0$ to ensure that the shape function satisfies the throat condition, we get

\begin{equation}
    \begin{split}
           \xi(r)=&\frac{1}{A (4 \beta -9)+8 \beta +3}\left[r^{-\frac{2 (A (4 \beta -9)+8 \beta +3)}{-13 A \beta +\beta +6}} r_0^{\frac{-5 A \beta -18 A+16 \beta +6}{-13 A \beta +\beta +6}}\right.\\&\left. \left((8 \beta +3) r_0^{\frac{\beta +6}{-13 A \beta +\beta +6}}-4 \beta  C_2 r_0^{\frac{13 A \beta }{-13 A \beta +\beta +6}}\right)\right.\\&\left.+A r^{-\frac{2 (A (4 \beta -9)+8 \beta +3)}{-13 A \beta +\beta +6}} \left\{(2 \beta -3) C_2 \left(r^{\frac{2 (4 A \beta -9 A+8 \beta +3)}{-13 A \beta +\beta +6}}-r_0^{\frac{2 (4 A \beta -9 A+8 \beta +3)}{-13 A \beta +\beta +6}}\right)\right.\right.\\&\left.\left.+(4 \beta -9) r_0^{\frac{-5 A \beta -18 A+17 \beta +12}{-13 A \beta +\beta +6}}\right\}+4 \beta  C_2\right].
    \end{split}
\end{equation}

From \eqref{eq:main} the expression for the shape function is given by

\begin{equation}
\begin{split}
        S_f(r)=&\frac{1}{C_2 (A (4 \beta -9)+8 \beta +3)}\left[r^{\frac{-21 A \beta +18 A-15 \beta }{-13 A \beta +\beta +6}} r_0^{\frac{-5 A \beta -18 A+16 \beta +6}{-13 A \beta +\beta +6}}\right.\\&\times\left. \left(C_2 (2 (A+2) \beta -3 A) r_0^{\frac{13 A \beta }{-13 A \beta +\beta +6}}\right.\right.\\&\left.\left.-(A (4 \beta -9)+8 \beta +3) r_0^{\frac{\beta +6}{-13 A \beta +\beta +6}}\right)+C_2 r (2 A (\beta -3)+4 \beta +3)\right].
\end{split}
\end{equation}

The derivative of this function at the wormhole throat is expressed as

\begin{equation}
    S_f'(r_0)=\frac{A (17 \beta -6)-\frac{3 r_0 (A (7 \beta -6)+5 \beta )}{C_2}+7 \beta -6}{(13 A-1) \beta -6}.
\end{equation}

\begin{figure}
            \centering
            \subfloat[$S_f$\label{fig:bsf1}]{\includegraphics[width=0.49\linewidth]{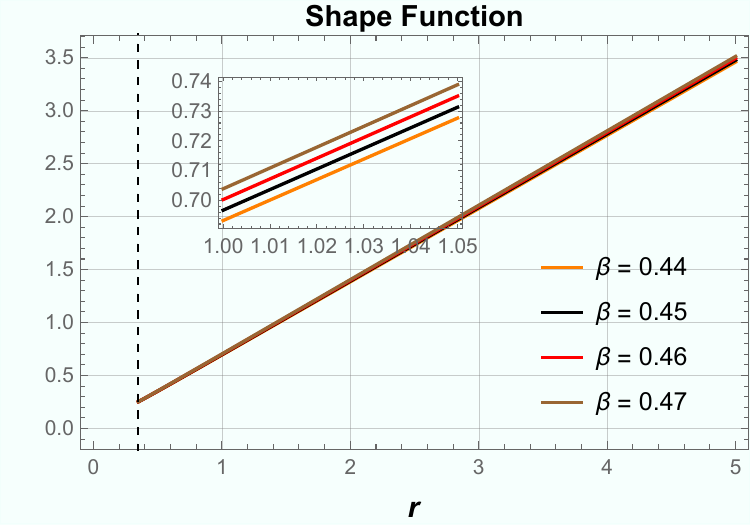}}
    	    \subfloat[$\frac{S_f}{r}$\label{fig:bsf2}]{\includegraphics[width=0.49\linewidth]{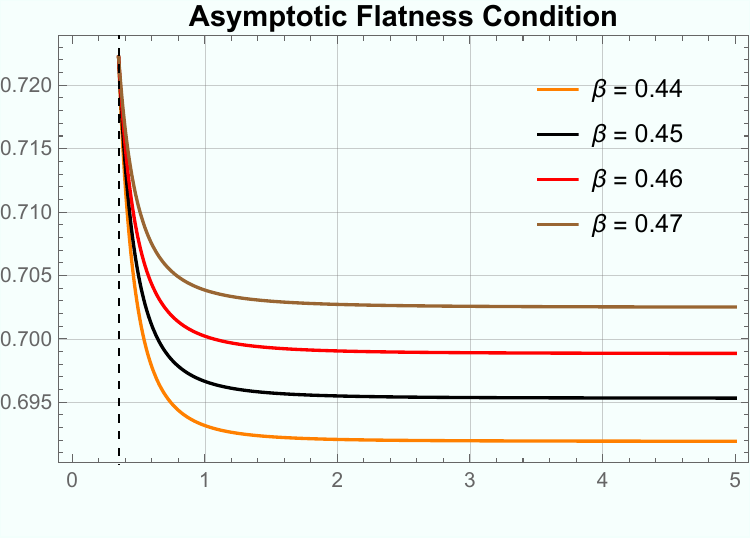}}\\
    	    \subfloat[$S_f'$\label{fig:bsf3}]{\includegraphics[width=0.49\linewidth]{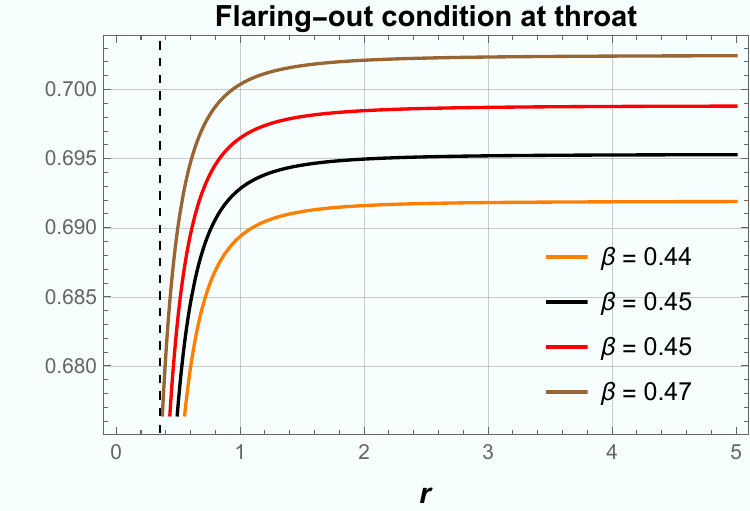}}
            \subfloat[$\frac{S_f-rS_f'}{S_f^2}$\label{chap5/fig:bsf4}]{\includegraphics[width=0.49\linewidth]{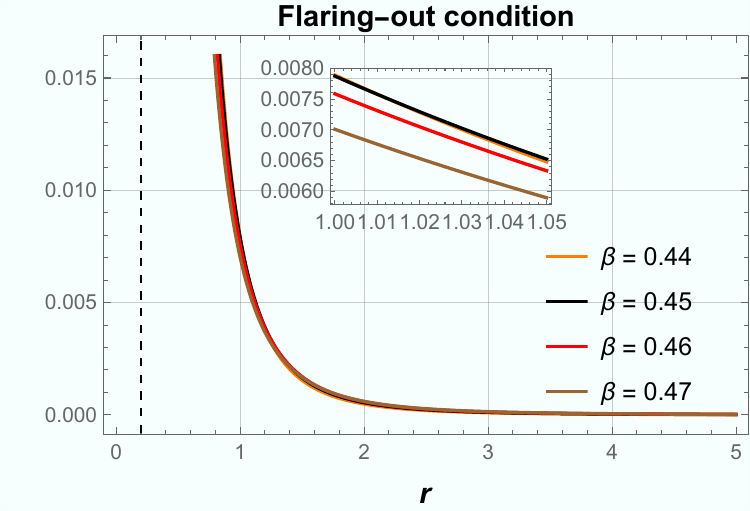}}
            \caption{Case 2: Plot showing the characteristics of $S_f$ for $r_0=0.35$, $\alpha=1.4$, $C_2=1.26$, $A=2.25$.}
            \label{fig:bsf}
        \end{figure}

To satisfy the flaring-out condition at $r_0$, the above expression should be less than 1. Consequently, we meticulously select the parameter values. By choosing $r_0=0.35$, $\alpha=1.4$, $C_2=1.26$, and $A=2.25$, the function $S_f$ becomes a monotonically increasing function and remains less than $r$ when $0.212<\beta<0.8$ [\figureautorefname~\ref{fig:bsf1}]. The condition $\frac{S_f-rS_f'}{S_f^2}>0$ is fulfilled for all $r$ within the parameter space $(0.23,0.518)$ of the model parameter $\beta$ [\figureautorefname~\ref{fig:bsf3} and \ref{chap5/fig:bsf4}]. The existence of conformal symmetry causes the function $S_f/r$ to approach the limiting value $\frac{2 A (\beta -3)+4 \beta +3}{A (4 \beta -9)+8 \beta +3}$ as $r$ tends to $\infty$. Notably, this value is less than 1 within the chosen parameter range [\figureautorefname~\ref{fig:bsf2}]. In this scenario, we can observe that the coupling constant $\beta$ has a significant influence on the existence of a viable wormhole solution.

For this solution, the energy density and pressures yields 

\begin{gather}
    \begin{split}
        \rho=&\frac{\alpha }{(\beta +1) r^2 (A (4 \beta -9)+8 \beta +3)}  \left[4 A \beta\right.\\&\left. -\frac{18 A (\beta +1) (2 (A+2) \beta -3 A) r^{-\frac{2 (A (4 \beta -9)+8 \beta +3)}{-13 A \beta +\beta +6}} r_0^{\frac{2 (A (4 \beta -9)+8 \beta +3)}{-13 A \beta +\beta +6}}}{(13 A-1) \beta -6}\right.\\&\left.-6 A+8 \beta +3\right]+\frac{18 \alpha  A r^{-\frac{18 (A-1) (\beta +1)}{(13 A-1) \beta -6}} r_0^{\frac{-5 A \beta -18 A+17 \beta +12}{-13 A \beta +\beta +6}}}{C_2 ((13 A-1) \beta -6)},
    \end{split}
\end{gather}
\begin{gather}
    \begin{split}
        p_r=&-\frac{3 \alpha }{(\beta +1) C_2 r^2 (-13 A \beta +\beta +6) (A (4 \beta -9)+8 \beta +3)}  \left[C_2 (-13 A \beta +\beta +6)\right.\\&\left.+6 (\beta +1) r^{-\frac{2 (A (4 \beta -9)+8 \beta +3)}{-13 A \beta +\beta +6}} r_0^{\frac{-5 A \beta -18 A+16 \beta +6}{-13 A \beta +\beta +6}} \left(C_2 (2 (A+2) \beta -3 A) r_0^{\frac{13 A \beta }{-13 A \beta +\beta +6}}\right.\right.\\&\left.\left.-(A (4 \beta -9)+8 \beta +3) r_0^{\frac{\beta +6}{-13 A \beta +\beta +6}}\right)\right],
    \end{split}
\end{gather}
\begin{gather}
    \begin{split}
         p_t=&\frac{\alpha }{6 (\beta +1) (2 \beta -1) C_2 r^2}  \left[\frac{18 A C_2}{((13 A-1) \beta -6) (A (4 \beta -9)+8 \beta +3)}\right.\\&\left. \left\{6 \left(2 \beta ^2+\beta -1\right) (2 (A+2) \beta -3 A)\right.\right.\\&\left.\left.\times r^{-\frac{2 (A (4 \beta -9)+8 \beta +3)}{-13 A \beta +\beta +6}} r_0^{\frac{2 (A (4 \beta -9)+8 \beta +3)}{-13 A \beta +\beta +6}}-(2 \beta -1) ((13 A-1) \beta -6)\right\}\right.\\&\left.-\frac{108 A \left(2 \beta ^2+\beta -1\right) r^{-\frac{2 (A (4 \beta -9)+8 \beta +3)}{-13 A \beta +\beta +6}} r_0^{\frac{-5 A \beta -18 A+17 \beta +12}{-13 A \beta +\beta +6}}}{(13 A-1) \beta -6}\right].
    \end{split}
\end{gather}

\begin{figure*}
	    \centering
	    \subfloat[$\rho$\label{fig:brho}]{\includegraphics[width=0.49\linewidth]{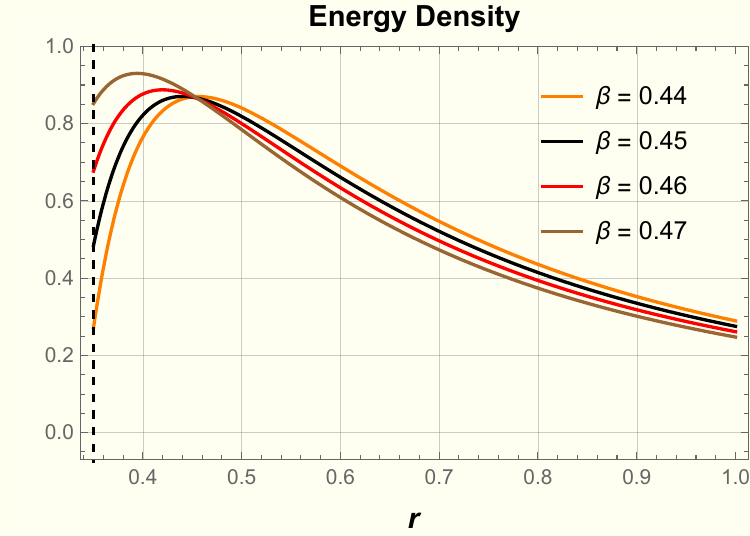}}
	    \subfloat[$\rho+p_{r}$\label{fig:be1}]{\includegraphics[width=0.49\linewidth]{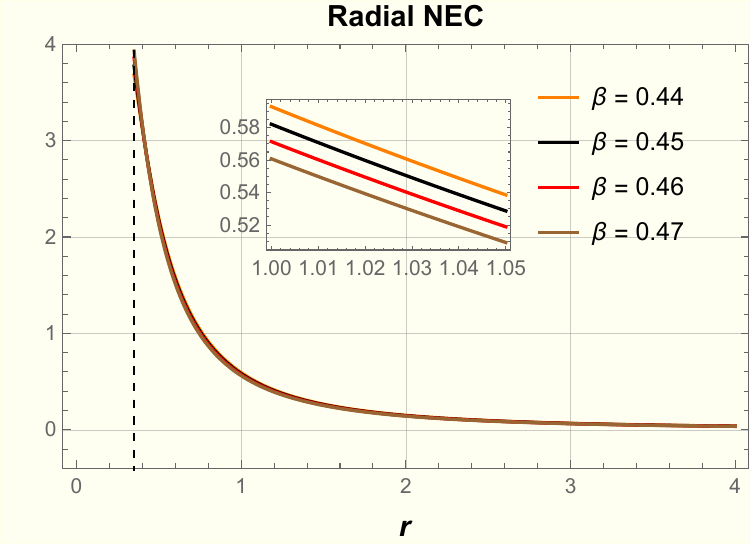}}\\
	    \subfloat[$\rho+p_{t}$\label{fig:be2}]{\includegraphics[width=0.49\linewidth]{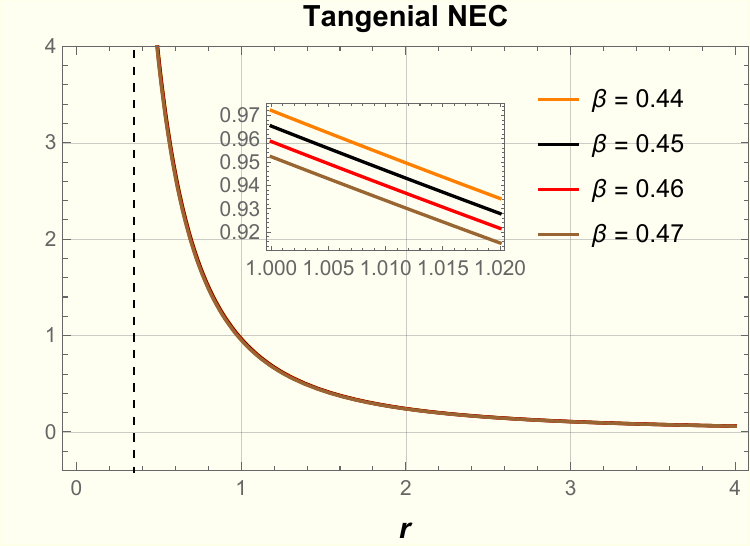}}
	    \subfloat[$\rho-|p_{r}|$\label{fig:be3}]{\includegraphics[width=0.49\linewidth]{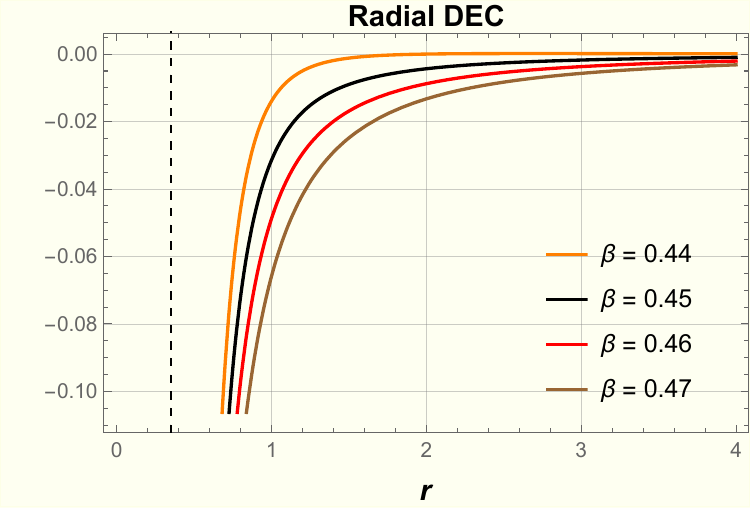}}\\
	    \subfloat[$\rho-|p_{t}|$\label{fig:be4}]{\includegraphics[width=0.49\linewidth]{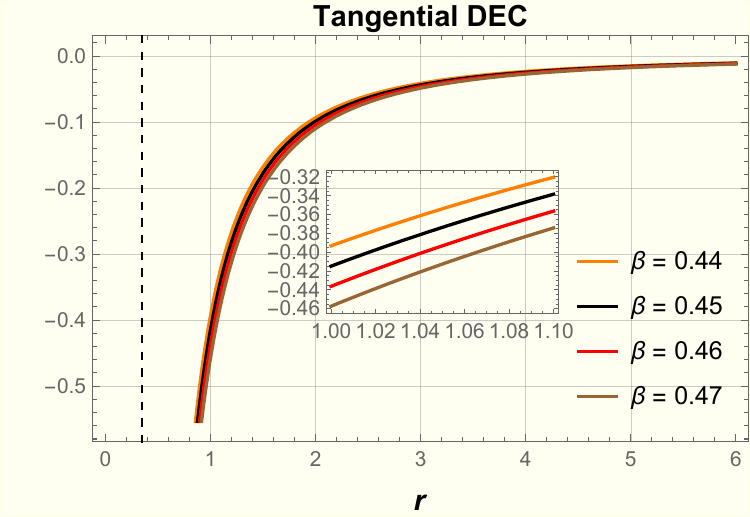}}
	    \subfloat[$\rho+p_{r}+2p_{t}$\label{fig:be5}]{\includegraphics[width=0.49\linewidth]{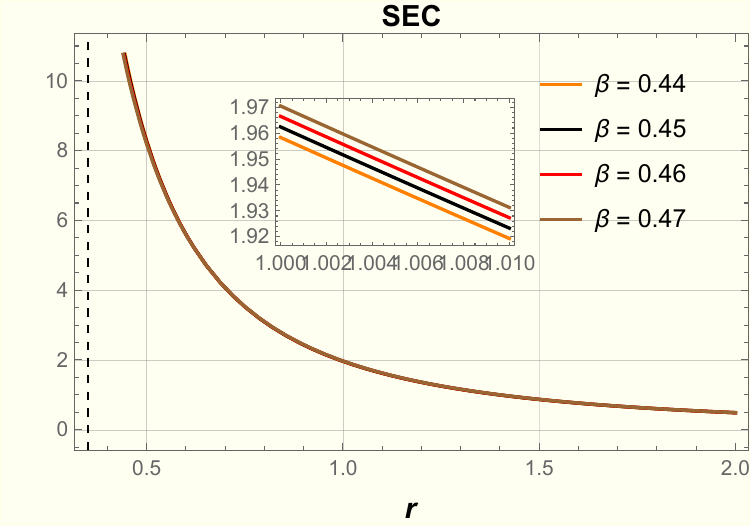}}
	    \caption{Case 2: Plot showing the profile of energy density and various ECs for $r_0=0.35$, $\alpha=1.4$, $C_2=1.26$, $A=2.25$.}
	    \label{fig:2caseec}
	\end{figure*}

For all values in the parameter space, the energy density remains positive throughout the domain [\figureautorefname~\ref{fig:brho}]. Both DECs are violated [\figureautorefname~\ref{fig:be3} and \ref{fig:be4}] and SEC is satisfied [\figureautorefname~\ref{fig:be5}]. It is to be noted that both NECs are satisfied clearly indicating the presence of non-exotic fluid [\figureautorefname~\ref{fig:be1} and \ref{fig:be2}].

 \subsection{Case 3: Wormhole admitting barotropic EoS}

In this context, we shall explore a specific situation where the EoS follows a barotropic relation, represented as $p_t = w \rho$. This equation expresses the relationship between the tangential pressure $p_t$ and the energy density $\rho$. By using equations \eqref{eq:rho} and \eqref{eq:pr}, we can obtain

 \begin{equation}
     \frac{\alpha  \left((w+1) \left((\beta +6) r \xi'(r)+2 (8 \beta +3) \xi(r)\right)+4 \beta  (w-2) C_2-6 w C_2\right)}{(\beta +1) (2 \beta -1) C_2 r}=0.
 \end{equation}

 Solving this equation for $\xi$ with initial condition $\xi(r_0)=0$, we have

\begin{equation}
    \xi(r)=-\frac{C_2 (-4 \beta +2 \beta  w-3 w) r^{-\frac{16 \beta }{\beta +6}-\frac{6}{\beta +6}} \left(r^{\frac{15 \beta }{\beta +6}+1}-r_0^{\frac{15 \beta }{\beta +6}+1}\right)}{(8 \beta +3) (w+1)}.
\end{equation}

This ensures that the wormhole solution satisfies the throat condition and the corresponding shape function is given by 

\begin{equation}
    S_f(r)=\frac{(2 \beta  (w-2)-3 w) r^{-\frac{15 \beta }{\beta +6}} \left(r^{\frac{15 \beta }{\beta +6}+1}-r_0^{\frac{15 \beta }{\beta +6}+1}\right)}{(8 \beta +3) (w+1)}+r.
\end{equation}

The constraining relation for the parameter values to satisfy the flaring-out condition at the wormhole throat is $\frac{-7 \beta +5 \beta w+6}{\beta +\beta w+6 w+6}<1$. In this case, we fix the value of the model parameter $\beta$ and study the influence of the EoS parameter $w$ on the characteristics of the wormhole solution. For $\alpha = 1.5$, $\beta=0.9$, $C_2=2$, and $r_0=1$, the shape function exhibits increasing behavior and remains positive for $w>0$ (in particular, we consider $w$ within the range $(0,1)$ to ensure physical viability) [\figureautorefname~\ref{fig:csf1}]. Throughout the radial coordinate $r$ domain, we have $r>S_f$, implying the finiteness of the proper radial distance function. The flaring-out condition is satisfied, ensuring the violation of the effective NEC, which is necessary for a traversable wormhole [see \figureautorefname~\ref{fig:csf3} and \ref{fig:csf4}]. The limiting value of the function $S_f/r$ for an infinitely large radial coordinate is $\frac{2 \beta (5 w+2)+3}{(8 \beta +3) (w+1)}$. As $w$ decreases, this limit approaches its lower value [\figureautorefname~\ref{fig:csf2}]. For the present shape function, the physical quantities \eqref{eq:rho}-\eqref{eq:pt} can be expressed as

\begin{align}
    \rho=&\frac{\alpha }{(\beta +1) (w+1) r^2},\\
    p_r=&-\frac{\alpha  r^{-\frac{18 (\beta +1)}{\beta +6}} \left((\beta +6) ((4 \beta -6) w+3) r^{\frac{16 \beta +6}{\beta +6}}-18 (\beta +1) (2 \beta  (w-2)-3 w) r_0^{\frac{16 \beta +6}{\beta +6}}\right)}{(\beta +1) (\beta +6) (8 \beta +3) (w+1)},\\
    p_t=&\frac{\alpha  w}{(\beta +1) (w+1) r^2}.
\end{align}

\begin{figure*}
	    \centering
	    \subfloat[ $\rho$\label{fig:crho}]{\includegraphics[width=0.49\linewidth]{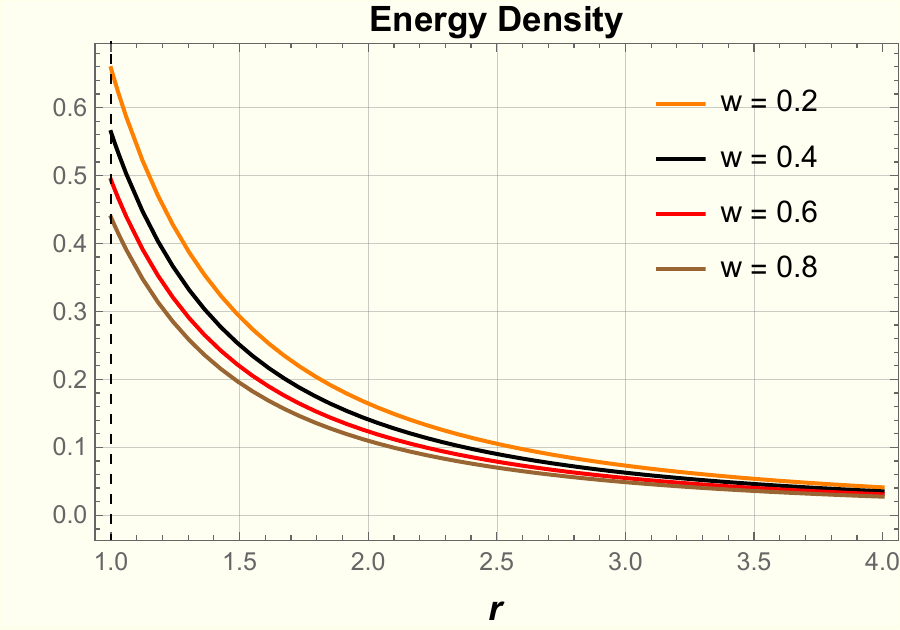}}
	    \subfloat[$\rho+p_{r}$\label{fig:ce1}]{\includegraphics[width=0.49\linewidth]{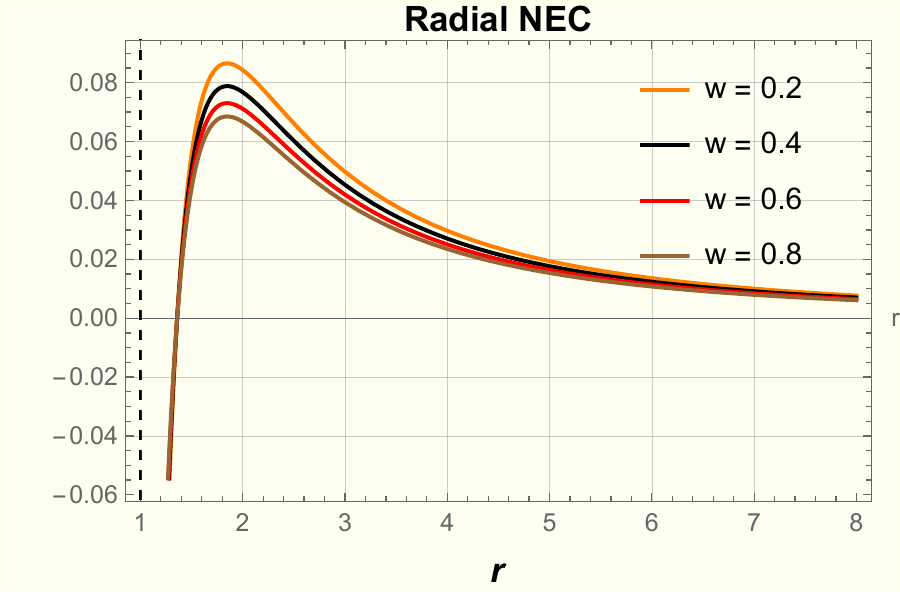}}\\
	    \subfloat[$\rho+p_{t}$\label{fig:ce2}]{\includegraphics[width=0.49\linewidth]{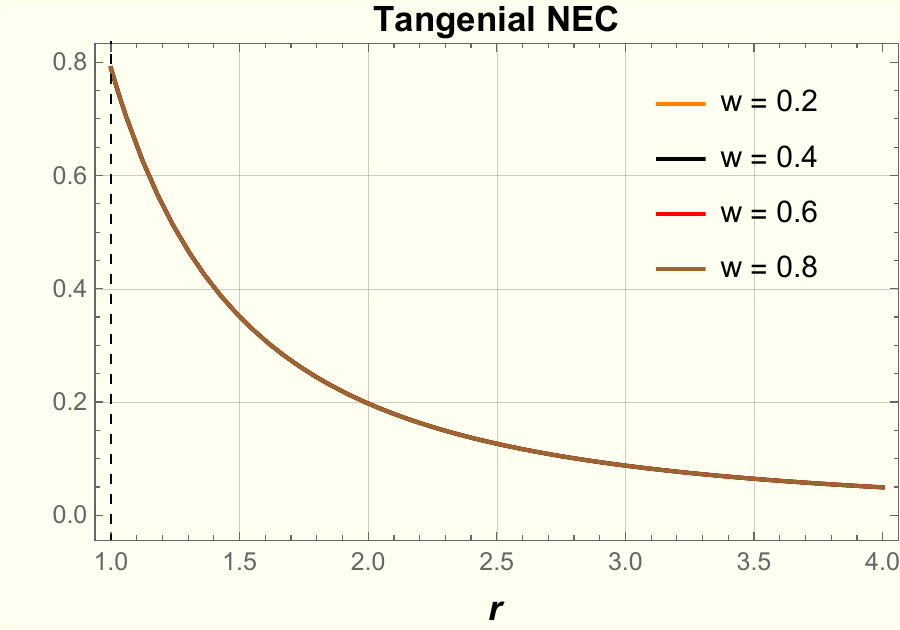}}
	    \subfloat[$\rho-|p_{r}|$\label{fig:ce3}]{\includegraphics[width=0.49\linewidth]{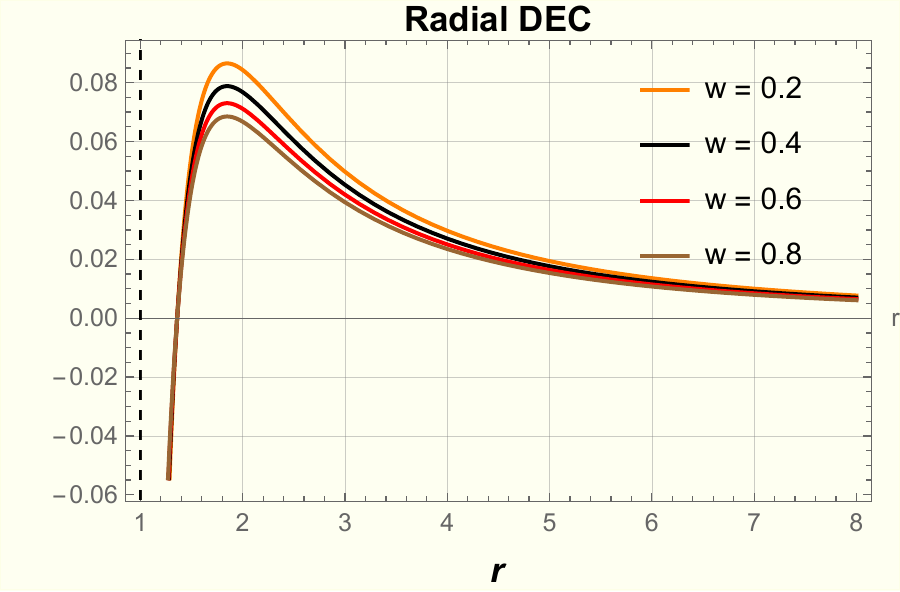}}\\
	    \subfloat[$\rho_*-|p_{t}|$\label{fig:ce4}]{\includegraphics[width=0.49\linewidth]{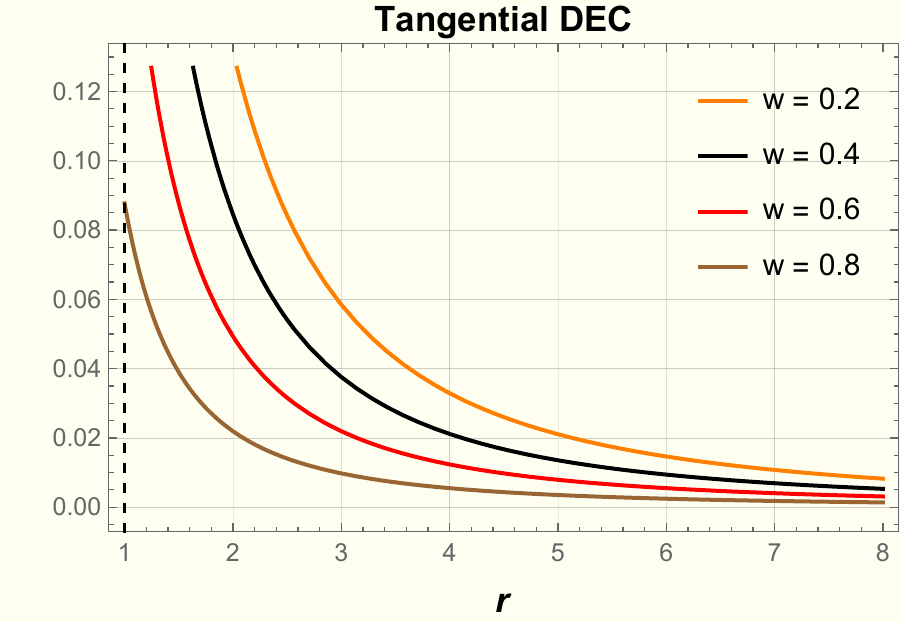}}
	    \subfloat[$\rho+p_{r}+2p_{t}$\label{fig:ce5}]{\includegraphics[width=0.49\linewidth]{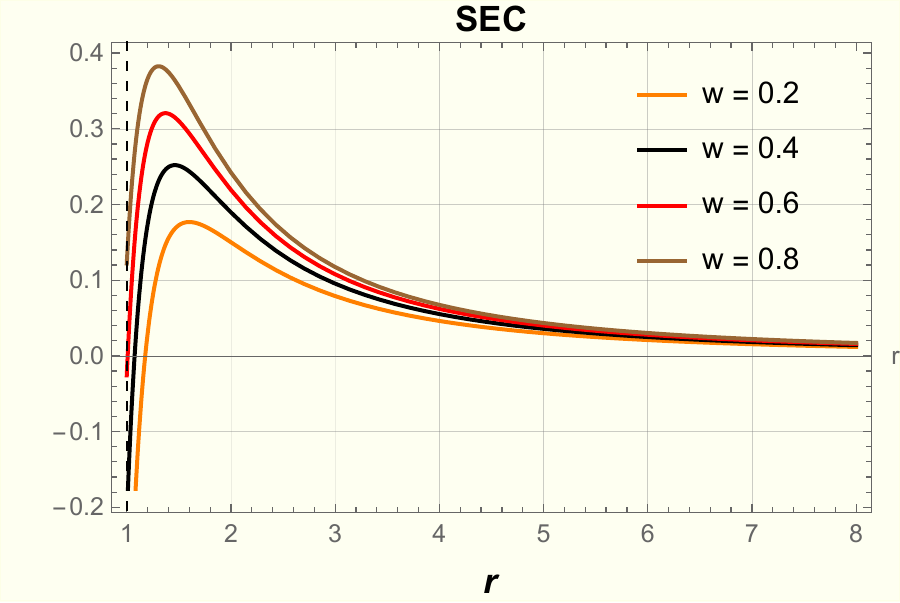}}
	    \caption{Case 3: Plot showing the profile of energy density and various ECs for $\alpha = 1.5, \beta=0.9, C_2=2$ and $r_0=1$.}
	    \label{fig:Cec}
	\end{figure*}

   The NEC and DEC are both satisfied when considering the tangential pressure. However, with the radial pressure, these conditions are violated, especially in the vicinity of the throat. The behavior of the SEC follows a similar pattern. These characteristics are illustrated in \autoref{fig:Cec}.

\section{Embedding Procedures}

Embedding diagrams serve as a crucial tool for visually conceptualizing wormhole structures. The selection of the shape function $S_f(r)$ assumes significance in determining the characteristics of these diagrams. In our examination, emphasis is given to the equatorial slice defined by $\theta=\frac{\pi}{2}$ and a fixed coordinate of time, denoted as $t= \text{constant}$. Under these specified conditions, the metric derived from \eqref{eq:whmetric} assumes the form

\begin{equation}\label{eq:metricembedding1}
    ds^2 = \dfrac{dr^2}{1-\frac{S_f(r)}{r}}+r^2 \; d\phi^2.
\end{equation}

Subsequently, this slice can be embedded into its hypersurface with cylindrical coordinates $(r,\phi,z)$, resulting in the metric

\begin{equation}\label{eq:metricembedding2}
    ds^2 = dz^2+dr^2+r^2 \; d\phi^2.
\end{equation}

By comparing \eqref{eq:metricembedding1} and \eqref{eq:metricembedding2}, an expression for $z(r)$ can be derived as

\begin{equation}
    \dfrac{dz}{dr}=\pm\left[ \dfrac{S_f(r)}{r-S_f(r)}\right] ^{1/2}.
\end{equation}

\begin{figure}
    \centering
    \includegraphics[width=0.7\linewidth]{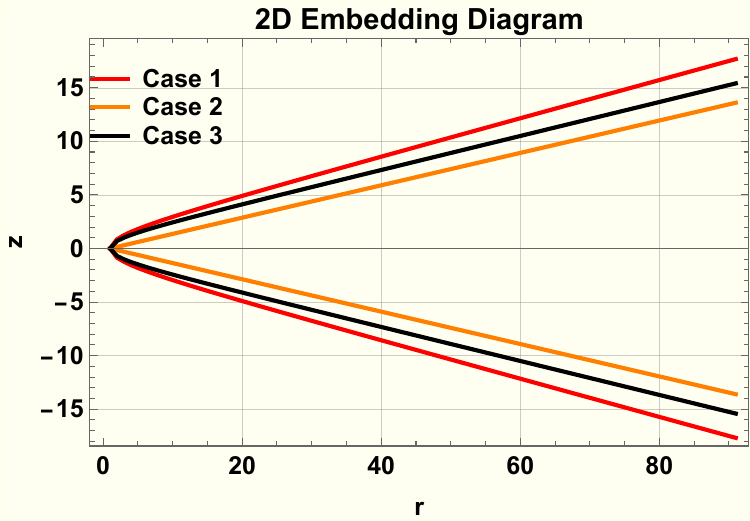}
    \caption{Plot showing the two-dimensional embedding diagram of wormhole solutions. Here $\alpha=1.5$, $C_2=4$, $r_0=1$ and $\beta=2$ for case 1, $\alpha=1.4$, $C_2=1.26$, $r_0=0.35$, $A=2.25$ and $\beta=0.45$ for case 2 and $\alpha = 1.5, \beta=0.9, C_2=2$, $r_0=1$ and $w=0.4$ for case 3.}
    \label{fig:2ded}
\end{figure}

For all three cases, the resultant two-dimensional embedding diagram is presented in  \autoref{fig:2ded} and the corresponding three-dimensional diagram is presented in \autoref{fig:ed}.

\section{Equilibrium Condition }\label{sec:chap5:V}

    Essentially, $f(Q,T)$ theory is a non-conservative theory of gravity i.e., in general,
    \begin{equation}
        \Tilde{\nabla}_a {T^a}_b\ne 0,
    \end{equation}
    where $\Tilde{\nabla}$ denotes the covariant derivative. In the seminal work by Xu et al \cite{Xu:2019sbp}, the corrected equation balancing the energy-momentum tensor is provided as

    \begin{equation}\label{eq:nonconservation}
    \begin{split}
        \Tilde{\nabla}_a {T^a}_b=\frac{1}{f_T-1}\left\{-\Tilde{\nabla}_a(f_T{\Theta^a}_b)-\frac{2}{\sqrt{-g}}\nabla_c\nabla_a {H_b}^{ca}+\nabla_a\left(\frac{1}{\sqrt{-g}}\nabla_c {H_b}^{ca}\right)\right.\\\left.-2\nabla_a {\mathcal{C}^a}_b+\frac{1}{2}f_T\partial_b T\right\},
    \end{split}
    \end{equation}
    where ${H_b}^{ca}$ is the hypermomentum tensor density and 
    \begin{equation}
        {\mathcal{C}^a}_b=\frac{1}{\sqrt{-g}}\nabla_c\left(\sqrt{-g}f_Q {P^{ca}}_b+\frac{{H_b}^{ca}}{2}\right).
    \end{equation}

    The right-hand side of the equation \eqref{eq:nonconservation} represents the covariant derivative of the energy-momentum tensor, while the left part is attributed to the coupling between geometry and matter.  As a result, the standard Tolman-Oppenheimer-Volkoff (TOV) equation is modified by incorporating a term associated with the coupling effect. For the present analysis, considering a static spherically symmetric metric with anisotropic matter distribution and adopting a linear $f(Q,T)$ model, the modified TOV \cite{Rahaman:2013xoa} equation can be expressed as

    \begin{equation}\label{eq:tov}
        p'_r+ \frac{\Phi'}{2}(p_r+\rho)+\frac{2}{r}(p_r-p_t)+\beta \left\{\frac{\Phi'}{2}(p_r+\rho)+\frac{2}{r}(p_r-p_t)+\frac{\rho'}{2}+\frac{p_r'}{2}-\frac{5p_t'}{3}\right\}=0,
    \end{equation}
    where $\Phi = 2\mathcal{R}_f$. The equation presented above characterizes the equilibrium conditions governing a stable state of the wormhole, considering the interactions of gravitational force $F_g$, hydrostatic force $F_h$, anisotropic force $F_a$, and the coupling force $F_c$. These forces are precisely defined by $F_g =  \frac{\Phi'}{2}(p_r+\rho)$, $F_h=p'_r$, $F_a =\frac{2}{r}(p_r-p_t)$, and $F_c=\beta \left\{\frac{\Phi'}{2}(p_r+\rho)+\frac{2}{r}(p_r-p_t)+\frac{\rho'}{2}+\frac{p_r'}{2}-\frac{5p_t'}{3}\right\}$. Thus, the balance of these forces is succinctly described by the expression $F_g + F_h + F_a +F_c = 0$, as derived from equation \eqref{eq:tov}. The \autoref{fig:eq} illustrates the behavior of interacting forces for all three cases.

\subsection*{Anisotropy Analysis }
The assessment of anisotropy holds a pivotal role in analyzing the internal geometry of a relativistic wormhole configuration. A well-established method for quantifying anisotropy in a wormhole involves the relation 

\begin{equation}
    \Delta = p_t - p_r.
\end{equation}

The determination of the wormhole's geometry is contingent on the anisotropic factor. When the tangential pressure surpasses the radial pressure $\Delta > 0$, it indicates a repulsive structure, with an anisotropic force acting outward. Conversely, if the radial pressure exceeds the tangential pressure $ \Delta < 0$, it implies an attractive geometric configuration. Furthermore, $\Delta\propto - F_a$, establishes the relationship between $\Delta$ and the anisotropic force $F_a$. By straightforwardly interpreting the behavior of $F_a$ for various wormhole solutions in \autoref{fig:eq}, one can observe that the tangential pressure exceeds the radial one, indicating the presence of a repulsive geometry in all three cases. 

\section{Concluding Remarks}\label{sec:chap5:VI}

\begin{figure}
            \centering
            \subfloat[$S_f$\label{fig:csf1}]{\includegraphics[width=0.49\linewidth]{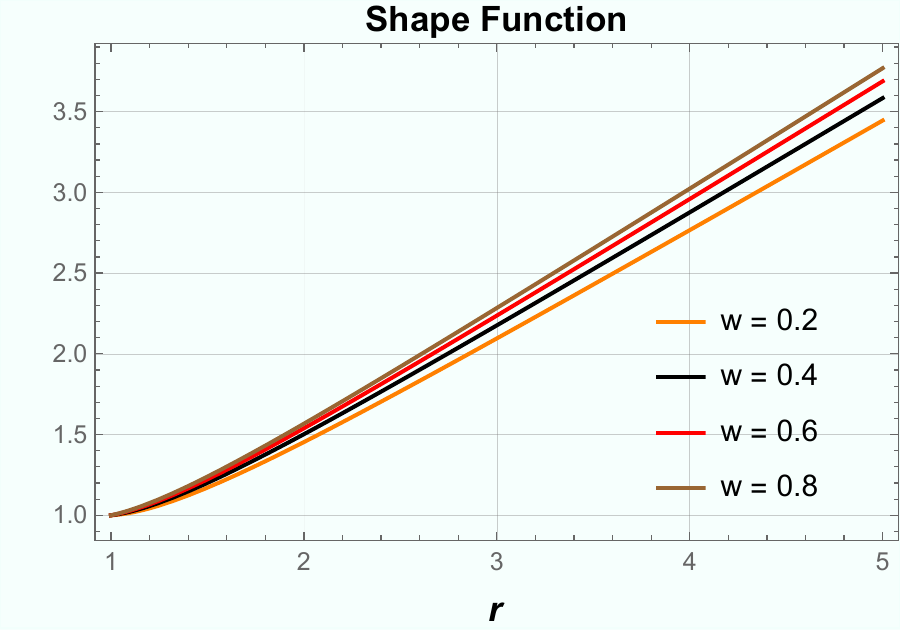}}
    	    \subfloat[$S_f/r$\label{fig:csf2}]{\includegraphics[width=0.49\linewidth]{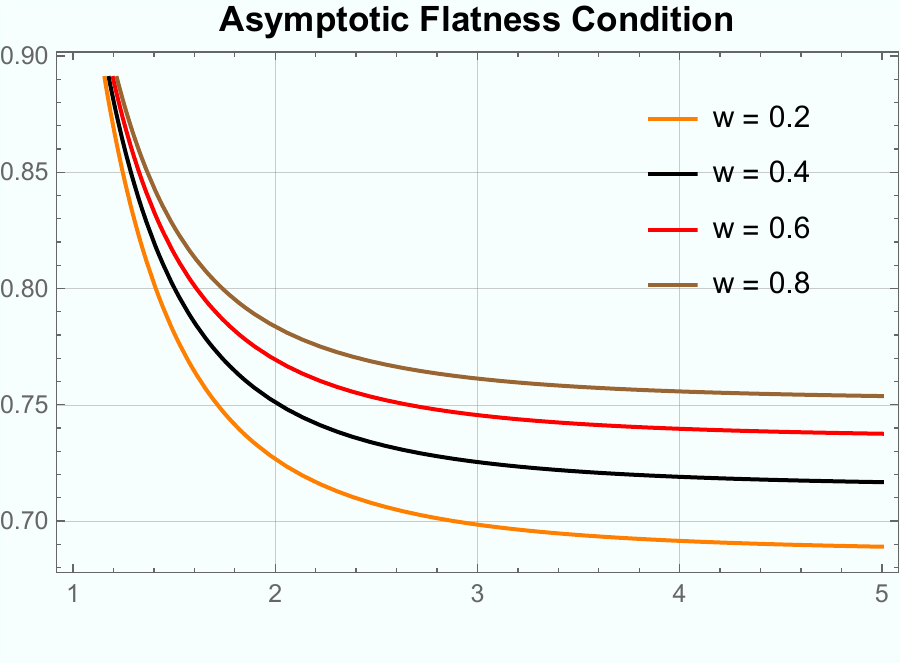}}\\
    	    \subfloat[$S_f'$\label{fig:csf3}]{\includegraphics[width=0.49\linewidth]{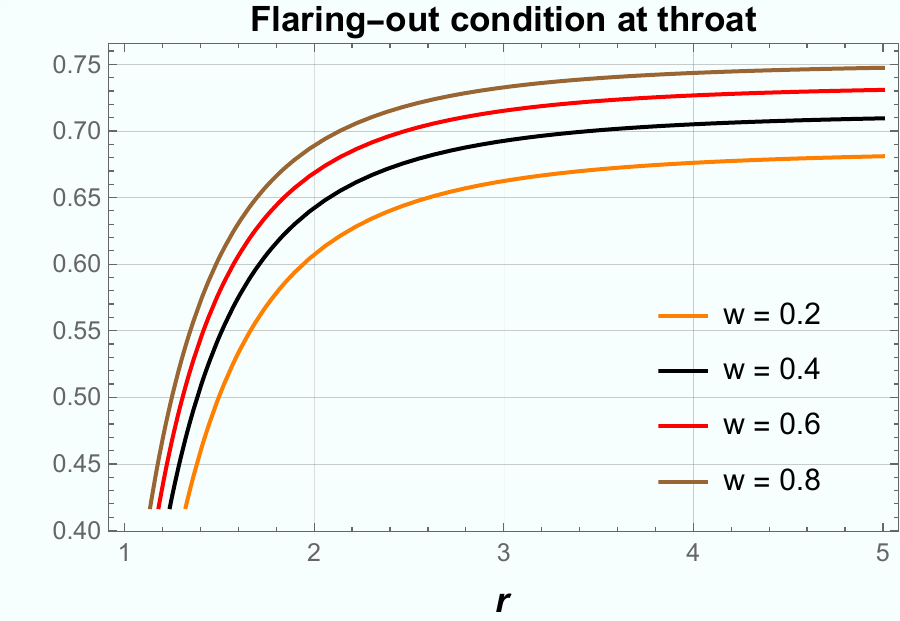}}
            \subfloat[$\frac{S_f-rS_f'}{S_f^2}$\label{fig:csf4}]{\includegraphics[width=0.49\linewidth]{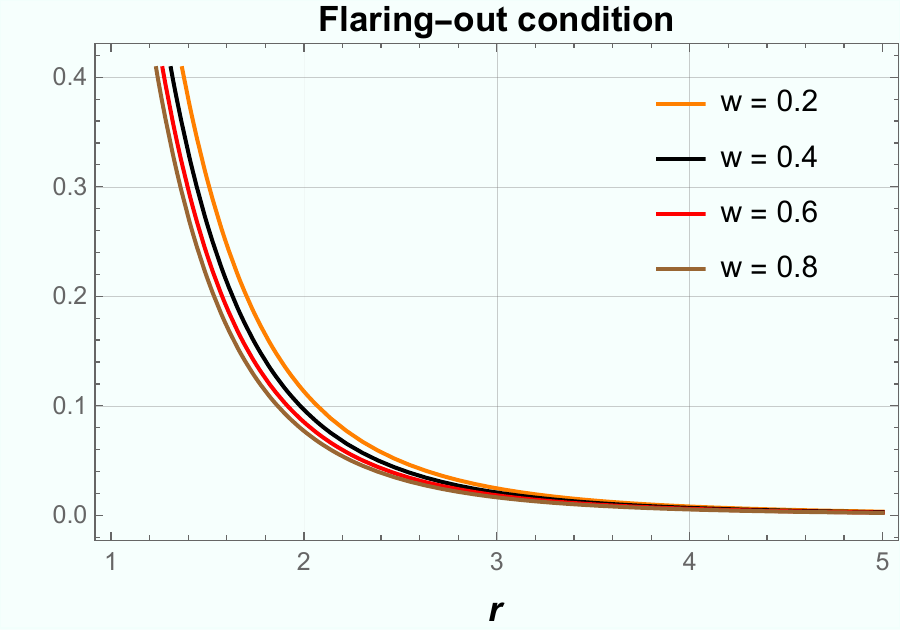}}
            \caption{Case 3: Plot showing the characteristics of $S_f$ for $\alpha = 1.5, \beta=0.9, C_2=2$ and $r_0=1$.}
            \label{fig:Csf}
        \end{figure}

\begin{figure*}
	    \centering
	    \subfloat[Case 1: 3D embedding for $\alpha=1.5$, $C_2=4$, $r_0=1$ and $\beta=2$.\label{fig:aed}]{\includegraphics[width=0.5\linewidth]{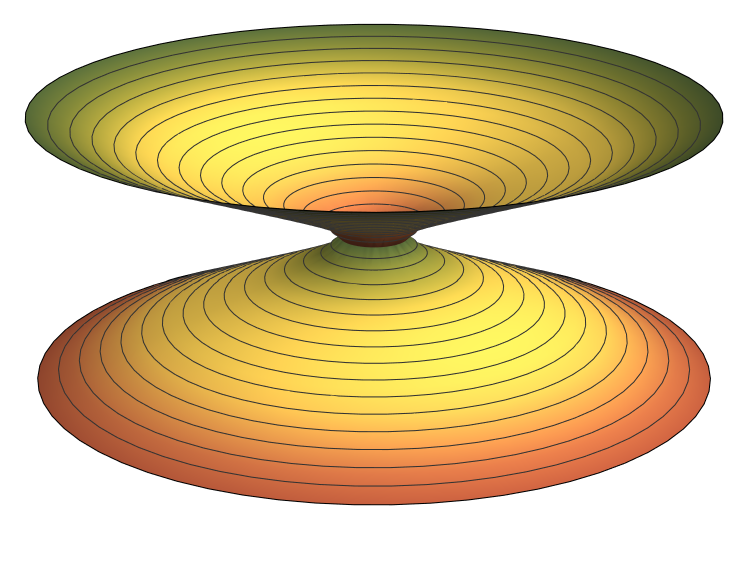}}\\
	    \subfloat[Case 2: 3D embedding for $\alpha=1.4$, $C_2=1.26$, $r_0=0.35$, $A=2.25$ and $\beta=0.45$. \label{fig:bed}]{\includegraphics[width=0.5\linewidth]{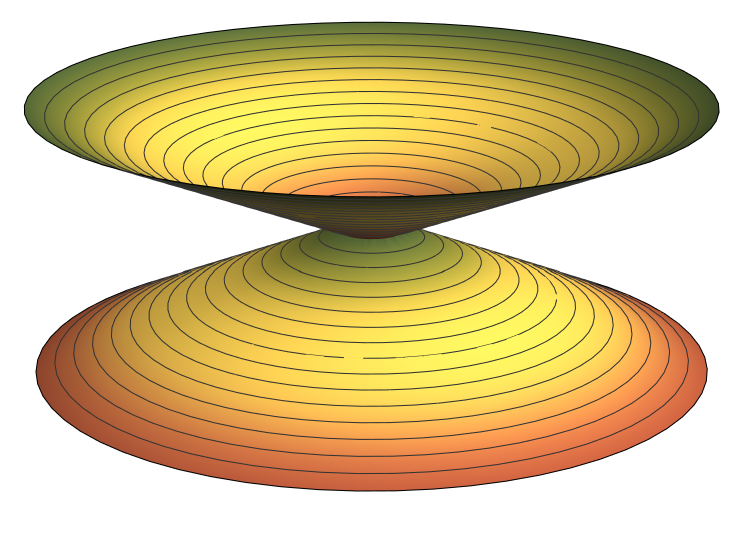}}\\
	    \subfloat[Case 3: 3D embedding for $\alpha = 1.5, \beta=0.9, C_2=2$, $r_0=1$ and $w=0.4$.\label{fig:ced}]{\includegraphics[width=0.5\linewidth]{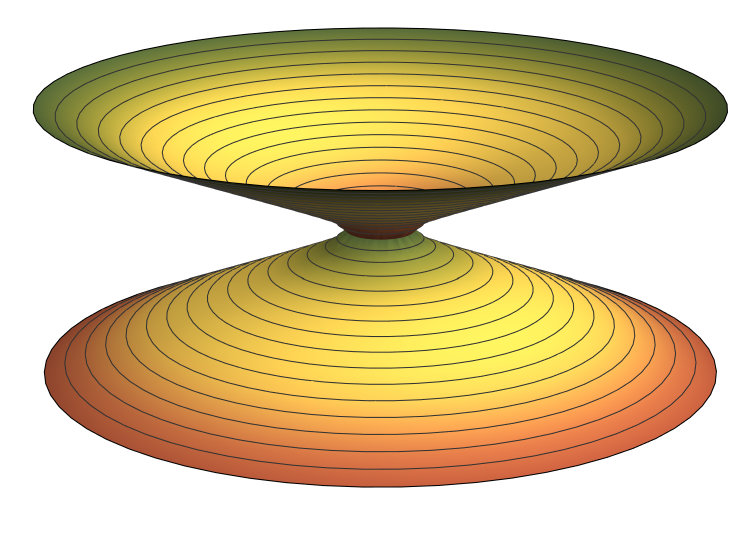}}
	    \caption{Plot showing the three-dimensional embedding diagram of wormhole solutions}
	    \label{fig:ed}
	\end{figure*}

\begin{figure*}
	    \centering
	    \subfloat[Case 1: Stability state of wormhole solution with $\alpha=1.5$, $C_2=4$, $r_0=1$ and $\beta=2$.\label{fig:aeq}]{\includegraphics[width=0.5\linewidth]{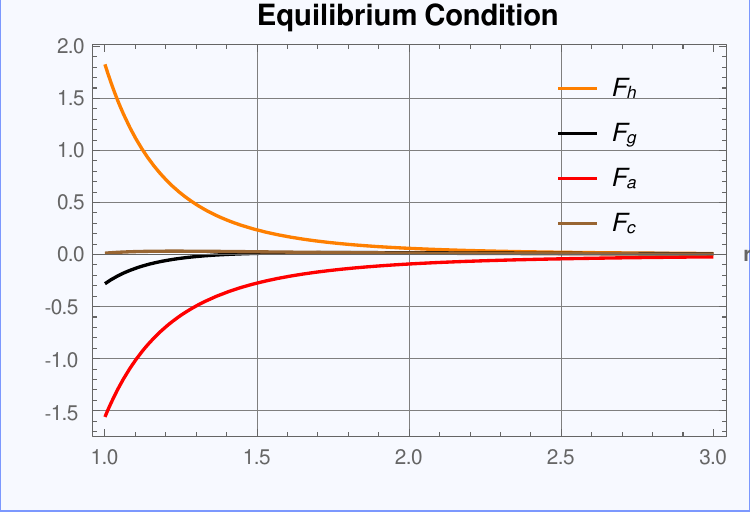}}\\
	    \subfloat[Case 2: Stability state of wormhole solution with $\alpha=1.4$, $C_2=1.26$, $r_0=0.35$, $A=2.25$ and $\beta=0.45$. \label{fig:beq}]{\includegraphics[width=0.5\linewidth]{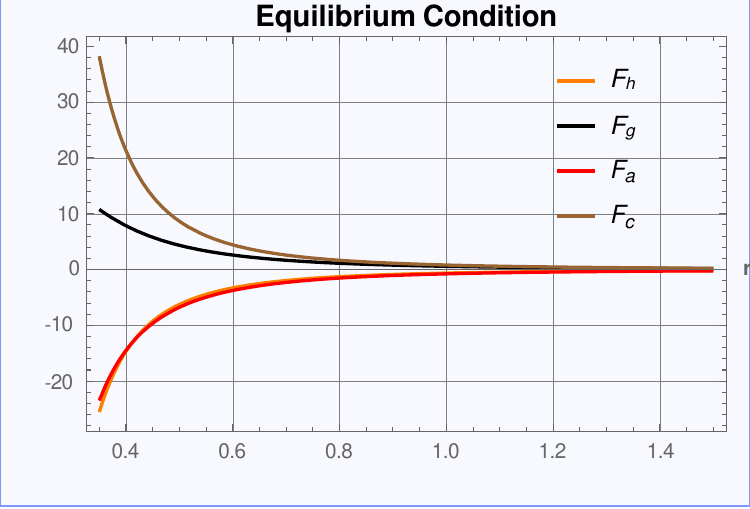}}\\
	    \subfloat[Case 3: Stability state of wormhole solution with $\alpha = 1.5, \beta=0.9, C_2=2$, $r_0=1$ and $w=0.4$.\label{fig:ceq}]{\includegraphics[width=0.5\linewidth]{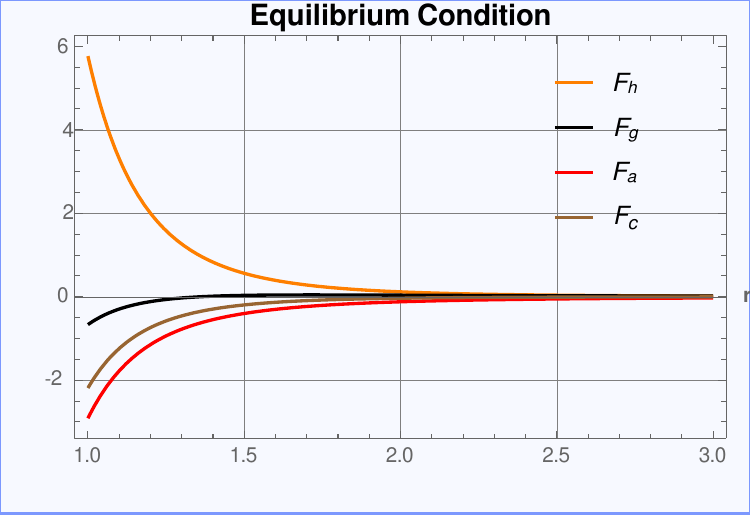}}
	    \caption{A plot showing the equilibrium state of different wormhole solutions}
	    \label{fig:eq}
	\end{figure*}

Wormholes with their distinctive nature allow for the possibility of linking distant regions through a bridge-like pathway. Of particular significance is the potential for spacetime travel, including the significant concept of time travel. In this chapter, we conducted a thorough examination of wormhole solutions within the framework of $f(Q,T)$ modified gravity theory, taking into account the presence of conformal symmetries. Throughout this investigation, we explored the feasibility of traversable wormholes under diverse scenarios, incorporating EoS and anisotropic relations, while considering the influence of CKVs. CKVs play a crucial role in establishing the mathematical connection between the geometry of spacetime and the matter it contains, particularly through the lens of Einstein's field equations. These vectors serve as indispensable tools for reducing the non-linearity order of field equations in various modified theories, contributing to the understanding and analysis of complex gravitational interactions. Moreover, the geometric background of $f(Q,T)$ theory offered prominent possibilities to address and potentially resolve exotic matter issues, a longstanding challenge in studying wormholes and their viability. By exploring the implications of this modified gravity theory, we sought to shed light on the underlying principles that govern the behavior of traversable wormholes. The key features of this chapter are discussed below:

\begin{itemize}
    \item The fundamental issue with the physical viability of a traversable wormhole in the context of GR is the presence of exotic fluids. Here, we have attempted to address this issue through the utilization of a non-metricity-based geometry-matter coupling approach.
    
    \item Considering the anisotropic form of matter distribution, we obtain the governing field equations and accomplish them with the aid of CKVs. These vectors are derived from the killing equations, utilizing the principles of Lie algebra. They exhibit the property of conformal invariance that allows the metric to undergo a conformal mapping to itself along the vector $\eta$ through the action of the Lie derivative operator.
    
    \item As our initial case, we examined the traceless EoS, which is often associated with the Casimir effect involving a massless field. This scenario has been previously explored by Lobo et al. \cite{Lobo:2009ip} in the context of $f(R)$ theories to investigate wormhole geometries. In this case, the radial NEC is violated only near the wormhole throat, while for $r \gg r_0$, it is satisfied. This observation suggests that exotic matter is confined to the region near the wormhole throat, without extending significantly to larger distances.

    \item In our subsequent case, we examined a wormhole solution with an anisotropic relation, where the tangential pressure is related to the radial pressure as $p_t = A p_r$. We derived the shape function and thoroughly examined the energy conditions. Remarkably, both the radial and tangential NECs were satisfied, eliminating the necessity for exotic matter. In \cite{MontelongoGarcia:2010xd}, authors exhibited similar outcomes in the context of nonminimal curvature-matter coupling gravity. Further, B\"{o}mer et al studied wormhole solutions in modified $f(T)$ theory and showed the satisfying behavior of NEC \cite{Boehmer:2012uyw}. Moreover, it is interesting to note that, under this circumstance, the wormhole solution is highly sensitive to the model parameter $\beta$. A small variation in the magnitude of $\beta$ has a significant influence on the shape function and the corresponding energy conditions.

    \item As our last case, we studied barotropic EoS relation with tangential pressure and analyzed the wormhole solution within this context. In \cite{Herrera:2013fja}, a more generalized version of this relation is considered. In this scenario, we studied the influence of the EoS parameter $w$ on the wormhole solution. The shape function is found to satisfy the required conditions for $w>0$. Based on the physical viability, we restricted the parameter space of $w$ to the interval $(0,1)$. The impact of the EoS parameter $w$ on the characteristics of the shape function $S_f$ is shown in the \autoref{fig:Csf}.

    \item The study is conducted by carefully considering a valid parameter space for the model parameters. The characteristics of the resulting wormhole solutions are thoroughly analyzed within this parameter range to ensure their physical viability. An embedding diagram for each case to visually interpret the solution is presented in \autoref{fig:ed}.
    
    \item In each of the three cases, the shape function exhibits satisfactory behavior that complies with all the necessary criteria within the specified parameter range. However, it should be noted that the asymptotic flatness conditions cannot be attained due to the presence of CVKs. In \cite{Mustafa:2019oiy}, authors examined a similar instance in the case of Rastall theory, in which they considered a traceless scenario along with CKVs.

    \item The geometry-matter coupling effect gave rise to an additional force $F_c$, prompting an investigation into the modified TOV equation. The resulting wormhole solutions are determined to be stable. The profile of the interacting forces is depicted in \autoref{fig:eq}.
\end{itemize}

In all, this chapter presents new wormhole solutions in the context of nonmetricity-based geometry-matter coupling gravity with CKVs exhibiting physically valid behavior.

\newpage
\thispagestyle{empty}

\vspace*{\fill}
\begin{center}
    {\Huge \color{NavyBlue} \textbf{CHAPTER 6}}\\
    \
    \\
    {\Large\color{purple}\textsc{\textbf{Wormhole Solutions in \texorpdfstring{$f(R,\mathcal{L}_m)$}{f(R,L m)} gravity with non-commutative geometry}}}\\
    \ 
    \\
    \textbf{Publication based on this chapter}\\
\end{center}
\textsc{On Possible Wormhole Solutions Supported by Non-commutative Geometry within $f(R, L_m)$ Gravity}, \textbf{NS Kavya}, V Venkatesha, G Mustafa, PK Sahoo, \textit{Annals of Physics} \textbf{455}, 169383, 2023 (Elsevier, Q1, IF - 3) DOI: \href{https://doi.org/10.1016/j.aop.2023.169383}{10.1016/j.aop.2023.169383}
\vspace*{\fill}

\pagebreak

\def\baselinestretch{1}
\chapter{\textsc{Wormhole Solutions in \texorpdfstring{$f(R,\mathcal{L}_m)$}{f(R,L m)} gravity with non-commutative geometry}}\label{chap6}
\def\baselinestretch{1.5}
\pagestyle{fancy}
\fancyhead[R]{\textit{Chapter 6}}
\textbf{Highlights:}
\textit{\begin{itemize}
    \item Non-commutativity is a key feature of spacetime geometry. The current chapter explores the traversable wormhole solutions in the framework of $f(R,\mathcal{L}_m)$ gravity within non-commutative geometry. 
    \item By using the Gaussian and Lorentzian distributions, we construct tideless wormholes for the nonlinear $f(R,\mathcal{L}_m)$ model, i.e., $f(R,\mathcal{L}_m)=\dfrac{R}{2}+\mathcal{L}_m^\alpha$. 
    \item For both cases, we derive shape functions and discuss the required properties with satisfying behavior. 
    \item For the required wormhole properties, we develop some new constraints. 
    \item The influence of the involved model parameter on energy conditions is analyzed graphically which provides a discussion about the nature of exotic matter.
    \item Further, we check the physical behavior regarding the stability of wormhole solutions through the TOV equation.
    \item An interesting feature regarding the stability of the obtained solutions via the speed of sound parameters within the scope of average pressure is discussed. 
\end{itemize}}

\section{Introduction}\label{I}

    \par In exploring the manifolds' features in different lights, the concept of non-commutative geometry is phenomenal. In the analysis of spacetime structure, non-commutativity can be presented via modified matter sources. Primarily, it unifies weak and strong forces with gravitational force. P. Aschieri et al \cite{Aschieri:2005zs} have constructed classical governing equations of gravity on non-commutative geometry.  Both the deformation of the spacetime geometry and the quantization can be effectively dealt with non-commutativity. In D-brane \cite{Seiberg:1999vs}, the coordinates of spacetime can be treated as non-commutative operators represented by $[y^a,y^b]=\mathit{i}\theta^{ab}$. Such operators lead to spacetime discretization. This is indicated by the second-order antisymmetric matrix $\theta^{ab}$ \cite{Doplicher:1994zv,Kase:2002pi,Smailagic:2003yb,Nicolini:2008aj}. One can remark that non-commutativity substitutes smear objects for point-like structures. The smearing phenomenon can be employed by replacing the Dirac delta function with a Gaussian and a Lorentzian distribution of minimal length $\sqrt{\theta}$. In \cite{Schneider:2020ysd}, Schneider et al. have discussed the Gaussian and Lorentzian distributions via simple smearing of a matter distribution within the black hole. With Gaussian distribution, Sushkov examined the wormholes supported by phantom energy \cite{Sushkov:2005kj}. Kuhfittig, in \cite{Kuhfittig:2012gi}, shows that certain thin-shell wormholes that are unstable in GR behave stable as a consequence of non-commutativity. The physical impact of the short separation of non-commutative coordinates can be seen in \cite{Nicolini:2009gw}. Rahaman et al. investigated wormhole geometry with Gaussian distribution and proved the feasibility of solution in four and five dimensions \cite{Rahaman:2012pg}.  
    
    \par For the static spherically symmetric point-like gravitational source having total mass $M$, Gaussian and Lorentzian distribution of energy densities are given by \cite{Nicolini:2005vd,Smailagic:2003yb},
    \begin{eqnarray}
        \label{gauss}\rho= \frac{M e^{-\frac{r^2}{4 \theta }}}{8 \pi ^{\frac{3}{2}} \theta ^{\frac{3}{2}}},\\
        \label{lorentz}\rho=\frac{\sqrt{\theta } M}{\pi ^2 \left(\theta +r^2\right)^2}.
    \end{eqnarray}
     
     These choices reflect the notion that the source is spread out or smeared rather than being concentrated at a single point. This is mainly because of the intrinsic uncertainty in the coordinate commutator. Further, the noncommutative correction becomes significant in a region near the origin, specifically when $r \lesssim \theta$. Within this neighborhood, the effects of noncommutativity regularize both the radial and tangential pressures, as well as the matter density. From the particular choice \eqref{gauss} and \eqref{lorentz}, the physical parameters (especially energy density) are finite and asymptotically vanish, supporting the vacuum solution at points far away from the origin. 
    \par In the present chapter, we attempt to study static spherical symmetric Morris-Thorne wormhole structure in the paradigm of the newly proposed $f(R,\mathcal{L}_m)$ gravity. This chapter is organized as follows:  In \autoref{sec:chap6:III}, we examine the wormhole model with Gaussian and Lorentzian distribution and derive the corresponding shape functions. Also, we analyze the influence of model parameter on the shape functions and energy conditions. Next, \autoref{sec:chap6:IV} assesses the stability of wormholes using the TOV equation. In \autoref{sec:chap6:V} we interpret the physical aspects of the wormhole by examining  average pressure and speed of sound. Finally, \autoref{sec:chap6:VI} gives the discussion of results and concluding remarks.
	

\section{Wormhole Geometry in \texorpdfstring{$f(R,\mathcal{L}_m)$}{f(R,L m)} } \label{sec:chap6:III}

\par The gravitational interaction of the wormhole geometry \eqref{eq:whmetric} with anisotropic matter distribution \eqref{eq:chap1:anisotropicT} in $f(R,\mathcal{L}_m)$ gravity can be described using the field equations given by \eqref{chap4:eq:fieldequation1},

		\begin{gather}
		    \label{fe1}4f_R\dfrac{S_f'}{r^2}-(f-f_{\mathcal{L}_m}\mathcal{L}_m)=(2\rho+p_r+2p_t)f_{\mathcal{L}_m},\\
      \begin{split}
          \label{fe2}	6f_R''\left(1-\dfrac{S_f}{r} \right)+3f_R'\left(\dfrac{S_f-rS_f'}{r^2} \right)+2f_R\left(\dfrac{3S_f-rS_f'}{r^3} \right) 
				-(f-f_{\mathcal{L}_m}\mathcal{L}_m)\\=(-\rho-2p_r+2p_t)f_{\mathcal{L}_m}, 
      \end{split}
		    \\
			\label{fe3}6\dfrac{f_R''}{r}\left(1-\dfrac{S_f}{r} \right)-f_R\left(\dfrac{3S_f-rS_f'}{r^3} \right)-(f-f_{\mathcal{L}_m}\mathcal{L}_m)=(-\rho+p_r-p_t)f_{\mathcal{L}_m}. 
		\end{gather}

\subsection{Non-commutative Wormhole models}\label{III}

		\par In this section, we shall consider a viable wormhole model to study the characteristics of wormhole geometry. In particular, we suppose the non-linear form given by,    
				\begin{equation}\label{mod}
		    f(R,\mathcal{L}_m)=\dfrac{R}{2}+\mathcal{L}_m^\alpha,
		\end{equation}
	    where $\alpha$ is a free parameter. For $\alpha=1$ the case reduces to GR. We presume that the matter Lagrangian density $\mathcal{L}_m$ depends on energy density $\rho$ i.e., $\mathcal{L}_m=\rho$ \cite{Bertolami:2008ab,Garcia:2010xb,Brown:1992kc,Hawking:1973uf,Faraoni:2009rk,Garcia:2010xb}. Now, comparing the equations \eqref{fe1} and \eqref{fe3} for $f(R,\mathcal{L}_m)$ model \eqref{mod} we can get the expressions for radial and tangential pressures as, 
		\begin{align}
		    \label{pr}p_r&=-\dfrac{\rho}{\alpha}\left[(\alpha-1)+\dfrac{S_f}{r^3\rho^\alpha}\right], \\
		    \label{pt}p_t&=\dfrac{r S_f'+S_f}{2\alpha r^3 \rho^{\alpha-1}} -\rho.
		\end{align}
	
      \subsubsection{Gaussian energy density}

	   \par The equation \eqref{gauss} describes the energy density for Gaussian distribution. With the physical parameters $\rho$ \eqref{gauss}, $p_r$ \eqref{pr} and $p_t$ \eqref{pt} the field equation \eqref{fe2} reduces to,

        \begin{equation}
            \frac{S_f'(r)}{r}=8^{-\alpha } \pi ^{-\frac{3 \alpha }{2}} r \left(\frac{M e^{-\frac{r^2}{4 \theta }}}{\theta ^{3/2}}\right)^{\alpha }.
        \end{equation}
        \par On solving the above ordinary differential equation, the shape function of the wormhole with Gaussian distribution can be obtained. This is given by,
	 \begin{equation}\label{Asf}
	      \begin{split}
	         S_f(r)=\frac{2^{1-3 \alpha } \pi ^{-\frac{3 \alpha }{2}} \theta  \left[\sqrt{\pi\theta }\; e^{\frac{\alpha  r^2}{4 \theta }} \text{erf}\left(\frac{\sqrt{\alpha } r}{2 \sqrt{\theta }}\right)-\sqrt{\alpha } r\right] \left(\frac{M e^{-\frac{r^2}{4 \theta }}}{\theta ^{3/2}}\right)^{\alpha }}{\alpha ^{3/2}}+k,
	         \end{split}
	 \end{equation}
    where, $\text{erf} (z)=\frac{2}{\sqrt{\pi}}\int_0^z e^{-t^2} dt$ is the Gauss error function and $k$ is the integrating constant. Now, to obtain the particular solution, we find the value of $k$ by imposing the throat condition $S_f(r_0)=r_0$. Then we have,
    \begin{equation}\label{k1}
           \begin{split}
               k=\frac{\sqrt{\alpha }\; r_0 \left[\left(8 \pi ^{\frac{3}{2}}\right)^{\alpha} \alpha +2 \theta  \left(\dfrac{M e^{-\frac{r_0^2}{4 \theta }}}{\theta ^{\frac{3}{2}}}\right)^{\alpha }\right]-2 \sqrt{\pi }\; \theta ^{\frac{3}{2}}\; e^{\frac{\alpha  r_0^2}{4 \theta }}\; \text{erf}\left(\frac{\sqrt{\alpha } r_0}{2 \sqrt{\theta }}\right) \left(\dfrac{M e^{-\frac{r_0^2}{4 \theta }}}{\theta ^{\frac{3}{2}}}\right)^{\alpha }}{\left(8 \pi ^{\frac{3}{2}}\right)^{\alpha}\alpha ^{\frac{3}{2}}}.
           \end{split}
           \end{equation}

   In order to achieve the traversability of a wormhole the shape function should satisfy the flaring-out condition. For the present scenario, $S_f(r)$ satisfies $S_f'(r_0)<1$ at the throat if the following inequality holds:
       \begin{equation}\label{ineq1}
       \left(\dfrac{M\;e^{-\frac{1}{4\theta}}}{8\pi^{\frac{3}{2}}\theta^{\frac{3}{2}}}\right)^\alpha<\frac{1}{r_0^2}.
    \end{equation}

    The above inequality is significant in determining the relation between $M$, $\theta$, $\alpha$, and $r_0$. For the GR scenario, with $r_0=1$ and $\theta=4$, we can get the constraining relation on the total mass $M$ as, $M<64\pi^{3/2} e^{1/16}$.  By taking the plot of shape function $S_f(r)$ with respect to $r$, we examined its behavior for $M=1.2, \theta=4$, and $r_0=1$. One can refer to \autoref{tab:table0} for detailed analysis. For different values of $\alpha$, \figureautorefname$\;$\ref{fig:Asf} shows the characteristics of $S_f(r)$. It is obvious from our choice of $k$ that $S_f(r)$ satisfies the throat condition. From \figureautorefname$\;$\ref{fig:Asf1}, it can be seen that $S_f(r)>0$ is a monotonically increasing function. Moreover, $S_f(r)<r$, implying the finiteness of the proper radial distance function. Also, the flaring-out condition is satisfied [\figureautorefname$\;$\ref{fig:Asf2a}, \ref{fig:Asf2b}]. The Lorentzian manifold becomes flat asymptotically as the value  $S_f(r)/r\to0$ for large $r$ and this can be interpreted from \figureautorefname$\;$\ref{fig:Asf3}.

    Substituting equations \eqref{gauss}, \eqref{Asf}, \eqref{k1} in \eqref{pr} and \eqref{pt}, the pressure elements take the form,
    \begin{align}
         \begin{split}
            p_r=&\frac{\left(\frac{M e^{-\frac{r^2}{4 \theta }}}{\theta ^{3/2}}\right)^{1-\alpha }}{8 \pi ^{3/2} \alpha ^{5/2} r^3} \left[-2 \sqrt{\pi } \theta ^{3/2} e^{\frac{\alpha  r^2}{4 \theta }} \text{erf}\left(\frac{\sqrt{\alpha } r}{2 \sqrt{\theta }}\right) \left(\frac{M e^{-\frac{r^2}{4 \theta }}}{\theta ^{3/2}}\right)^{\alpha }\right.\\&\left.+2 \theta  \left(\sqrt{\pi } \sqrt{\theta } e^{\frac{\alpha  r_0^2}{4 \theta }} \text{erf}\left(\frac{\sqrt{\alpha } r_0}{2 \sqrt{\theta }}\right)-\sqrt{\alpha } r_0\right) \left(\frac{M e^{-\frac{r_0^2}{4 \theta }}}{\theta ^{3/2}}\right)^{\alpha } \right.\\&\left.+\sqrt{\alpha } \left(-r \left((\alpha -1) \alpha  r^2-2 \theta \right) \left(\frac{M e^{-\frac{r^2}{4 \theta }}}{\theta ^{3/2}}\right)^{\alpha }-8^{\alpha } \pi ^{\frac{3 \alpha }{2}} \alpha  r_0\right)\right],
        \end{split}\\
        \begin{split}
            p_t=&\frac{\left(\frac{M e^{-\frac{r^2}{4 \theta }}}{\theta ^{3/2}}\right)^{1-\alpha } }{16 \pi ^{3/2} \alpha ^{5/2} r^3} \left[2 \sqrt{\pi } \theta ^{3/2} e^{\frac{\alpha  r^2}{4 \theta }} \text{erf}\left(\frac{\sqrt{\alpha } r}{2 \sqrt{\theta }}\right) \left(\frac{M e^{-\frac{r^2}{4 \theta }}}{\theta ^{3/2}}\right)^{\alpha }\right.\\&\left.+2 \theta  \left(\sqrt{\alpha } r_0-\sqrt{\pi } \sqrt{\theta } e^{\frac{\alpha  r_0^2}{4 \theta }} \text{erf}\left(\frac{\sqrt{\alpha } r_0}{2 \sqrt{\theta }}\right)\right) \left(\frac{M e^{-\frac{r_0^2}{4 \theta }}}{\theta ^{3/2}}\right)^{\alpha }\right.\\&\left.+\sqrt{\alpha } \left(8^{\alpha } \pi ^{\frac{3 \alpha }{2}} \alpha  r_0-r \left(2 \theta +\alpha  (2 \alpha -1) r^2\right) \left(\frac{M e^{-\frac{r^2}{4 \theta }}}{\theta ^{3/2}}\right)^{\alpha }\right)\right].
        \end{split}
    \end{align}
       
    \par Furthermore, with the above pressure elements, we examined the energy conditions NEC, DEC, and SEC for Gaussian distribution [ref \figureautorefname$\;$\ref{fig:Aec}]. In this case, NEC is not satisfied for radial pressure [\figureautorefname$\;$\ref{fig:Ae1}] but for tangential pressure it holds [\figureautorefname$\;$\ref{fig:Ae2}]. Also, both DECs are violated and SEC is satisfied [\figureautorefname$\;$\ref{fig:Ae3},\ref{fig:Ae4},\ref{fig:Ae5}]. In addition, $\text{NEC}_{eff}\equiv \frac{rS_f'-S_f}{r^3}$ is violated as a consequence of satisfied flaring-out condition.

	 \begin{table*}[htbp]
		\caption{Gaussian Distribution: Summarizing the nature of shape function $S_f$ with $M=1.2,\theta=4$ and $r_0=1$.}
		    \label{tab:table0}
		    \centering
		    \begin{tabular}{c c c}
		        \hline\hline
		        \textit{Function }           & \textit{Result}  & \textit{Interpretation} \\
		        \hline
		        $S_f(r)$  & \makecell{ $0<S_f(r)<r\;\forall\;r>r_0$ \\and  $\; S_f(r_0)=r_0$\\ for $\alpha>0.41$}  & \makecell{Viable form of shape function\\ and throat condition is satisfied}\\
		        \hline
		        $\frac{S_f(r)-rS_f'(r)}{S_f(r)^2}$ & \makecell{$>0$, for $\alpha\ge0.58$\\ and $S_f'(r_0)<1$} & Flaring-out condition is satisfied \\
		       \hline
		       $\dfrac{S_f(r)}{r}$& \makecell{approaches to 0\\ for large value of $r$\\ and $\alpha>0$}& Asymptotic flatness condition is satisfied\\
		       \hline\hline
			\end{tabular}
		\end{table*}

    \textbf{Wormhole Solutions:} In the context of Gaussian distribution, we have verified the criteria satisfied by the wormhole, such as finite redshift, throat condition, asymptotic condition, flaring-out condition, and the violation of effective null energy condition. Based on the constraining relation \eqref{ineq1}, we chose $M=1.2, \theta = 4$ and $r_0=1$ and analyzed the influence of model parameter $\alpha$ on the wormhole solution. For different values of $\alpha$ (say, $\alpha_i=0.645,0.650,0.655,09.660,0.656$), corresponding wormhole metrics read:
    \begin{equation}
        ds^2=e^{c}dt^2-\psi_i dr^2  - r^2\left(d\theta^2+\text{sin}^2\theta \,d\phi^2\right),
    \end{equation}
    where $c$ is some constant. For shape functions $S_{f_i}$ corresponding to $\alpha_i$, $\psi_i\equiv\left(1-\frac{S_{f_i}}{r}\right)^{-1}$, with

    \begin{align}
        \psi_1=&\frac{r}{\left(e^{-\frac{r^2}{16}}\right)^{0.645} \left(0.315227 r-1.39139 e^{0.0403125 r^2} \text{erf}(0.20078 r)\right)+r-0.99173},\\
        \psi_2=&\frac{r}{\left(e^{-\frac{r^2}{16}}\right)^{0.65} \left(0.304023 r-1.33676 e^{0.040625 r^2} \text{erf}(0.201556 r)\right)+r-0.991964},\\
        \psi_3=&\frac{r}{\left(e^{-\frac{r^2}{16}}\right)^{0.655} \left(0.293234 r-1.2844 e^{0.0409375 r^2} \text{erf}(0.20233 r)\right)+r-0.992191},\\
        \psi_4=&\frac{r}{\left(e^{-\frac{r^2}{16}}\right)^{0.66} \left(0.282845 r-1.23419 e^{0.04125 r^2} \text{erf}(0.203101 r)\right)+r-0.992411},\\
        \psi_5=&\frac{r}{\left(e^{-\frac{r^2}{16}}\right)^{0.665} \left(0.272839 r-1.18605 e^{0.0415625 r^2} \text{erf}(0.203869 r)\right)+r-0.992626}.
    \end{align}
    
    \subsubsection{Lorentzian energy density}

    \par The non-commutative geometric distribution is an intrinsic aspect of a Lorentzian manifold \cite{Nicolini:2005vd}. It is independent of the spacetime properties such as curvature. In this section, we study the scenario of the traversable wormhole with Lorentzian energy density distribution \eqref{lorentz}. Substituting \eqref{lorentz}, \eqref{pr} and \eqref{pt}, the field equation \eqref{fe3} becomes,

    \begin{equation}\label{ode2}
        \frac{S_f'(r)}{r}=\pi ^{-2 \alpha } r \left(\frac{\sqrt{\theta } M}{\left(\theta +r^2\right)^2}\right)^{\alpha }.
    \end{equation}

    \par The aforementioned equation is significant in determining the desired shape function. It is known that the shape function $S_f(r)$ at the throat should have a fixed point i.e., $S_f(r_0)=r_0$. Therefore, the ordinary differential equation \eqref{ode2} is an initial value problem. The particular solution of this equation is obtained as,
    
     \begin{equation}\label{Bsf}
	      \begin{split}
	         S_f(r)=\frac{r^3}{3\pi ^{2 \alpha }}   \left( M\theta ^{-\frac{3}{2}}\right)^{\alpha } \, _2F_1\left(\frac{3}{2},2\alpha ;\frac{5}{2};-\frac{r^2}{\theta }\right) +k,
	         \end{split}
	 \end{equation}
    where, ${}_2F_1(a,b;c;z)$ is the hypergeometric function and $k$ is the constant of integration given by,
    \begin{equation}
           \begin{split}
               k=r_0-\frac{r_0^3}{3\pi ^{2 \alpha }}   \left( M\theta ^{-\frac{3}{2}}\right)^{\alpha } \, _2F_1\left(\frac{3}{2},2\alpha ;\frac{5}{2} ;-\frac{r_0^2}{\theta }\right).
           \end{split}
           \end{equation}
   Additionally, we consider a constraining relation
   \begin{equation}\label{ineq2}
       \left(\dfrac{M\sqrt{\theta}} {\pi^2(1+\theta)^2}\right)^\alpha<\dfrac{1}{r_0^2}, 
   \end{equation}   
   in order to satisfy the flaring-out condition at the throat. For $\alpha=1$ we can retain the inequality for GR. With $r_0=1$ and $\theta=4$ the inequality \eqref{ineq2} reads, $M<25\pi^2/2$.

    \par Now we have to choose the range of values for the model parameter for which the obtained shape function satisfies all the necessary requirements. To this end, with the help of the plot of shape function versus radial coordinate, we studied the behavior of $S_f(r)$. The value of $\alpha$ is constrained to get the viable form of the shape function with Lorentz distribution [refer Table$\;$\ref{tab:table1}]. The effect of model parameter $\alpha$ on $S_f(r)$ for $M=1.2, \theta=4$ and $r_0=1$ is depicted in \figureautorefname$\;$\ref{fig:Bsf}. It can be observed that $S_f(r)$ is a non-negative monotonically increasing function in the domain of radial coordinate $r$ [\figureautorefname$\;$\ref{fig:Bsf1}] and satisfies the condition $S_f(r)<r$. Further, \figureautorefname$\;$\ref{fig:Bsf2a}, \ref{fig:Bsf2b} reveals that the shape function obeys the flaring-out condition. For an infinitely large value of the radial coordinate $S_f(r)/r$ approaches to zero [\figureautorefname$\;$\ref{fig:Bsf3}]. Thus, we can say that the shape function so obtained for the Lorentz distribution satisfies all the essential conditions.

    Further, with the shape function \eqref{Bsf} and energy density \eqref{lorentz}, the radial and tangential pressures can be rewritten as, 

    \begin{align}
        \begin{split}
            p_r=&\frac{\left(\frac{\sqrt{\theta } M}{\left(\theta +r^2\right)^2}\right)^{1-\alpha }}{3 \pi ^2 \alpha  \theta  r^3} \left[-r^3 \left(\theta +r^2\right) \, _2F_1\left(1,\frac{5}{2}-2 \alpha ;\frac{5}{2};-\frac{r^2}{\theta }\right) \left(\frac{\sqrt{\theta } M}{\left(\theta +r^2\right)^2}\right)^{\alpha }\right.\\&\left.-3 (\alpha -1) \theta  r^3 \left(\frac{\sqrt{\theta } M}{\left(\theta +r^2\right)^2}\right)^{\alpha }-3 \pi ^{2 \alpha } \theta  r_0\right.\\&\left.+r_0^3 \left(\theta +r_0^2\right) \, _2F_1\left(1,\frac{5}{2}-2 \alpha ;\frac{5}{2};-\frac{r_0^2}{\theta }\right) \left(\frac{\sqrt{\theta } M}{\left(\theta +r_0^2\right)^2}\right)^{\alpha }\right].
        \end{split}
        \end{align}
        
        \begin{align}
        \begin{split}
            p_t=&\frac{\left(\frac{\sqrt{\theta } M}{\left(\theta +r^2\right)^2}\right)^{1-\alpha }}{6 \pi ^2 \alpha  \theta  r^3} \left[r^3 \left(\theta +r^2\right) \, _2F_1\left(1,\frac{5}{2}-2 \alpha ;\frac{5}{2};-\frac{r^2}{\theta }\right) \left(\frac{\sqrt{\theta } M}{\left(\theta +r^2\right)^2}\right)^{\alpha }\right.\\&\left.-3 (2 \alpha -1) \theta  r^3 \left(\frac{\sqrt{\theta } M}{\left(\theta +r^2\right)^2}\right)^{\alpha }+3 \pi ^{2 \alpha } \theta  r_0\right.\\&\left.-r_0^3 \left(\theta +r_0^2\right) \, _2F_1\left(1,\frac{5}{2}-2 \alpha ;\frac{5}{2};-\frac{r_0^2}{\theta }\right) \left(\frac{\sqrt{\theta } M}{\left(\theta +r_0^2\right)^2}\right)^{\alpha }\right].
        \end{split}
    \end{align}

    \par In addition, energy conditions interpret the characteristics of motion of energy and matter. Here, we studied the behavior of various energy conditions for Lorentz distribution with $M=1.2, \theta=4$, and $r_0=1$. The NEC is violated for radial pressure, supporting the requirement of the exotic fluid. Further, SEC and tangential NEC are obeyed. There is a violation of both the DECs. 

    \par In the next section, we shall analyze the physical aspects of the Gaussian and Lorentzian wormholes.

 \begin{table*}[h!]
		\caption{Lorentzian Distribution: Summarizing the nature of shape function $S_f$ with $M=1.2,\theta=4$ and $r_0=1$.}
		    \label{tab:table1}

		    \centering
		    \begin{tabular}{c c c}
		        \hline\hline
		        \textit{Function }           & \textit{Result}  & \textit{Interpretation} \\
		        \hline
		        $S_f(r)$  & \makecell{ $0<S_f(r)<r\;\forall\;r>r_0$ \\ and  $\; S_f(r_0)=r_0$\\ for $\alpha>0.5$}  & \makecell{Viable form of shape function\\ and throat condition is satisfied}\\
		        \hline
		        $\frac{S_f(r)-rS_f'(r)}{S_f(r)^2}$ & \makecell{$>0$, for $\alpha>0.52$\\ and $S_f'(r_0)<1$} & Flaring-out condition is satisfied \\
		       \hline
		       $\dfrac{S_f(r)}{r}$& \makecell{approaches to 0 \\for large value of $r$\\ when $\alpha>0.5$}& Asymptotic flatness condition is satisfied\\
		       \hline\hline
			\end{tabular}
		\end{table*}

  \textbf{Wormhole solutions:} Within the framework of the Lorentzian distribution, we have examined the criteria that a traversable wormhole should satisfy. These include finite redshift, throat condition, asymptotic condition, flaring-out condition, as well as the violation of the effective null energy condition. By utilizing the constraining relation denoted by \eqref{ineq2}, we have selected specific values for the parameters $M=1.2$, $\theta=4$, and $r_0=1$, enabling us to investigate the impact of the model parameter $\alpha$ on the wormhole solution. For various values of $\alpha$ (i.e., $\alpha_i=0.645, 0.650, 0.655, 0.660, 0.656$), the corresponding metrics describing the wormhole are as follows:

  \begin{equation}
        ds^2=e^{c}dt^2-\psi_i dr^2  - r^2\left(d\theta^2+\text{sin}^2\theta \,d\phi^2\right),
    \end{equation}

    where $\psi_i$'s are given by,

    \begin{align}
        \psi_1=& -\frac{44.6546 r}{1. \left(r^2+4\right)^{1.29} \left(\frac{1}{\left(r^2+4\right)^2}\right)^{0.645} r^3 \, _2F_1\left(1.29,\frac{3}{2};\frac{5}{2};-\frac{r^2}{4}\right)-44.6546 r+43.8154},\\
        \psi_2=& -\frac{45.5992 r}{1. \left(r^2+4\right)^{1.3} \left(\frac{1}{\left(r^2+4\right)^2}\right)^{0.65} r^3 \, _2F_1\left(1.3,\frac{3}{2};\frac{5}{2};-\frac{r^2}{4}\right)-45.5992 r+44.7611},\\
        \psi_3=& -\frac{46.5637 r}{1. \left(r^2+4\right)^{1.31} \left(\frac{1}{\left(r^2+4\right)^2}\right)^{0.655} r^3 \, _2F_1\left(1.31,\frac{3}{2};\frac{5}{2};-\frac{r^2}{4}\right)-46.5637 r+45.7268}, \\
        \psi_4=&-\frac{47.5487 r}{1. \left(r^2+4\right)^{1.32} \left(\frac{1}{\left(r^2+4\right)^2}\right)^{0.66} r^3 \, _2F_1\left(1.32,\frac{3}{2};\frac{5}{2};-\frac{r^2}{4}\right)-47.5487 r+46.7129}, \\
        \psi_5=& -\frac{48.5545 r}{1. \left(r^2+4\right)^{1.33} \left(\frac{1}{\left(r^2+4\right)^2}\right)^{0.665} r^3 \, _2F_1\left(1.33,\frac{3}{2};\frac{5}{2};-\frac{r^2}{4}\right)-48.5545 r+47.7199}.
    \end{align}
\section{Equilibrium condition}\label{sec:chap6:IV}
\begin{figure}[htbp]
        \centering
        \subfloat[Equilibrium picture for Gaussian distribution\label{fig:Aeq}]{\includegraphics[width=0.6\linewidth]{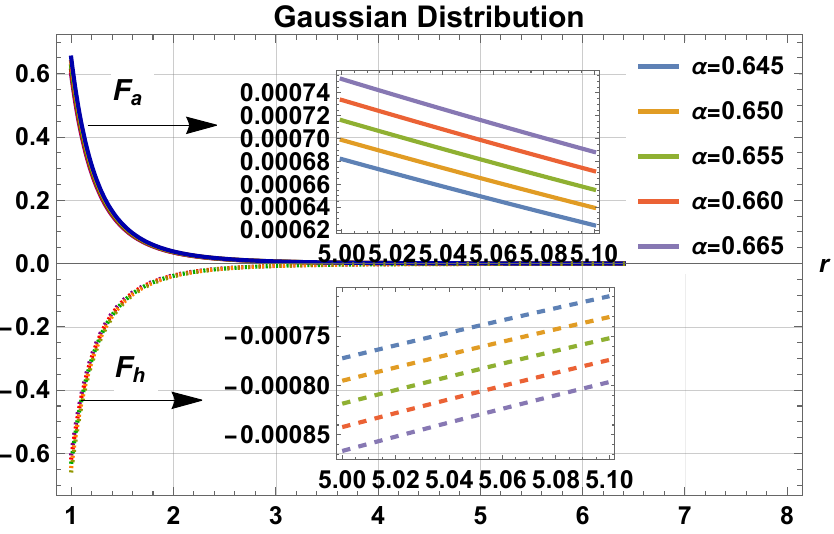}}\\
	    \subfloat[Equilibrium picture for Lorentzian distribution\label{fig:Beq}]{\includegraphics[width=0.6\linewidth]{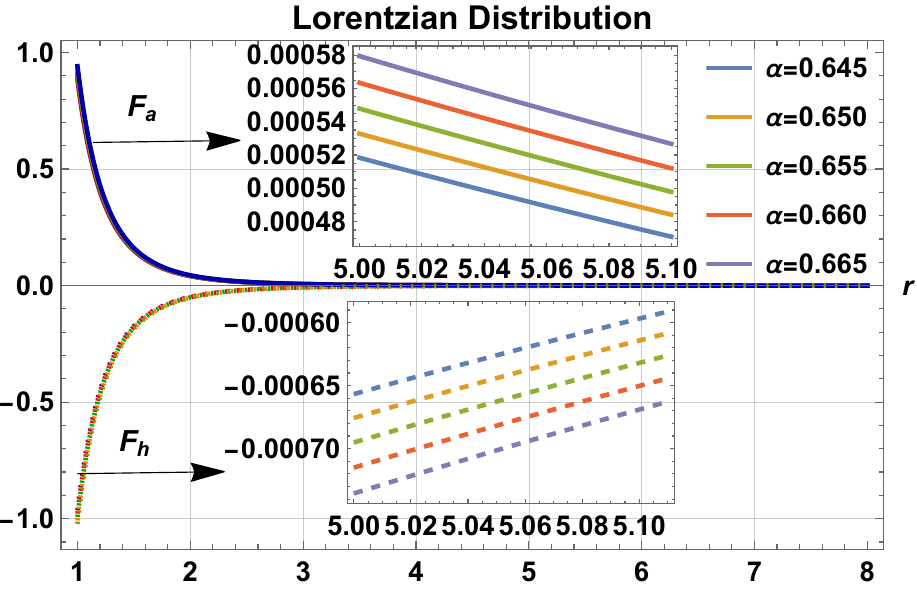}}
        \caption{The profile of hydro-static and anisotropic forces for different values of model parameter $\alpha$ with the total gravitational mass $M=1.2$ \textit{(mass)}, the square of minimal length $\theta=4$ \textit{(length$^{2}$)} and the throat radius $r_0=1$ \textit{(length)}.}
        \label{fig:chap6:eq}
    \end{figure}

    In this section, we shall analyze the stability of Gaussian and Lorentzian wormhole models. For this purpose, we use the TOV equation \eqref{eq:tov} by setting $F_c=0$ (since the coupling effect vanishes for the choice of Lagrangian in hand for $f(R,\mathcal{L}_m$ theory). Thus, \eqref{eq:tov} can be rewritten as, $F_g+F_h+F_a=0$. Since we have considered the tideless scenario, we have $\mathcal{R}_f=0$ implying  $F_h+F_a=0$. \figureautorefname$\;$\ref{fig:Aeq} and \figureautorefname$\;$\ref{fig:Beq} illustrate the behavior of hydro-static and anisotropic forces for Gaussian distribution and Lorentzian distribution. From these plots, one can assess the influence of the model parameter on the equilibrium condition. It can be interpreted from \figureautorefname$\;$\ref{fig:chap6:eq} that the nature of both hydro-static and gravitational forces are similar but opposite to one another.

\section{Interpretation of wormhole}\label{sec:chap6:V}
In this section, we shall discuss the physical aspects of the wormhole models.

\subsection{Average Pressure}
\par The average pressure $p$ can be described as,
$$p=\dfrac{1}{3}(p_r+2p_t).$$
For Gaussian distribution, the expression for average pressure reads,
    \begin{equation}
        p=\frac{(2-3 \alpha ) M e^{-\frac{r^2}{4 \theta }}}{24 \pi ^{3/2} \alpha  \theta ^{3/2}},
    \end{equation}
and for the Lorentzian distribution, it is given by,
    \begin{equation}
        p=-\frac{(3 \alpha -2) \sqrt{\theta } M}{3 \pi ^2 \alpha  \left(\theta +r^2\right)^2}.
    \end{equation}
    
\subsection{Speed of Sound}

\par The speed of sound parameter $v_s^2$ determines the stability of a wormhole. A wormhole is said to be stable if $0<v_s^2<1$ \cite{Capozziello:2020zbx,Poisson:1995sv,Usmani:2010cd,Mandal:2021qhx,Boshkayev:2020vrg}. The speed of sound parameter is expressed as, 
$$v_s^2=\dfrac{dp}{d\rho}.$$
For both Gaussian and Lorentzian distributions, this physical quantity is given by, 
\begin{equation}\label{sound}
    \dfrac{dp}{d\rho}=\frac{2}{3 \alpha }-1.
\end{equation}
One can note that $0\leq\dfrac{dp}{d\rho}<1$ is satisfied for $1/3\leq\alpha<2/3$. Therefore, non-commutative wormhole models are stable for $1/3\leq\alpha<2/3.$
\section{Concluding Remarks}\label{sec:chap6:VI}
\par
\begin{figure}[htbp]
            \centering
            \subfloat[$S_f(r)$\label{fig:Asf1}]{\includegraphics[width=0.49\linewidth]{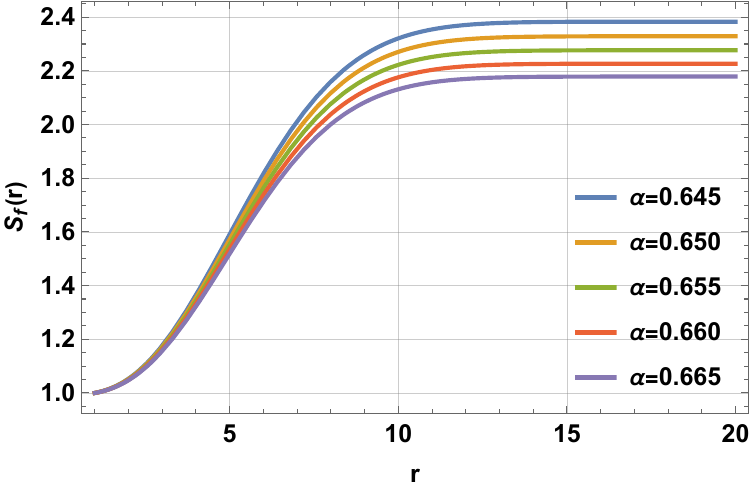}}
    	    \subfloat[$\frac{S_f(r)-rS_f'(r)}{S_f(r)^2}$\label{fig:Asf2a}]{\includegraphics[width=0.49\linewidth]{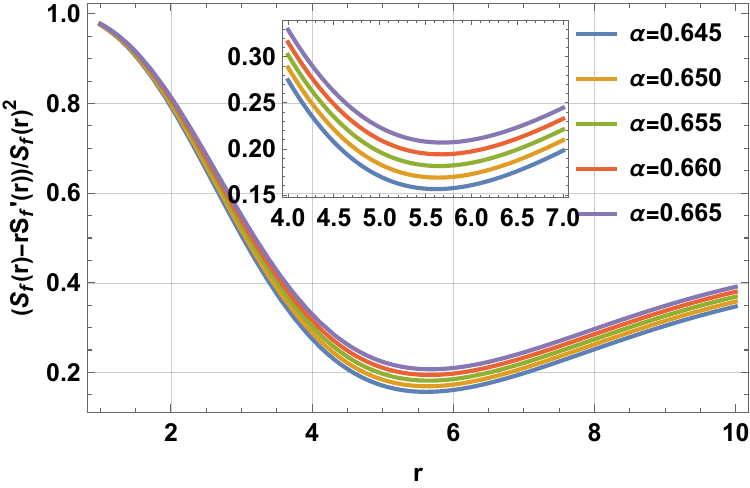}}\\
    	    \subfloat[$S_f'(r)$\label{fig:Asf2b}]{\includegraphics[width=0.49\linewidth]{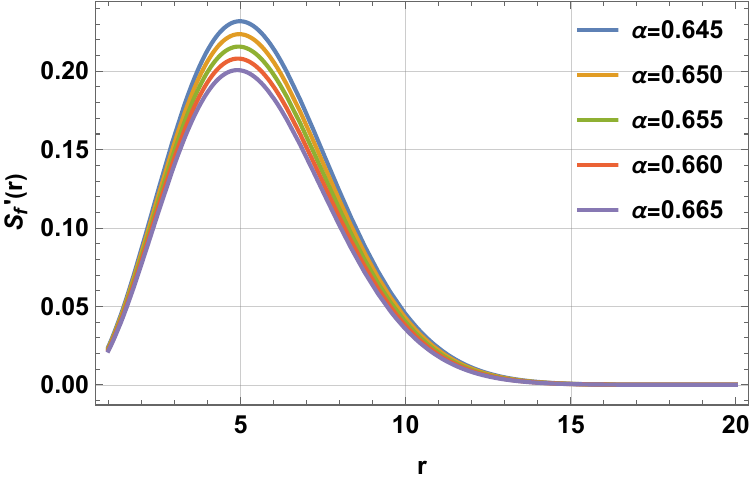}}
            \subfloat[$S_f(r)/r$\label{fig:Asf3}]{\includegraphics[width=0.49\linewidth]{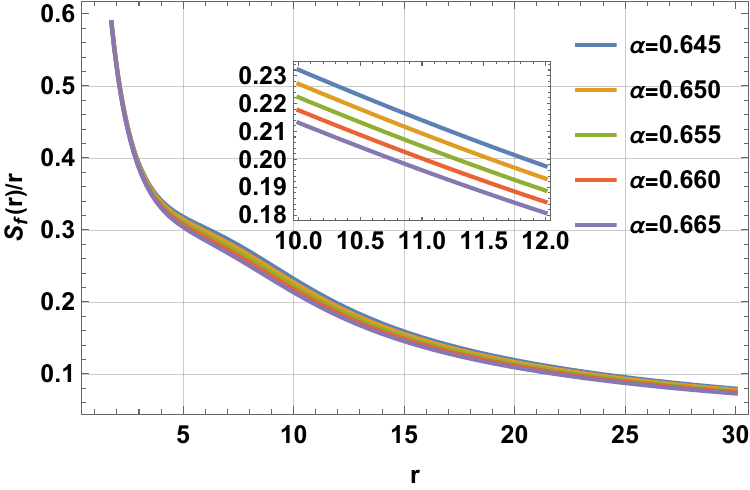}}
            \caption{Gaussian Distribution: The influence of model parameter $\alpha$ on the behavior of shape function $S_f$ with the total gravitational mass $M=1.2$ \textit{(mass)}, the square of minimal length $\theta=4$ \textit{(length$^{2}$)} and the throat radius $r_0=1$ \textit{(length)}.}
            \label{fig:Asf}
        \end{figure}

\begin{figure*}[htbp]
	    \centering
	    \subfloat[Energy density $\rho$\label{fig:Arho}]{\includegraphics[width=0.45\linewidth]{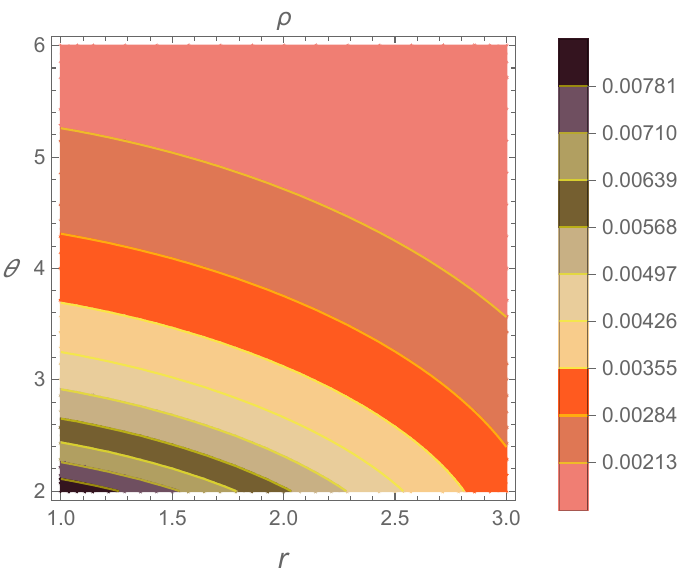}}
	    \subfloat[NEC $\rho+p_r$\label{fig:Ae1}]{\includegraphics[width=0.45\linewidth]{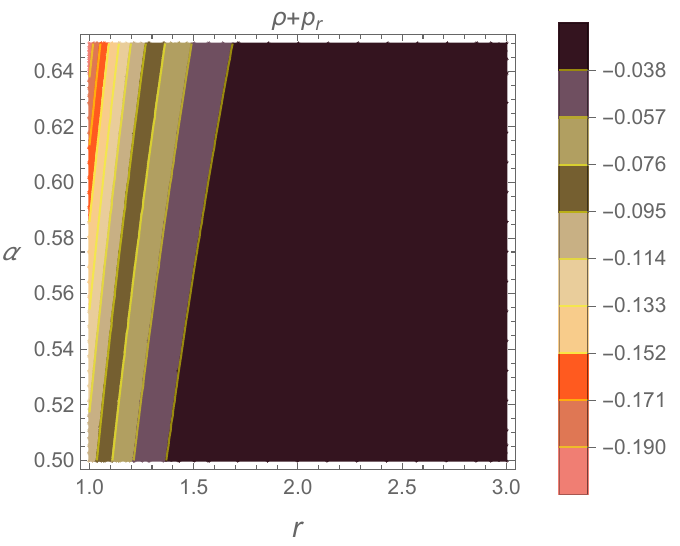}}\\
	    \subfloat[NEC $\rho+p_t$\label{fig:Ae2}]{\includegraphics[width=0.45\linewidth]{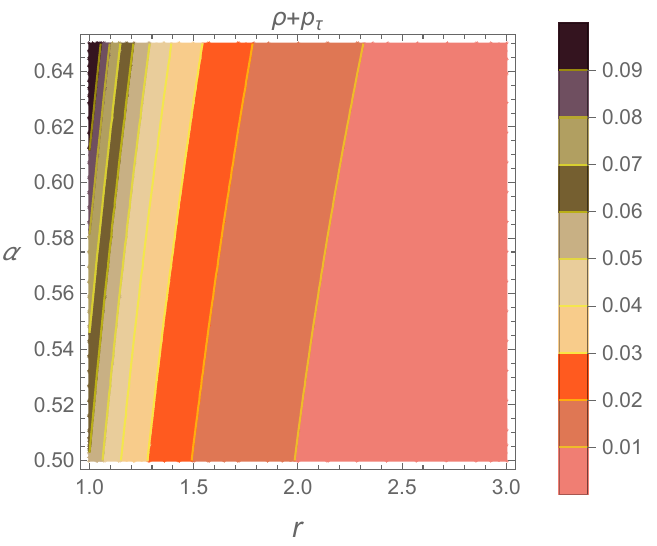}}
	    \subfloat[DEC $\rho-|p_r|$\label{fig:Ae3}]{\includegraphics[width=0.45\linewidth]{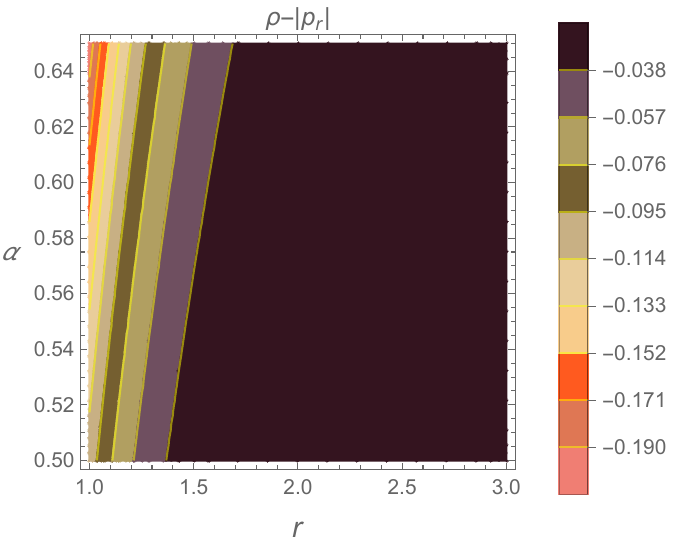}}\\
	    \subfloat[DEC $\rho-|p_t|$\label{fig:Ae4}]{\includegraphics[width=0.45\linewidth]{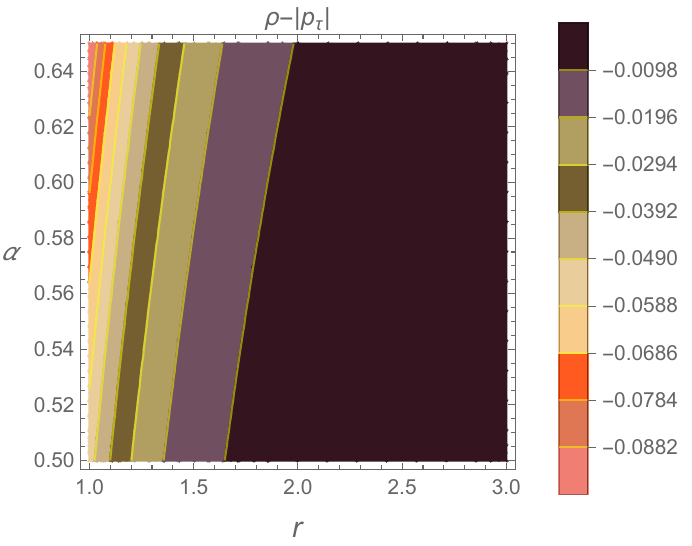}}
	    \subfloat[SEC $\rho+p_r+2p_t$\label{fig:Ae5}]{\includegraphics[width=0.45\linewidth]{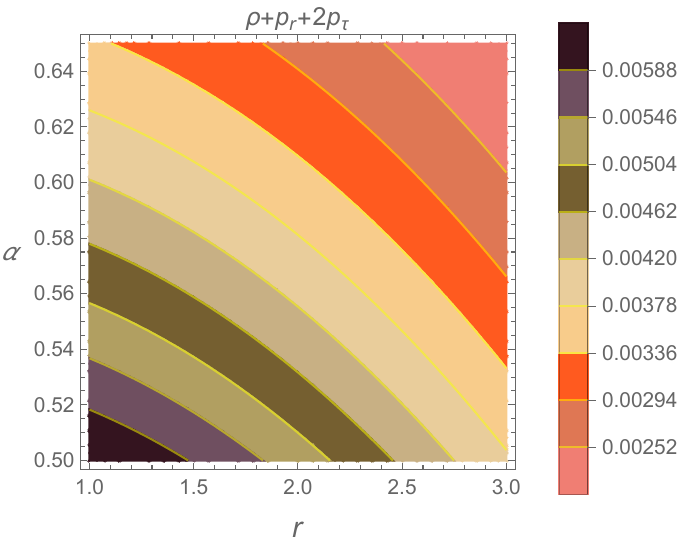}}
	    \caption{Gaussian Distribution: plot showing the profile of (a) energy density varying w.r.t radial coordinate $r$ and the square of minimal length $\theta$ \textit{(length$^{2}$)} with the total gravitational mass $M=1.2$ \textit{(mass)}, (b)-(f) different energy conditions varying w.r.t radial coordinate $r$ and model parameter $\alpha$ with the total gravitational mass $M=1.2$ \textit{(mass)}, the square of minimal length $\theta=4$ \textit{(length$^{2}$)} and the throat radius $r_0=1$ \textit{(length)}.}
	    \label{fig:Aec}
	\end{figure*}

\begin{figure}[htbp]
        \centering
        \subfloat[$S_f(r)$\label{fig:Bsf1}]{\includegraphics[width=0.49\linewidth]{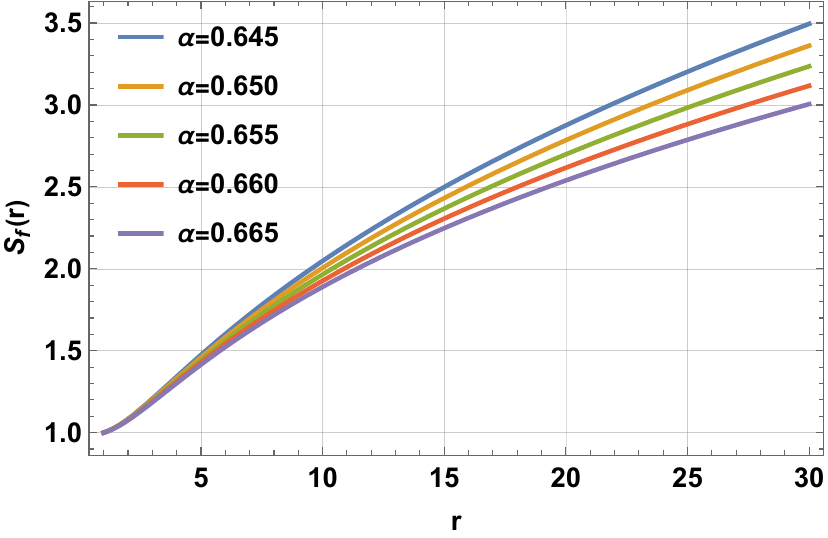}}
	    \subfloat[ $\frac{S_f(r)-rS_f'(r)}{S_f(r)^2}$\label{fig:Bsf2a}]{\includegraphics[width=0.49\linewidth]{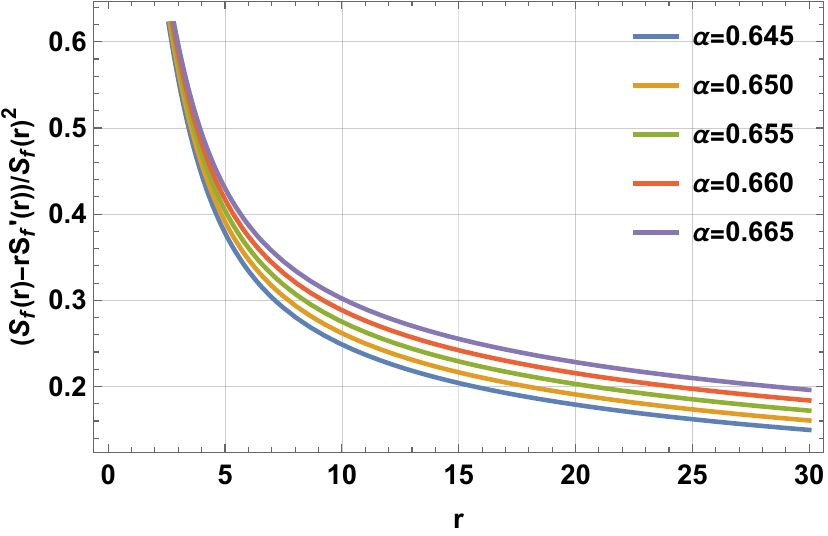}}\\
        \subfloat[$S_f'(r)$\label{fig:Bsf2b}]{\includegraphics[width=0.49\linewidth]{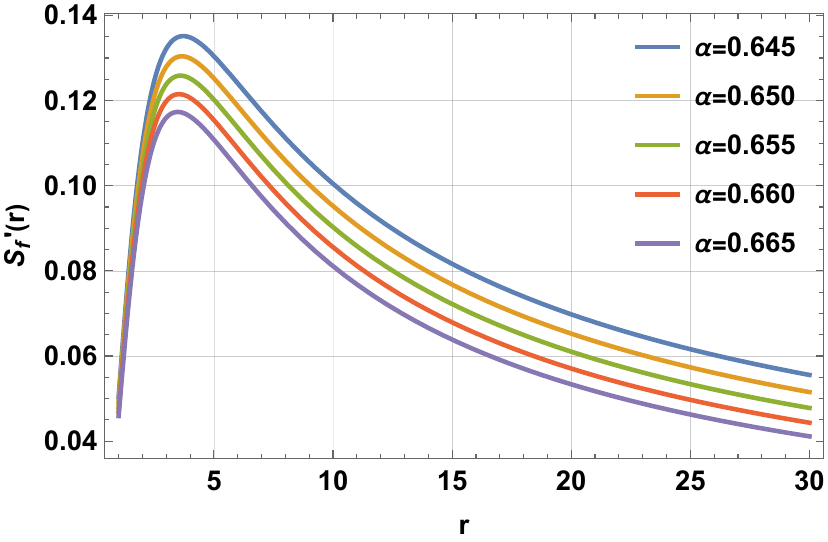}}
	    \subfloat[$S_f(r)/r$\label{fig:Bsf3}]{\includegraphics[width=0.49\linewidth]{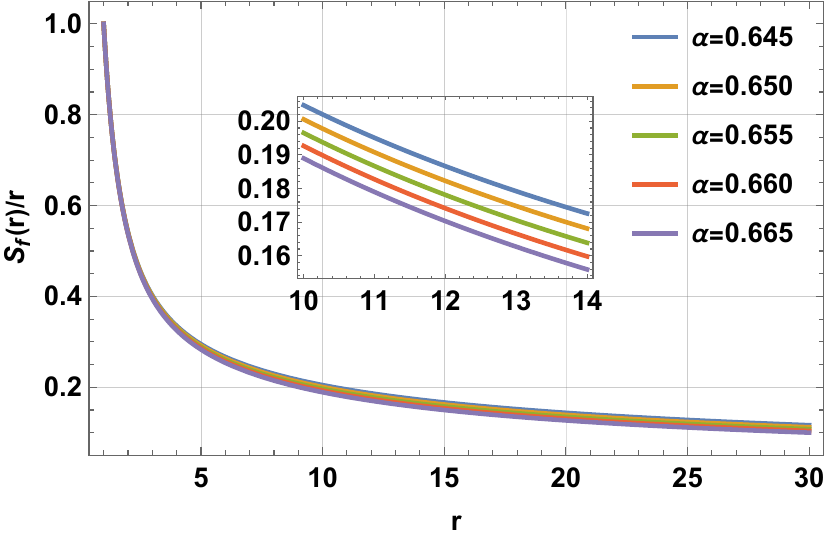}}
        \caption{Lorentzian Distribution: The influence of model parameter $\alpha$ on the behavior of shape function $S_f$ with the total gravitational mass $M=1.2$ \textit{(mass)}, the square of minimal length $\theta=4$ \textit{(length$^{2}$)} and the throat radius $r_0=1$ \textit{(length)}.}
        \label{fig:Bsf}
    \end{figure}

\begin{figure*}[htbp]
	    \centering
	    \subfloat[Energy density $\rho$\label{fig:Brho}]{\includegraphics[width=0.45\linewidth]{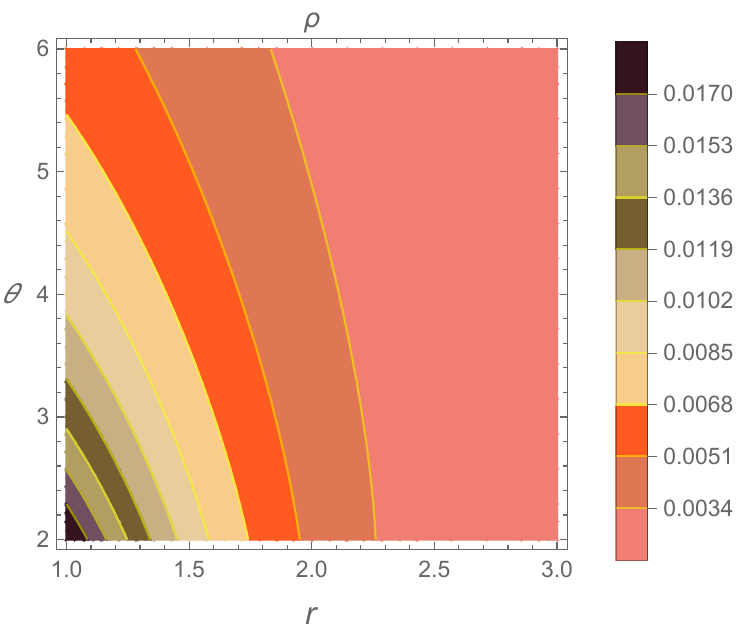}}
	    \subfloat[NEC $\rho+p_r$\label{fig:Be1}]{\includegraphics[width=0.45\linewidth]{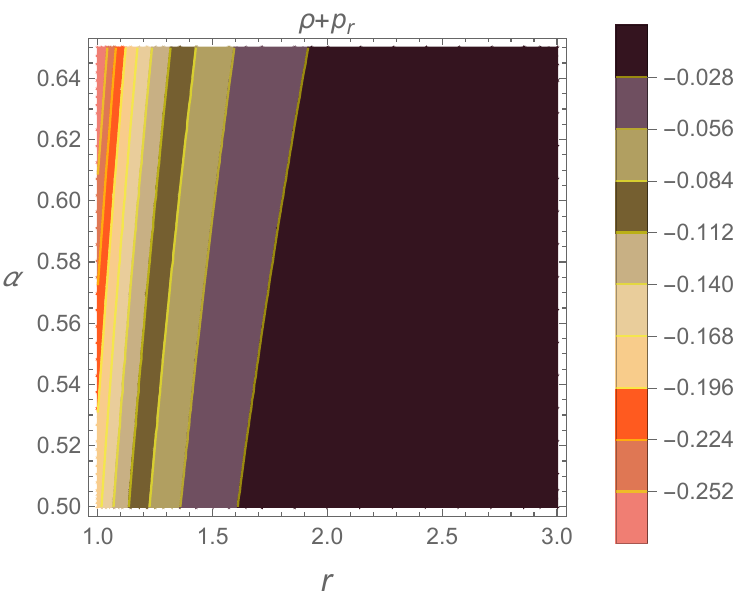}}\\
	    \subfloat[NEC $\rho+p_t$\label{fig:Be2}]{\includegraphics[width=0.45\linewidth]{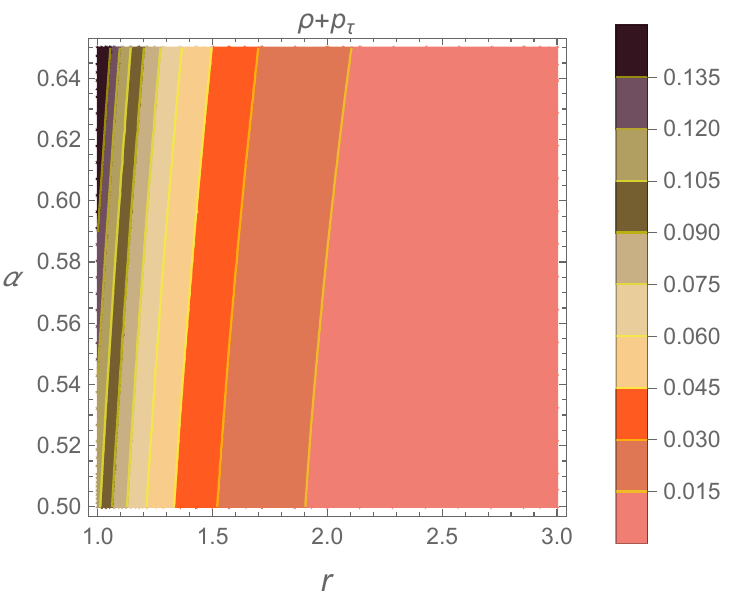}}
	    \subfloat[DEC $\rho-|p_r|$\label{fig:Be3}]{\includegraphics[width=0.45\linewidth]{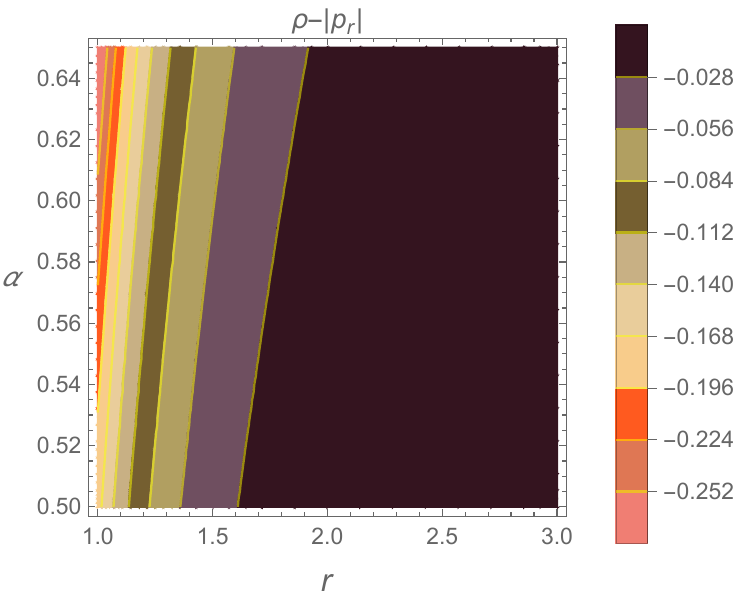}}\\
	    \subfloat[DEC $\rho-|p_t|$\label{fig:Be4}]{\includegraphics[width=0.45\linewidth]{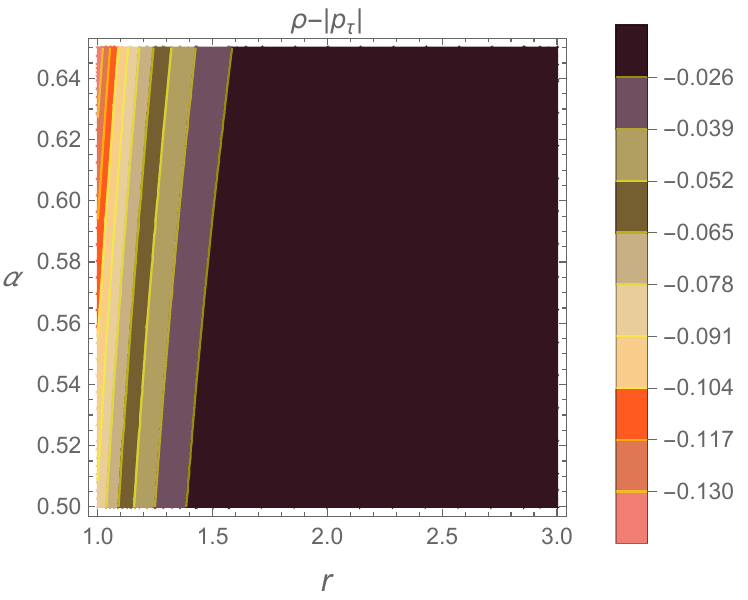}}
	    \subfloat[SEC $\rho+p_r+2p_t$\label{fig:Be5}]{\includegraphics[width=0.45\linewidth]{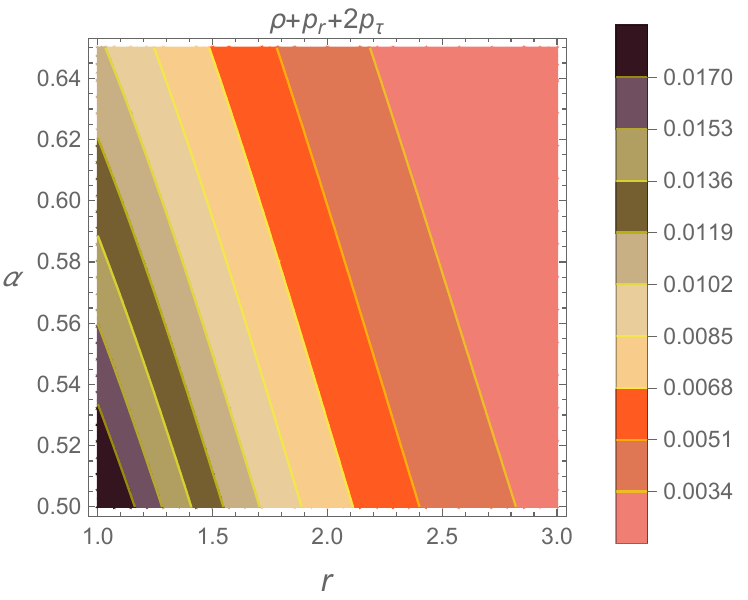}}
	    \caption{Lorentzian Distribution: plot showing the profile of (a) energy density varying w.r.t radial coordinate $r$ and the square of minimal length $\theta$ \textit{(length$^{2}$)} with the total gravitational mass $M=1.2$ \textit{(mass)}, (b)-(f) different energy conditions varying w.r.t radial coordinate $r$ and model parameter $\alpha$ with with the total gravitational mass $M=1.2$ \textit{(mass)}, the square of minimal length $\theta=4$ \textit{(length$^{2}$)} and the throat radius $r_0=1$ \textit{(length)}.}
	    \label{fig:Bec}
	\end{figure*}

 In our current chapter, we used an explicit coupling of matter with geometry which replaces the Ricci scalar in the Einstein-Hilbert action with an arbitrary function of the Ricci scalar and the matter Lagrangian. The geometry-matter coupling theory such as $f(R,\mathcal{L}_m)$ gravity can remarkably address the issue of exotic matter \cite{MontelongoGarcia:2010xd,Garcia:2010xb}. On the other hand, non-commutative geometry with modified matter sources provides a mathematical approach to dealing with physical phenomena.
\begin{itemize}
    \item With the anisotropic matter distribution, we studied wormholes with zero tidal force. 
    \item Further, we presumed a non-linear $f(R,\mathcal{L}_m)$ model $f(R,\mathcal{L}_m)=\dfrac{R}{2}+\mathcal{L}_m^\alpha$ with $\alpha$ being a free parameter. From \cite{Jaybhaye:2022gxq,Kavya:2022dam,Pradhan:2022msm}, we can see that the model is capable to explain the present scenario of the Universe. 
    \item In the first case, we examined the wormhole scenario with Gaussian distribution. Here, we derived a shape function satisfying the throat condition. We verified the range of values of the model parameter $\alpha$ for which the obtained shape function obeys the flaring-out and asymptotic flatness conditions [ref Table$\;$\ref{tab:table0}].
    \item In the second case, we considered Lorentzian distribution to analyze wormhole properties. The obtained shape function obeys all the necessary conditions for a traversable wormhole. The parameter values of $\alpha$ for which the shape function fulfills the criteria are represented in the Table$\;$\ref{tab:table1}.
    \item We verified the flaring-out condition at the throat for both non-commutative geometries. The obtained inequalities $\left(\frac{M\;e^{-\frac{1}{4\theta}}}{8\pi^{\frac{3}{2}}\theta^{\frac{3}{2}}}\right)^\alpha<\frac{1}{r_0^2}$ and $ \left(\frac{M\sqrt{\theta}}          {\pi^2(1+\theta)^2}\right)^\alpha<\frac{1}{r_0^2}$ constrained the parameter values $M$, $r_0$, and $\theta$. In addition, with the aid of the speed of sound parameter, we checked the stability of the wormhole. It is inferred from equation \eqref{sound} that these non-commutative wormholes are stable for $1/3\le\alpha<2/3$. Further, the equilibrium condition is verified through the TOV equation. 
    \item In \figureautorefname$\;$\ref{fig:Asf} and \ref{fig:Bsf}, we analyzed the impact of model parameter $\alpha$ on the behavior of shape functions. It is necessary to note that a minute variation in the value of $\alpha$ can impact the nature of shape functions. The nature of both shape functions is similar to that of the results obtained by Shamir and his collaborators \cite{Shamir:2021xwh}, in the context of exponential gravity coupled with the matter.
    \item In addition, throughout the chapter, we analyzed the influence of this parameter $\alpha$ on the physical properties of the wormhole. Also, in all figures, we plotted the profile for those values of $\alpha$ for which we obtain a physically plausible wormhole solution. Further, in our analysis, we ignored those values of the model parameter leading to the negatively defined energy density. \figureautorefname$\,$\ref{fig:Arho} and \ref{fig:Brho} show the density profile for both Gaussian and Lorentzian distribution.
    \item In both cases, there is a violation of the NEC, implying the existence of an exotic matter source. This agrees with the result obtained in GR and various modified theories with non-commutativity. In \cite{Sokoliuk:2022efj},  for the linear $f(Q)$ model with charge, NEC is violated. Also, in \cite{Mustafa:2021bfs}, the result is obtained in the absence of charge. A similar instance can be seen in \cite{Mustafa:2021vqz} with matter coupling in teleparallel gravity, which indicates the presence of exotic matter. In the context of Rastall gravity, Mustafa et al. \cite{Zubair:2019qqz} obtained the wormhole solutions violating NEC. 
\end{itemize}
\par In all, the chapter has presented a stable viable wormhole model in the framework of $f(R,\mathcal{L}_m)$ gravity with non-commutative distributions.

\newpage
\thispagestyle{empty}

\vspace*{\fill}
\begin{center}
    {\Huge \color{NavyBlue} \textbf{CHAPTER 7}}\\
    \
    \\
    {\Large\color{purple}\textbf{\textsc{Big Bang Nucleosynthesis in \texorpdfstring{$f(T)$}{f(T)} Gravity}}} \\
    \
    \\
    \textbf{Publication based on this chapter}\\
\end{center}
\textsc{Can Teleparallel $f(T)$ Models Play a Bridge Between Early and Late Time Universe?}, \textbf{NS Kavya}, SS Mishra,  PK Sahoo, V Venkatesha, \textit{Monthly Notices of the Royal Astronomical Society} (accepted) (Oxford University Press, Q1, IF - 4.8, H-index - 372) DOI: \href{https://doi.org/10.1093/mnras/stae1723}{10.1093/mnras/stae1723}
\vspace*{\fill}

\pagebreak

\def\baselinestretch{1}
\chapter{\textsc{Big Bang Nucleosynthesis in \texorpdfstring{$f(T)$}{f(T)} Gravity}}\label{chap7}
\def\baselinestretch{1.5}
\pagestyle{fancy}
\fancyhead[R]{\textit{Chapter 7}}

\textbf{Highlights:}\\
\textit{\begin{itemize}
    \item In the present chapter, we consider two highly motivated hybrid $f(T)$ models and constrain them using the observations from the Big Bang Nucleosynthesis era.
    \item In addition, using late-time observations of Cosmic Chronometers and Gamma-Ray-Bursts, the ranges of the model parameters are confined which are in good agreement with early time bounds.
    \item Subsequently, we investigate the viability of the models in the primordial-intermediating-present epochs.
\end{itemize}}
\section{Introduction} 

The concrete affirmation and validity of the standard cosmological model comes from the very beginning of the timeline covering primordial nucleosynthesis to the present-time acceleration. Based on the cosmological principle and the high energy interactions of fundamental elements, the standard cosmological model is capable of predicting the events that can reach up to nearly $10^{-43}$ s of the post-Big Bang era.  Currently, the standard model of particle physics, represented by the $SU(3)_C \otimes SU(2)_L \otimes U(1)_Y$ gauge theory of strong and electroweak interactions, presents a fundamental framework for understanding quarks and leptons which has been experimentally verified at energies up to approximately 1000 GeV. 

Numerous observational surveys, with several teams working together, probe the evolution of the Universe at different stages, covering a wide range of observational factors. For instance, the Hubble parameter $H_0$, the present mass density $\rho_0$, the deceleration parameter $q_0$, and the distance modulus $\mu$ provide insight into late-time phenomena of the Universe \cite{WMAP:2006bqn,WMAP:2010qai,Demianski:2016zxi}, and observations of the transition redshift $z_t$ are crucial for understanding changes in the Universe's phase. However, observations of the intermediate epoch of the Universe remain under explored to date. Interestingly, there are several significant experimental observations of the primordial Universe to center on, such as the abundance of $^3$He, $^4$He, D, $^7$Li, the baryon-to-photon ratio, and the CMBR, which help us predict dynamics as early as 0.01 s after the Big Bang. Since the Universe undergoes expansion and cooling in its early stages, the rates of weak interactions, pivotal for maintaining thermal equilibrium between neutrinos and matter, decrease progressively. Likewise, the rates of reactions converting neutrons and protons decline. Eventually, these rates dip below the expansion rate, leading to a freeze-out. Simultaneously, the strong and electromagnetic nuclear reaction rate also decreases and undergoes freeze-out. These epochs contribute to the nucleosynthesis era of the Big Bang model \cite{Fields:2019pfx,Barrow:2020kug,Anagnostopoulos:2022gej,Kang:2008zi,Mishra:2023onl,Capozziello:2017bxm}. 

Though the standard cosmological model provides viable explanations for phenomena from early to late times and aligns well with observational data, it has some limitations \cite{Joyce:2014kja,Tsujikawa:2003jp,Velten:2014nra,Astesiano:2022ozl,Katsuragawa:2016yir,Zaregonbadi:2016xna,Joudaki:2016kym}. This necessitates exploring models beyond $\Lambda$CDM. While quintessence models, which incorporate scalar fields with potential, can explain cosmological data with certain adjustments, the absence of strongly motivated scalar field models from theoretical particle physics is significant. Consequently, the effects of cosmic acceleration might be better explained within the framework of modified gravity theories, where extensions of GR are considered.

In this chapter, we explore an extension of gravitational theory TEGR \autoref{sub:chap1:TEGR}, which was initially proposed by Einstein himself in \cite{Unzicker:2005in, Maluf:2002zc, Obukhov:2002tm, Ferraro:2006jd, Linder:2010py}. TEGR describes the gravitational field using the torsion tensor. Although many results from TEGR match those from GR, extensions of TEGR can yield different outcomes compared to extensions of GR \cite{Cai:2015emx} (Also see \autoref{fig:cha1:trinity1}). As a result, torsion-based modified theories have garnered significant attention and have been thoroughly investigated in recent years \cite{Wu:2010mn,Chen:2010va,Dent:2010nbw,Sudharani:2023hss,Bahamonde:2021srr,Moreira:2021xfe,Mishra:2023khd,Kofinas:2014daa,Gonzalez-Espinoza:2021mwr}. One can also see some interesting works that adopt different techniques to constrain $f(T)$ gravity models \cite{Nunes:2016qyp,Nunes:2016plz,Ren:2022aeo,Briffa:2020qli}. Motivated by this, we study two teleparallel $f(T)$ models and their influence on the dynamics of the Universe from early to late times. 

This chapter is organized in the following manner: In \autoref{sec2} we provide a detailed geometric foundation of $f(T)$ teleparallel gravity. The \autoref{sec3} presents the formalism of BBN. Using BBN constraints we analyze two hybrid gravity models in \autoref{sec4}. The early Universe constraints are obtained from BBN, while the late time constraints are obtained from recent observational data sets of Cosmic chronometers and Gamma-Ray-Bursts. The methodology and results of this procedure are explained in \autoref{sec5} and in \autoref{sec6}, we conclude our results. 

\section{Geometric Formulation of \texorpdfstring{$f(T)$}{f(T)} Gravity}\label{sec2}
The simplest departure from the TEGR is considering an arbitrary functional form of the torsion scalar gives rise to the so-called "$f(T)$ theory". This theory uses the Weitzenb\"{o}ck connection instead of the usual torsion-less Levi-Civita connection of GR. Using the connection, the non-null torsion tensor can be obtained as,
\begin{equation}\label{eq:torsiontensor}
    T_{\mu \nu}^{\lambda}=\overset{w}{\Gamma}_{\nu \mu}^\lambda-\overset{w}{\Gamma}_{\mu \nu}^\lambda=e_{\gamma}^\lambda(\partial_{\mu}e_{\nu}^\gamma-\partial_{\nu}e_{\mu}^\gamma).
\end{equation}
Here $\overset{w}{\Gamma}_{\nu \mu}^\lambda$ is the Weitzenb\"{o}ck connection and the tetrad fields $e_{\mu}^\gamma$ or $e_{\mu}(x^\gamma)$ create a tangent space at each point $x^\gamma$ of the manifold. The metric tensor of the manifold can be written in terms of the tetrad fields as $g_{\mu \nu}(x)=\eta_{\alpha \beta} \, e_\mu^{\alpha}(x) \, e_\nu^{\beta}(x)$, with the pseudo-Riemannian metric $\eta_{\alpha \beta}=diag(1,-1,-1,-1)$.

 The torsion scalar can be defined as 
\begin{equation}\label{eq:torsionscalar}
    T\equiv {S_{\lambda}}^{\mu \nu} T_{\mu \nu}^\lambda,
\end{equation}
where the superpotential tensor is defined as ${S_{\lambda}}^{\mu \nu} \equiv \frac{1}{2}({K^{\mu \nu}}_{\lambda}+\delta_\lambda^\mu {T^{\alpha \nu}}_{\alpha}-\delta_\lambda^\nu {T^{\alpha \mu}}_{\alpha})$ and the contorsion tensor is defined as ${K^{\mu \nu}}_{\lambda} \equiv -\frac{1}{2}({T^{\mu \nu}}_{\lambda}-{T^{\nu \mu}}_{\lambda}-{T_{\lambda}}^{\mu \nu})$. Using the torsion scalar one can modify the GR action to teleparallel action as
\begin{equation}\label{eq:chap7:action}
    S=\frac{1}{2k^2} \int d^4xe[T+f(T)],
\end{equation}
where $k=\sqrt{8\pi G}$ and $e=det(e_\mu^A)=\sqrt{-g}$. 

Variation of the above action with respect to the tetrads gives rise to the motion equation
\begin{multline}
 \label{eq:field}
   (1+f_T)\left[e^{-1}\partial_\mu\,(e\,{e^\lambda_{\,A}} \,S_\lambda^{\,\, \nu \mu})-e^\alpha_{\,A} \,T^{\lambda}_{\,\, \mu \alpha}\, S_{\lambda}^{\,\, \mu \nu} \right]+\\
   e^{\lambda}_{\,A}\, S_\lambda^{\,\, \nu \mu}\, \partial_\mu T \, f_{TT}+\frac{1}{4} \,e^\nu_{\,A} \,[T+f]=4 \pi G \,e^\lambda_{\,A} \stackrel{em}{T}_\lambda^{\,\,\, \nu},
\end{multline}
 where $f_T, \, f_{TT}$ are the first and second derivative of $f$ with respect to the torsion scalar, respectively and $\stackrel{em}{T}_\lambda^{\,\,\, \nu}$ denotes the usual energy momentum tensor. For further discussion, we consider a spatially flat Friedmann-Lemaitre-Robertson-Walker (FLRW) metric of the form $ds^2=dt^2-a^2(t)\,\delta_{ij}\,dx^i\, dx^j$ with $a(t)$ being the scale factor in terms of time. The vierbein ansatz, corresponding to the metric, considered of the form $e_\mu^A=diag(1,a(t),a(t),a(t))$. The vierbein along with \eqref{eq:torsionscalar} gives rise to the relation $T=-6H^2$ where the Hubble parameter $H=\dot{a}/a$.

Inserting the vierbein in the motion equation \eqref{eq:field}, the modified Friedmann equations can be obtained as
\begin{equation}
\label{eq:motion1}
    12 H ^2 (1+f_T)+T+f=2k^2 \rho,
\end{equation}
\begin{equation}\label{eq:motion2}
    48 H ^2 \dot{H} f_{TT}-(1+f_T)(4\dot{H}+12H^2)+f-T=2 k^2 p,
\end{equation}
where $\rho=\rho_m+\rho_r$ and $p=p_m+p_r$. The above motion equations can be rewritten in the GR equivalent form as
\begin{gather}\label{eq:motiongr1}
    H^2=\frac{k^2}{3}(\rho +\rho_{DE}),\\ \label{eq:motiongr2}
    3H^2+2\dot{H}=-\frac{k^2}{3}(p +p_{DE}).
\end{gather}
 The DE components in the above equations are expressed in the context of $f(T)$ gravity as 
\begin{gather}\label{eq:rhode}
    \rho_{DE}=\frac{3}{k^2} \left(\frac{T f_T}{3}-\frac{f}{6}\right),\\\label{eq:pde}
    p_{DE}=\frac{1}{2k^2} \frac{f-T f_T +2 T^2 f_{TT}}{1+f_T+2 T f_{TT}}.
\end{gather}

\section{Background of BBN}\label{sec3}

Recent studies have suggested that BBN occurred within the initial fractions of seconds following the Big Bang, approximately around 0.01 s, extending to several hundred seconds after. During this period, the Universe was extremely hot and dense. BBN, along with the CMBR, provides firm validation for the high temperatures that characterized the primordial Universe. Here, we shall explore BBN physics in the context of $f(T)$ teleparallel gravity. The occurrence of BBN takes place in the era in which radiation was dominating. The relativistic elements (as well as mass-less radiation) in this stage have the energy density $\rho_r$ given by 

\begin{equation}\label{eq:rhor}
    \rho_r=\frac{\pi^2}{30} g_* \mathcal{T}^4.
\end{equation}

Here, $\mathcal{T}$ is the temperature, and $g_*$ represents the effective number of degrees of freedom, which is nearly equal to 10. The abundance of neutrons is calibrated by considering the proton-to-neutron conversion rate \cite{Kolb:1990vq}

\begin{equation}\label{eq:lambdapn}
    \lambda_{pn}(\mathcal{T}) = \lambda_{(n + \nu_e \rightarrow p + e^-)} + \lambda_{(n + e^+ \rightarrow p + \bar{\nu}_e)} + \lambda_{(n \rightarrow p + e^- + \bar{\nu}_e)}.
\end{equation}

The total rate is given by
\begin{equation}\label{lambdatot}
    \lambda_{tot}(\mathcal{T}) = \lambda_{pn}(\mathcal{T}) + \lambda_{np}(\mathcal{T}),
\end{equation}

where $\lambda_{np}(\mathcal{T})$ is the inverse of $\lambda_{pn}(\mathcal{T})$. In view of \eqref{eq:lambdapn}, the above equation takes the form \cite{Anagnostopoulos:2022gej}

\begin{equation}
   \lambda_{tot}(\mathcal{T})= 4 \mathcal{A} T 3(4!T 2 + 2 \times 3!QT + 2!Q2).
\end{equation}

In the above equation,  $Q$  represent the neutron-proton mass difference, given by  $Q = m_n - m_p = 1.29 \times 10^{-3} \, \text{GeV}$ . Additionally, $\mathcal{A}$ is taken as  $\mathcal{A} = 1.02 \times 10^{-11} \, \text{GeV}^{-4}$. Using the relation 
\begin{equation}
    Y_p \equiv e^{-\frac{t_n - \mathcal{T}_F}{\tau}}
 \frac{2e^{-\frac{Q}{T(\mathcal{T}_F)}}
}{1 + e^{-\frac{Q}{T(\mathcal{T}_F)}}
},
\end{equation}
the primordial mass fraction of $^4$He can be calibrated with $t_n$ and $\mathcal{T}_F$ being the freeze-out time and freeze-out temperature for nucleosynthesis and weak interactions, respectively, and $\tau$ refers to the mean lifetime of neutrons \cite{Barrow:2020kug}. 

In the context of GR, the first Friedmann equation can be written as $H^2=\frac{k^2}{3}\rho_{eff}$ \eqref{eq:motiongr1}, where $\rho_{eff}$ is the effective energy density of the system. Since radiation dominates in the BBN era, observational evidence indicates that any other contribution is negligible compared to radiation when the concordance model is considered. Thus, we have
\begin{equation}\label{eq:HGR}
    H^2 \approx \frac{k^2}{3} \rho_r \equiv H^2_{GR},
\end{equation}
where $k=1/M_p$, with $M_p = 1.22 \times 10^{19} \, \text{GeV}$ being the plank mass.

Here, we provide the convention adopted in the later part of this chapter: to distinguish the Hubble rate in GR from that obtained in $f(T)$ theory, we denote the former as $H_{GR}$ and the latter simply as $H$. Thus, equation \eqref{eq:motiongr1} along with the equation \eqref{eq:HGR} reads

\begin{equation}
    H = H_{GR} \sqrt{1 + \frac{\rho_{DE}}{\rho_r}}.
\end{equation}

Expanding the quantity within the square root to first order is justified by the dominance of DE density being significantly smaller than that of radiation during the radiation-dominated era and thus obtaining 

\begin{equation}\label{eq:deltaH}
    \Delta H = H - H_{GR} \approx \frac{\rho_{DE}}{\rho_r} \frac{H_{GR}}{2}.
\end{equation}

The equation \eqref{eq:rhor} in the above equation yields 

\begin{equation}\label{eq:H(T)}
    H(\mathcal{T}) = \sqrt{\frac{\pi^2 g_*}{90}} \frac{\mathcal{T}^2}{M_{p}}.
\end{equation}

Thus, the scale factor evolves as a $\sim t^{1/2}$, with $t$ being the cosmic time. The relation between temperature and time is therefore given by

\begin{equation}
    \frac{1}{t} \sim \left( \frac{32\pi^3 g_*}{90} \right)^{1/2} \frac{\mathcal{T}^2}{M_p}
\end{equation}

or in other words $\mathcal{T}(t) \sim (t/\text{sec})^{-1/2} \, \text{MeV}$. 

Central to BBN is the concept of the freeze-out temperature, $\mathcal{T}_F$. Neutrons and protons interconvert via weak interactions. When the temperature exceeds 1 MeV, much higher than the expansion rate, these conversions remain in equilibrium. However, as the temperature falls below 1 MeV, the neutron-to-proton ratio 'freezes out' at approximately 1/6, gradually declining due to the decay of free neutrons. The freeze-out temperature in hand is related to the specific Hubble rate  $H(\mathcal{T}_F) = \lambda_{tot}(\mathcal{T}_F) \approx c_q \mathcal{T}_F^5$, where  $c_q = 4\mathcal{A}4! \approx 9.8 \times 10^{-10} \, \text{GeV}^{-4}$ . Utilizing equation \eqref{eq:H(T)} at  $\mathcal{T} = \mathcal{T}_F$ , we find
\begin{equation}
     \mathcal{T}_F = \left( \frac{\pi^2 g_*}{90 M^2_{p} c^2_q} \right)^{1/6} \approx 0.6 \, \text{MeV}
\end{equation}

Given $H(\mathcal{T}_F) = c_q \mathcal{T}^5_F$ , we derive  $\Delta H = 5c_q T^4_f \Delta \mathcal{T}_F$. Substituting $\Delta H$  from equation \eqref{eq:deltaH}, we obtain

\begin{equation}\label{eq:deviation}
     \frac{\Delta \mathcal{T}_F}{\mathcal{T}_F} \approx \frac{\rho_{DE}}{\rho_r} \frac{H_{GR}}{10c_q \mathcal{T}^5_F}. 
\end{equation}

Taking into account the observational constraint, we arrive at the expression constraining the freeze-out temperature as

\begin{equation}\label{eq:bbnbound}
    \left| \frac{\Delta \mathcal{T}_F}{\mathcal{T}_F} \right| < 4.7 \times 10^{-4}.
\end{equation}

\section{Constraining \texorpdfstring{$f(T)$}{f(T)} models} \label{sec4}
 From theoretical and observational aspects of BBN, one can derive constraints on a given cosmological model. A cosmological model is presumed to be viable only if it meets the conditions set by BBN. Therefore, in this section, we shall compare various $f(T)$ gravity models with BBN calculations.
 
\subsection{Hybrid exponential model}
We begin considering the Hybrid exponential $f(T)$ model having the form
\begin{equation}\label{eq:model1}
    f(T)=T \left(e^{\frac{n  T_0}{T}}-1\right),
\end{equation}
with $n$ being the free parameter. The model, in similar lines with \cite{Anagnostopoulos:2021ydo}, has proved its worth by providing a better fit to the observational datasets than the $\Lambda CDM$ model. The model reduces to TEGR with the limit $n=0$. The DE density \eqref{eq:rhode} for this model can be calculated as
\begin{equation}\label{eq:rhodem1}
\rho_{DE}=\frac{1}{2 \,{M_p}^2} \left( e^{\frac{n  H_0^2}{H^2}} \left(12 \, n \, H_0^2-6 H^2\right)+6 H^2 \right).
\end{equation}
Using the above expression in the first motion equation \eqref{eq:motion1}, we obtain 
\begin{equation}
e^{\frac{n  H_0^2}{H^2}} \left(H^2-2 n  H_0^2\right)=H_0^2 \Omega_{m0}  (z+1)^3
\end{equation}
where $\Omega_{m0}= \rho_{m0}/3{H_0}^2$, is the present density parameter. 

Again we use \eqref{eq:rhodem1} to find the deviation in freeze out temperature from \eqref{eq:deviation},

\begin{equation}\label{eq:Tfm1}
\frac{\Delta \mathcal{T}_F}{\mathcal{T}_F}=\frac{e^A \left(180 \,n \, H_0^2 \,{M_p}^2-\pi ^2 \,g \,{\mathcal{T}_F}^4\right)+\pi ^2 \,g\, {\mathcal{T}_F}^4}{30 \sqrt{10\,g} \,\pi \, c_q\, {M_p}^5 \,{\mathcal{T}_F}^7 },
\end{equation}
where $A={\frac{90 n  H_0^2 {M_p}^2}{\pi ^2 g {\mathcal{T}_F}^4}}$. To constrain the model using the bound on deviation, the values for the fixed parameters are incorporated as $M_p=1.22 \times 10^{19}\,\,GeV$ , $\mathcal{T}_F=0.0006\,\,GeV$, $ c_q=9.8\times 10^{-10} GeV^{-4} $ and $g \sim 10$. Moreover, the present value of the Hubble parameter is considered as $H_0 = 1.47\times 10^{-42\;}GeV$ which is equivalent to the value from Planck 2018 results i.e., $H_0=67.2$ km $s^{-1}$ Mp$c^{-1}$.

\begin{figure}
    \centering
    \includegraphics[width=0.7\linewidth]{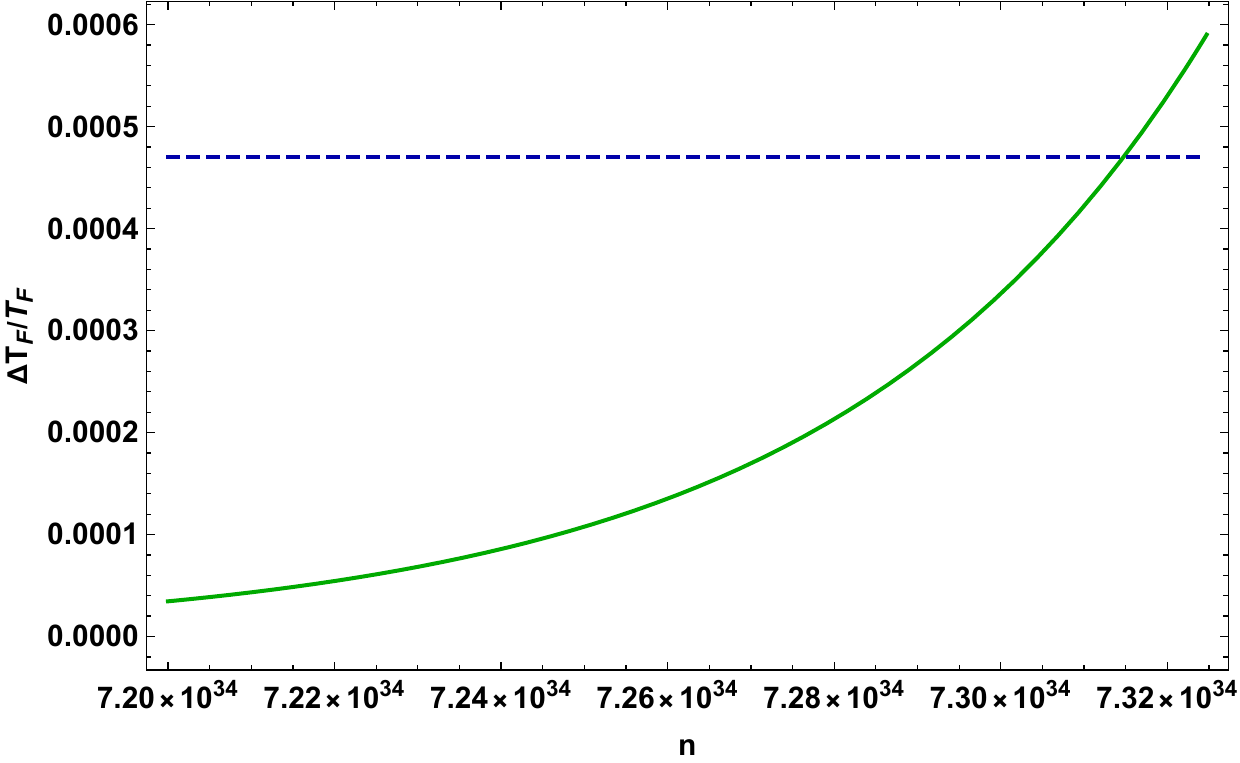}
    \caption{$n$ vs $\Delta \mathcal{T}_F/\mathcal{T}_F$ for the Hybrid exponential model. The blue dashed line represents the upper bound of $\Delta \mathcal{T}_F/\mathcal{T}_F$. }
    \label{fig:bbn1}
\end{figure}

We depict from \autoref{fig:bbn1} that the model satisfies the BBN bound \eqref{eq:bbnbound} for $n \leq 7.3147 \times 10^{34}$. The blue dashed curve represents the upper bound of the deviation $\left(\frac{\Delta \mathcal{T}_F}{\mathcal{T}_F}=0.00047\right)$. As the curve is asymptotic to zero towards the left, the domain we obtain for the model parameter is unbounded below. 

\subsection{Hybrid tangent hyperbolic model}
The Hybrid tangent hyperbolic model is motivated from \cite{Wu:2010av} because it allows phantom divide crossing line for the effective EoS parameter and corroborates with the recent observational datasets. The model consists of a tangent hyperbolic and a power law term that has  the following functional form 

\begin{equation}\label{eq:model2}
    f(T)=\lambda T_0 \,\left(\frac{T}{T_0}\right)^a \tanh\left(\frac{T_0}{T}\right),
\end{equation}
with $a$ and $\lambda$ being the free parameters. The model reduces to TEGR with the limit $\lambda=0$. The DE density \eqref{eq:rhode} for this model reads
\begin{multline}\label{eq:rhodem2}
    \rho_{DE}=\frac{\lambda \, M_p^2}{2} \, E ^{2a-2}\left(12\,H_0^2 \, \text{sech}\left(\frac{1}{E}\right)^4+
    \right. \left.
    6H^2\,(1-2a)\, \tanh\left(\frac{1}{E}\right)^2\right),
\end{multline}
where $E=H/H_0$. Using the above expression in the first motion equation \eqref{eq:motion1}, we obtain 
\begin{multline}
  H^2=\lambda \left(\frac{H}{H_0}\right)^{2a-2} \left(2 {H_0}^2 \text{sech}\left(\frac{H_0}{H}\right)^{4}+ \right.\\\left.
  H^2(1-2a) \tanh \left(\frac{H_0}{H}\right)^{2}   \right)+\Omega_{m0}(1+z)^3.
\end{multline}

Considering the present time scenario in the above equation, one can obtain the dependency between the model parameters as
\begin{equation}\label{eq:lambda}
    \lambda=\frac{1- \Omega_{m0}}{2 \, \text{sech}(1)+(1-2a)\, \tanh(1)}.
\end{equation}
Using \eqref{eq:rhodem2} in the theoretical expression for the deviation in freeze-out temperature \eqref{eq:deviation}, we get

\begin{equation}\label{eq:Tfm2}
\begin{split}
    \frac{\Delta \mathcal{T}_F}{\mathcal{T}_F}= &\frac{1}{c_q \,g\, {\mathcal{T}_F}^9 \sqrt{\frac{90}{C}} (-2 a \,\tanh (1)+\tanh (1)+2 \, \text{sech}(1))}\times \\& \left[{H_0} \,3^{1-2 a} \,10^{-n-\frac{1}{2}} \,\pi ^{2 a-4}\, (1-\Omega_{m0})\left(\frac{90}{\pi^2 C}\right)^a \times \right.\\&\left.\left(\pi ^2 g (1-2 a)\, {\mathcal{T}_F}^4\, \tanh \left(C\right)+180 {H_0}^2 {M_p}^2 \text{sech}^2\left(C\right)\right)\right],
\end{split}
\end{equation}
where $C=\frac{90 \, {H_0}^2 \, {M_p}^2}{\pi ^2 \, g \, {\mathcal{T}_F}^4}$.

\begin{figure}
    \centering
    \includegraphics[width=0.7\linewidth]{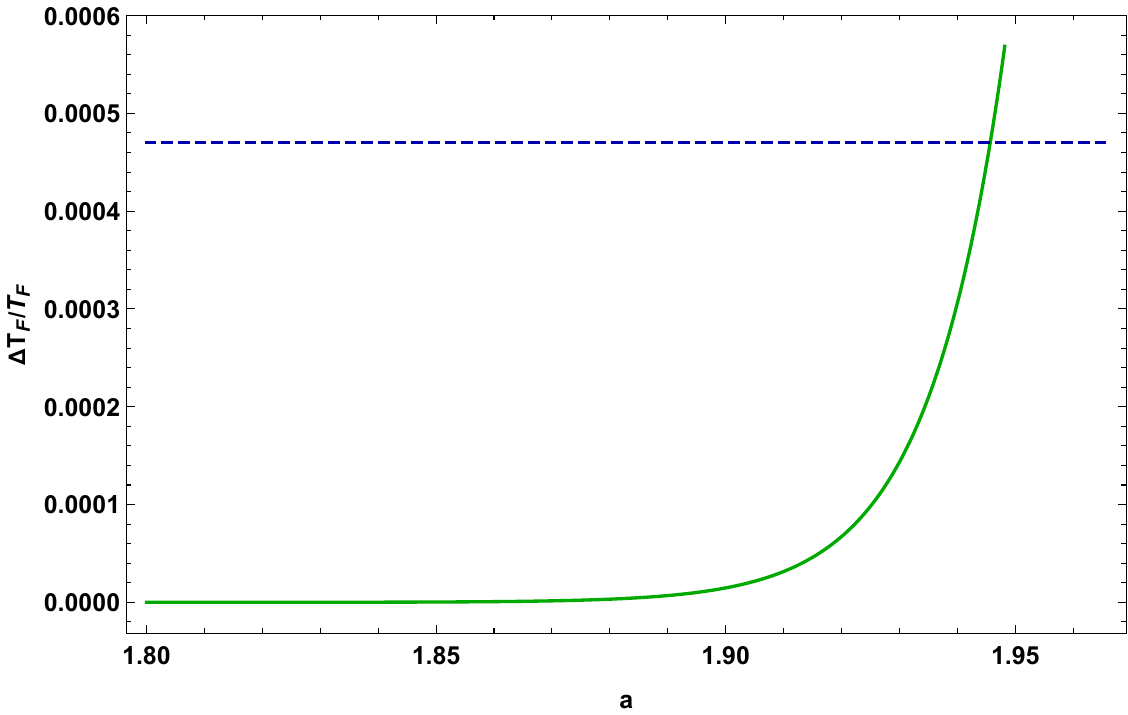}
    \caption{$a$ vs $\Delta \mathcal{T}_F/\mathcal{T}_F$ for the Hybrid tangent hyperbolic model. The blue dashed line represents the upper bound of $\Delta \mathcal{T}_F/\mathcal{T}_F$.}
    \label{fig:bbn2}
\end{figure}

We depict from \autoref{fig:bbn2} that the model satisfies the BBN bound \eqref{eq:bbnbound} for $a \leq 1.94566$. The values of fixed parameters are considered the same as the first model. By straightforward calculation, one can obtain the range for the other model parameter $\lambda$ from \eqref{eq:lambda}. The blue dashed curve represents the upper bound of the deviation $\left(\frac{\Delta \mathcal{T}_F}{\mathcal{T}_F}=0.00047\right)$. Similar to the first model, the curve is asymptotic to zero towards the left, hence, the domain for the model parameter is unbounded below.

\section{Verification of the models in late-time using observational datasets}\label{sec5}

To develop a cosmological model that viably explains the evolution of the Universe, we need to find valid ranges of space for free parameters. This is achieved through the Bayesian approach along with the MCMC methodology. For this purpose, in the present chapter, we use Cosmic Chronometers and Gamma-Ray-Bursts datasets.

In the previous section, we restricted the model parameters of two highly motivated models through BBN constraints. This section is dedicated to test the credibility of the said models in various epochs while staying within the limits.

\begin{figure}
    \centering
    \includegraphics[width=0.56\linewidth]{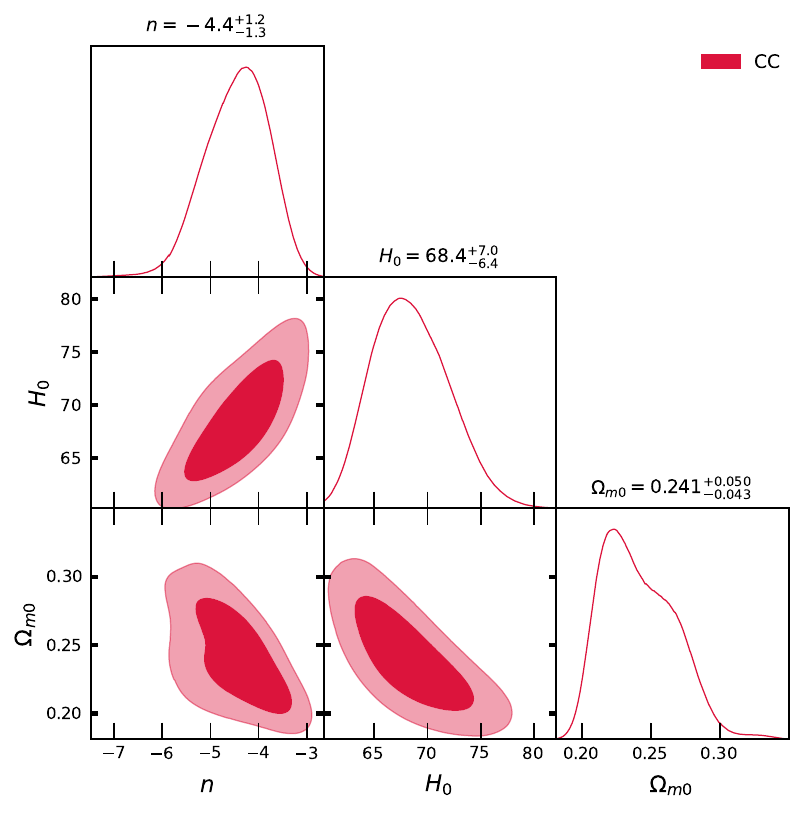}
    \caption{Likelihood probability distributions and 2D contours for the Hybrid exponential model obtained from MCMC analysis of the CC dataset. The dark-shaded region represents $1 \sigma$ CL and the light-shaded region represents $2 \sigma$ CL. }
    \label{fig:c1}
\end{figure}

\begin{figure}
    \centering
    \includegraphics[width=0.56\linewidth]{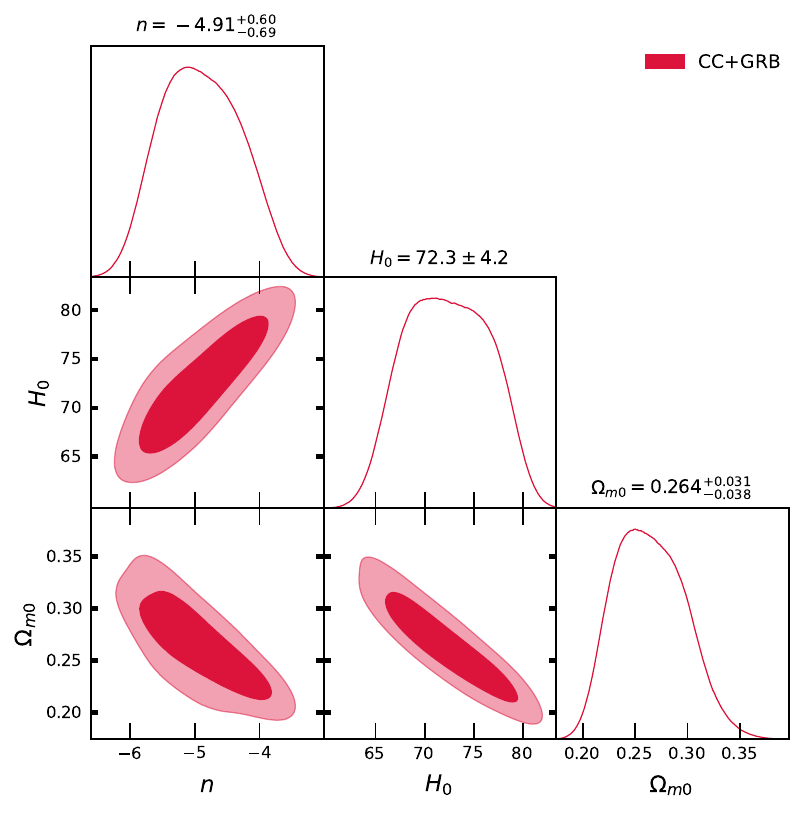}

        \caption{Likelihood probability distributions and 2D contours for the Hybrid exponential model obtained from MCMC analysis of the CC + GRB dataset. The dark-shaded region represents $1 \sigma$ CL and the light-shaded region represents $2 \sigma$ CL.}
    \label{fig:c2}
\end{figure}

\begin{figure}
    \centering
    \includegraphics[width=0.56\linewidth]{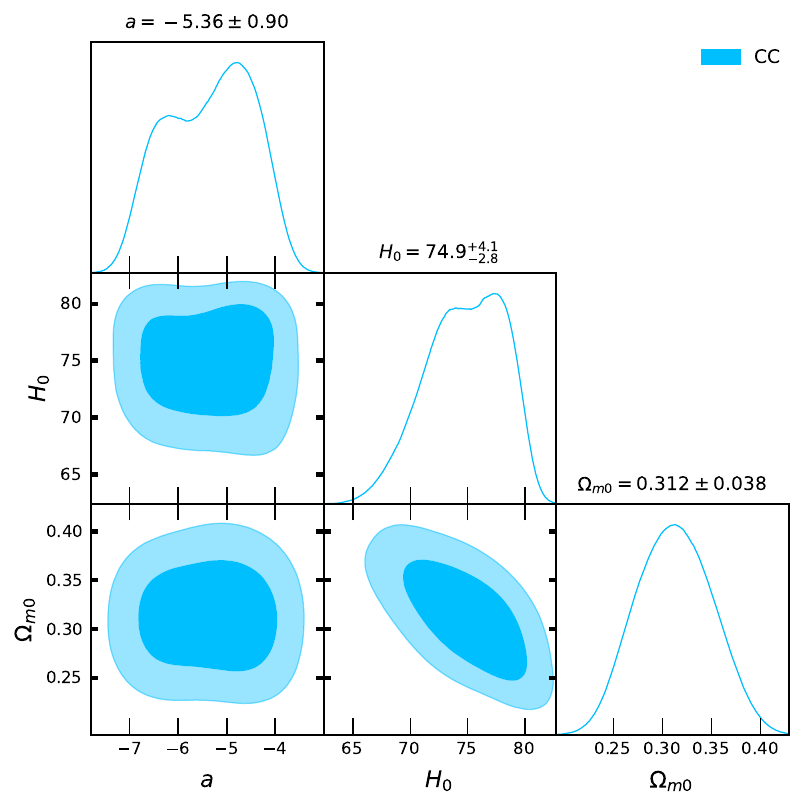}
    \caption{Likelihood probability distributions and 2D contours for the Hybrid tangent hyperbolic model obtained from MCMC analysis of the CC dataset. The dark-shaded region represents $1 \sigma$ CL and the light-shaded region represents $2 \sigma$ CL.}
    \label{fig:c3}
\end{figure}

\begin{figure}
    \centering
    \includegraphics[width=0.56\linewidth]{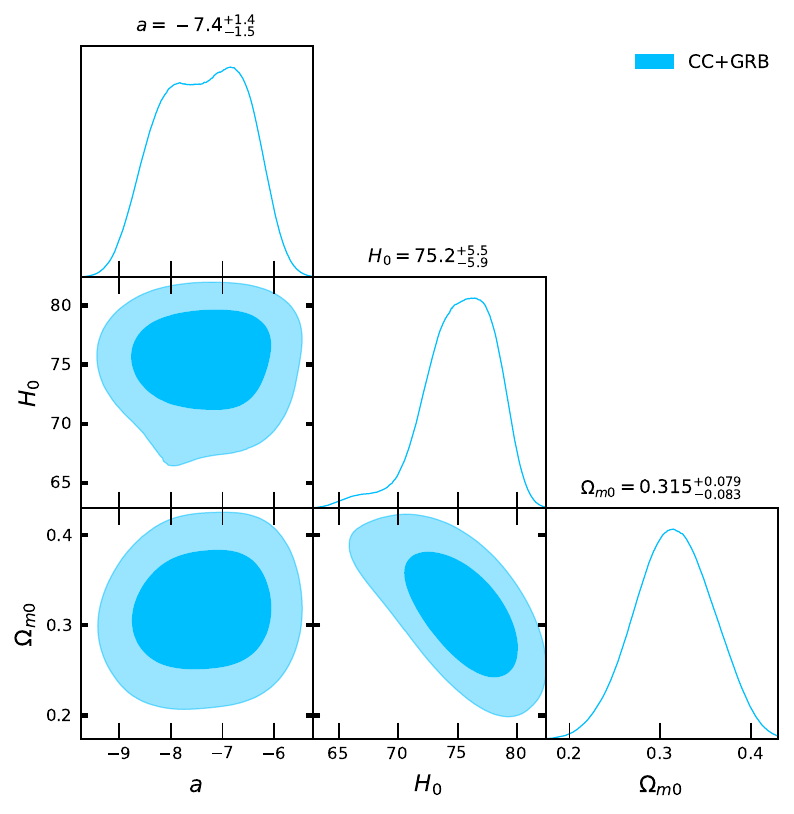}
    \caption{Likelihood probability distributions and 2D contours for the Hybrid tangent hyperbolic model obtained from MCMC analysis of the CC + GRB dataset. The dark-shaded region represents $1 \sigma$ CL and the light-shaded region represents $2 \sigma$ CL.}
    \label{fig:c4}
\end{figure}

\begin{figure}
    \centering
    \includegraphics[width=0.7\linewidth]{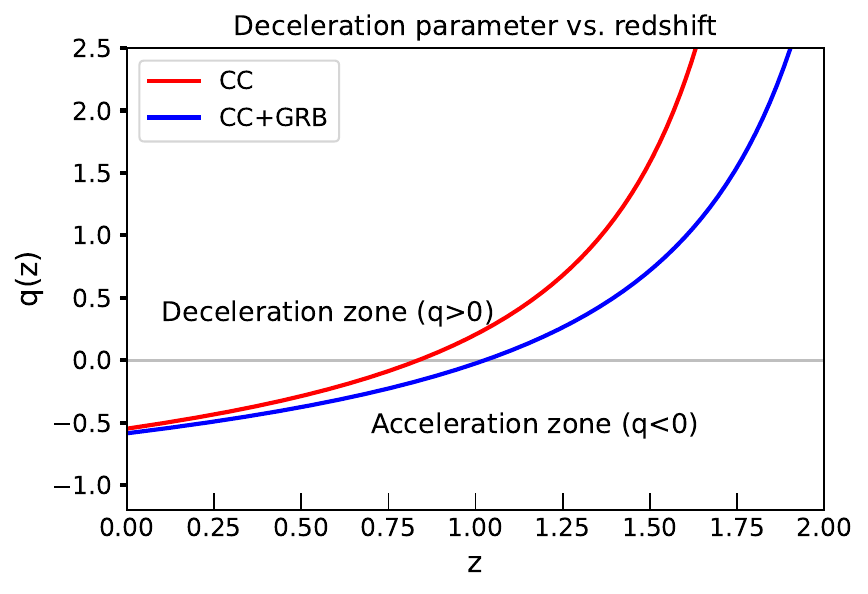}
    \caption{Deceleration parameter (constrained from CC and CC+GRB) vs redshift for the Hybrid exponential model, indicating the phase transition.}
    \label{fig:DP1}
\end{figure}

\begin{figure}
    \centering
    \includegraphics[width=0.7\linewidth]{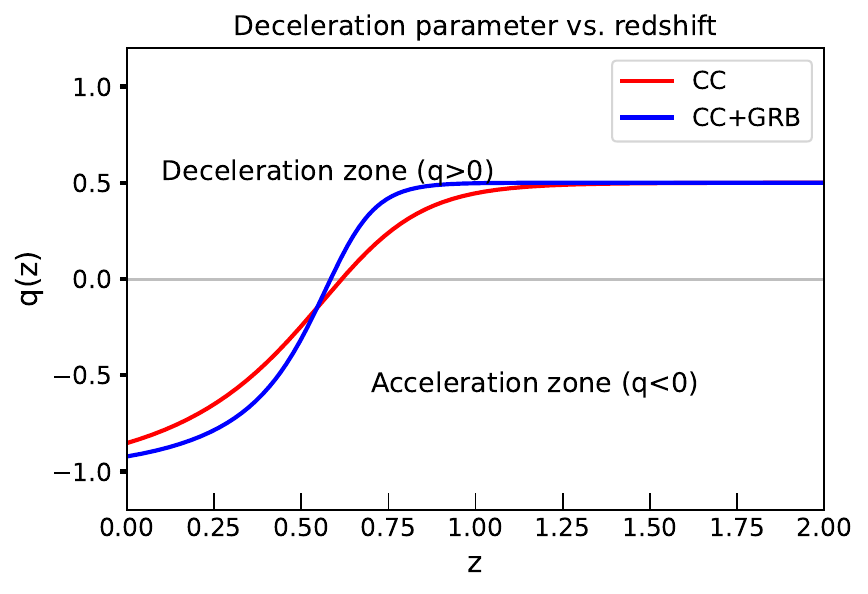}
    \caption{Deceleration parameter (constrained from CC and CC+GRB) vs redshift for the Hybrid tangent hyperbolic model, indicating the phase transition.}
    \label{fig:DP2}
\end{figure}

We start by performing an MCMC analysis to constrain the free parameters using the CC and GRB datasets. It can be observed in \autoref{fig:c1}-\ref{fig:c4} that the model parameter ranges obtained from the sampling lie well within the BBN range which confirms that the models are suitable candidates to explain the early time as well as late-time behavior of the Universe. However, agreement with all the inter mediating epochs is another essential requirement for any theory to be considered as an alternative to GR. In this context, we study the behavior of the deceleration parameter for both our models. For the hybrid exponential model, one can observe in \autoref{fig:DP1} that the parameter transits from the deceleration to the acceleration zone at redshifts mentioned in \autoref{table1}. On the other hand, for the hybrid tangent hyperbolic model in \autoref{fig:DP2}, the $\it{q}$-profile shows a better trend in the phase transition with more accurate redshifts. The present values of the deceleration parameter and the transition redshifts summarized in \autoref{table1} are found to be suitably aligned with the recent observational data. 

\begin{figure}
    \centering
    \includegraphics[width=0.7\linewidth]{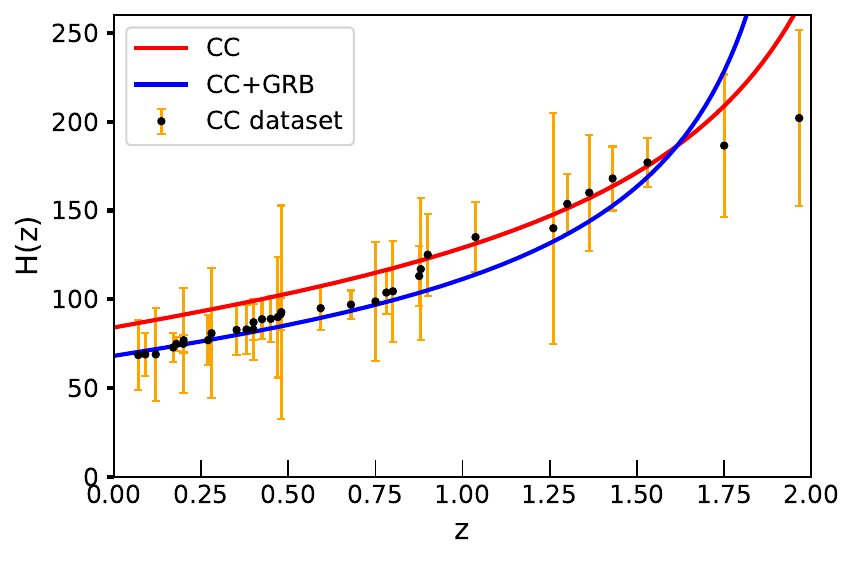}
    \caption{The Hubble function (constrained from CC and CC+GRB) 
    against redshift and error bars of CC dataset for the Hybrid exponential model.}
    \label{fig:HP1}
\end{figure}

\begin{figure}
    \centering
    \includegraphics[width=0.7\linewidth]{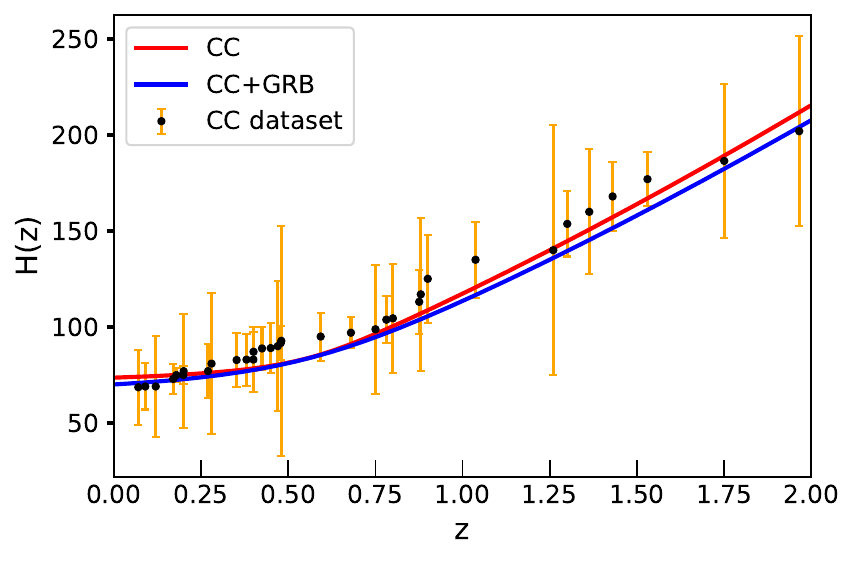}
    \caption{The Hubble function (constrained from CC and CC+GRB) 
    against redshift and error bars of CC dataset for the Hybrid tangent hyperbolic model.}
    \label{fig:HP2}
\end{figure}

\begin{figure}
    \centering
    \includegraphics[width=0.7\linewidth]{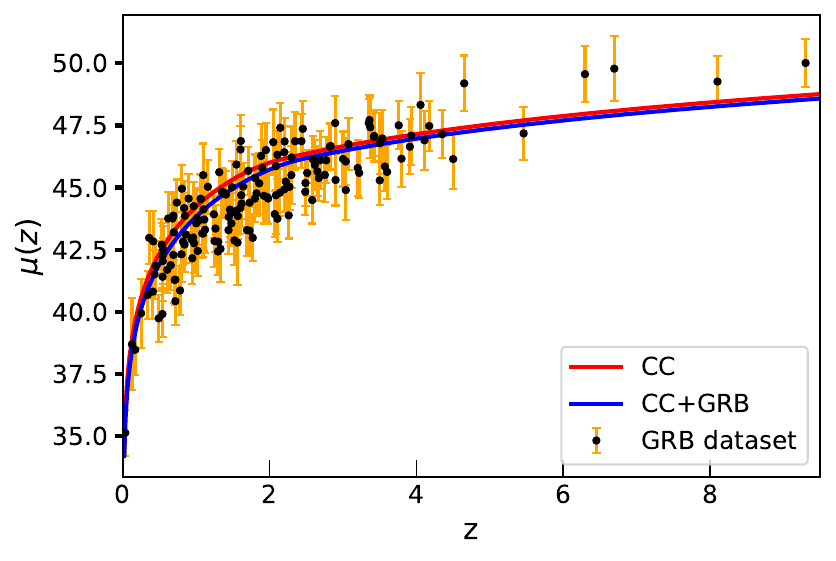}
    \caption{The distance modulus function (constrained from CC and CC+GRB) 
    against redshift and error bars of GRB dataset for the Hybrid exponential model.}
    \label{fig:mu1}
\end{figure}

\begin{figure}
    \centering
    \includegraphics[width=0.7\linewidth]{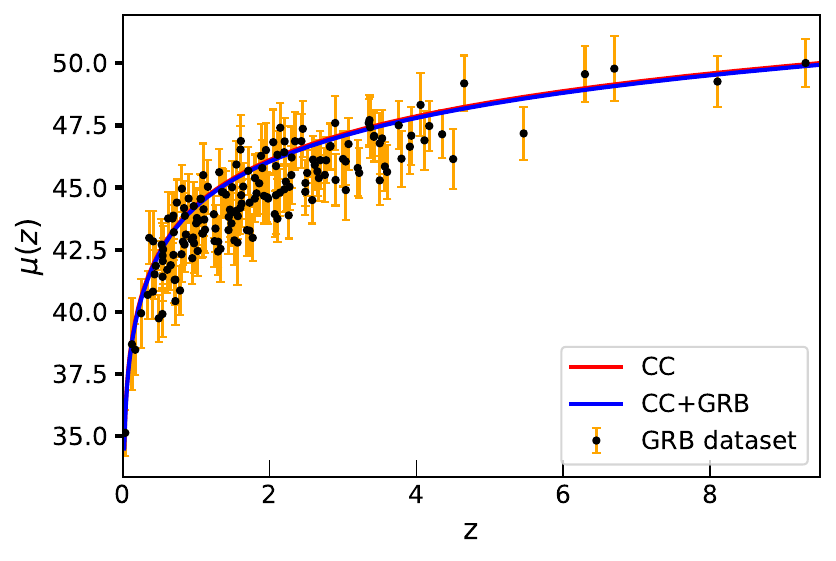}
    \caption{The distance modulus function (constrained from CC and CC+GRB) 
    against redshift and error bars of GRB dataset for the Hybrid tangent hyperbolic model.}
    \label{fig:mu2}
\end{figure}

Furthermore, the Hubble and distance modulus functions are confronted with the 34 data points of CC and 162 data points of GRB datasets, respectively. The GRB data is particularly considered because of its capability to explore the higher redshifts. For both models, we find the functions are in nice agreement with the data points (See \autoref{fig:HP1}-\ref{fig:mu2}). The common ranges obtained for the model parameters from considering early and late-time analysis are summarized in \autoref{table2}.

\begin{table}
 \centering
 \caption{Summary of the transition redshift $(z_t)$ and present value of deceleration parameter $(q_0)$ obtained from $q(z)$ of both models.
 }
 
 \label{table1}
 \setstretch{1.5}
    \begin{tabular}{c c c c}
    
 \hline\hline
\multicolumn{1}{c} {\it{\textbf{Model}}} & \it{\textbf{Dataset}} & {$z_t$} & $q_0$ \\ \hline
\multicolumn{1}{ c  }{Hybrid Exponential } &
\multicolumn{1}{ c }{CC} & 0.838 & -0.548     \\ 
\multicolumn{1}{ c  }{}                        &
\multicolumn{1}{ c }{CC + GRB} & 1.026 & -0.585   \\ \cline{1-4}
\multicolumn{1}{ c  }{Hybrid Tangent Hyperbolic } &
\multicolumn{1}{ c }{CC} & 0.617 & -0.852  \\ 
\multicolumn{1}{ c  }{}                        &
\multicolumn{1}{ c }{CC + GRB} & 0.585 & -0.921 \\  \hline \hline  
    \end{tabular}
\end{table}

\begin{table}
\setstretch{1.4}
 \centering
 \caption{Summary of the ranges obtained from both early and late time for the $f(T)$ models. In the first row, the results of model parameter $n$ of the Hybrid Exponential model are presented while in the second row, the results of model parameter $a$ of the Hybrid Tangent Hyperbolic model are presented.}
 
 \label{table2}

    \begin{tabular}{c c c c}
    
\hline\hline
\multicolumn{1}{c} {\it{\textbf{Model Parameter}}} & \it{\textbf{Observation}} & \it{\textbf{Range}} & \it{\textbf{Common Range}}  \\ \hline
\multicolumn{1}{ c  }{$n$} &
\multicolumn{1}{ c }{CC} & $(-5.7,-3.2)$ &      \\ 
\multicolumn{1}{ c  }{}  
             &
\multicolumn{1}{ c }{CC + GRB} & $(-5.6,-4.31)$ &  $(-5.6,-4.31)$  \\ 
\multicolumn{1}{ c  }{} 
             &
\multicolumn{1}{ c }{BBN} & $(-\infty,7.3147 \times 10 ^{34})$  \\ \hline
\multicolumn{1}{ c }{$a$} &
\multicolumn{1}{ c }{CC} & $(-6.26,-4.46)$   \\ 
\multicolumn{1}{ c  }{}                        &
\multicolumn{1}{ c }{CC + GRB} & $(-8.9,-6)$ & $(-6.26,-6)$ \\ 
\multicolumn{1}{ c  }{}
              &
\multicolumn{1}{ c }{BBN} & $(-\infty,1.94566)$  \\ \hline\hline 

    \end{tabular}
\end{table}

\section{Concluding Remarks}\label{sec6}

To execute this, we adopted a new approach that includes different evolutionary phases of the Universe starting from the nucleosynthesis era. The challenging outcomes of this result raise curiosity to know more about the assumed teleparallel models. The presence of $\it{T}$ term in the denominator makes both the functions approach zero in early times $(T \rightarrow \infty)$. This indicates that without any limits on the model parameters, the models reduce to TEGR at the time of the Big Bang, which solves the cosmological constant problem. A similar line of reasoning can be followed for the late-time as well, the coupling with polynomial $\it{T}$ term makes it GR-like when $T \rightarrow 0$. 

To find a proper justification for whether the models are eligible candidates to connect the early-time and late-time epochs, we commenced from the primordial era by constraining them using BBN observations. The model parameters are restricted as $n \leq 7.3147 \times 10^{34}$ for the exponential model and $a \leq 1.94566$ for the tangent hyperbolic model. Two widely used datasets CC and GRB have been used to constrain the models further by performing the MCMC technique. It is to be noted that the ranges of the model parameters align with the range obtained from BBN and the other cosmological parameters are in agreement with the observational data \cite{Planck:2018vyg,Mandal:2023bzo}. 

\textit{Deceleration parameter}, the most important tool to show phase changes in the intermediate epochs, is then studied to observe the behavior of our models in the transitional times. We find that both models show the transition from the deceleration zone to the acceleration zone. The redshift at which the alteration is occurring is slightly higher \cite{Rani:2015lia,Naik:2023yhl} for the Hybrid exponential model constrained by CC+GRB, while it is around the most widely accepted value for the remaining three \cite{Cunha:2008ja}. For the hybrid tangent hyperbolic model, the deceleration parameter approaches $0.5$ for higher redshift which coincides with the behavior of the $\Lambda$CDM model. However, for the hybrid exponential model, a definitive intermediate epoch cannot be observed as it deviates after a certain redshift. Nevertheless, in both models, the present values of the deceleration parameter confirm the ongoing accelerated expansion of the Universe. In addition, the constrained teleparallel $f(T)$ models are found to be viable in the early phase of the Universe and they align with the data points of the CC and GRB dataset, which is verified using $H(z)$ and $\mu(z)$ functions. 

Thus by joining the pieces, we conclude that both the hybrid exponential and hybrid tangent hyperbolic models can fairly explain the physical aspects of the dynamics of the Universe. In particular, the hybrid tangent hyperbolic model can be considered a fine alternative to GR. 

\newpage



%

\def\baselinestretch{1}
\markboth{}{\small{{\it \hfill Conclusion}}}
\fancyhead[R]{ }
\chapter*{\textsc{Conclusion and Future Perspective}}
\addcontentsline{toc}{chapter}{\textsc{Conclusion}}
\def\baselinestretch{1.5}

The main objective of this thesis is to study modified theories of GR. The thesis is structured in such a way that the focus is given to several aspects of the spacetime manifold, such as probing the dynamics and nature of the accelerating Universe, analyzing wormhole geometries, and exploring the primordial behavior of the universe. This is carried out in the context of different modified theories, primarily, $f(T)$, $f(Q)$, $f(Q,T)$, $f(Q,L_m)$, and $f(R,L_m)$ gravities. Recently, the exploration of modified theories of gravity has entered an exciting phase, driven by advances in observational surveys and the growing demand for explanations beyond the scope of GR. As cosmologists and astrophysicists strive to unravel the mysteries of DM, DE, and the early universe, modified theories of gravity provide promising frameworks. The significant ongoing research and a likely new era in the near future are dedicated to gravitational waves (GWs). The detection of GWs by LIGO and Virgo has opened a new window into the universe. Future observations, especially with next-generation detectors like the Einstein Telescope and LISA, will provide stringent tests for modified gravity models. These models can predict deviations in the waveforms of GWs from binary mergers, and any observed discrepancies from GR predictions could offer evidence for or against specific modified theories of gravity. This could provide a platform to explore these theories, their potential modifications, and the dynamics within their frameworks to a greater extent. 

In recent years, advancements in technology have led to significant developments in observational surveys. 
The future of modified gravity research lies in the synergy between theoretical advancements and observational breakthroughs. Theoretical physicists are developing more sophisticated models that can be directly tested by upcoming observations. As data precision improves, these models will either gain support or require refinement. The interconnection between theory and observation is expected to lead to the development of more accurate models of gravity, potentially exploring new aspects of the universe's fundamental forces. The CMB and LSS are crucial observational pillars for cosmology. Upcoming surveys like the Simons Observatory and CMB-S4 will provide high-precision data on the early universe, allowing for tighter constraints on modified gravity theories. These theories often predict different growth rates of cosmic structures, which can be contrasted against data from galaxy surveys (like DESI and Euclid) and CMB measurements. Any detected anomalies in the distribution or evolution of cosmic structures could point toward modifications in the underlying gravitational framework. One of the emerging trends is the cross-correlation of data from different observational probes, such as combining GW observations with electromagnetic counterparts or correlating LSS with CMB data. This approach could enhance the sensitivity to subtle effects predicted by modified gravity theories. As datasets from different missions and surveys become more comprehensive, the potential for discovering new physics through cross-correlation increases significantly. A primary reason for developing modified gravity theories is to explain the accelerated expansion of the universe without invoking a cosmological constant. Future observations from missions like the James Webb Space Telescope (JWST) and the Rubin Observatory will provide deeper knowledge into the nature of DE.

The future of modified theories of gravity is closely intertwined with the progress of observational cosmology. With a wealth of new data expected from upcoming surveys and experiments, the next decade promises to be a transformative period for our understanding of gravity. Whether these theories will lead to a paradigm shift or further cement the dominance of GR remains to be seen, but the journey promises to be as enlightening as it is challenging.

\setlinespacing{1.}

\addcontentsline{toc}{chapter}{Bibliography}
\fancyhead[R]{\textit{Bibliography}}
\printbibliography

\setlinespacing{1.5}

\def\baselinestretch{1}
\markboth{}{\small{{\it \hfill Appendix}}}
\chapter*{\textsc{Appendix}}
\addcontentsline{toc}{chapter}{\textsc{Appendix}}
\def\baselinestretch{1.5}
\fancyhead[R]{\textit{Appendix}}
\setstretch{1.5}

\section*{Cosmological Developments and Surveys}

Throughout the 20th and early 21st centuries, remarkable milestones have shaped our understanding of the Universe. In 1915, Albert Einstein introduced the GR, illustrating how energy density warps spacetime. Just two years later, Willem de Sitter proposed models of the Universe, one static with a cosmological constant and another that described an empty, expanding Universe. The theory's predictions were validated in 1919 when Arthur Eddington observed the bending of light during a solar eclipse. By 1922, Alexander Friedmann identified solutions to Einstein's equations that implied the Universe was expanding, a concept further supported by Edwin Hubble's 1929 discovery of a redshift-distance relationship. 

The detection of the 2.7 K cosmic microwave background radiation in 1965 by Arno Penzias and Robert Wilson provided crucial evidence of the Big Bang, an achievement later recognized with a Nobel Prize. Andrei Sakharov in 1967 offered a theoretical explanation for the observed matter-antimatter asymmetry. In 1969, Charles Misner and Robert Dicke tackled the horizon and flatness problems of the Big Bang model. The introduction of the cosmic inflation theory by Alan Guth and Alexei Starobinsky in 1980 addressed these issues. 

The 1990s saw significant advancements, including the discovery of the Universe's accelerating expansion in 1998 and further refinement of the cosmic microwave background's properties by the COBE satellite in 1992 and subsequent instruments. Entering the 21st century, the 2dF Galaxy Redshift Survey and the Cosmic Background Imager provided strong evidence for DE and offered unprecedented insights into the Universe's structure. NASA's WMAP in 2003 provided a highly detailed image of the cosmic microwave background, supporting the idea of a 13.7 billion-year-old Universe. 

The detection of baryon acoustic oscillations by SDSS and 2dF in 2005 further confirmed predictions of dark matter models. The direct detection of gravitational waves by LIGO and Virgo in 2016 opened a new frontier in astronomy. In 2019, the first image of a black hole at the center of the M87 galaxy was captured, reinforcing Einstein's theories. Finally, the launch of the James Webb Space Telescope in 2021 marked the beginning of a new era in space exploration and cosmology.

\section*{Statistical Methodology}
To estimate the viable range for model parameters we use the Bayesian approach with MCMC analysis. 
The Bayesian technique interprets probability as a degree of belief, where parameters are treated as random variables shaped by the data and prior knowledge. Starting with an initial prior distribution for the parameters, the data is used to update and refine these beliefs. This process leads to the creation of a posterior distribution, which combines all the information from the data with the prior beliefs. The posterior distribution thus provides a complete summary of what is known about the parameters, integrating both prior knowledge and new evidence. This posterior distribution can be determined with the help of prior and likelihood which can be expressed using the following relation
\begin{equation}
    P(\Theta \mid \mathcal{D}) = \frac{P(\mathcal{D} \mid \Theta) \cdot \pi(\Theta)}{P(\mathcal{D})}\implies P(\Theta \mid \mathcal{D})\propto P(\mathcal{D} \mid \Theta) \cdot \pi(\Theta).
\end{equation}
The above equation is a result of Bayes theorem.  MCMC sampling provides a sophisticated method for estimating model parameters, especially in situations where the posterior distribution is analytically intractable. By generating a series of samples, MCMC allows the empirical distribution of these samples to converge to the actual posterior distribution as the sequence grows. This means that any question regarding the posterior distribution of the parameters can, in principle, be answered by analyzing the associated Markov chain. A Markov chain is a stochastic process that produces a sequence of states, with each state depending solely on the one immediately before it. The Metropolis-Hastings algorithm is the most basic and widely used MCMC method. In our study, we utilized MCMC for both the original dataset, containing measured redshifts and distance modulus/Hubble parameter, and the predicted dataset, which includes measured redshifts and predicted distance modulus/Hubble parameter derived from our ensemble learning model.

To determine the best-fit parameters of our cosmological model, we adopt the approach of minimizing the $\chi^2$ function. It is important to note that minimizing the $\chi^2$ is equivalent to maximizing the likelihood, which in turn is equivalent to minimizing the negative log-likelihood. This approach allows us to explore the parameter space and obtain a statistical distribution of the parameter values that are consistent with the observational constraints. To find the best-fit parameter space, we maximize the likelihood $\mathcal{L}\propto e^{ -\frac{\chi^2}{2} }$, where $\chi^2$ function is defined by,
\begin{equation}\label{eq:xi}
    \chi^2=\Delta V^T (C^{-1} )\Delta V.
\end{equation}
Here, $C$ denotes the covariance matrix of uncertainties, and $\Delta V$ is the difference between the theoretical and observational values of the key physical quantity. In our analysis, we utilize the widely-used MCMC package called \texttt{emcee} \cite{Foreman-Mackey:2012any}, which provides efficient and reliable sampling techniques.

\section*{Data sets}

\subsection*{Pantheon+}
The Pantheon$+$ dataset contains distance moduli estimated from 1701 light curves of 1550 SNeIa with a redshift range of $0.001\leq z\leq 2.2613$, acquired from 18 distinct surveys. Notably, 77 of the 1701 light curves are associated with Cepheid-containing galaxies.  It holds a central role in explaining the expanding Universe. Significantly, the spectroscopically collected SNeIa data such as, SuperNova Legacy Survey (SNLS), Sloan Digital Sky Survey (SDSS), Hubble Space Telescope (HST) survey, Panoramic Survey Telescope and Rapid Response System (Pan-STARRS1) provide a solid evidence in this regard. In \autoref{chap4} we have made use of 2018 Planck data that consists of 1048 data points and we set $H_0=69$ km/s/Mpc \cite{Pan-STARRS1:2017jku}. Pantheon$+$ has the benefit of being able to constrain $H_0$ in addition to the model parameters. Pantheon+ is used in \autoref{chap2} and \ref{chap3}, in which $H_0$ is constrained along with other model parameters. We extremize the $\chi^2$ function \eqref{eq:xi} as shown below to fit the parameter of the model from the Pantheon$+$ samples.
\begin{equation}\label{Eq:ChiSN}
    \chi^2_{SNeIa}= \Delta\mu^T (C_{stat+sys}^{-1})\Delta\mu,
\end{equation}
where $C_{stat+sys}$ is the covariance matrix of Pantheon$+$ dataset formed by adding the systematic and statistic uncertainties and $\Delta\mu$ is the distance residual given by
\begin{equation}
    \Delta\mu_i=\mu_i-\mu_{th}(z_i),
\end{equation}
where $\mu_i$ is the distance modulus of the $i^{th}$ SNeIa. Note that $\mu_i=m_{Bi}-M$, where $m_{Bi}$ is the apparent magnitude of $i^{th}$ SNeIa and $M$ is fiducial magnitude of an SNeIa. The theoretical distance modulus $\mu_{th}$ can be calculated from the following expression
\begin{equation}\label{Eq:mu}
    \mu^{th}(z,\Theta)=5\log_{10}\left(\frac{d_L(z,\Theta)}{1Mpc} \right)+25,
\end{equation}
where $d_L$ is the model-based luminosity distance in Mpc given by
\begin{equation}
    d_L(z,\Theta)=\frac{c(1+z)}{H_0}\int_0^z \frac{d\xi}{E(\xi)}, 
\end{equation}
where $c$ is the speed of light and $E(z)=\frac{H(z)}{H_0}$. 

When analyzing SNeIa data alone, there exists a degeneracy between the parameters $H_0$ and $M$. To address this issue, a modification is made to the SNeIa distance residuals presented in equation \eqref{Eq:ChiSN} as shown in previous studies \cite{Perivolaropoulos:2023iqj,Brout:2022vxf}. Specifically, the modified residuals $\Delta\Tilde{\mu}$ are defined as
\begin{equation}
    \Delta\Tilde{\mu}=\begin{cases}
			\mu_i-\mu_i^{Ceph}, & \text{if $i\in $ Cephied hosts}\\
            \mu_i-\mu_{th}(z_i), & \text{otherwise}
		 \end{cases}
\end{equation}
where $\mu_i^{Ceph}$ indicates Cepheid host of the $i^{th}$ SNeIa which is provided by SH0ES. It is to be noted that $\mu_i-\mu_i^{Ceph}$ is sensitive to the Hubble constant $H_0$ and $M$. In our analysis, we take $M=-19.253$ which has been determined from SH0ES Cepheid host distances (see \cite{Riess:2021jrx}).

\subsection*{Cosmic Chronometers (CC):}
Using the differential aging technique, the CC method deduces the Hubble rate by studying ancient, quiescent galaxies that are closely positioned in redshift. This approach is based on defining the Hubble rate $ H = -\frac{1}{1+z}\frac{dz}{dt}$  for a FLRW metric. The CC method is significant because it can determine the Hubble parameter independently of any specific cosmological model. Based on various surveys \cite{Jimenez:2003iv,Simon:2004tf,Stern:2009ep,Moresco:2012jh,Zhang:2012mp,Moresco:2015cya,Moresco:2016mzx,Ratsimbazafy:2017vga}, we utilized 34 CC data points in the study within the redshift range of  $0.1 <z< 2$. In \autoref{chap2} and \ref{chap4}, 31 CC data points are used (readers may refer to Table IV in \cite{Naik:2023ykt}).  Now, as mentioned earlier, we consider the chi-square function $\chi^2$ to evaluate the unknown parameters. For CC dataset with $n$ number of data points, it is given by
            \begin{equation}
             \chi^2_{CC} =\sum\limits_{k=1}^{n}\dfrac{\left[H_{th}(z_k,\Theta)-H_{ob}(z_k)\right]^2}{\sigma^2_{H(z_k)}}.
            \end{equation}
            Here, $H_{th}$ indicates the theoretically obtained value of the Hubble parameter and $H_{ob}$ represents its observed value and $\sigma$ is the standard deviation.

\subsection*{Baryonic Acoustic Oscillations (BAO):}
BAO serves as a significant cosmological probe for studying the large-scale structure of the Universe. These oscillations originate from acoustic waves in the early Universe, which compress baryonic matter and radiation in the photon-baryon fluid. This compression leads to a distinctive peak in the correlation function of galaxies or quasars, providing a standard ruler for measuring cosmic distances. The comoving size of the BAO peak is determined by the sound horizon at the time of recombination, which relies on the baryon density and the temperature of the cosmic microwave background. 

At a given redshift $z$, the position of the BAO peak in the angular direction determines the angular separation $\Delta \theta= r_d/((1+z)D_A(z))$, while in the radial direction, it determines the redshift separation $\Delta z=r_d/D_H(z)$. Here, $D_A$ represents the angular distance, $D_H = c/H$ corresponds to the Hubble distance, and $r_d$ denotes the sound horizon at the drag epoch. By accurately measuring the position of the BAO peak at different redshifts, we can constrain combinations of cosmological parameters that determine $D_H/r_d$ and $D_A/r_d$. By selecting an appropriate value for $r_d$, we can estimate $H(z)$. In this study, we employ a dataset comprising 26 non-correlated data points obtained from line-of-sight BAO measurements \cite{BOSS:2014hwf,Gaztanaga:2008xz,Blake:2012pj,Chuang:2013hya,Chuang:2012qt,Busca:2012bu,Oka:2013cba,BOSS:2013rlg,BOSS:2013igd,Bautista:2017zgn,BOSS:2016zkm,Ross:2014qpa} (see Table V of \cite{Naik:2023ykt})). Similar to the CC method, the BAO data is incorporated into the analysis through the computation of the chi-square function \eqref{eq:xi} as 
\begin{equation}
        \chi^2_{BAO}(\Theta)=\sum_{k=1}^{26}\left[\frac{(H_{th}(z_i,\Theta)-H_{ob}^{BAO}(z_i))^2)}{\sigma_{H^2(z_k)}}\right],
\end{equation}
where $H_{th}$ represents the theoretical values of the Hubble parameter for a specific model with model parameters $\Theta$. On the other hand, $H_{ob}^{BAO}$ corresponds to the observed Hubble parameter obtained from the BAO method, and $\sigma_H$ denotes the error associated with the observed values of $H^{BAO}$.

\subsection*{Gamma-Ray-Bursts (GRBs):}
GRBs are the most intense explosions in the Universe, having enormous energy. It is possible to observe these cosmic phenomena at comparatively higher redshifts. The correlation between $\nu F_{\nu}$, $E_{p,i}$, and $E_{\text{iso}}$ is given by
\begin{equation}
    \log_{10} \left( \frac{E_{\text{iso}}}{1 \, \text{erg}} \right) = b + a \log_{10} \left( \frac{E_{p,i}}{300 \, \text{keV}} \right),
\end{equation}
where $\nu F_{\nu}$ represents the rest-frame spectrum peak energy, $E_{\text{iso}}$ is the isotropic-equivalent radiated energy \cite{Amati:2002ny}, $a$ and $b$ are constants, and $E_{p,i}$ denotes the spectral peak energy in the cosmological rest frame of the GRBs. $E_{p,i}$ correlates with the observer-frame quantity $E_p$ through its relation being derived from $E_p$ as $E_{p,i} = E_p(1 + z)$. $E_{iso}$ can be calibrated using the luminosity distance $d_L$ and bolometric fluence $S_{\text{bolo}}$ which is related as
\begin{equation}
    E_{\text{iso}} = 4\pi d_L^2(z, \text{cp}) S_{\text{bolo}} (1 + z)^{-1}.
\end{equation}
Here, we have utilized $162$ data points for the redshift ranging from $0$ to $9.3$ \cite{Demianski:2016zxi}. The $\chi^2$ function \eqref{eq:xi} for GRBs data takes the form 
\begin{equation}
        \chi^2_{GRB}(\Theta)=\sum_{k=1}^{162}\left[\frac{(\mu_{th}(z_i,\Theta)-\mu_{ob}^{GRB}(z_i))^2)}{\sigma_{\mu^2(z_k)}}\right].
\end{equation}

\newpage
\thispagestyle{empty}

\vspace*{\fill}
\begin{center}
    {\Huge \color{NavyBlue} \textbf{PUBLICATIONS}}
\end{center}
\vspace*{\fill}

\pagebreak

\markboth{}{\small{{\it \hfill Publications}}}
\chapter*{\textsc{Publications}}
\addcontentsline{toc}{chapter}{\textsc{Publications}}
\def\baselinestretch{1}
\fancyhead[R]{\textit{Publications}}

\section*{Thesis Publications:}
\begin{enumerate}[label=(\arabic*),leftmargin=*]
		\setlength{\itemsep}{10pt}
		\setlength{\parskip}{1pt}
		\setlength{\parsep}{1pt}
		
            \item  \textbf{NS Kavya}, V Venkatesha, \textsc{Embedding the $\Lambda$CDM Framework in Non-minimal $f(Q)$ Gravity with Matter-Coupling}, \textit{Physics Letters B} \textbf{856}, 138927, 2024 (Elsevier, Q1, IF – 4.3) 
            \item  \textbf{NS Kavya}, V Venkatesha, S Mandal, PK Sahoo, \textsc{Constraining Anisotropic Cosmological Model in $f(R, L_m)$ Gravity}, \textit{Physics of the Dark Universe} \textbf{38}, 101126, 2022 (Elsevier, Q1, IF - 5.5)
            \item  \textbf{NS Kavya}, V Venkatesha, G Mustafa, PK Sahoo, \textsc{On Possible Wormhole Solutions Supported by Non-commutative Geometry within $f(R, L_m)$ Gravity}, \textit{Annals of Physics} \textbf{455}, 169383, 2023 (Elsevier, Q1, IF - 3)
            \item  \textbf{NS Kavya}, G Mustafa, V Venkatesha, \textsc{Probing the Existence of Wormhole Solutions in $f(Q,T)$ Gravity with Conformal Symmetry}, \textit{Annals of Physics} textbf{468}, 169723, 2024 (Elsevier, Q1, IF - 3)
            \item  \textbf{NS Kavya}, SS Mishra, PK Sahoo, V Venkatesha, \textsc{Can Teleparallel $f(T)$ Models Play a Bridge Between Early and Late Time Universe?}, \textit{Monthy Notices of Royal Astronomical Society} \textbf{532},  3126–3133, 2024  (Oxford University Press, Q1, IF - 4.8, H-index - 372)
	       \item  DM Naik, \textbf{NS Kavya}, L Sudharani, V Venkatesha, \textsc{Impact of a Newly Parametrized Deceleration parameter on the Accelerating Universe and the Reconstruction of $f(Q)$ Non-metric Gravity Models}, \textit{The European Physical Journal C} \textbf{83}, 9, 1-15, 2023 (Springer, Q1, IF - 4.4) 
\end{enumerate}
\subsection*{Other Publications during Ph.D:}
\begin{enumerate}[label=(\arabic*),leftmargin=*]
		\setlength{\itemsep}{10pt}
		\setlength{\parskip}{1pt}
		\setlength{\parsep}{1pt}
		
		\item  V Venkatesha, \textbf{NS Kavya}, PK Sahoo, \textsc{Geometric structures of Morris -Thorne wormhole metric in $f(R, L_m)$ gravity and energy conditions}, \textit{Physica Scripta} \textbf{98} (6), 065020, 2023 (IOP Science, Q2, IF - 2.9)
		\item  \textbf{NS Kavya}, V Venkatesha, G Mustafa, PK Sahoo, SVD Rashmi, \textsc{Static traversable wormhole solutions in $f(R, L_m)$ gravity},\textit{Chinese Journal of Physics} \textbf{84}, 1-11, 2023 (Elsevier, Q2, IF - 5)
            \item  DM Naik, \textbf{NS Kavya}, V Venkatesha, \textsc{Observational insights into the accelerating universe through reconstruction of the deceleration parameter}, \textit{Chinese Physics C} \textbf{47} (8), 085107, 2023 (IOP Science, Q1, IF - 3.9)
            \item  DM Naik, \textbf{NS Kavya}, L Sudharani, V Venkatesha, \textsc{Model-independent cosmological insights from three newly reconstructed deceleration parameters with observational data}, \textit{Physics Letters B} \textbf{844}, 138117, 2023 (Elsevier, Q1, IF - 4.4)
            \item  L Sudharani, \textbf{NS Kavya}, DM Naik, V Venkatesha, \textsc{Hubble parameter reconstruction: A tool to explore the acceleration of the universe with observational constraints}, \textit{Chinese Journal of Physics} \textbf{85}, 250-263, 2023 (Elsevier, Q2, IF - 5)
            \item  CC Chalavadi, \textsc{Wormhole solutions supported by non-commutative geometric background in  gravity},, V Venkatesha, \textsc{Wormhole solutions supported by non-commutative geometric background in  gravity}, \textit{The European Physical Journal Plus} \textbf{138} (10), 1-14, 2023 (Springer, Q2, IF - 2.8) 
            \item  V Venkatesha, CC Chalavadi, \textbf{NS Kavya}, PK Sahoo, \textsc{Wormhole geometry and three-dimensional embedding in extended symmetric teleparallel gravity}, \textit{New Astronomy} 105, 102090, 2024 (Elsevier, Q3, IF - 2.096 )
            \item L Sudharani, \textbf{NS Kavya}, DM Naik, V Venkatesha, \textsc{Probing accelerating cosmos via reconstructed Hubble parameter and its influence on f (T) gravity models},  \textit{Nuclear Physics B} 998, 116410, 2024 (Elsevier, Q1, IF - 2.5, H-index - 268)
            \item  CC Chalavadi, V Venkatesha, \textbf{NS Kavya}, SVD Rashmi, \textsc{Conformally symmetric wormhole solutions supported by noncommutative geometry in gravity},\textit{Communications in Theoretical Physics} \textbf{76} (2), 025403, 2024
            \item  \textbf{NS Kavya}, G Mustafa, V Venkatesha, PK Sahoo, \textsc{Exploring wormhole solutions in curvature–matter coupling gravity supported by noncommutative geometry and conformal symmetry}, \textit{Chinese Journal of Physics} \textbf{87}, 751-765, 2024 (Elsevier, Q2, IF - 5)
            \item  M Koussour, \textbf{NS Kavya}, V Venkatesha, N Myrzakulov, \textsc{Cosmic expansion beyond  $\Lambda$CDM: investigating power-law and logarithmic corrections}, \textit{The European Physical Journal Plus} \textbf{139} (2), 1-13,2024 (Springer, Q2, IF - 2.8) 
            \item  L Sudharani, K Bamba, \textbf{NS Kavya}, V Venkatesha, \textsc{Governing accelerating Universe via newly reconstructed Hubble parameter by employing empirical data simulations}, \textit{Physics of the Dark Universe} \textbf{45}, 101522,2024 (Elsevier, Q1, IF - 5.5)
            \item  SS Mishra, \textbf{NS Kavya}, PK Sahoo, V Venkatesha, \textsc{Constraining extended teleparallel gravity via cosmography: A model-independent approach}, \textit{Astrophysical Journal} (accepted) (IOP Science, Q1, IF - 4.9, H-index - 479)
\end{enumerate}

\newpage
\thispagestyle{empty}

\vspace*{\fill}
\begin{center}
    {\Huge \color{NavyBlue} \textbf{ACADEMIC EVENTS}}
\end{center}
\vspace*{\fill}

\pagebreak
\fancyhead[R]{\textit{Academic Events}}
\subsection*{Participation in Workshops/Seminars/Conferences during Ph.D:}

\begin{enumerate}[label=(\arabic*),leftmargin=*]
		\setlength{\itemsep}{10pt}
		\setlength{\parskip}{1pt}
		\setlength{\parsep}{1pt}
        \item Presented a research paper in the international conference ICMAA - 2024 held from 02-04 August 2024, organized by Department of Mathemtics, Karnatak University, Dharwad.
        \item Presented a research paper in conference CosmoGravitas: 1st International Conference in Cosmology and Gravitation held from 10-14 June 2024 hosted by the Centre for Theoretical Physics and Natural Phhilosophy. Mahidol University, Nakhonsawan Campus, Thailand.
        \item Obtained IMS Prize for the year 2023 - the best research paper for presenting at 89th Annual Conference of the Indian Mathematical Society - An International Meet during 22-25 December 2023 organised by BITS-Pilani, Hyderabad Campus. Hyderabad, Telangana. 
        \item Presented a research paper in the international conference Differential Geometry and its Applications (DGA-22), from 4th-5th Mar 2022, Jointly organized by the Department of Mathematics, Kuvempu University, Shankaraghatta and The Tensor Society, Lucknow.
        \item Presented a research paper in the international conference ICMSA - 2022, during 28-30 July 2022, organised by School of Mathematical Sciences, Swami Ramananda Theerth Marathwada University, Nanded, Maharashtra.
        \item Participated in the Summer School on GWA during 01-12 July 2024, organised by ICTS-TIFR, Bengaluru.
        \item Participated in the workshop ADAP - 2023, during 05-08 September 2023 at MANUU, Hyderabad, Telangana.
        \item Presented a research paper online in the international conference RAMSIA - 2023, during 22-24 June 2023, organized in hybrid mode by Department of Mathematics, Institute of Applied Sciences and Humanities, GLA University, Mathura, Uttar Pradesh.
        \item Presented a research paper in the 16th international conference MSAST 2022 held during 21-23 December 2022, organised online by Institute IMBIC, India. 
        \item Presented a research paper in the National conference NCMAS - 2022, organised online by Department of Mathematics, Uttarakhand Open University, Haldwani, Uttarakhhand.
        \item Participated in the 27th International Academy of Physical Sciences (CONIAPS – XXVII) on Advances in relativity and Cosmology (PAEC-2021), from 26-28 October 2021, organized online by Deparment of Mathematics, BITS-Pilani, Hyderabad Campus, Hyderabad.
        \item Participated in the webinar Emerging Trends in Computational Mathematics, on 26 November 2021, organised by Department of Mathematics, School of Engineering, Presidency University, Bengaluru.
        \item Participated in the international e-conference ICND-2021, during 20-24 December 2021, on the eve of National Mathematics Day, to celebrate the 134th Birth Anniversery of Mathematical Genius Srinivasa Ramanujan, organized by Department of Mathematics, School of Physical Sciences, Central University of Karnataka, Kalaburagi.
        \item Participated in the 5 days International online workshop ANTDE - 2022, during 06-10 June 2022, organised by the Department of Mathematics, Malaviya National Institute of Technology, Jaipur. 
        \item Participated virtually in the international conference ICOMAA - 2022, during 28-29 June 2022, organized by Department of Mathematics, University of Kalyani, India.
        \item Participated in the online workshop WRPA - 2022, on 17th November 2022, organised by Amity University, Uttar Pradesh.
        \item Participated in the online international workshop GFEACT - 2022, during 13-14 December 2022, organised by Siksha Bhavana, Visva-Bharati, Santiniketan, West Bengal.
        \item Participated in the webinar ISIMC - 2022 during 21-22 December 2022, organized by Department of Mathematics, School of Physical Sciences, Central University of Karnataka, Kalaburagi.
        \item Participated in the seven days online workshop during 15-21 December 2023, organized by ISAR, Chennai. 
        \item Participated in the online workshop NAR 2024, during 20-22 March 2024, organized by Deparment of Applied Sciences, IIIT, Allahabad.
\end{enumerate}

\newpage
\thispagestyle{empty}

\vspace*{\fill}
\begin{center}
    {\Huge \color{NavyBlue} \textbf{REPRINTS}}
\end{center}
\vspace*{\fill}

\pagebreak

\includepdf[pages={1}]{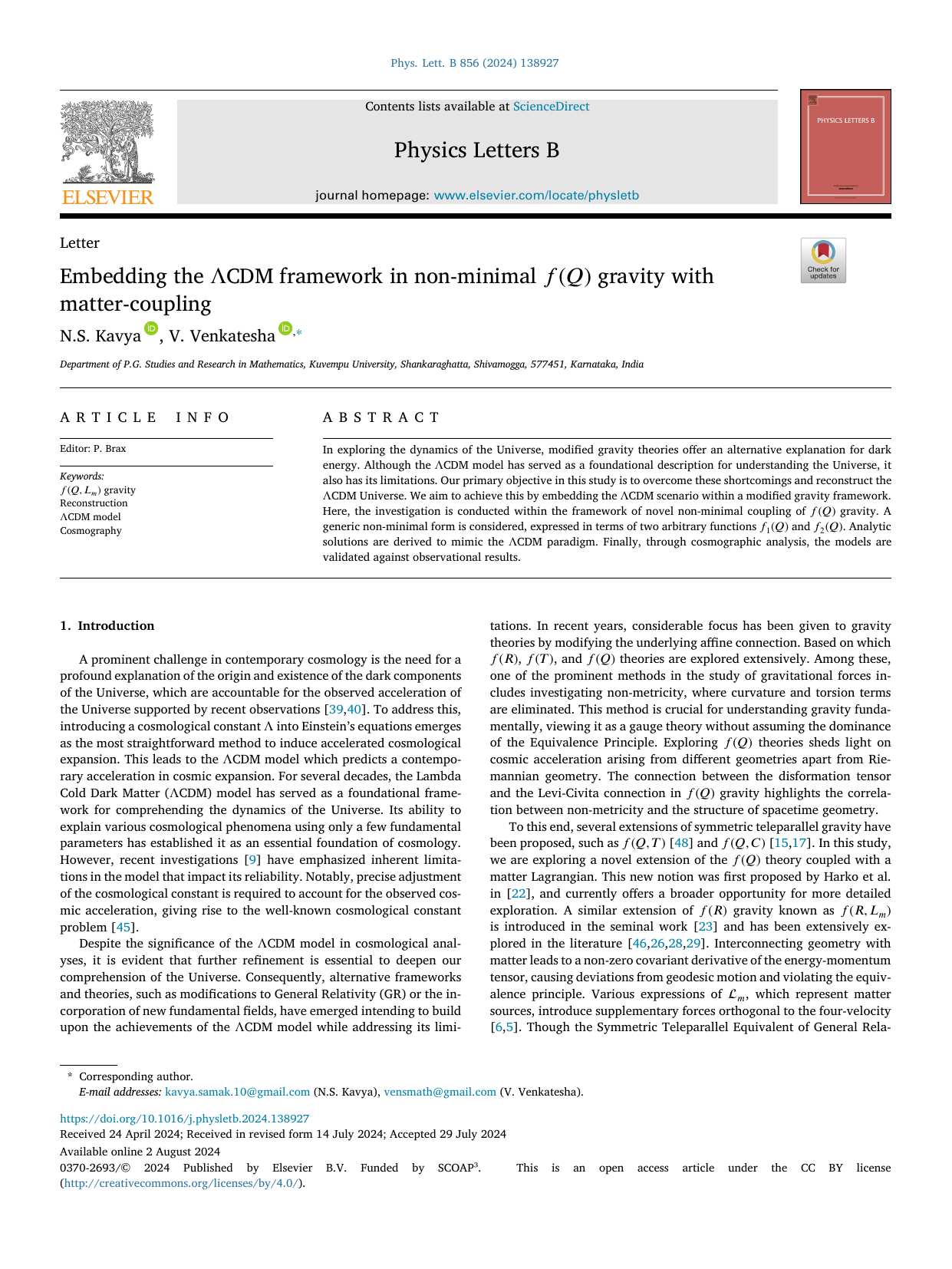}
\includepdf[pages={1}]{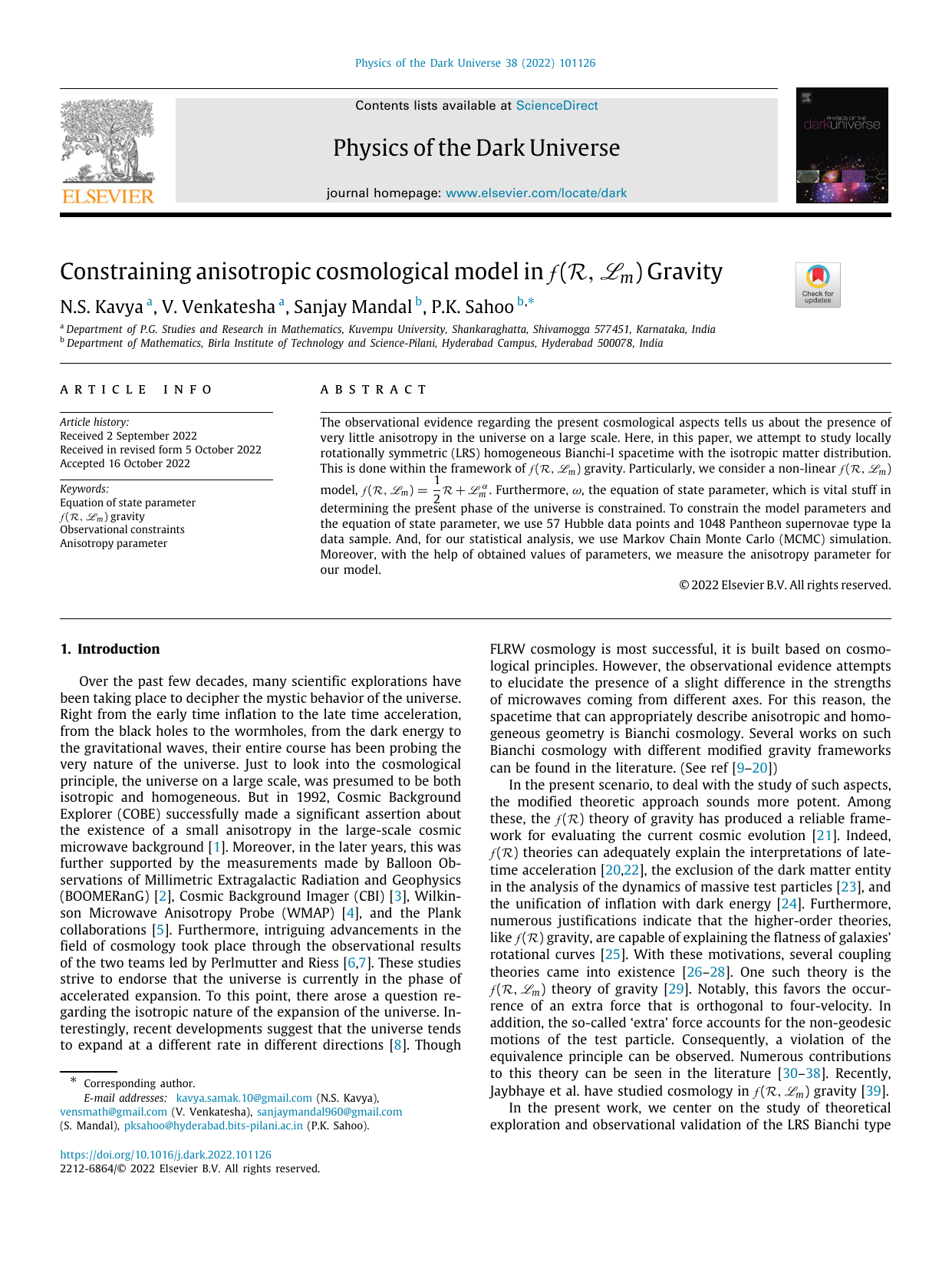}
\includepdf[pages={1}]{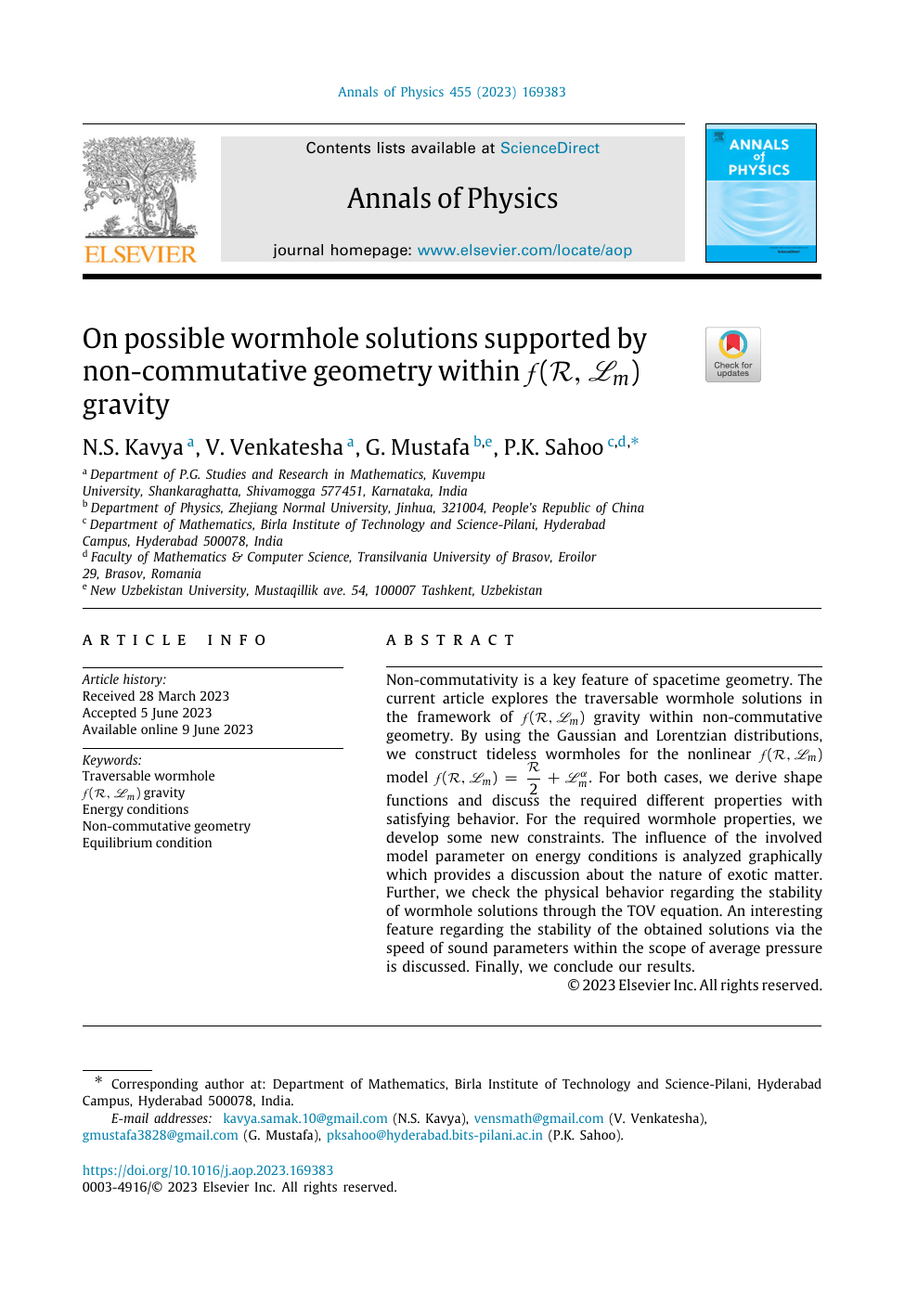}
\includepdf[pages={1}]{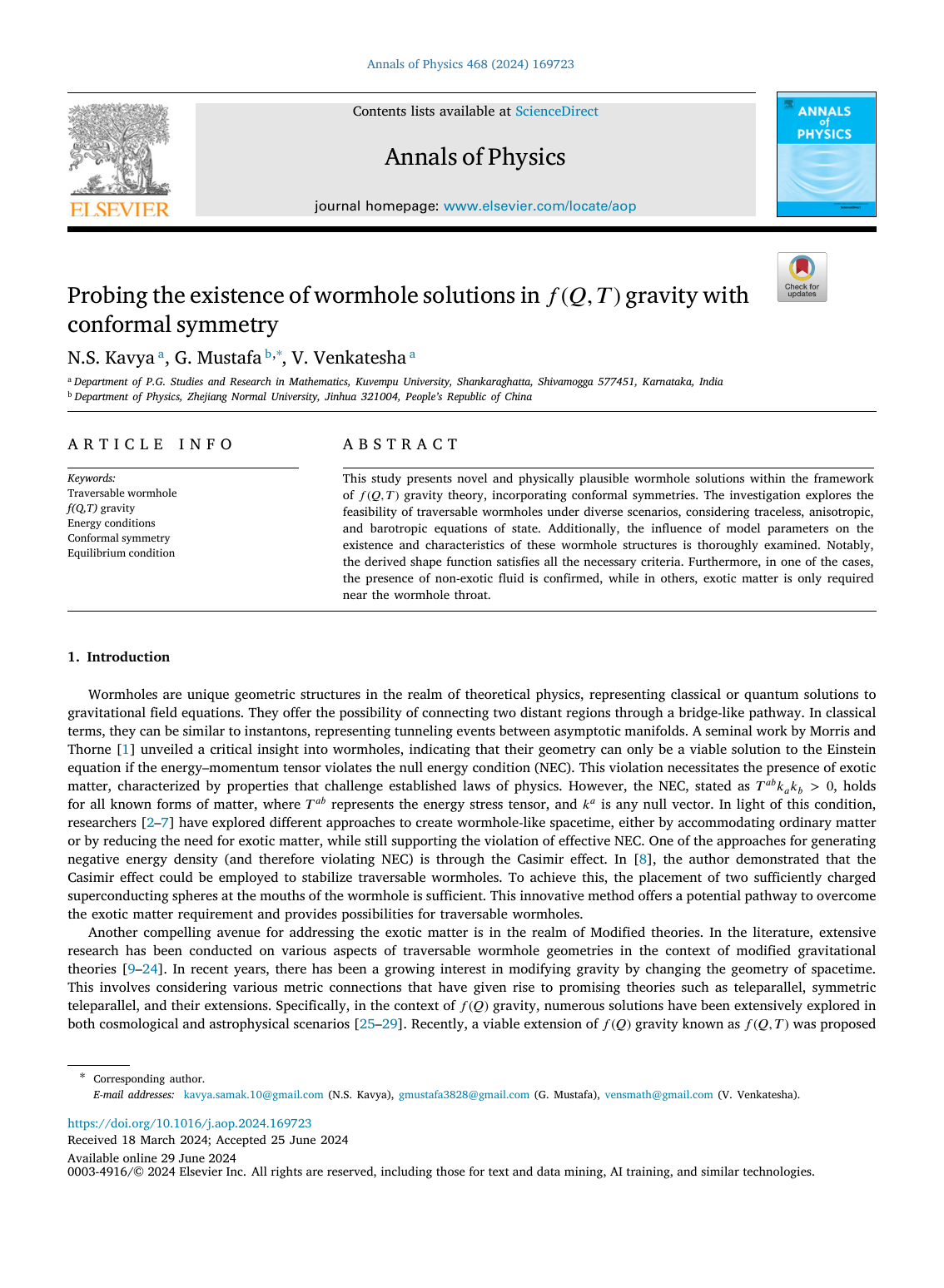}
\includepdf[pages={1}]{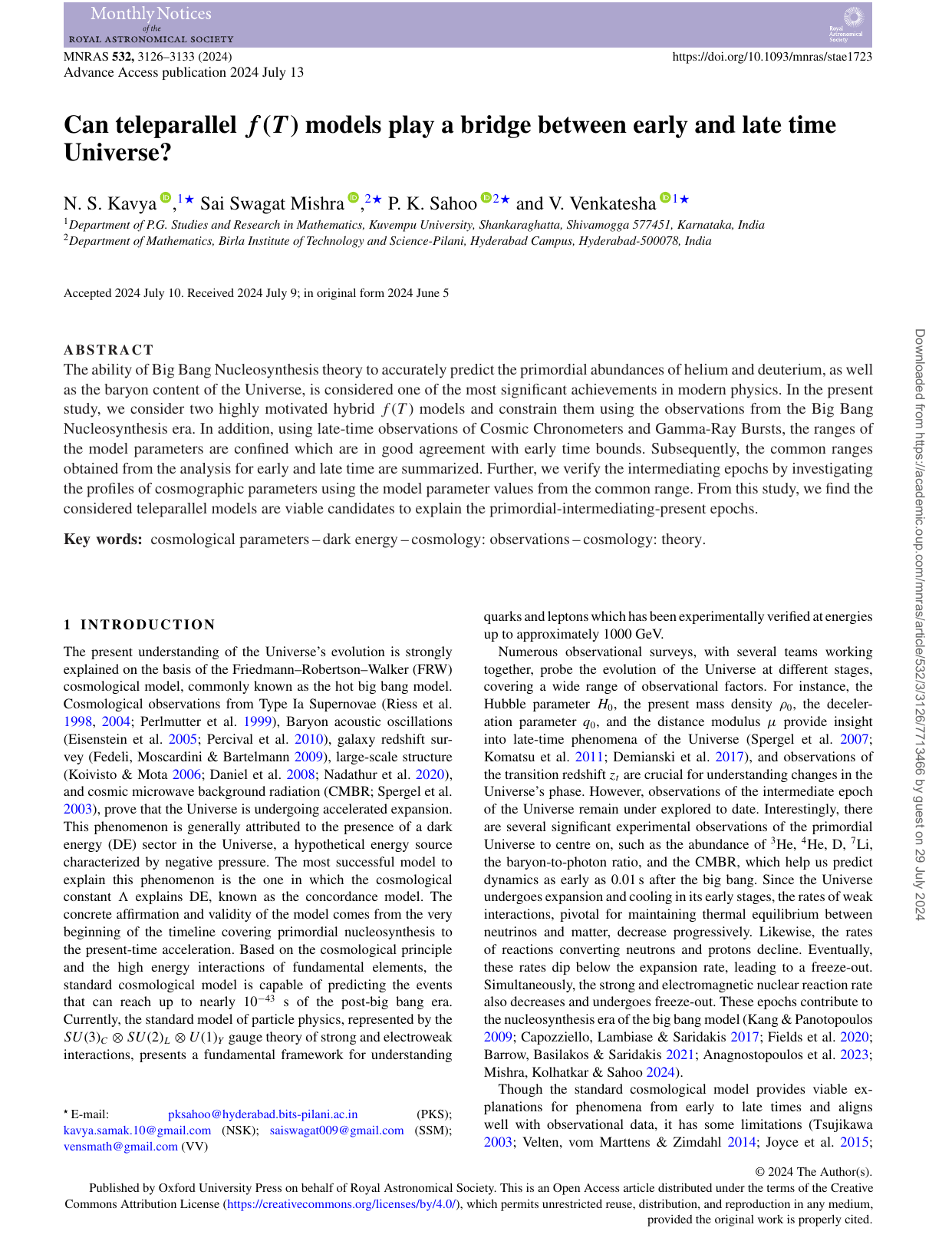}
\includepdf[pages={1}]{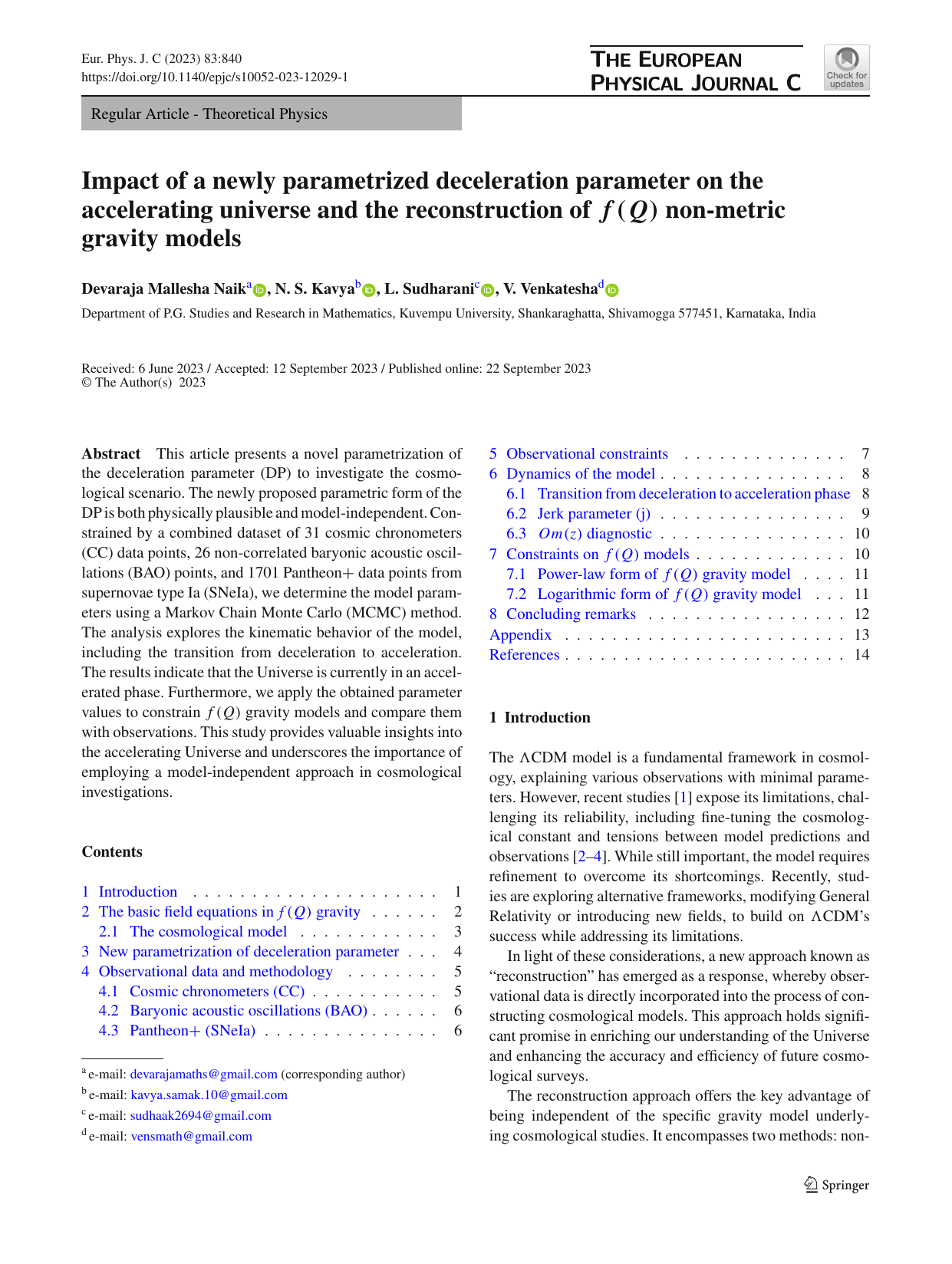}





\end{document}